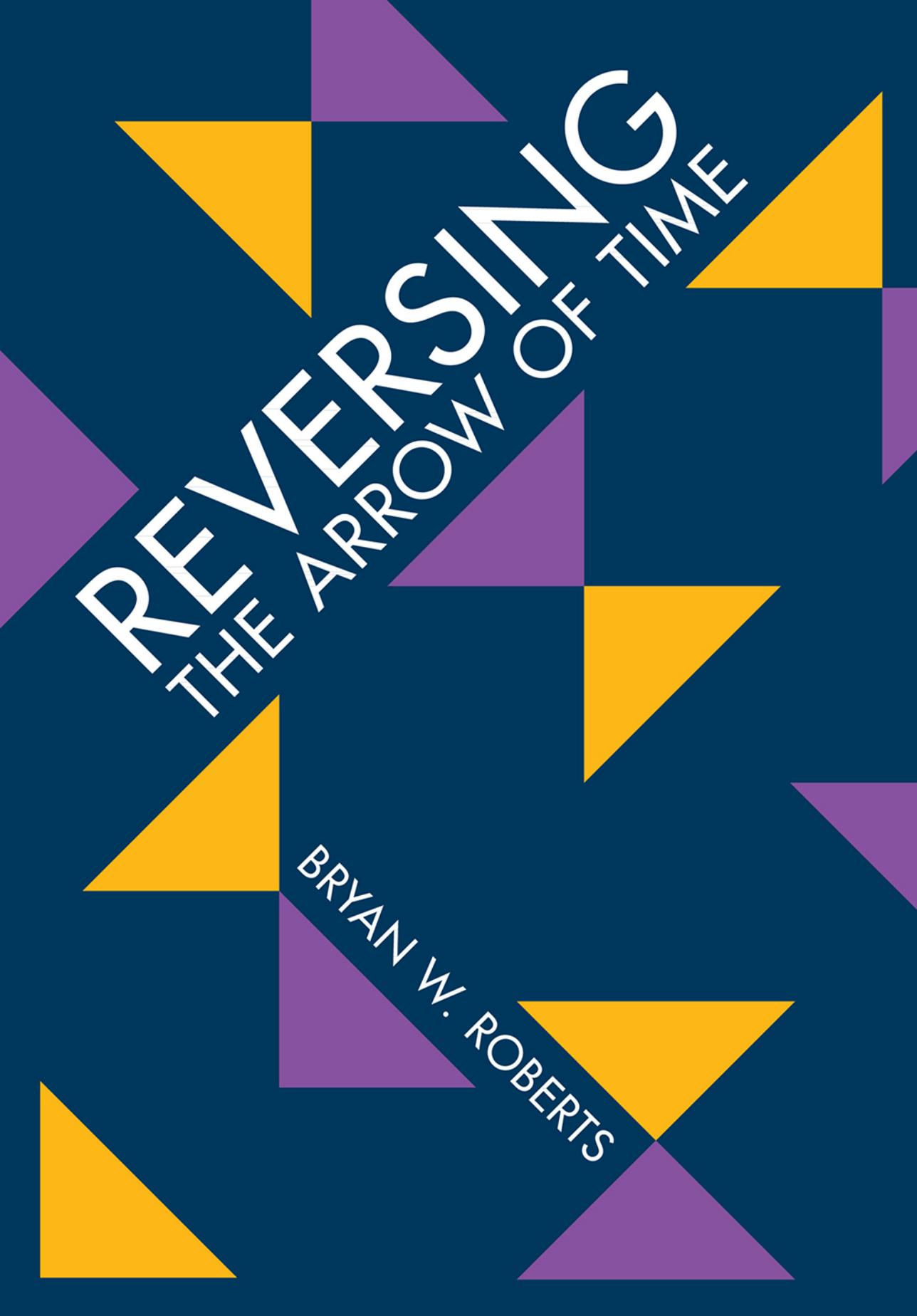

# REVERSING
## THE ARROW OF TIME

### BRYAN W. ROBERTS

# REVERSING THE ARROW OF TIME

'The arrow of time' refers to the curious asymmetry that distinguishes the future from the past. *Reversing the Arrow of Time* argues that there is an intimate link between the symmetries of 'time itself' and time reversal symmetry in physical theories, which has wide-ranging implications for both physics and its philosophy. This link helps to clarify how we can learn about the symmetries of our world, how to understand the relationship between symmetries and what is real, and how to overcome pervasive illusions about the direction of time. Roberts explains the significance of time reversal in a way that intertwines physics and philosophy, to establish what the arrow of time means and how we can come to know it. This book is both mathematically and philosophically rigorous yet remains accessible to advanced undergraduates in physics and the philosophy of physics. This title is also available as Open Access on Cambridge Core.

BRYAN W. ROBERTS is a philosopher of physics and Associate Professor at the London School of Economics and Political Science, where he directs the Centre for Philosophy of Natural and Social Sciences (CPNSS). He won the Leverhulme Prize in 2017 for his work on observables.

# REVERSING THE ARROW OF TIME

### BRYAN W. ROBERTS

*London School of Economics and Political Science*

CAMBRIDGE
UNIVERSITY PRESS

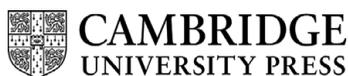

# Contents

# What Is in This Book

This is a book about the arrow of time, that curious asymmetry that distinguishes the future from the past. My aim is to answer two philosophical questions: what is the arrow of time, and how can we come to know about it? These questions depend on a remarkable concept of twentieth-century physics called 'time reversal'.

My central thesis is that there is an intimate link between the symmetries of 'time itself' and the concept of time reversal symmetry in our best physical theories, which has wide-ranging implications for both physics and its philosophy. This link helps to clarify how we can learn about the symmetries of our world; how to understand the relationship between symmetries and what's real; and how to overcome pervasive illusions about the direction of time. It also helps to establish a sense in which time has an arrow. I will argue that, although the world is filled with illusory arrows of time, there is still an asymmetry in the structure of time itself, which survives the idealisations of our best scientific theories and which holds independently of the contingent facts about matter and energy. I will explain how to understand what time asymmetry means and how to identify the experimental evidence for it.

To prepare you for the writing style, this work intertwines mathematics, philosophy, and physics, as natural philosophy often does. It would be a mistake to follow some commentators in thinking that 'philosophy' is a synonym for 'speculation'. The philosophy in this book rather consists in a fascination with big questions, together with an insistence on sharp conceptual distinctions and clear, empirically informed thought. This is not a work of popular science: many discussions will assume some understanding of modern philosophy and physics, at about the level of an advanced undergraduate. That said, I have tried to keep things easy-going where possible and have included a large number of diagrams. Chapter 1 in particular is readable by any audience and includes a compact discussion of some classic philosophy of time as well as a history of time reversal in physics.

Each of the eight chapters states and argues for a separate conclusion, and each chapter can be read independently of the others. However, there is also a single, over-arching proposal that informs every chapter – that the 'structure' of time is linked to the structure of our dynamical theories, in a way that allows one to empirically check whether time itself has an arrow. The core of that proposal is described and argued for in Chapter 2. I then use this perspective in Chapter 3 to try to settle a debate about what time reversal means. Chapter 3 can also be used to guide a general course on the philosophy of physics for advanced undergraduate or master's students: it includes detailed reviews of the structure of classical and quantum mechanics, noting many philosophical debates along the way. Chapter 4 then introduces what 'time reversal symmetry' means and connects it to the philosophy of symmetry more broadly. This chapter can be used to guide an introductory course on symmetry.

Chapter 5 debunks a number of purported 'arrows' that are not true asymmetries of time: in electromagnetism, statistical mechanics, cosmology, quantum theory, and causation. Perhaps the most heterodox claim of the book appears in Chapter 6, where I argue that classical thermodynamics, despite being widely viewed as a paradigmatic example of a time-asymmetric theory, is in fact temporally symmetric in all but a trivial way. This chapter also serves as an introduction to the philosophical foundations of thermodynamics, including a systematic presentation of its geometric structure, which is perhaps not yet widely appreciated by philosophers or physicists.

Finally, in Chapter 7, I argue that time reversal symmetry violation in particle physics provides evidence for an arrow of time itself and explain how it avoids some possible pitfalls. The time asymmetry of electroweak theory is, as far as I can tell, the only compelling experimental evidence for an arrow of time, but it is very compelling indeed. Chapter 8 then turns to the study of other time reversing symmetries like CT (time reversal with charge conjugation) and CPT (charge conjugation plus parity and time reversal) symmetry in relativistic quantum field theory. Here I defend the arrow of time from the claim that it is erased by these more general symmetries. In particular, I argue against a claim of Richard Feynman that reversing time necessarily exchanges matter and antimatter.

To cover all this ground, I have built on the efforts of many extraordinary philosophers, physicists, and friends. In the few places that I disagree with them, I have tried to meet them half-way. I hope the result is some progress in the physics and philosophy of time.

# Acknowledgements

I experience philosophy and physics the way Cummings experiences poetry: groping about in the blackest night, for ever clearer, dearer light. For their patience, support, and inspiration when I was lost, I thank the speakers and participants in the Cambridge–LSE 'Philosophy of Physics Bootcamp', all of whom gave me generous and helpful feedback. These include Andreas Achen, Emily Adlam, Jeremy Butterfield, Erik Curiel, Neil Dewar, Juliusz Doboszewski, Henrique Gomes, Josh Hunt, Klaas Landsman, Joanna Luc, Ruward Mulder, Noel Swanson, and James Wills.

The writing started during the first summer of the COVID-19 global pandemic. I'm grateful to the people who stayed close to me during that unusual time, but especially to my family, to Katarina Prkačin and Jeremy Butterfield for their endless love and encouragement, to Anita Roberts for her committed copy-editing, to the Fenland Fusion, and to Barbara Prkačin for pushing me to the finish line.

Much of this research was done during a magical time while I was a visitor at the Inter-University Centre in Dubrovnik, Croatia, and then a Visiting Fellow Commoner at Trinity College, Cambridge. I thank both of these institutions for their support. Both visits were made possible thanks to the generosity of the Philip Leverhulme Foundation through the Philip Leverhulme Prize, of Trinity College, and of my wonderful colleagues in the LSE Department of Philosophy, Logic and Scientific Method.

I also owe thanks for the feedback and lessons I received from many people in conversations over the years, who are too many to list but especially include Harvey Brown, Jeremy Butterfield, Craig Callender, Adam Caulton, John Earman, James Ladyman, David Malament, Wayne Myrvold, John Norton, Laura Ruetsche, Giovanni Valente, and David Wallace. Special thanks to Gerhard Hegerfeldt for his correspondence about Chapter 4 and to James Wills for discussions about the material in Chapter 6. Mary Chase wrote that "in this world, Elwood, you must be oh so smart, or oh so pleasant – well, for years I was smart – I recommend pleasant". I am lucky to have been supported by people who were both. I hope you all enjoy this work.

# 1

# A Brief History of Time Reversal

---

***Précis.*** *The symmetries of time can be understood through the symmetries of motion, both in a sense that is familiar to philosophers and in the history of physics.*

---

Can time be accurately described in an undirected way, like a great eternal string with no preference for one direction over the other? Or, is it directed like an arrow, with two distinct ends? Philosophers often point out that human experience is vividly directed: we remember the past and not the future; we age towards the future and not the past. But, does time have a direction beyond such facts about human psychology and physiology? This chapter will introduce the main thesis of this book, that the answer is yes: time really is directed like an arrow, in a sense given by what physicists call 'time reversal' asymmetry. In particular, this asymmetry can be detected empirically through our experience of the motion of matter-energy. This asymmetry will be familiar to philosophers, but the evidence for it was developed over the course of two centuries in the history of physics. In this chapter, I will explain both the philosophy and the history behind these claims.

The majority of this book will be cast in the language of physics, which is best-suited to capturing our empirical evidence about the structure of time. However, I would also like to point out a connection between this evidence and the broader philosophy of time. So, Section 1.1 connects my argument to the asymmetries of time that are perhaps most familiar to philosophers, known as the 'A series' and the 'B series' of John McTaggart. The remaining sections then show how the symmetries of time have played a prominent role in two centuries of physics. Section 1.2 points out that the origins of time reversal can be traced to Carnot's theory of engines. Section 1.3 reviews its role in the famous reversibility paradox of statistical physics. Section 1.4 describes how time reversal invariance rose to prominence in the first half of the twentieth century, and Section 1.5 recounts the great shock that physicists felt when they discovered the first evidence of time asymmetry in electroweak interactions.

## 1.1 On the A Series and the B Series

John McTaggart, an eccentric Cambridge philosopher of Trinity College who was known to salute cats as he met them[1], gave an account of time's arrow that has been influential amongst philosophers: call an undirected description of time a *C series*, and a directed description a *B series* (we will shortly have an A series too). The C series provides language to say whether or not an event falls between two others, or a 'betweenness' relation, while the B series adds the language of an ordering relation. The ordering relation allows one to say something that goes beyond the C series: that an event stands in a before-after relation with respect to others, and (ordinarily[2]) not vice versa. In this language, our question "Is time directed?" becomes "Beginning with a C series description, is there reason to think that time is accurately described by a B series?"

McTaggart believed that it would take a special sort of process to produce a B series from a C series description. Inspired by Hegel's categories, he took this process to involve causality. He also proposed a candidate: the characteristics of being past, present, and future, which he called *A series* descriptions, seem to "pass along" the C series as the future becomes present, and the present becomes past. This 'passage' would determine the kind of ordering required for the B series: say that one event occurs 'after' another event if and only if it happens, during passage – that one is in the future while the other is in the past (or present), but not the reverse. The schema is illustrated in Figure 1.1. Unfortunately, McTaggart himself found it hard to make sense of his A series notion of 'change' from future to present to past, and he ultimately rejected it, as well as the reality of time more generally, as incoherent.[3]

---

[1]  As reported by Dickinson (1931, p.68).
[2]  Following Lewis (1979), one might make an exception for closed timelike curves and the cyclic histories of Nietzsche (1974). But, as we will see in Chapter 2, this is no barrier to defining temporal asymmetry.
[3]  Mellor (1998, Chapter 7) provides a classic discussion, and Ingthorsson (2016) at book-length.

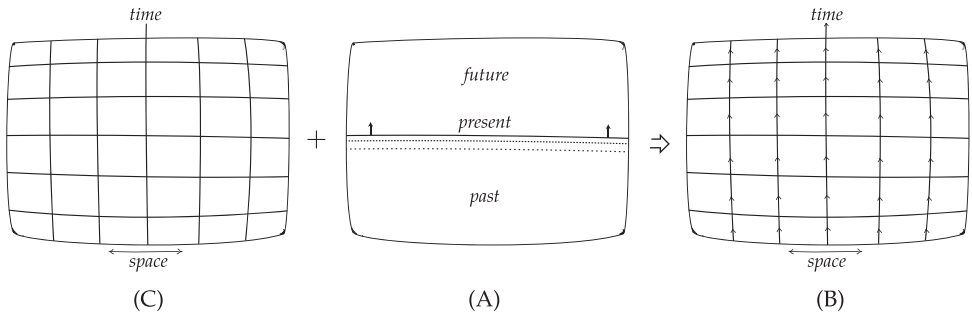

Figure 1.1  McTaggart took his C series plus A series to determine a B series.

McTaggart inspired a voluminous metaphysics of time literature that I'm afraid I won't breach. My aim here is rather to bring that metaphysics a little closer to the physics: McTaggart's A, B, and C series each have a natural expression in physics, so long as we are willing to replace his notions of time and change with more modern ones. For example, Earman (2002a) construes McTaggart's B series as a spacetime with a temporal orientation.[4] The C series is then just a spacetime without a temporal orientation. But, according to McTaggart, the A series is supposed to be linked to the B series and the C series, through what metaphysicians after Broad (1923) now variously interpret as 'passage' or 'becoming'. Many philosophers of physics have despaired of finding an A series in modern physics.[5] Others, such as Maudlin (2002a, 2007), are more optimistic.

It is not my purpose to take a position on this debate here. However, I would like to draw out a different aspect of McTaggart's picture that I think helps to maintain good, clear thinking about the nature of time. Namely, we should begin with a clean, clear separation of the concepts of 'time' and 'change'. Of course, these concepts must be intimately linked, as McTaggart suggests. But, let us not tether a concept as rich as time to just one conceptual framework. Like McTaggart, I would like to 'pull apart' two concepts of time, in order to examine their relationship.

I will pull these concepts apart in a way that is natural in the practice of physics. In physics, we sometimes analyse time using spacetime structure, as when we describe a relativistic spacetime in special or general relativity.

---

[4] See Section 2.5.3 for a more detailed discussion about temporal orientations.
[5] Callender (2017) and Earman (2002a) both identify the A series 'Becoming' as an aspect of the Manifest Image rather than the Scientific Image – adopting the nomenclature of Sellars (1962) – which led Earman to call for metaphysicians of Becoming to "remain locked in their mutual embrace of Becoming and sink from view into the metaphysical mire" (Earman 2002a, p.2). Maudlin (2002b) responded with a defence of the concept of change in modern physics.

Other times we analyse a concept that is perhaps more appropriately called 'change', when we imagine the replacement of one state of the world with another. The latter can be described using a structure commonly called state space, or configuration space, or phase space, as in classical or quantum mechanics. When change is described this way in physics, it is often referred to as a *dynamical system*, whose selection of possible changes is called a *law of motion*. So, let me make the distinction in this way: 'time itself' will refer to spacetime structure, while 'change' will refer to the changing state in a state space.

The overarching idea that will be carried through every chapter in this book can be put in these terms: that time and change are linked in a way that allows one to learn about the structure of time by studying the structure of change. In particular, in order to learn whether time has an asymmetry or 'arrow', one can study the asymmetries of change in the material world.

In Chapter 2, I will show how to make this idea precise, beginning with a concept called *time reversal:* we can understand an 'undirected' description of time to mean that the structure of time does not change when it is 'reversed'. I will then show how this concept can be used to determine whether time itself has an arrow. Disclaimer: my aim with this proposal is not to reanimate McTaggart, nor to argue that he would endorse any such view.[6] If one likes, it may be possible to associate the B series with spacetime structure and the A series with change in dynamical systems. Indeed, if one does so, then there are certain kinds of change that provide evidence for a direction of time: not all change, but just a special kind of change that is called 'time reversal violating', and which is discussed in Chapter 7.

The framework threading through this book finds its origins in the pioneering work of Wigner (1939) on the representation theory of relativistic quantum mechanics. It can be distilled down into two postulates:

1. If changing states are interpreted as occurring in spacetime, then those changes must share a common structure with spacetime.
2. Given this, the asymmetries of spacetime can be inferred from asymmetries of those changing states.

The mathematical tool that Wigner used to describe the 'common structure' in the first postulate is called a *representation:* roughly speaking, it is a structure-preserving map, from a spacetime structure to a dynamical system. So, to keep that clearly in mind, I will refer to the first postulate as the 'Representation View'. This view will be motivated and developed in detail in Chapter 2. A special case will be of particular interest to me: that if states are described as changing *with respect to time*, then that change must share some common structure with time itself.

---

[6] As it happens, the great 'Space and Time' address of Hermann Minkowski (1908) was given in the same year that McTaggart (1908) published his famous article, but I know of no evidence that either one knew of the other's work at the time.

Wigner used the first postulate to determine the possible dynamical systems of quantum theory, given that they are formulated in the context of Minkowski spacetime. My proposal throughout this book will reverse this thinking and instead use the structure of dynamical change to draw inferences about the structure of spacetime. This leads to the second postulate: by drawing on our observations of change in dynamical systems, I will argue that one can determine whether time has an arrow – and indeed, that there is extremely strong evidence that it does.

The way that this inference works can be illustrated using a toy theory. Suppose the changing state of an animal is described by the metamorphosis of a caterpillar into a butterfly. There is an asymmetry in this theory of change, which is that the reverse metamorphosis cannot occur. In other words, the 'time reversed' description is impossible, as illustrated in Figure 1.2. This is an asymmetry in a description of change. However, if time shares the symmetries of this particular change, then it might provide evidence that time itself has an asymmetry too.

This toy theory takes place at a level that omits a great deal of information about change. For example, the interaction of the animal with its environment is completely ignored. Once that hidden information is restored, it is not so clear that the change being described really is asymmetric. I call such erroneous inferences 'misfiring' arrows of time and discuss them in detail in Chapter 5. However, a first step in avoiding them is to move from theories of biology to theories of fundamental physics. If we describe motion on a fundamental level, by drilling down to the most basic description of change that we can find in the nature of matter and energy, then we might manage to avoid misfiring arrows and identify a true asymmetry in time. In Chapter 7, I will argue that we have evidence for time asymmetry in this sense.

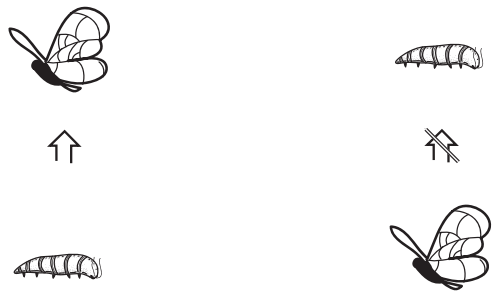

Figure 1.2  Time asymmetry: a possible description (left) whose time reverse is not possible (right).

The situation is perhaps similar to a claim of McTaggart (1908, p.464), that, "[i]t is only when the A series, which gives change and direction, is combined with the C series, which gives permanence, that the B series can arise". If the A series is a description of change in a dynamical system, and if that description shares the symmetries of time, then an asymmetry in time itself can arise, which one might interpret as the B series. This helps to dispel a well-known concern about how the laws of motion can be used to make inferences about the direction of time itself, rather than just motion.[7] In this book I will cleanly separate time and change. But, like Wigner, I will argue that the two are linked through a representation. It is this link that allows one to make inferences about the nature of time on the basis of observations about motion.

McTaggart (1908, p.474) himself asks, near the end of his article, whether events in the C series might have some quality that gives them order, writing, "[w]hat is that quality, and is it a greater amount of it which determines things to appear as later, and a lesser amount which determines them to appear as earlier, or is the reverse true?" One way to understand the argument I will make over the course of this book is that time does have a quality somewhat like this. It is not a quality of any one event but rather of the structure of time as a whole: its symmetries are linked to the symmetries of dynamical change in a way that establishes an asymmetry. As to which direction is truly 'later' and which is 'earlier', my account say very little. The arrow of time is as Wittgenstein (1958, §454) described the drawing, '↣': "[t]he arrow points only in the application that a living being makes of it". In my view, this makes it no less remarkable that time in our world has an arrow.

The remaining chapters will develop the argument for this view, through an analysis of temporal symmetry under the time reversal transformation. Time reversal is a thoroughly modern concept, and so I will analyse its meaning using the language of modern physics. However, I would also like to convey the charming way that temporal symmetry came to be so important, through an easy-going history of time reversal. That history begins, in the next section, with engines.

---

[7] A version of this concern can be found in Black (1959), with more sophisticated statements found in Earman (1974), Gołosz (2017), and Sklar (1974, §F).

## 1.2  Ingenuity and Engines

In the summer of 1816, the French physicist Jean-Baptiste Biot convinced the owners of a former church to let him use its boiler to study the polarisation of light passing through turpentine vapour. Not something to be left unattended near an open flame, the experiment detonated in a great explosion that sent the boiler's cover flying and set the roof of the church on fire. Undeterred, Biot advised anyone repeating his experiment to place the boiler behind an impenetrable wall, since

"the explosion of the vapor, its ignition and that of the liquid, could cause miserable death, and in the most inevitable and cruel manner, to people located at quite a distance."[8]

Explosions aren't always an inconvenience: that flying boiler cover might have more helpfully been used to push an object along a track, like a train. It is really most useful when it can be repeated in a controlled manner to keep the train going, as had been achieved by British inventors like Newcomen and Watt in the eighteenth century.[9] Indeed, soon after the boiler incident, Biot (1817) published a textbook describing a burgeoning class of machines that were powered by vapour explosions. What held these ingenious machines or 'engines' back was a lack of understanding as to what distinguishes a useless explosion from an optimally useful one.

Answering this question made use of a proto-concept of time reversal, introduced by Sadi Carnot in his 1824 *Reflections on the Motive Power of Fire*. Writing while on duty in the French army, Carnot stumbled on a crucial observation, that a useful engine would have to cycle back to its initial state so that the explosive motion could be repeated. This was the ingenuity that ultimately led to modern engines: that all processes that produce motion from heat "can be executed in a reverse sense and in a reverse order" (Carnot 1824, p.19). Carnot's 'reverse sense' and 'reverse order' introduced the concept of time reversal for the first time but applied in a way that is subtly different from its modern usage. Let me review it in a little more detail, using Carnot's most famous example.

Carnot began with the Carnot cycle, which he describes in terms of a stunningly simple example of a gas in a cylinder that expands and contracts, and so can be used to force a piston and drive motion. A model of such a gas is illustrated in the pressure–volume diagram of Figure 1.3. This model is well-known to physicists: the cylinder is initially in contact with a source of heat, which allows it to expand while retaining constant temperature (isothermally) along the top path in the diagram. It then continues to expand in isolation from any heat exchange (adiabatically) along the right path, resulting in a drop in temperature. A reverse process then follows: the pressure on the piston is increased to drive the volume back down, maintaining a constant temperature by losing the same amount of heat to a cold source, as along the bottom path. The compression then continues adiabatically until the temperature is elevated back to its initial value along the left path. As a result, the pressure, volume, and heat of the system are all restored to their original values, and the process can be repeated. There has also been a total amount of work done by the engine, $W := \int P \, dV$, which is equal to the area of the shape traced out by the curves.

---

[8]  My translation of Biot (1819, p.133); this curious article provided one of the first studies of optical rotation in turpentine. Happily, no one was injured in the accident.

[9]  A classic history of this development is Dickinson (1939).

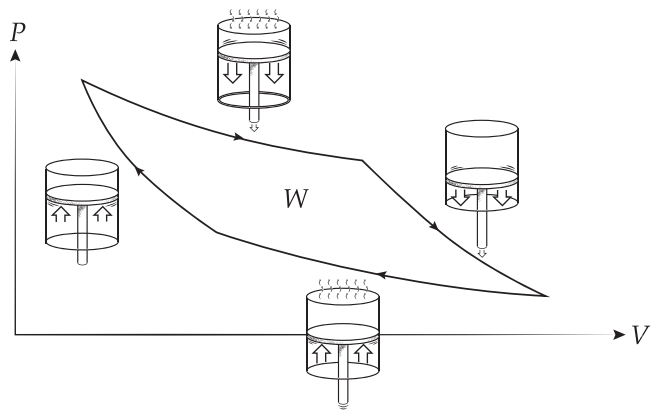

Figure 1.3 Carnot's heat engine: isothermal expansion (top), adiabatic expansion (right), isothermal compression (bottom), and adiabatic expansion (left).

From an engineering perspective, the Carnot cycle aims to do two things: to do as much work as possible and to return back to where it started so that the process can repeat. These are both achieved with the help of what Carnot took to be his central conceptual insight, the pairing of two processes with two 'inverse' processes:

The operations that we have just described could have been done in the inverse order and sense. . . . In our first operations, there was at the same time a production of motive power and a transport of caloric [heat] from body *A* to body *B*; in the inverse operations, there was at the same time an expense of motive power and a return of caloric from body *B* to body *A*. (Carnot 1824, pp.10–11)[10]

---

[10] My translation. Carnot's successful use of 'caloric' here, a chemical element postulated to characterise heat before the kinetic theory that was assumed to be conserved, has been the subject of much debate in the philosophy of science (cf. Chang 2003; Laudan 1981; Myrvold 2020a; Psillos

Table 1.1. *An expansion in the Carnot cycle is the time reverse of some compression, though these specific pairings are not the time reverse of each other.*

| Process | | Inverse Process |
|---|---|---|
| Isothermal expansion | ↔ | Isothermal compression |
| Adiabatic expansion | ↔ | Adiabatic compression |

These pairings, shown in Table 1.1, relate each process to some 'time reversed' process, in that each compression corresponds to an expansion described in the reverse time direction. In particular, an isothermal compression with heat flowing in is the time reverse of some isothermal expansion with heat flowing out; and, an adiabatic expansion is the time reverse of some adiabatic compression.

This is a subtle variation on typical modern usage of time reversal: strictly speaking, Carnot's pairings are not the time reverse of each other, since they take place at entirely different pressures and volumes. In fact, if one were to carry out the 'strict' time reversal of the first two parts of the cycle (the top and right paths in Figure 1.3), one would just trace back along the same lines to the original state. This produces a cycle with zero area, and which thus does zero work. How then does Carnot choose the right compression process to follow the expansion?

It is the natural choice of an engineer: choose the inverse processes that are 'optimal', in the sense of maximising the amount of work done by the engine. After following the top and right paths in the diagram, there are various ways of zig-zagging back to restore the original amounts of pressure, volume, and heat. But, since the work done is given by the area inscribed by the paths, these will always be less than or equal to the work done in Carnot's cycle. Assuming that the first two paths in the cycle achieve the engine's maximum and minimum temperatures, the unique work-maximising cycle is the Carnot cycle. That is how Carnot selects the 'inverse' operations: he does not pair expansions with their strict time reverses but rather chooses those 'inverse operations' that produce the best possible engine.[11]

---

1999). The subsequent development of equilibrium thermodynamics discussed in Chapter 6 is often viewed as a response to the discovery that no such chemical element exists!

[11] A reading of Carnot along these lines is set out in much more careful detail by Uffink (2001, §4), who duly cautions that Carnot himself does not make any explicit connection between 'inverse operations' and 'time inversion'.

This is the proto-version of time reversal that appeared in Carnot's theory of heat and work, on the road to identifying the behaviour of an optimal engine. Unfortunately, all of this discussion took place with a rather rough idea of what 'time reversal' actually means. That was a side-effect of the limited language of thermodynamics that was available at the time of Carnot. Fortunately, more precise thinking about time reversal would become available in the next episode in our story, the development of statistical mechanics.

### 1.3  Well, You Just Try to Reverse Them!

The appearance of time asymmetry is commonly associated with the phenomena of classical thermodynamics, like an exploding boiler or a realistic mechanical engine that dissipates heat. We tend to experience these processes as unfolding in one way but not the other: the boiler explodes but does not 'un-explode'; the engine dissipates the heat it generates but does not spontaneously heat up. That sort of time asymmetry is often said to be a consequence of the second law of thermodynamics, that in at least some contexts, entropy does not decrease. In Chapter 6, I will argue that the situation is more subtle. But, for this story, the more important difficulty is that classical thermodynamics makes no mention of a system's underlying constituents. With growing interest in the nature of the material that makes up a gas or an engine, the natural next step was to use a theory of fundamental matter to try to explain thermodynamic behaviour.

One prominent perspective on fundamental matter in the nineteenth century was the atomist one, commonly attributed to Democritus, Boyle, and Bošković. On this view, all physical phenomena can be reduced to "the particular sizes, shapes, and situations of the extremely little bodies that cause them" (Boyle 1772, p.680). The possible motions of these phenomena would then be described by the laws of a dynamical theory, in the sense of Section 1.1.

What does it mean to 'time reverse' these structureless little bodies? We could imagine a film of the particles played back in reverse. One would at least expect to see their positions occur in the reverse time-order and with velocities in the opposite directions. This provides a rough, preliminary way to think about the time reversal transformation, which will be clarified in Chapters 2–3. For now, following the discussion of Section 1.1, we can take time reversal symmetry in a dynamical system to mean that there is a possible trajectory of particle positions and velocities such that, if we consider the trajectory in the reverse time-order and with reversed velocities, then the resulting curve is a possible trajectory as well. Time *asymmetry* would then be a dynamical system that is not time symmetric. In a time asymmetric system, the time reverse of at least one trajectory describes motion that is impossible according to the laws.

This concept had a dramatic effect on the work of Ludwig Boltzmann, who proposed to reduce all of thermodynamics to the statistics of huge numbers of little particles (Boltzmann 1872). Boltzmann's explanation of their apparent time asymmetry was given in his famous '*H*-theorem', where he seems to conclude that generic classical mechanical systems are likely to approach a 'stationary state', meaning one that does not change over time and which attains the maximum possible entropy.[12] The reverse 'entropy-decreasing' process, according to Boltzmann, would be highly unlikely.

A simple way to understand Boltzmann's thinking is in terms of a *counting argument:* roughly, that high-entropy states can happen in such an enormous variety of ways that they occupy the 'greatest volume' of possibilities. It is like imagining a house with a thousand blue rooms and one room that is red.[13] Think of the blue rooms as analogous to high-entropy states. Now, suppose that you were to leave a red room and enter an arbitrary new room (with a uniform probability of entering a given room); it is overwhelmingly likely that the new room will be blue. Moreover, it is likely that if you continue to repeat this process, your room colour will (with high probability) be unchanging or 'stationary' over time.[14]

This sort of counting argument has the potential to explain many asymmetries: an exploded boiler and a dissipated gas are both descriptions that belong to the overwhelming majority of possible states. So, it is natural that we should expect a system to end up in such a state. Unfortunately, this is not enough to explain the time-asymmetric behaviour of such systems. Returning to the house analogy, suppose we find a person in a red room and ask what colour room they are most likely to have come from? The very same volume argument concludes: a blue one. That is, the counting argument by itself provides equally good evidence for a high entropy state to the future and to the past.

---

[12] There is some debate about whether this is actually what Boltzmann meant to conclude by this theorem and about whether the resulting state is really 'stationary'; see Klein (1973, p.73) vs. Von Plato (1994, p.81), and Uffink (2007, §4) for a convincing clarification.

[13] In fact, given around $10^{26}$ molecules of air in your kitchen, the analogy there would require a house with around $10^{10^{26}}$ rooms, with all but one coloured blue.

[14] Here, the supposition that you enter an 'arbitrary' room with uniform probability is contentious: there is little agreement among philosophers of physics on the extent to which it works (cf. Uffink 2001).

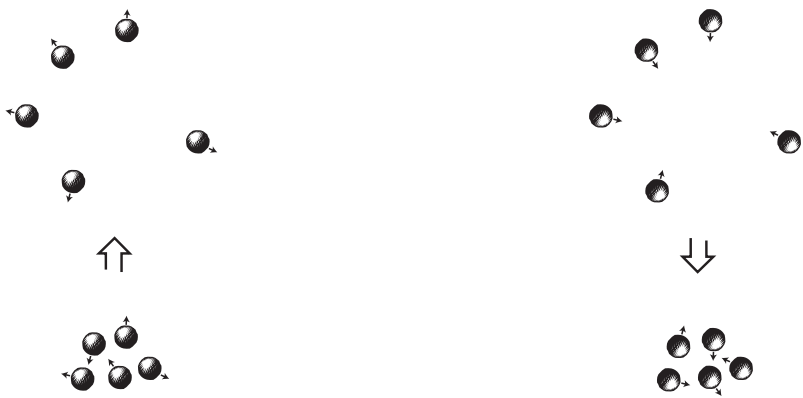

Figure 1.4 Reversing the velocities of a thermodynamic process (left) produces an antithermodynamic process (right).

One might hope that the apparent time asymmetry of thermodynamics could be explained by some other aspect of the laws of classical particle mechanics. This hope was famously dashed by the Austrian chemist Josef Loschmidt,[15] using the concept of time reversal:

Indeed, if in the above case, after a time $\tau$ which is long enough to obtain the stationary state, one suddenly assumes that the velocities of all atoms are reversed, we would obtain an initial state that would appear to have the same character as the stationary state. For a fairly long time this would be appropriate, but gradually the stationary state would deteriorate, and after passage of the time $\tau$ we would inevitably return to our initial state. (Loschmidt 1876, p.139)

Loschmidt's observation was that all known possible motions in classical particle mechanics correspond to a time-reversed motion that is also possible, with particle positions occurring in the reverse time-order and with their velocities reversed. Boltzmann had described a physical system in terms of a time-symmetric system of classical particles. So, if such a system evolves classically from a low-entropy state to a high-entropy one, then there is always a counterpart system that evolves 'anti-thermodynamically', from a stationary state to a non-stationary one, as in Figure 1.4. This dramatically contradicts Boltzmann's conclusion that a system of molecules generically evolves towards a stationary state.

In discussion with Loschmidt about this reversibility paradox, Boltzmann is rumoured to have replied, "Well, you just try to reverse them!" (Brush 1976a, p.605). However, he took the problem very seriously and developed a collection of creative responses to Loschmidt's reversibility paradox.[16] One popular textbook response today is to restrict Boltzmann's argument to those descriptions that begin in a special low-entropy initial state. It is then sometimes claimed that, as an economical way of explaining all time asymmetry at once, one can postulate that the universe as a whole began with very low entropy. The meaning and status of this postulate, dubbed the 'Past Hypothesis' by Albert (2000), is a matter of ongoing debate.[17] Price (1996, 2004) has argued that such arguments do not provide evidence for time asymemtry. I agree, and will give this argument from the perspective of my account in Chapter 5.

---

[15] This 'reversibility paradox' was anticipated by Maxwell, Tait, and Thomson between 1867 and 1870, and revived by Ehrenfest and Ehrenfest-Afanassjewa (1907); see Brush (1976a, pp.82–3) and Brush (1976b, pp.602–5).

However, all is not lost for the arrow of time: developments after Boltzmann led to an entirely new kind of time asymmetry. To appreciate how this happened, it helps to first review how time reversal came to play a more centre-stage role in physics.

## 1.4 The Rise of Time Reversal

Physicists of the nineteenth century, like many of us, were fascinated by cats. Witness the French mathematician Jules Richard:

The problem of the cat who falls back on its paws has preoccupied scholars for several years. Here is what the problem amounts to. A cat launched into the air always falls back on its paws; how can this be done? It seems that the cat launched in the air and with no point on which to press could not modify its motion in any way.[18]

The problem was widely studied throughout the nineteenth and twentieth centuries, including by James Clerk Maxwell, who in a letter to his wife described a widespread rumour about his time spent in Cambridge: "There is a tradition in Trinity that when I was here I disovered a method of throwing a cat so as not to light on its feet, and that I used to throw cats out of windows".[19]

The origin of this problem, according to Paul Painlevé (1904, p.1171–2), is an assumption of time reversal symmetry: when a system begins at rest, the equations of motion "do not change, and neither do the initial conditions, when one changes $(t - t_0)$ into $-(t - t_0)$". From this, Painlevé argued that a rigid cat cannot rotate as it falls. Earman (2002b) calls this the "Bad News for Cats Theorem". As far as I can tell, some further assumptions are needed to make Painlevé's argument work, but his insight that time reversal can give rise to an informative physical argument still broke new ground.[20] Fortunately for cats, there is a ready solution to the problem: the cat's motion need not be rigid (Figure 1.5). Simple models have even been proposed that clarify the non-rigid motion of a falling cat, and the problem remains of interest to modern mathematicians.[21]

---

[16] See Uffink (2001) for an overview.
[17] Cf. Earman (2006); Wallace (2010, 2017).
[18] My translation of Richard (1903, p.183).
[19] Reported in Maxwell's biography by Campbell and Garnnett (1882, p.499). The present author recommends that you be nice to your cat.

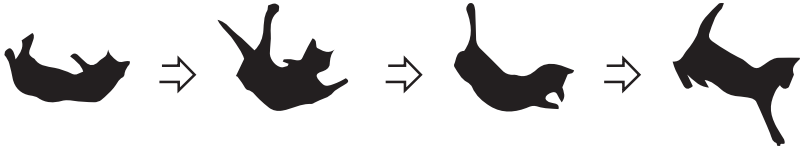

Figure 1.5 Non-rigid motion avoids the Bad News for Cats theorem, which was derived by Painlevé (1904) on the assumption of time reversal symmetry.

Arguments underwritten by time reversal invariance became increasingly common after Wigner (1931) introduced the first precise definition in his celebrated book, *Group Theory and its Application to the Quantum Mechanics of the Atomic Spectra*. This book simultaneously introduced a central role for group theory in modern physics, as well as a central role for time reversal. I will examine Wigner's definition in precise detail in Chapter 3; but here is the passage in which time reversal made its debut, translated as 'time inversion' in Wigner's writing. After considering a system in which all the translation and velocity boost symmetries have been suppressed, Wigner writes:

The transformation $t \rightarrow -t$ remains an additional symmetry element. It transforms a state $\varphi$ into the state $\theta\varphi$ in which all velocities (including the 'spinning' of the electrons) have opposite directions to those in $\varphi$. (Hence, 'reversal of the direction

---

[20] A Painlevé-like statement that *can* be proved is that the falling cat's motion is periodic. Let $S$ be a set of states, and let $t \mapsto \varphi_t$ be a one-parameter group of bijections on $S$ satisfying $\varphi_{t+t'} = \varphi_t\varphi_{t'}$ for all $t, t' \in \mathbb{R}$, representing the dynamics. Suppose $\tau$ is a representation of time reversal symmetry (see Chapter 4), so $\tau \circ \varphi_t = \varphi_{-t} \circ \tau$ for all $t \in \mathbb{R}$, which satisfies $\tau \circ R = R \circ \tau$ for some non-trivial bijection $R$, e.g., $R$ could be a rotation. *Proposition*: a state $s \in S$ can only satisfy $\tau(s) = s$ (initially at rest) and $\varphi_t(s) = R(s)$ for some $t \neq 0$ (finally changed) if it is periodic, in that $\varphi_{t'}(s) = s$ for some $t' \neq t$. *Proof*: Being 'finally changed' is equivalent to $\varphi_{2t}s = \varphi_t \circ R(s)$. Our assumptions thus imply that $\varphi_{2t}(s) = \varphi_t \circ R(s) = \varphi_t \circ R \circ \tau(s) = \tau \circ \varphi_{-t} \circ R(s) = \tau(s) = s$. The conclusion then follows with $t' = 2t$.

[21] The first mechanical solution to the falling cat problem was given by Kane and Scher (1969); for later mathematical developments, see Montgomery (1993) and the references therein.

of motion' is perhaps a more felicitous, though longer, expression than 'time inversion'.) The relation between time inversion and the change which the lapse of time induces in a system is of great importance. (Wigner 1931, p.325)

Wigner's comment about the importance of time reversal symmetry was entirely correct.[22] In addition to Bad News for Falling Cats, time reversal symmetry was used in the development of low-temperature physics, through Wigner's analysis of energetic degeneracy (Wigner 1932). It was used to show why non-trivial superpositions of bosons and fermions are never observed in nature (Wick, Wightman, and Wigner 1952). Time reversal symmetry was even shown by Wald (1980) to be incompatible with the possibility of pure-to-mixed state transitions in quantum theories of gravity. Today, the concept of time reversal is a cornerstone of modern physics.

The idea that time is symmetric might even have been absorbed as a central axiom of modern physics, from which a great number of conclusions could be derived, had nature not conspired to have it otherwise. At the time of Wigner's writing, all known laws of elementary motion appeared to exhibit manifest time reversal symmetry. Shockingly, this pattern eventually failed: in 1964, time reversal symmetry was dramatically ejected from the axioms of physics, and a new origin for an arrow of time became available.

## 1.5 The Great Shock

In the first half of the twentieth century, the importance of time reversal in fundamental physics had come to be appreciated. However, few seemed to consider the possibility that it might fail to be a symmetry. The earliest exception that I know of is a comment by one of the founders of quantum mechanics, Paul Dirac. With characteristic foresight, Dirac wrote:

A transformation . . . may involve a reflection of the coordinate system in the three spacial dimensions and it may involve a time reflection, the direction . . . in space-time changing from the future to the past. I do not believe there is any need for physical laws to be invariant under these reflections, although all the exact laws of nature so far known do have this invariant. (Dirac 1949, p.393)

Dirac's comment reveals a remarkable puzzle for anyone interested in the status of time asymmetry: we are often in a situation of not knowing the correct laws of motion for a physical system. But, the definition of a symmetry of motion involves a statement of how the system can change over time. So, how can we possibly determine whether or not any transformation like time reversal is a symmetry without knowing the laws of motion?

---

[22] I have suggested in Section 1.1 that an asymmetry of motion allows us to infer an asymmetry of time. So, for now let me set aside Wigner's comment about motion; I will return to it in Chapter 2.

Progress towards a solution came in the form of a 'symmetry principle', originally formulated by Pierre Curie, which has been the subject of a great deal of study in the philosophy of physics (and which I will review in Chapter 6):

> When certain effects reveal a certain asymmetry, this asymmetry must be found in the causes that have given rise to it. (Curie 1894, p.401)

One interpretation of this principle takes a 'cause' to be a law of motion together with an initial state, an 'effect' to be a final state.[23] So, if a final state has an asymmetry that is not found in the initial state, then that asymmetry must somehow be in the dynamical law itself. This provides a way to evaluate certain symmetries of motion: if two successive states of a system do not share a unitary symmetry, *then that symmetry must be violated by the equations of motion*.

Curie's principle was first applied in particle physics to show that the parity transformation (or what Dirac called 'spatial reflection') is not a symmetry of the equations of motion, although the principle was not referred to by that name. This application arose out of a problem in particle physics known as the $\theta$–$\tau$ puzzle. Two in-going particles, denoted $\theta$ and $\tau$, were known to have the same masses and lifetimes but to decay into different outgoing states: the former into two pions (positive and neutral), and the latter into three pions (two positive and one negative). Since one of these decay products was invariant under the parity transformation and the other was not, parity invariance implies that they must have originated from different particles $\theta$ and $\tau$, by an application of Curie's principle. So, given that $\theta$ and $\tau$ were not the same particle, the puzzle was to explain why nature appeared to conspire to give them the same masses, and indeed the same lifetimes in a decay interaction. Lee and Yang put the puzzle in a controversial way:

> One might even say that the present $\theta$-$\tau$ puzzle may be taken as indication that parity conservation is violated in weak interactions. This argument is, however, not to be taken seriously because of the paucity of our present knowledge concerning the nature of the strange particles. (Lee and Yang 1956, p.254)

---

[23] Cf. Belot (2003), Earman (2004), and Roberts (2013a).

The experimentalist Norman Ramsey was intrigued and asked Richard Feynman if he thought that it was worth designing an experiment to test for parity violation. Feynman said yes, but that he was almost sure that the result would be that the world is parity invariant. When Ramsey asked if Feynman was confident enough to bet 100 dollars to one, Feynman is said to have replied "No, but 50 dollars I will!" (Gardner 1991, p.91). Feynman soon lost his money when the experimentalists Chien-Shiung Wu, Ambler, et al. (1957) proved that the decay of the Cobalt-60 atom violates parity. In addition, her discovery immediately solved the $\theta-\tau$ puzzle: the two particles were in fact one and the same – now known as a positive $K^+$ meson or kaon, which is subject to parity-violating interactions.

With parity symmetry violated, the physics community came to widely believe that another symmetry must hold instead. The transformation that exchanges ordinary matter with an exotic substance called 'antimatter', called 'charge conjugation' (denoted 'C'), turns out to have been violated by the Wu experiment as well.[24] However, the combined transformation consisting of C together with parity P produces a transformation denoted CP, which was thought to remain a symmetry. And, by a very general theorem known as the CPT theorem, discussed more in Chapter 8, CP symmetry is equivalent to time reversal symmetry. In this way, the physics community hung on to the assumption of time reversal symmetry, in spite of the recent fate of parity.

Why did they assume this? James Cronin gave a colourful explanation some years later:

It just seemed evident that CP symmetry should hold. People are very thick-skulled. We all are. Even though parity had been overthrown a few years before, one was quite confident about CP symmetry. (Cronin and Greenwood 1982, p.41)

In fact, it was not just the absurdity of the human condition that led physicists to replace parity symmetry with CP symmetry. After Wu's experiment, a number of simple and powerful theoretical models were quickly developed to explain her result, including one developed by the young Stephen Weinberg (1958). These elegant models did a compelling job of describing the behaviour of these new interactions. As it happened, they also strongly suggested both CP invariance and time reversal invariance. So, *pace* Cronin, this was not so much a matter of thick-skulled intuition but of the hard struggle to find any alternative.[25]

---

[24] This was pointed out by Garwin, Lederman, and Weinrich (1957), who independently verified Wu's experiment.

It is difficult to understate the great shock to the community when James Cronin and Val Fitch discovered that CP symmetry fails and therefore that time reversal symmetry fails too. The feeling was summarised here:

> It came as a great shock that microscopic $T$ invariance is violated in nature, that 'nature makes a difference between past and future' even on the most fundamental level. We might feel that such a statement is sensationalist rather than scientific; yet there is indeed something very special about a violation of the invariance under T or CP. (Bigi and Sanda 2009, p.5)

This story is worth describing in some detail, although we will revisit it in Chapter 7.

Cronin identifies a deep paper by Gell-Mann and Pais (1955) as a central influence on his thinking, writing that, "you get shivers up and down your spine, especially when you find you understand it" (Cronin and Greenwood 1982, p.40). Gell-Mann and Pais had proposed a means of detecting the violation of C symmetry, but it was a means of detecting CP symmetry violation as well. Cronin and Fitch had an accelerator at the Brookhaven National Laboratory on Long Island, New York, which could set an accurate bound on both kinds of symmetry violation. So, to test CP symmetry, they designed an experiment firing a long-lived neutral kaon state $K_L$ into a spark chamber and taking photographs of thousands of particle decay events to see what came out.

At the time, neutral kaons were characterised by their decay into three pions, one neutral and two of opposite charges. This decay is compatible with CP invariance, by another application of Curie's principle: the neutral kaon $K_L$ and the three-pion state are reversed by CP. In contrast, a two-pion state is left unchanged by CP, and so a decay into just two pions would imply CP violation (Figure 1.6). Cronin and Fitch set out to check whether they could show that, to a high degree of accuracy, no CP violating two-pion decay events could be found.

After a long analysis of all the photographs, they found that to the contrary, a small but unmistakable number of long-lived neutral kaons decayed into two pions, violating CP and time reversal symmetry. They immediately began checking their result and discussing it with colleagues at Brookhaven. After explaining it to their colleague Abraham Pais over coffee,

---

[25] Weinberg describes one phase relation in his model: "This phase relation would follow from CP invariance, but is difficult to understand on any other basis" (Weinberg 1958, p.783).

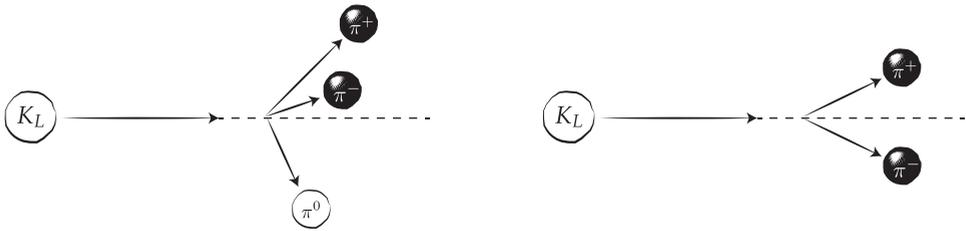

Figure 1.6 Neutral kaon decay into three pions (left) and two pions (right). The neutral pion is invisible to the spark chamber, but its trajectory is calculated by conservation of momentum. The two-pion decay implies CP-violation.

Pais reported that, "[a]fter they left I had another coffee. I was shaken by the news" (Pais 1990).

Cronin and Fitch were awarded the 1980 Nobel Prize for their discovery.[26] In the award ceremony speech, Gösta Ekspong described how the physics community viewed the implications for the direction of time in poetic terms:

> The laws of physics resemble a canon by Bach. They are symmetric in space and time. They do not distinguish between left and right, nor between forward and backward movements. For a long time everyone thought it had to be like that.... [Cronin and Fitch's] discovery ... implied consequences for time reflection. At least one theme is played more slowly backwards than forwards by Nature. (Ekspong 1980)

Theoreticians caught up with the detection of CP and time asymmetry over the next decade. In order to describe these interactions in terms of the developing theory of non-abelian gauge fields due to Yang and Mills (1954), the previous three-quark gauge theory had to be adjusted. This led to the modern six-quark theory of flavour mixing and paved the way for the unified quantum theory of strong and electroweak interactions known as the Standard Model.

As I will argue in Chapter 7, there are subtleties in Curie's principle that prevent its use in the detection of time asymmetry without appeal to CPT symmetry. However, new symmetry principles were soon developed that have been less discussed by philosophers but which enabled the more direct detection of time asymmetry. These principles were successfully applied to produce the first evidence of time reversal symmetry violation without appeal to CPT symmetry, by the CPLEAR Collaboration (1998) at CERN.

---

[26] Cronin describes this history in a delightful University of Chicago lecture transcribed by Margaret Greenwood (Cronin and Greenwood 1982).

This was followed by a number of creative tests of CP symmetry and time reversal symmetry violation in other sectors, including the *B*-meson and, quite recently, in the lepton sector through muon–electron neutrino oscillation.[27] This provided evidence that time reversal symmetry violation is here to stay, and in increasing quantities.

## 1.6 Summary

The dramatic discovery of time reversal symmetry violation has not yet been widely embraced by philosophers, at least as evidence for an arrow of time. Some have found it helpful to set aside the philosophical analysis of time asymmetry in particle physics, focusing their analysis on other important issues,[28] while others have expressed flat-out scepticism about its existence (Horwich 1989, p.56). Maudlin (2007, p.118) has suggested that this latter response has "a certain air of desperation" to it, insofar as Nobel prizes have already been awarded. But, I think there is also a conceptual issue to be overcome: these discoveries may establish an asymmetry in the way particle states change, but what reason is there to think that this establishes an arrow of time itself?

Answering this question requires some philosophical analysis. McTaggart's separation of time into 'A series' and 'B series' components is a start. However, in this book, I will make the separation using the language of asymmetries in dynamical systems on the one hand, and asymmetries in spacetime structure on the other. My central postulate will be that these two kinds of asymmetries are intimately linked. However, this link must be spelled out and motivated in more detail, if we are to have a convincing account of what time reversal in particle physics has to do with the reflection of time itself. The project of the next chapter is to develop one such account.

---

[27] CP-violation by $B_0$-mesons was detected independently by the BaBar Collaboration (2001) and the Belle Collaboration (2001); time reversal symmetry violation was later detected by the BaBar Collaboration (2012). New evidence for CP violation to a much larger extent was recently discovered in the lepton sector by the T2K Collaboration (2020).

[28] Cf. Callender (2000, p.249) and Price (1996, p.18).

# 2

# What Time Reversal Means

*Précis.* *The meaning of time reversal can be reconstructed from the structure of time translations. 'Instantaneous' time reversal is just its proxy in a state space representation.*

The Russian physicist Vladimir Fock is rumoured to have remarked:

Physics is essentially a simple science. The main problem is to understand which symbol means what.[1]

What does time reversal mean? Its textbook presentation is a little mysterious. On the one hand, time reversal is presented as a simple mapping $t \mapsto -t$, with the explanation that it 'reverses the arrow of time'. On the other hand, it is treated as a richly structured concept that reverses all sorts of quantities, like momenta, magnetic fields, and spin. It even conjugates wave functions. Where does all this structure come from? Received opinion is that it comes from 'motion reversal', not 'time reversal'. But 'motion reversal' is rarely defined, and the relationship between the map $t \mapsto -t$ and all of that rich structure is rarely made clear. Philosophers and physicists have made some progress in clearing the fog, but disagreement remains.

   This chapter will show how elementary considerations from representation theory explain the mystery. I will argue that it is firstly a matter of determining which symbol means what: the time reversal transformation $t \mapsto -t$ is not a map on time coordinates like $t =$ two o'clock, but rather a map on time *translations* like $t =$ time-shift by two hours. All the rich structure of time reversal that gets referred to as 'motion reversal' arises out of how time translations are represented on a state space, which I will derive in Chapter 3. All that structure is not just motion reversal: it is a representation of time reversal.

---

[1]  Khriplovich and Lamoreaux (1997, p.53).

The central idea that I will present is a general perspective on how spacetime symmetries appear in state space, which I call the *Representation View*. In short, I will propose that we separate our discussions of time into two components: time translations in spacetime and time translations in state space, which are related by a representation. This simple idea is so powerful that it will carry us through several of the arguments in this book. In this chapter, I will use it to clarify how time reversal gets its meaning.

I begin by setting out the problem of defining time reversal in Section 2.1 and then recounting in Section 2.2 how two competing camps have responded. Section 2.3 then presents a first look at the Representation View, which I claim mediates between these camps in a way that dissolves the debate. Section 2.4 presents the philosophy of time that will underlie my response: that time is not a mere ordered set but has 'relational' or 'structural' properties like a Lie group of time translations. Drawing on structuralist and functionalist perspectives for inspiration, Section 2.5 argues that time reversal is a reversal of time translations; and Section 2.6 recovers its structure as a group element from the structure of time translations. Finally, the meaning of time reversal in state space is presented in Section 2.7, as the transformation corresponding to $\tau : t \mapsto -t$ in a state space representation.

## 2.1  The Definability Problem

Some favourite interpretative strategies in the philosophy of physics appear unhelpful when applied to time reversal. For example, take Percy Bridgman's operationalism. Bridgman performed experiments at higher pressures than had ever been achieved before; in 1922, one exploded, killing two people and destroying his basement laboratory in the Jefferson physics building at Harvard – even shattering the windows of the nearby law school.[2] In high-pressure environments where existing gauges blow, it wasn't clear to Bridgman how 'pressure' should be defined. His response, inspired by Einstein's (1905) account of simultaneity, was to interpret all meaning as deriving from physical operations:

---

[2]  As recounted by Walter (1990, pp.37–8). Sadly, Bridgman's research assistant Atherton Dunbar and carpenter William Connell were both killed by the explosion, and eight graduate students in the room above were injured.

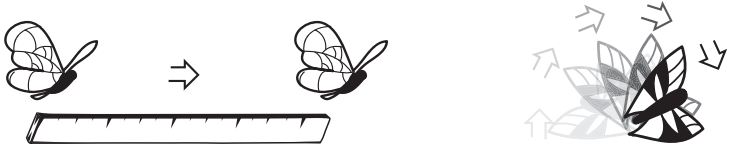

Figure 2.1  Operational definitions of spatial translation and rotation.

We may illustrate by considering the concept of length: what do we mean by the length of an object? ... To find the length of an object, we have to perform certain physical operations. The concept of length is therefore fixed when the operations by which length is measured are fixed: that is, the concept of length involves as much as and nothing more than the set of operations by which length is determined. (Bridgman 1927, p.5)

Setting aside the radical implications of operationalism for physics,[3] it appears that spatial translation and rotation can be defined operationally: a spatially translated subsystem is one resulting from the operation of rigid motion along a ruler, while a spatially rotated one results from rigid rotation about an axis, as shown in Figure 2.1. Time translation can be defined operationally too, by comparison of a subsystem to the state of a clock.

But, operationalism seems at best unhelpful when it comes to time reversal. What operation, short of science fiction, would reverse time? Physicists and philosophers tend to reply with analogies to films played in reverse. But this strategy is imprecise, and its relevance is questionable. This is especially true for phenomena like the time reversal violating neutral kaon: apart from its inconvenient invisibility in a spark chamber, the typical lifetime of even the 'long-lived' neutral kaon state $K_L$ is around $5 \times 10^{-8}$ seconds. In this case, the film analogy would require an impossibly high frame-rate of over 20 million frames per second.

As a matter of convention, one can give the phrase 'time reversal' whatever definition one wishes. Many physicists do. But, if no justification were given, there would be no connection between time reversal and any non-conventional arrow of time. We would lose the physics community's poetic conclusion that, following the experimental discovery of time reversal violation, we learned that "at least one theme is played more slowly" backwards in time than forwards.[4] Call this the *definability problem:* How can one justify a definition of time reversal in a way that makes it relevant to the arrow of time in our world?

---

[3]  As Bridgman himself notes, operationalism complicates the meaning of simple concepts, since (for example) measuring stellar lengths in light years and measuring laboratory lengths with rulers require different operations, resulting in multiple concepts of 'length'.

[4]  See Section 1.5.

## 2.2  Two Competing Camps

Philosophers and physicists have responded to the definability problem by separating it into two parts: an order reversing part and an instantaneous part. David Albert (2000, §1) explains these two parts using the term *instantaneous state* to refer to a point in a typical state space, which describes the world at an instant: a distribution of particles, a probability distribution, a description of an electric field on a spacelike surface, and so on.[5] Examples of a state space include a Hilbert space or a phase space, as discussed in more detail in Chapter 3. Albert uses *temporal sequence* to refer to any one-parameter set of these states, parametrised by time. Then the definition of 'time reversal' could include one or both of the following:

1. *Time order reversal:* transformation of a temporal sequence of instantaneous states to a sequence with the very same elements, but with the reverse ordering.
2. *Instantaneous reversal:* transformation of each instantaneous state to a 'reversed' instantaneous state.

Viewing a sequence of instantaneous states as a stack of pancakes, the first transformation reverses the order of items in the stack, while the second transformation turns around the individual pancakes, as shown in Figure 2.2.

Although standard textbook treatments take time reversal to involve both, the first is much easier to understand. Sachs even begins his classic book, *The Physics of Time Reversal*, with the proposal that "[t]he qualitative meaning of the time variable is that of an ordering parameter in one-to-one correspondence to a sequence of events" (Sachs 1987, p.4). Defining 'time' to be an instance of an ordering makes the statement 'Time reversal is ordering reversal' into a trivial analytic truth, analogous to 'A bachelor is unmarried'. But, this definition is plausible whether or not we share Sachs' quoted view on time. Time order reversal can even be operationally defined: take two clocks, one that counts down and the other that counts up, and associate each clock-time with a physical event, as in Figure 2.3. The melting ice cubes might suggest that time 'goes up the page'. But, operationally, each clock determines a sequence of states ordered by increasing clock-times, and each sequence can be defined as the time order reverse of the other.

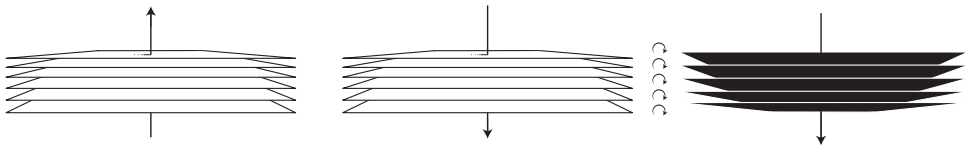

Figure 2.2  A time ordering (left), its time order reversal (middle), and time order reversal together with instantaneous reversal (right).

---

[5] Albert further asks that these descriptions be 'complete' and 'genuinely instantaneous', as I discuss in Sections 2.5.1 and 3.1.1. See also Butterfield (2006a,b) for a similar view. I view my account as compatible with both.

In contrast, instantaneous reversal appears a bit more mysterious. Of course, a time order reversal of the form $t \mapsto -t$ induces a transformation of velocity $v = dx/dt$ to its negative, $v \mapsto -v$. But, textbooks treat many other quantities as transforming under time reversal as well: spin, momentum, and magnetic fields all reverse sign under time reversal, even though none of these are typically written as a rate of change of some parameter. Even a magisterial treatise on time reversal like the book *CP Violation* by Bigi and Sanda (2009, p.5), for all its virtues, leaves room for confusion when they say that time reversal "can be viewed as reversal of motion", and then – in the same paragraph – that it provides evidence that "nature makes a difference between past and future". What then is the link between these two?

Physicists and philosophers have responded to this situation in two ways. The first camp, containing a majority of physicists and philosophers, emphasises the first claim of Bigi and Sanda: they accept instantaneous reversal, while still insisting that time reversal is 'really' motion reversal. This tradition goes right back to Wigner's founding discussion of time reversal introduced in Section 1.4, although Wigner uses the phrase 'time inversion' in place of time reversal: "'reversal of the direction of motion' ['Bewegungsumkehr'] is perhaps a more felicitous, though longer, expression than 'time inversion'" (Wigner 1931, p.325).[6] The second camp rejects instantaneous reversal: as Albert (2000, p.18) exclaims, "What can it possibly mean for a single instantaneous physical situation to be happening '*backward*'?" He answers: "gibberish", and Callender (2000, p.254) agrees. They conclude that time reversal can only mean the conceptually less-problematic concept of time order reversal.[7]

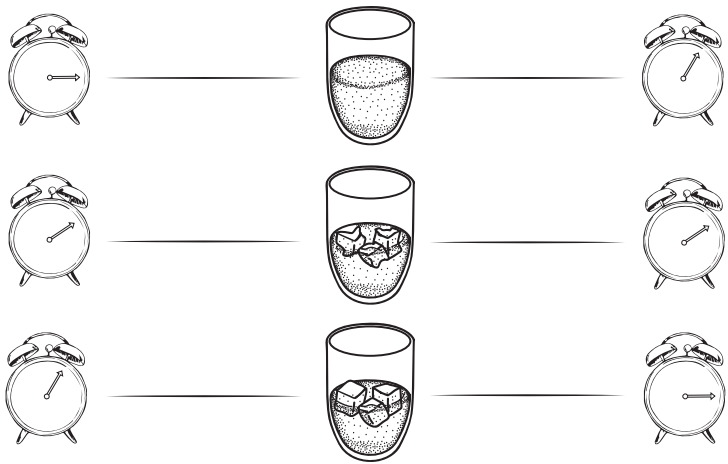

Figure 2.3 Operational definition: the orderings determined by increasing clock times for the left and right clocks are the time order reverse of each other.

The first camp has been defended in a variety of ways. Sachs (1987, p.23) suggests that the hidden velocities associated with "microscopic current structure" explain the reversal of the magnetic field. For other quantities, such as those in quantum theory, it is common to appeal to rules about a classical-to-quantum correspondence: if classical momentum transforms as $p \mapsto -p$, then it is not so far-fetched to imagine that quantum momentum transforms in the same way. John Earman gives an argument like this, in his explanation of why time reversal ought to conjugate position wavefunctions:

Since $\psi(\mathbf{x}, 0)$ encodes information about the momentum of the particle it must be 'turned around' or non-trivially transformed by the time reversal operation so that the reversed state at $t = 0$ describes the wave packet propagating in the opposite direction. So instead of making armchair philosophical pronouncements about how the state cannot transform, one should instead be asking: How can the information about the direction of motion of the wave packet be encoded in $\psi(\mathbf{x}, 0)$? Well (when you think about it) the information has to reside in the phase relations of the components of the superposition that make up the wave packet. And from this it follows that the time reversal operation must change the phase relations. (Earman 2002b, p.248)

Although some find it intuitive to motivate time reversal on the basis of classical 'motion reversal', it's hard not to see this as a change of subject, from 'time reversal' to 'motion reversal'. It is also not always clear how to apply it. For example, the term 'motion reversal' alone offers little advice about how internal degrees of freedom transform under time reversal. However, not all such defences appeal to motion reversal: Malament (2004) and Roberts (2017) give derivations of time reversal in electromagnetism and quantum theory, respectively, which appeal directly to facts about time.

---

[6] After Wigner (1931), two prominent examples are the wonderful textbooks by Sakurai and by Ballentine: "*time reversal* is a misnomer; it reminds us of science fiction. Actually what we do in this section can be more appropriately characterized by the term *reversal of motion*" (Sakurai 1994, p.266); and, "the term 'time reversal' is misleading, and the operation that is the subject of this section would be more accurately described as *motion reversal*" (Ballentine 1998, p.377). Among philosophers, see Earman (2002b, p.247): "the reversal of motion prescription for particles forces out the standard account of the time reversal behaviour of electromagnetic fields"; and Uffink (2001, p.314): "the term 'time reversal' is not meant literally. That is to say, we consider processes whose reversal is or is not allowed by a physical law, not a reversal of time itself".

[7] See Albert (2000, §1) and Callender (2000) for statements of this view, which both physicists and philosophers have endorsed; cf. Allori (2015), Castellani and Ismael (2016), and Rachidi, Rubinstein, and Paolone (2017, §§1.3.6 and 1.3.7).

Still, all this does seem to support Albert's summary of the textbook account:

(1) that in the case of Newtonian mechanics the procedure is 'obviously' to reverse the velocities of all the particles, and to leave everything else untouched; and (2) that the question needs to be approached afresh (but with the Newtonian case always somehow in the back of one's head) in each new theory one comes across; and (3) that what it is *in all generality* for one physical situation to be the time reverse of another is (not surprisingly!) an obscure and difficult business. (Albert 2000, p.18)

To those who would conclude that time reversal is obscure or difficult, Albert responds: "It isn't, really". A conceptually clearer alternative, according to him, is to say that time reversal is just the reflection of time, $t \mapsto -t$.

I agree with this alternative, although I will soon argue that this does not settle the debate in favour of the first camp. We should all agree that, at the end of the day, it must be possible to view time reversal as a conceptually simple operation, which does nothing more than reverse time. If we are to continue referring to it as 'time reversal' with any honesty, then all its bells and whistles must ultimately derive from this touchstone. And yet, as Callender (2000, p.262) says, "many arguments attempt to blur the difference" between temporal reflection $t \mapsto -t$ and the transformation of other degrees of freedom associated with motion. In an effort to restore clarity, Callender even refers to the former as 'time reversal' and the latter 'Wigner reversal'; may each be worthy of the name.

I will argue that Callender's distinction is unnecessary, because the two amount to the same thing. The first camp's beautifully simple intuition is correct, that time reversal is a map that transforms time to its negative. However, more is needed to understand it in state space, where there is room for confusion about what Albert and Callender say. Referring to the highly structured state space of quantum mechanics, associated with a Hilbert space $\mathcal{H}$, Callender considers a curve $\psi(t)$ indexed by a parameter $t$ associated with time, and says:

It does not logically follow, as it does in classical mechanics, that the momentum or spin must change signs when $t \mapsto -t$. Nor does it logically follow from $t \mapsto -t$ that one must change $\psi \mapsto \psi^*$. (Callender 2000, p.23)

Here I disagree. Similarly, I think Albert goes too far when he writes about electromagnetism that,

Magnetic fields are *not* the sorts of things that any proper time-reversal transformation can possibly turn around. Magnetic fields are not – either logically or conceptually – the *rates of change* of anything. (Albert 2000, p.20)

On the face of it, these authors seem to reject the standard dogma that time reversal conjugates wavefunctions and reverses magnetic fields. Some physicists have even followed suit in this observation, distinguishing the 'strict' time reversal of Albert and Callender, which is not a symmetry of most physical theories, from the 'soft' time reversal of most physicists which often is a symmetry (cf. Rachidi, Rubinstein, and Paolone 2017, §1.3.4).

That sort of conclusion goes too far, and these statements have correspondingly been met with widespread criticism.[8] But, this should not distract from the basic insight of Albert and Callender, which I will argue is correct: time reversal simply reverses time. Albert and Callender are even right to say that the transformation rules for time reversal on state space do not logically follow from this alone; however, once the meaning of 'time's passage' in state space has been specified, I will argue that they do.

Thus, to be clear: I also endorse the instantaneous reversal transformation of the second camp. However, I would like to offer a path of reconciliation. Let me characterise their essential claims as follows:

**Time Reflection Camp:** Time reversal is just time order reversal: a reflection of time's arrow.
**Instantaneous Camp:** Time reversal is time order reversal, together with a non-trivial transformation of instantaneous states in a space of states.

In this chapter and the following, I will show a sense in which these two views are in fact one and the same. This has only been obscured because our perspective on time has been too simple. In the next section, I will outline a more careful view that corrects this.

## 2.3 The Representation View

The Time Reflection Camp describes time reversal as a reflection of time, $t \mapsto -t$. The Instantaneous Camp describes it as including a transformation $T$ on state space. In this section, I will present a general view of spacetime symmetries that mediates between the two camps and argue that it settles the debate. My discussion here will be relatively informal; the details will then follow in the remainder of this chapter and the next.

---

[8] Cf. Arntzenius (2004), Arntzenius and Greaves (2009), Butterfield (2006b, §3.3), Earman (2002b), Malament (2004), North (2008), Peterson (2015), and Roberts (2012, 2017, 2021).

In physics and its philosophy, we often speak loosely of a symmetry as if its meaning were fixed in all contexts. For example, we speak of rotations and time translations in spacetime, and then again of rotations and time translations in a state space, like a Hilbert space. But, mathematically and conceptually, spacetime and state space are two different things. When we define a concrete symmetry transformation, we must specify what it transforms: Is it a transformation of spacetime, or a transformation of a state space? We cannot have it both ways.

When we do use the same language for symmetries in two contexts, it is justified only because there is a systematic relationship between them. To make that explicit, we can imagine taking the concept of a symmetry and 'pulling it apart' into (at least) two separate concepts: spacetime symmetries on the one hand and state space symmetries on the other. What is the relationship between these concepts?

This question is answered by a framework for characterising dynamical systems that is well-known in physics, originally appearing in the great work of Wigner (1939) on the representation of Minkowski spacetime symmetries on Hilbert space.[9] The framework has two components: first, there is a structure characterising a set of symmetry transformations, such as those of spacetime or of a gauge theory; second, there is a structure describing the symmetries of a space of states. I will discuss the meaning of both in more detail in what follows. The relationship between them is sometimes called a *group action*, but I will call it a *representation:* a homomorphism (or 'structure-preserving map') from the first structure to the other. This ensures that a 'structure-preserving copy' of the first symmetries are living inside the symmetries of state space.

All that I would like to add to this well-known view is an interpretive proposal, that a representation is what gives meaning to the concept of a

---

[9] Although Bargmann (1954) and Wigner (1939) together form the *locus classicus* for this framework, the basic idea is implicit in Sophus Lie's (1893) analysis of differential equations using what are now called Lie groups. Wigner himself traces the idea to the work of Ettore Majorana (1932) on representing rotations of fermions; Wigner's paper was submitted just months before Majorana's mysterious disappearance in March 1938 at age 32, on a boat leaving Palermo, never to be seen again (Recami 2019).

'spacetime symmetry' in the context of a state space. This simple, guiding symmetry principle is what underpins the argument of virtually every chapter in this book. So, let me set it out clearly for reference:

*The Representation View:* A symmetry of a state space can be interpreted as a 'spacetime symmetry' only if it is an element of a *representation* of a spacetime symmetry structure.

The Representation View is a wide-ranging perspective, which can be applied to virtually any spacetime symmetries and any state space structure. Among other things, the spacetime symmetries can be Newtonian or Minkowski; state space can be Newtonian configuration space or quantum theory. The view can even be given a local expression, using the concept of a local Lie group, in order to be applied in general relativity.[10] However, for most purposes in this book, and especially this chapter, the particular symmetries of interest will be the *time translations*.

In the special case of time translations, the Representation View says that a symmetry of state space only deserves to be called a 'time translation' if it is a representation of the time translation group. This requires us to 'pull apart' the concept of a time translation and identify two separate concepts: time translations in spacetime and time translations as state space symmetries. The connection between them is then given by a representation, which ensures there is a homomorphic copy of the spacetime time translations in state space, as in the state space diagram for a harmonic oscillator illustrated in Figure 2.4.

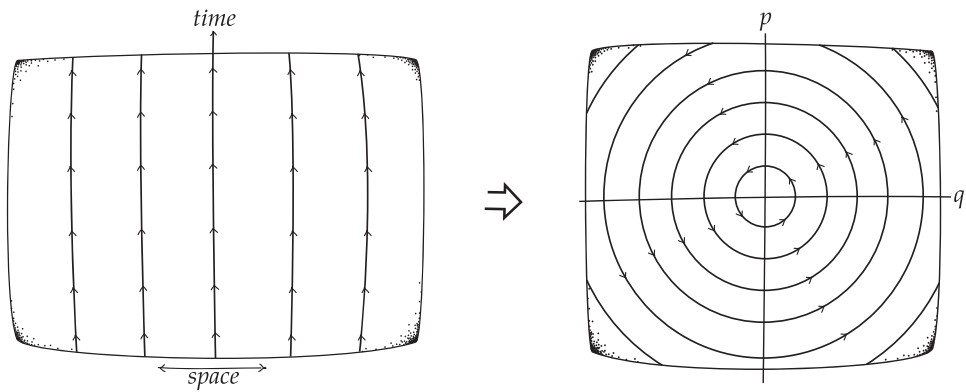

Figure 2.4 A group of time translations on spacetime (left) represented amongst the symmetries of a state space (right) for the harmonic oscillator.

---

[10]  See Olver (1993, p.18) and also Section 2.7.

Sometimes we make the Representation View explicit, as when one derives the dynamics of quantum theory from a group describing time translations like $G = (R, +)$, which are then associated with a strongly continuous representation amongst the symmetries of Hilbert space.[11] Other times it is left implicit, as when one says things like: "Let time evolution be represented by the solutions to this differential equation on a state space". There is nothing wrong with such a statement. However, it is important to notice what it leaves out: nothing is said about why these solutions represent changes over time, as compared to changes with respect to any other parameter.

For example, in the path of a particle traced out along a line in Figure 2.5, the picture alone says nothing about why the line represents a time translation as opposed to a spatial translation. A sociologist might say that a description of time in state space is one in which the physicist uses the letter '*t*'. A philosopher of physics should say that a description of time in state space is a representation of time translations.

The Representation View provides an elegant solution to the debate about the meaning of time reversal in dynamical systems, described in Section 2.2. Instead of thinking of time and time reversal as a single concept, we can pull it apart into two: time reversal is a spacetime symmetry of the form $\tau : t \mapsto -t$, but it is also a state space symmetry $T$, where the two are related by a representation associating the former with the latter: $\tau \mapsto T$. After all, it is hard to justify calling a theory 'dynamical' without a representation of time's passage! So, Albert (2000) and Callender (2000) are right to view time reversal as a transformation that simply reverses time. And, when Earman (2002b) proposes that time reversal is a state space transformation that "turns around" the phase relations of a wave packet at an instant, he is right too: this is the effect of the state space transformation $T$. The two views are not inconsistent; they simply describe two different sides of the same representation relation.

In the remaining sections of this chapter, I will develop the philosophical and mathematical understandings of time that underpin the Representation

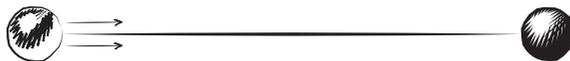

Figure 2.5  A translation in time, or space?

View. Of course, as the physics of time advances, we can expect our understanding of time to change. But, whatever structure turns out to more accurately describe temporal relations, we will now have a strategy for determining the associated meaning of time reversal, by viewing time in physics as a representation of that more accurate structure instead. What is perhaps surprising is that, when one adopts this view, the structure of time translations essentially determines the meaning of the time reversal operator $T$ on state space, with all its bells and whistles. To see this, one must first recognise various senses in which time is a highly structured concept, which includes group theoretic and continuity properties. That is the subject of the next section.

## 2.4 Temporal Structure

### 2.4.1 Structuralist and Functionalist Inspiration

The composer Hector Berlioz once wrote, "Time is a great master, it is said; the misfortune is that it is an inhumane master who kills the pupils".[12] Like Berlioz, I would like to identify some relevant properties of time without settling the question of what time actually is. Bracketing this question is an important part of understanding the broader nature of time, and it is the strategy I will adopt for the analysis of time reversal.

There are signs that neither physics nor philosophy is currently prepared to say what time is. The spacetimes of general relativity and later twentieth-century physics are widely believed to be an approximation to reality, which emerge on familiar scales from some hidden underlying substrate governed by quantum theory. There is currently little agreement about what that substrate is. However, abstract properties of time, like its symmetries, may hold some insight into its nature.

Philosophical debates about spacetime realism and antirealism have similarly led to little agreement on what time is. This is in spite of many important clarifications following classic works like Sklar (1974) and Earman (1989), and following the debate over Einstein's 'hole argument'.[13] The *substantivalist* takes spacetime to exist independently of its contents, as when Newton characterises Absolute Time and Absolute Space without reference to anything 'external'. Antirealism denies this, usually coupled with some statement of how facts about spacetime are reducible to something else, as when Leibniz suggests that temporal relations reduce to causal relations.[14] A modern relationist account was articulated and defended by Brown (2005) and by Brown and Pooley (2006), and remains the subject of much debate.[15]

---

[12] "Le temps est un grand maître, dit-on; le malheur est qu'il soit un maître inhumain qui tue ses éléves" (Berlioz 1989, p.390, 27 November 1856 letter to the playwright Saint-George).

[13] For an introduction, see Norton (2019) and the references therein.

However, recent developments have shown that it is possible to set aside the question of realism and still say something interesting. My discussion takes two developments as inspiration. In the first place, following John Worrall (1989), I find that a pleasant way to face controversy is to focus on our points of agreement: in the absence of a complete account of time, we can instead focus our attention on the more abstract structure of time, which is better-confirmed by experiment and less philosophically controversial.[16] In the second place, following Eleanor Knox (2013, 2019), I find that a natural way to identify that structure is through the functional role that time plays in more widely-agreed contexts. Butterfield and Gomes (2020) have placed this approach in its correct historical context: functionalism about a problematic concept like 'time' is an interpretation that identifies the concept with the occupant of a pattern of relations (a 'functional role') in a less problematic context, such as the behaviour of matter-energy – and, if we can, we also establish that the occupier of this role is unique.[17]

I will not take a position on either of these debates. Besides inspiration, the lesson that I would like to take from them is that the concept of 'time' in physics, whatever it actually refers to, must include rich relational or structural properties. As Callender (2017) points out, one simply cannot develop a physical theory without it:

> This demands scores of decisions about time. Are there instants? Ordered? Partially or totally? What is the topology ...? Is time continuous, dense, or discrete? Open or closed? One-dimensional? (Callender 2017, p.20)

I will shortly make this more concrete and identify particular 'structural patterns' expressed by a Lie group as providing the relevant 'functional role' for time. The pay-off for insisting on this, I will argue, is a clear account of where time reversal gets its meaning: the standard textbook definitions of time reversal can be viewed as fundamentally arising out of the Lie group structure of time, viewed here as the unique occupant of a well-understood pattern of relations (to use the language of Butterfield and Gomes 2020).

---

[14] This is a common reading of Leibniz 1715, p.18, expanded by Grünbaum (1973, §12), Reichenbach (1928, §43), and Winnie (1977), among others. Newton's statement of substantivalism can be found in his Scholium to the Definitions in the second edition of the *Principia* (Newton 1999, pp.54–5) and Clarke's defence of Newton's position in the famous Leibniz–Clarke correspondence (Leibniz and Clarke 2000).

[15] Cf. Gryb and Thébault (2016), Myrvold (2019), Norton (2008b), and Pooley (2013).

[16] Group structure has indeed been proposed as a foundation for the ontology of physics, following work on the ontology of physics by French (2014), Ladyman (1998), French and Ladyman (2003), and Ladyman and Ross (2007). I will not take a position on that debate here; for my position, see Roberts (2011).

[17] Butterfield and Gomes locate the origins of functionalism in the theories of mind due to Armstrong (1968), Lewis (1966, 1972), and Putnam (1960). They conclude that spacetime functionalism should be viewed as "a species of reduction (in particular: reduction of chronogeometry to the physics of matter and radiation)", where reduction is interpreted in a Nagelian sense (Butterfield and Gomes 2020, p.1).

### 2.4.2 The Lie Group of Time Translations

How should one characterise temporal structure? A substantivalist about spacetime might identify one structural property as given by a 'time translation', described as a function from the set of all temporal instants to itself; in contrast, a relationist might take that same structural property to be a function of matter-energy states. I would like to postulate something that both sides should agree on, that the temporal structure at least includes a structure describing time translations, illustrated in Figure 2.6.[18] In many applications this forms a 'Lie group': roughly speaking, a group that is also a differentiable manifold, associated with a collection of coordinate charts called an 'atlas'. More precisely, I mean the following (cf. Landsman 2021; Olver 1993):

> **Definition 2.1** A *Lie group* is a set $G$, a binary operation $+$, and an atlas such that:
>
> (i) (*group*) $G$ is a group with respect to the binary operation; and
> (ii) (*manifold*) $G$ is a smooth, connected, paracompact real manifold with respect to the atlas, and the group operation $(t, t') \mapsto t + t'$ and the group inverse map $t \mapsto -t$ are both smooth maps on $G$.

Postulate (i) (*group*) captures the sense in which time translations can be composed to form new time translations, through an associative map $+ : (t, t') \mapsto t + t'$. The identity element $0 \in G$ can be interpreted as 'no translation'; the existence of an inverse translation $-t$ such that $t - t = -t + t = 0$ can be interpreted as meaning that translations can 'cancel each other out' and produce no translation. The postulate that time translations satisfy (ii) (*manifold*) captures the claim that they form a connected continuum. Of course, time translations might not have these properties, for example in models that treat time as discrete. I see no reason why my proposal cannot be generalised to these contexts, following my comments at the end of the previous section. Moreover, on scales where theories like general relativity and quantum theory are well-confirmed, both (i) and (ii) are ubiquitous.

---

[18] Belot (2007, p.171) proposes an interpretation of time in the spirit of this view.

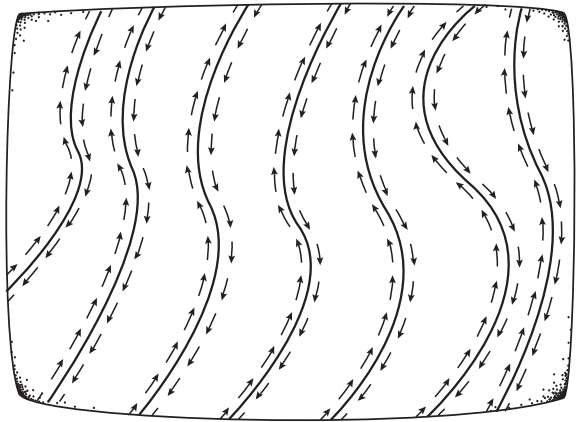

Figure 2.6 A group of time translations along curves in spacetime.

The following postulate is often true of time translations as well, and we will generally adopt it:

(iii) (*one-parameter*) As a manifold, *G* is one-dimensional.

When (iii) (*one-parameter*) is true, one may treat temporal relations like those of a one-dimensional timeline. This is of course not generally true in physics: even time translations in Minkowski spacetime fail to form a one-parameter group, owing to the absence of absolute time. It also fails in branching-time scenarios associated with multiverse and Everettian scenarios.[19] However, neither context poses a problem for us. In a relativistic spacetime with a reference frame associated with a timelike vector field, time translations threading the vector field form a one-parameter Lie group; this is a standard strategy in (3 + 1) approaches to relativity theory. Similarly, in Everettian branching time, the emergent experience of the measuring physicist is still described by the one-parameter Lie group translating along a 'thin red line' through the 'tree', whose branches are the worlds of a multiverse.[20]

Reverting to the language of Section 2.4.1, suppose that the 'pattern of relations' or 'functional role' of time is associated with these properties: time translations are a one-parameter Lie group. Then the occupier of this role is nearly unique: there are only two (connected, inextendible, paracompact in the standard topology) one-parameter Lie groups, the translations of the reals $(\mathbb{R}, +)$ and the group of rotations of the circle $SO(2)$ (Olver 1993, §1.2).

---

[19] Compare e.g. Wallace (2012), as well as Belnap (1992), Belnap, Perloff, and Xu (2001), and Belnap, Müller, and Placek (2021a,b).

[20] I take the 'thin red line' metaphor from Belnap and Green (1994).

If we speak 'locally', then we get uniqueness exactly: these two groups are isomorphic (as 'local' Lie groups) in a neighbourhood of every point. And, speaking globally, since modelling time translations as the circle group is appropriate only for exotic scenarios like closed timelike curves and eternal recurrence, I will set this possibility aside here.[21] Thus, now that we have more clarity on the philosophical underpinnings of this hypothesis, we can say in many applications: *let time translations be characterised by the Lie group* $(\mathbb{R}, +)$.

## 2.5 Time Translation Reversal

### 2.5.1 *Not Just Reversal of a Set of Instants*

Whatever time is, its structure includes time translations. The basic postulate of this section is that time reversal is the reversal of those time translations, in a sense that I will soon make precise. It will soon become clear that lack of clarity about this point is perhaps the single biggest source of confusion in discussions of the meaning of time reversal.

The postulate means, for example, that if $t \mapsto \Sigma_t$ is a sequence of spacelike hypersurfaces ordered by the real line, each representing an instant $t$, then time reversal cannot just be an order reversal of that set, such as,

$$\Sigma_t \mapsto \Sigma_{-t}. \tag{2.1}$$

Focusing on this transformation alone would ignore the rich structure of time. Time is not just a set of instants but also a set of translations describing temporal structure, like the relations between instants, and in particular how one can 'slide' each instant forward or backward by a duration $t$, via the transformation of each instant $\Sigma_{t_0}$ by a map, $\varphi_t : \Sigma_{t_0} \mapsto \Sigma_{t_0+t}$.

This structure can go some way towards clarifying the debate over the meaning of time reversal. The account of Albert and Callender, which I have called the Time Reflection Camp (Section 2.2), begins with a description of how a "complete description" of a physical situation would characterise instants:

There would seem to be two things you want from a description like that:
a. that it be genuinely *instantaneous* ...; and b. that it be *complete* (which is to say, that all the physical facts about the world can be read off from the full temporal set of its descriptions). (Albert 2000, pp.9–10)

---

[21] The analysis that I give in this chapter and the next would suggest the possibility of a theory of 'time travel' dynamical systems, given by unitary or symplectic representations of $SO(2)$; however, I am not aware of any such analysis.

But, there is an ambiguity in what could be meant here by a "full temporal set". A conservative reading might take it to mean a collection of instantaneous descriptions having nothing more than set structure. However, a more liberal interpretation might also include structure on the set of instants, such as the Lie group of time translations. Which one is it? Albert appears to adopt the conservative reading in his presentation of time reversal: "any physical process is necessarily just some infinite sequence $S_I, \ldots, S_F$ of instantaneous states. . . . And what it is for that process to happen backward in just for the sequence $S_F, \ldots, S_I$" (Albert 2000, p.111). But, on a more liberal reading, this might not be all there is to say about it: Albert has, after all, equipped time with an ordering structure to make it a sequence; so, this transformation might plausibly include a transformation of the structure of time translations too. In what follows, I will adopt the latter interpretation.

The more liberal interpretation appears more clearly in the account of Callender, which on the surface has a similar ambiguity:

Relative to a co-ordinisation of spacetime, the time reversal operator takes the objects in spacetime and moves them so that if their old co-ordinates were $t$, their new ones are $-t$, assuming the axis of reflection is the co-ordinate origin. . . . As I understand it here, '$T$' switches the temporal order by switching the sign of $t$. It also switches the sign of *anything logically supervenient* [my emphasis] upon switching the sign of $t$, e.g., the velocity $d\mathbf{x}/dt$. But that is all '$T$' does. (Callender 2000, pp.253–4)

On a conservative interpretation, the structures that are supervenient[22] on time might not include the structural facts like the time translations; but on a more liberal reading, they do. There is also a sense in which the liberal reading is already implicit in Callender's remark: time translation by $t$ is the map $\varphi_t : \Sigma_{t_0} \mapsto \Sigma_{t_0+t}$, where each $\Sigma_{t_0}$ is a spacelike surface. As a result, the transformation $t_0 \mapsto -t_0$ on instants induces a transformation on time translations, $\varphi_t \mapsto \varphi_{-t}$. In particular, if we write 'flip' to denote Callender's time reversal operation, then the succession of transformations 'flip-translate-flip' is equivalent to translating in the opposite direction,[23] as shown in Figure 2.7. In what follows, I will adopt a reading of both Albert and Callender according to which time reversal is a reversal of time translations, $\varphi_t \mapsto \varphi_{-t}$.

---

[22] Supervenience is a philosophical term of art: a set of properties $M$ supervenes on a set of properties $S$ if and only if a difference in $M$-properties implies a difference in $S$-properties (cf. McLaughlin and Bennett 2018). For example, statistical macro-states would usually be said to supervene on the underlying micro-states, but not conversely. As the philosopher Nicholas Rescher once told me: "Supervenience is a gift horse that one should stare in the mouth" (private communication).

[23] More formally, if $T$ is Callender's time reversal transformation defined by $T(\mathbf{x}, t) = (\mathbf{x}, -t)$, then the induced transformation on time translations given by $\varphi_t \mapsto \varphi_t^T := T \circ \varphi_t \circ T^{-1}$ is just equal to $\varphi_t^T = \varphi_{-t}$ for all $t$.

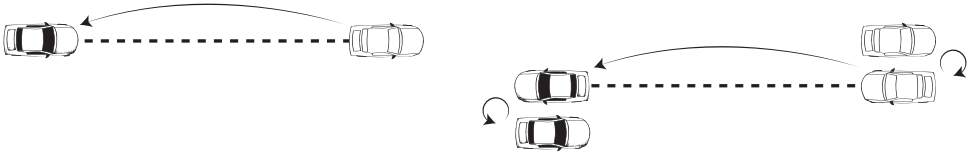

Figure 2.7 'Flip-translate-flip' (right) is equivalent to translating in the reverse direction (left).

Of course, one could insist on ignoring the structural properties of time when defining time reversal. This would be plausible in a naïve theory of time that consists entirely of an unstructured set of sentences or models describing what's happening at each instant $\Sigma_t$. But, philosophers have recently argued convincingly that this naïve perspective on theories is disastrous (cf. Dewar 2022; Halvorson 2012, 2019).

Thus, it is wrong to describe time as nothing more than an unstructured set of instants $\Sigma_t$. When physicists seem to suggest otherwise, I would like to interpret this as shorthand – as when Von Neumann (1932, p.354) says that in a dynamical theory, "there corresponds to the time an ordinary number-parameter $t$", or when Sachs (1987, p.4) says that, "[t]he qualitative meaning of the time variable is that of an ordering parameter in one-to-one correspondence to a sequence of events". Both immediately go on to attribute a great deal of further structure to time, including statements about its group of time translations, and thus they do not really mean that time is an ordered set and nothing more. Similarly, when philosophers like Albert (2000, p.11) write,[24] "any physical process is necessarily just some infinite sequence of states", I would like to read this as shorthand, so as not to deny the rich structural properties of time.

### 2.5.2  *Reversing Structure Too*

Maudlin (2007, §4) has noted an essential connection between time translations and time reversal in his defence of the passage of time:

The passage of time is deeply connected to the problem of the direction of time, or time's arrow. If all one means by a 'direction of time' is an irreducible intrinsic asymmetry in the temporal structure of the universe, then the passage of time

---

[24] See also Callender (2017), who writes of a semiclassical approach to canonical quantum gravity: "The claim may have other problems . . . but one of them is not the identification of time with a metaphysically rich batch of properties. Instead here the role time plays couldn't be more spare. Time is simply identified as that parameter with respect to which the quantum matter fields evolve". (Callender 2017, p.110)

implies a direction of time. But the passage of time connotes more than just an intrinsic asymmetry: not just any asymmetry would produce passing. (Maudlin 2007, p.109)

Let me focus on Maudlin's observation that time translations are 'directed'. Each time translation is indeed associated with another time translation in the 'reverse' direction. From this perspective, it is natural to view time reversal as a transformation relating oppositely-directed time translations. Confusingly, one typically writes $t \mapsto -t$ for both the transformation of time coordinates and the automorphism of a group of time translations; but, unless otherwise specified, I will always use the latter notation and interpret time reversal as an automorphism that reverses time translations.

Why is time reversal associated with the particular map $t \mapsto -t$ and not some other, such as $t \mapsto -t + t_0$? North has pressed this objection,[25] that there may be some freedom in "choice of temporal origin" (North 2008, p.218). But we avoid this issue here by defining time reversal as a map on time translations rather than on time coordinates: $t = 0$ is not a point on a time axis but rather the identity translation by a duration of zero time. No choice of temporal origin is needed to express this.

Perhaps a more challenging concern is: even viewing time reversal as a map on time translations, there are many order reversing transformations besides $t \mapsto -t$. For example, the two transformations of a line depicted in Figure 2.8 are two different order reversing transformations. One could impose the restriction that time reversal $\tau$ be a 'reversal', understood mathematically as an *involution*; this means that applying it twice produces the identity transformation, $\tau^2 =$ identity.[26] But, there are many involutions besides $f : t \mapsto -t$, such as the map $g : t \mapsto (1 - t^3)^{1/3}$; both satisfy $f \circ f = g \circ g = \mathbb{1}$, where $\mathbb{1}$ is the identity transformation $\mathbb{1}(t) = t$.

The inspiration of structuralists and functionalists helps here, as it did in Section 2.4.1. To use the nomenclature of Butterfield and Gomes (2020): although time reversal is a 'problematic' concept, in that it is not obvious how to define it, we can still identify its functional role and hope for a uniqueness result. That would break the underdetermination as to which order reversing involution is time reversal.

---

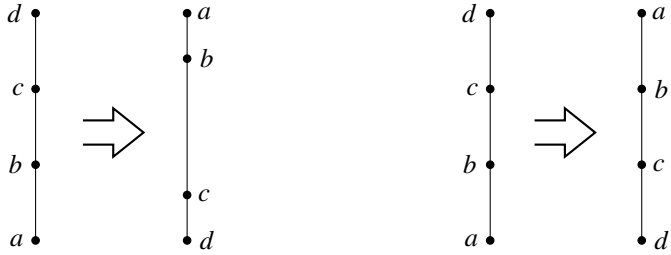

Figure 2.8 Two ways to reverse order.

I propose we make three such observations about its typical functional role in physical theorising:

1. (*Involution*) Time reversal is a 'reversal', in that applying it twice to the group of time translations must produce the identity transformation.
2. (*Automorphism*) Time reversal does nothing more than reverse time translations, and in particular it does not stretch, scale, or discontinuously transform the Lie group structure. This is enforced by requiring it to be an automorphism of $(\mathbb{R}, +)$.
3. (*Non-triviality*) Time reversal does something interesting, in that it is not the identity transformation.

These three properties can be shown to provide a functional definition of time reversal, in that they uniquely determine a transformation of the Lie group $(\mathbb{R}, +)$:

**Proposition 2.1** *If $\tau$ is a non-trivial automorphism of the one-parameter Lie group $(\mathbb{R}, +)$ satisfying $\tau(\tau(t)) = t$, then $\tau(t) = -t$.*

*Proof* First note that every continuous automorphism $\tau$ of $(\mathbb{R}, +)$ satisfies $\tau(t) = \tau(1)t$ for all $t \in \mathbb{R}$, where 1 is the element $t = 1$ and $\tau(1)t$ is ordinary multiplication. This is proved for natural numbers $t$ by induction: the base case is obvious, and if $\tau(t) = \tau(1)t$, then since $\tau$ is an automorphism of $(\mathbb{R}, +)$,

$$\tau(t + 1) = \tau(t) + \tau(1) = \tau(1)t + \tau(1) = \tau(1)(t + 1). \tag{2.2}$$

A second induction proves it for the rationals, and so it holds of the unique continuous extension of $\tau$ to $\mathbb{R}$. Applying $\tau(\tau(t)) = t$, we now get $t = \tau(\tau(t)) = \tau(\tau(1)t) = \tau(1)^2 t$, so $\tau(1) = \pm 1$. Thus $\tau(t) = \tau(1)t = \pm t$, and so by non-triviality we get $\tau(t) = -t$. ∎

I take this fact to settle the meaning of time reversal as a transformation of the Lie group of time translations $(\mathbb{R}, +)$: time reversal is the map $\tau : t \mapsto -t$. Notably, the proof immediately generalises to discrete time translations, such as those given by the group of integers or rationals under addition. However, one should make no mistake – the result that time translations admit a non-trivial automorphism is substantial. Chapter 4 will show that it implies every representation of this structure in state space is time reversal invariant.

### 2.5.3 *Generalising Malament's Picture*

Viewing time reversal as a map on the Lie group of time translations given by $t \mapsto -t$ is a natural generalisation of the account of time reversal given by Malament (2004). Showing this requires the language of relativity theory, although I will soon return to a more general discussion.

Spacetime in relativistic physics is represented by a Lorentzian manifold $(M, g_{ab})$, where $M$ is a smooth real manifold and $g_{ab}$ is a Lorentz-signature metric. A *temporal orientation* is an equivalence class of timelike vector fields defined by the equivalence relation of being co-directed.[27] If a temporal orientation exists, then there are exactly two. Co-directed vector fields 'point' into the same light cone lobe at every point, as illustrated in Figure 2.9, and so it is common practice to treat a temporal orientation as defining a temporal asymmetry (cf. Earman 1974). In describing a temporal orientation, one typically selects a representative vector field from the equivalence class, while keeping in mind that any other in the same equivalence class would serve equally well.

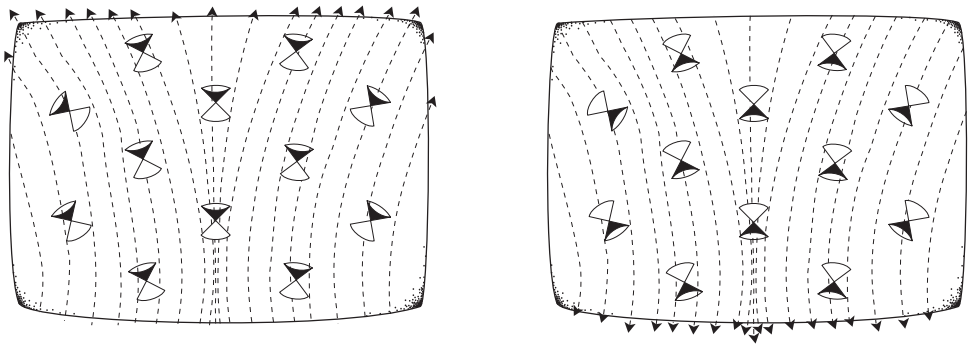

Figure 2.9 Time translations are associated with a temporal orientation.

---

[27] A pair of smooth timelike vector fields $\xi^a$ and $\phi^a$ are *co-directed* if and only if $\xi^a \phi_a > 0$.

Malament's basic proposal is then to view time reversal as the reversal of a temporal orientation:[28]

The time reversal operation is naturally understood as one taking fields on $M$ as determined relative to one temporal orientation to corresponding fields on $M$ as determined relative to the other. (Malament 2004, p.306)

I agree: this is a special case of the account I have given above. A spacetime $(M, g_{ab})$ can be equipped with the structure of a Lie group, given a smooth timelike vector field $\xi^a$: in some local region[29] around a point $p \in M$, there is a one-parameter group of diffeomorphisms 'threading' the vector field that is locally isomorphic to $(\mathbb{R}, +)$ in a neighbourhood of $0 \in \mathbb{R}$. This group is naturally interpreted as a group of 'time translations' in a reference frame associated with the timelike vector field $\xi^a$. The resulting set of integral curves $\gamma(t) := \varphi_t(p)$ has $\xi^a$ as its tangent vector field. Thus, the transformation $t \mapsto -t$ of the group $(\mathbb{R}, +)$ gives rise to a reversal of the tangent vector field $\xi^a \mapsto -\xi^a$ at every point in the region, and vice versa. In short: in the context of Malament's argument, treating time reversal as reversing a local Lie group of time translations is equivalent to treating it as reversing time orientation.

There are two advantages to viewing Malament's proposal as a special case of the reversal of time translations. First, time translation reversal can be expressed in any spacetime theory, and in any interpretation of that theory, so long as we agree on the (e.g. group) structure of time translations. This includes non-standard representations of time translation on a spacetime $(M, g_{ab})$, for example with exotic tachyon motion that cannot be described by a smooth timelike vector field, or in alternative spacetime theories. Second, as we will see in Chapter 3, time translation reversal allows one to more directly interpret time reversal in dynamical theories, when the latter admit a representation of time translations.

Malament views his proposal as a critical response to Albert (2000), and in particular Albert's claim that classical electromagnetism (and most other physical theories) are not time reversal invariant. I agree with this critique and that Albert has not given any reason to think that classical electromagnetism is time reversal violating.

---

[28] A similar but less-detailed proposal is found in Earman (1974, p.25) and Wald (1984, p.60).
[29] This 'local region' requirement allows one to avoid incomplete timelike vector fields, whose integral curves cannot be parametrised by all of $\mathbb{R}$, for example in black hole spacetimes that are (timelike) geodesically incomplete.

However, this conclusion does not necessarily follow from Albert's account if one includes relational or structural properties in his description of time reversal. Recall that the idea motivating both Albert and Callender is that time reversal has the simple form

$$t \mapsto -t. \tag{2.3}$$

As I have suggested, by interpreting this $t$ as a time *translation* rather than a time *coordinate*, we are led to the very same specific definition of time reversal that Malament has formulated in this context. All these accounts of time reversal and time reversal invariance amount to the same thing on the Representation View. To make this precise, let me now illustrate how the group theoretic structure of time reversal arises directly out of the structure of time translations.

## 2.6 Constructing the Group Element

When one postulates a concrete group of spacetime symmetries, such as the Lorentz group or the Galilei group, one often postulates a set of 'discrete' group elements that include time reversal. This group structure turns out to be crucial for the Representation View, in which we will take the meaning of time reversal on state space to be determined by such a group element. So far, we have only been viewing time reversal as an automorphism of a group, and not a group element itself. In this section I will show that, remarkably, the time reversal group element can be recovered directly from the structure of those time translations, or more generally of the group of spacetime symmetries.

In the discussion above, I argued that time translations are an essential part of the structure of time. We have seen in particular that, when that group is $(\mathbb{R}, +)$, the reversal of time translations $t \mapsto -t$ is the unique non-trivial, involutive automorphism. In this section I will show that it is always possible to extend $(\mathbb{R}, +)$ in a way that 'adds in' time reversal as a group element $\tau$, which was hidden all along in the structure of time translations. More generally, when time translations form part of a continuous spacetime symmetry group like the restricted Poincaré group, the same technique can be applied to construct the complete Poincaré group of continuous and discrete symmetries. This will allow us to apply the Representation View and to interpret the instantaneous time reversal operator on state space as the representative of this time reversal element $\tau$.

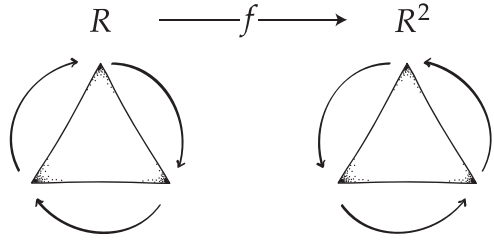

Figure 2.10 The flip automorphism exchanges the rotations of $C_3$, or equivalently reverses the direction of the rotations.

To understand how time translations can be extended to include time reversal, we will need to make use of the elementary group theory of semidirect products.[30]

**Definition 2.2** Given a group $(G, \cdot)$ together with a group $S$ of automorphisms $s : G \to G$, the *semidirect product operation* is the binary operation on $(G \times S)$ given by

$$(g, s)(g', s') := (g \cdot s(g'), s \circ s'). \tag{2.4}$$

The resulting group, denoted $G \rtimes S$, is called the *semidirect product* of $G$ with $S$.

As a warm-up for how we will use semidirect products to understand time reversal, consider the example of the cyclic group of order three, $C_3 = \{1, R, R^2\}$. It describes the symmetries of an equilateral triangle with $1 = R \cdot R^2$. This group has two automorphisms: the identity transformation $\iota$ and the exchange of the two rotations defined by $f : R \mapsto R^2$ and $f : R^2 \mapsto R$. The automorphism $f$ is equivalent to applying a 'flip' to the triangle before each rotation, which is again equivalent to a rotation in the reverse direction (Figure 2.10).

Now, suppose we want to create a larger group, which 'adds in' the flip automorphism $f$ as a group element. A semidirect product allows us to do this: writing the automorphism group as $S = \{\iota, f\}$, we take our new group elements to be the pairs $(g, s)$ for each $g \in C_3$ and $s \in S$ and define our new group operation using Eq. (2.4). This new group now has six elements. Three of them form a subgroup that just implements the original rotations, which we can write in shorthand as

$$1 := (1, \iota) \qquad R := (R, \iota) \qquad R^2 := (R^2, \iota). \tag{2.5}$$

But, we now have three more elements, $(1, f)$, $(R, f)$, and $(R^2, f)$, which apply a flip. Adopting the shorthand that $f := (1, f)$, one can check[31] that this implies that $f R f^{-1} = R^2$ and $f R^2 f^{-1} = R$. That is, conjugation by the 'flip element' $g \mapsto f g f^{-1}$ is equivalent to an application of our automorphism $f$. The result is a larger group describing both the rotational and flip symmetries of the triangle, isomorphic to the symmetric group $S_3$.

As the 'flip' analogy suggests, a semidirect product is exactly what we need to construct a time reversal group element. We begin with the group of time translations $G = (\mathbb{R}, +)$, which we have shown has a unique automorphism that is appropriate for time reversal, $\tau : t \mapsto -t$ (Proposition 2.1). Let $S = \{\iota, \tau\}$ be our automorphism group, with $\iota$ the identity transformation on $\mathbb{R}$. We can now construct a new, larger group with elements of the form $(t, s)$ for each $t \in \mathbb{R}$ and $s \in S$. Adopting additive notation for the binary operation $+$, a semidirect product operation is given by $(t, s)(t', s') := (t + s(t'), s \circ s')$.

As in our example of $C_3$, this new group contains a subgroup of elements of the form $(t, \iota)$ with $t \in \mathbb{R}$, which is isomorphic to the original group of time translations. So, as before, we can write $t := (t, \iota)$ as shorthand for these group elements. But, we now have a further set of elements $(t, \tau)$ with $t \in \mathbb{R}$ that includes time reversal. In particular, we adopt the shorthand $\tau := (0, \tau)$ and refer to this special element as the *time reversal group element*.

To check that this element behaves like time reversal, we can simply observe that it implements the automorphism $t \mapsto -t$ when it acts on time translations by conjugation, $\tau t \tau^{-1} = -t$. That is, by our definitions,

$$\tau t \tau^{-1} = (0, \tau)(t, \iota)(0, \tau^{-1}) = (0, \tau)(t, \tau^{-1}) = (-t, \iota) = -t. \qquad (2.6)$$

The resulting semidirect product group is denoted $(\mathbb{R}, +) \rtimes S$, which I will refer to as the *group of time translations with time reversal*. It has very little structure beyond the structure of time translations, except for the inclusion of a time reversal group element implementing the automorphism $t \mapsto -t$. In this sense, the group of time translations already contains the structure of a time reversal group element, through the extension of those time translations to one with time reversal as well.

Although I have restricted attention here to the simple time translation group $G = (\mathbb{R}, +)$, this procedure for introducing a time reversal element is much more general. Suppose we begin with the group $G = \mathcal{P}_+^\uparrow$ of rigid translations, rotations, and Lorentz boosts on Minkowski spacetime, called the (restricted) Poincaré group. This group has four automorphisms, $D = \{\iota, p, \tau, p\tau\}$, representing the identity, spatial translation reversal, time translation reversal, and the composition of the last two. Just as above, one can extend $P_+^\uparrow$ to a group that 'adds in' these transformations as group elements using a semidirect product in exactly the same way. The result is what is commonly known as the (complete) Poincaré group $\mathcal{P} := \mathcal{P}_+^\uparrow \rtimes D$.

---

[31] By our definitions, $f R f^{-1} = (1, f)(\iota, R)(1, f^{-1}) = (1, f)(f(R), f^{-1}) = (f(R), \iota) = (R^2, \iota) = R^2$. The remaining arguments are similar.

This procedure is indeed a standard technique in the construction of the Poincaré group (cf. Varadarajan 2007, §IX.2). It is perhaps not so widely studied. However, the discussion here shows that it provides a useful technique for establishing the group structure of discrete symmetries. We will return to its broader applications in Chapter 8, and especially Section 8.3, on the meaning of CPT symmetry.

## 2.7 The Representation View of Time Reversal

The last two sections show that the structural properties of time, and in particular the time translations, give rise to a group structure for time translations with time reversal. This provides a foundation for the meaning of time reversal on spacetime. To carry its meaning over to state space, we now apply the Representation View, set out in Section 2.3: a symmetry of state space is called 'time reversal' only if it is a representative of the time reversal group element. That representative, I claim, is just what is commonly referred to as $T$, the 'instantaneous time reversal transformation'. Remarkably, the Representation View allows one to more or less determine the meaning of this transformation. In this section I will outline how that works. Chapter 3 is then devoted to deriving its meaning in a variety of physical state spaces.

In the debate between the two camps, Albert (2000) and Callender (2000) pointed out that time reversal should be characterised by the simple property that $\tau : t \mapsto -t$. We have seen how that is true, and indeed that time reversal is an element of the spacetime symmetry group, when each $t$ is interpreted as a time translation. However, this does not imply that time reversal on the state space of a dynamical theory just 'reverses the little-$t$' parameter in each trajectory. A dynamical theory is hardly deserving of the name 'dynamical' unless it admits a representation of time translations. Once such a representation is chosen, 'time reversal' on state space simply refers to the representation of $\tau$, the time reversal group element.[32] Let me set out this aspect of the Representation View in more formal terms, so that we can apply it in physics.

---

[32] Struyve (2020) has made a similar observation about this very debate, diagnosing the disagreement about the meaning of time reversal as arising from different choices of 'ontology'. I agree, insofar as each 'ontology' is associated with a representation.

**Definition 2.3** Let $G$ contain a substructure representing time translations (typically a Lie group) as well as a time reversal element $\tau \in G$. Let $\mathcal{A}$ be state space structure (typically an object in some category), with $\mathrm{Aut}(\mathcal{A})$ its automorphism group. A *representation of $G$* is a homomorphism $\phi : G \to \mathrm{Aut}(\mathcal{A})$, whose restriction to the substructure of time translations is also a homomorphism. The *instantaneous time reversal transformation $T$* is the image of time reversal in this representation, $T := \phi(\tau)$.

I have intentionally tried to formulate this definition with a high degree of generality. It is indeed in the same spirit as a recent approach to algebraic quantum field theory, viewed as an embedding of a category of relativistic spacetimes into a category of algebras.[33] More generally, the theorist might take time translations to be associated with a discrete group, or even a semigroup (without inverses). Or, in the context of general relativistic spacetimes, we might restrict time translations to local regions by requiring $G$ to be a *local Lie group*; this restricts the definition of group operations to local regions (Olver 1993, p.18). On the Representation View, this perspective can be applied whenever our spacetime structure and our state space structure are defined by a functor between categories.

However, the application of Definition 2.3 in the simplest cases of dynamical systems is straightforward: suppose time translations are given by the Lie group $(\mathbb{R}, +)$ and that we construct the group with time reversal $G = (\mathbb{R}, +) \rtimes \{\iota, \tau\}$ as above, which satisfies $\tau t \tau^{-1} = -t$. Then we can take the symmetries of any state space, like a Hilbert space or classical configuration space, and find a representation of $G$ given by a group homomorphism $\phi(G)$ amongst those symmetries, and where the restriction of $\phi$ to the time translations $(\mathbb{R}, +)$ is a Lie group homomorphism. We can write time translations on state space as $\phi_t$ for each $t \in \mathbb{R}$, commonly called a 'phase flow'. Time reversal on state space is then the element $T := \phi_\tau$. Since a representation is a homomorphism, the fact that $\tau t \tau^{-1} = -t$ carries over to state space too, as the statement that for $T$ reverses the direction of each time translation:[34]

$$T \phi_t T^{-1} = \phi_{\tau t \tau^{-1}} = \phi_{-t}. \tag{2.7}$$

---

[33] Cf. Brunetti, Fredenhagen, and Verch (2003) and Rédei (2014).

[34] Experts may recognise this statement as indicative of time reversal invariance, and indeed it is. But this does not mean that time reversal invariance is assured: as we will see in Chapter 4, a time reversal violating system is one in which a representation of the group of time translations with time reversal does not exist.

Formally speaking, this transformation $T$ is just the standard instantaneous time reversal operator; it is 'instantaneous' only in the sense that it is represented by a transformation on (instantaneous) state space. But, this by itself is no more paradoxical than the fact that a time translation $\phi_t : \mathcal{S} \to \mathcal{S}$ by $t$ is a transformation on instantaneous state space, too. There is no special philosophical problem with either one: both represent structural facts about time and do not need to be interpreted as 'turning around' a spatial slice at an instant.

The objections of Albert and Callender about what it means to 'reverse the arrow of time at an instant' are thus dissolved. And, when Earman (2002b) does propose that time reversal must "turn around" the phase relations of a wave packet at an instant, we can always view this as shorthand for certain facts about the representative of the time reversal group element. But, the ultimate interpretation of $T$ is as we have said above: it is the representative of the time reversal group element, which reverses the direction of time translations as in Eq. (2.7). We will see more about how such 'shorthand' facts arise in Chapter 3.

## 2.8 Summary

Time reversal faces a problem of definability: to justify its meaning in a way that makes it relevant to the direction of time. This gives rise to two camps of interpretation. The Instantaneous Camp introduces a richly structured time reversal operator $T$ on state space, while the Time Reflection Camp favours a simpler picture in which time reversal just 'reverses time', $\tau : t \mapsto -t$.

I have argued that these statements are not incompatible. To clarify how this is the case, we must separate the concept of time into 'spacetime' and 'state space' components, where the latter gets its meaning from the fact that it is a structure-preserving copy of the former. This is the Representation View, a powerful approach to interpreting spacetime symmetries that we will use throughout this book. As we have seen above, the meaning of time reversal is essentially encoded in the structure of time, and in particular the time translations. The richly structured time reversal operator $T$ is not a mystery in this view: it is just the representative of time reversal $\tau$ on a state space representation.

Thus, Albert and Callender are right that time reversal is a simple concept at its foundation: it simply reverses time translations. But, just as Earman has proposed, an instantaneous time reversal transformation is also a natural part of state space: it arises whenever the dynamics of a theory is associated with a representation of time translations. Of course, much more would need to be said about what exactly the time reversal operator $T$ winds up looking like, theory by theory. That is the subject of Chapter 3.

# 3

# Time Reversal in Physical Theory

---

*Précis. The meaning of time reversal on state space is determined by a representation of time translations.*

---

When the time reversal transformation $T$ appears on the state space of a physical theory, it is usually required to do a number of idiosyncratic things, such as those listed in Figure 3.1. For example, it conjugates wavefunctions in quantum mechanics, and it reverses magnetic fields in electromagnetism. To explain this, its advocates often refer to it as 'motion reversal' and appeal to a wide variety of physical facts about how this affects instantaneous states. This led Albert (2000, p.18) to remark that in the textbooks, "for one physical situation to be the time reverse of another is (not surprisingly!) an obscure and difficult business". Even Earman (2002b) admits that the going gets tough:

> I do not mean to suggest by the above examples that fixing the properties of the time reversal operation is always such an easy or straightforward matter. It is not, and in some instances the quest may not end in any clear answer. (Earman 2002b, p.249)

This chapter aims to make the meaning of time reversal in state space a more straightforward matter. We now have the results of Chapter 2 in hand: on the Representation View (Section 2.3), a dynamical theory is by definition a representation of time translations on state space. We have seen how to construct a time reversal group element $\tau : t \mapsto -t$ that reverses those time translations (Section 2.6); and, when the representation of time translations extends to this larger structure, then we can immediately understand the time reversal transformation $T$ to be the image of $\tau$ in state space (Section 2.7). So, the meaning of time reversal is not so haphazard after all: it is determined by a representation of time translations. What remains is to verify that the familiar time reversal operator in textbooks is what we recover by this procedure.

| Reversed | | Preserved | |
|---|---|---|---|
| Momentum: | $p \mapsto -p$ | Position: | $q \mapsto q$ |
| Magnetic Field: | $B \mapsto -B$ | Electric Field: | $E \mapsto E$ |
| Spin: | $\sigma \mapsto -\sigma$ | Kinetic Energy: | $p^2/2m \mapsto p^2/2m$ |
| Position wavefunction: | $\psi(x) \mapsto \psi(x)^*$ | Transition probability: | $|\langle \psi, \phi \rangle|^2 \mapsto |\langle \psi, \phi \rangle|^2$ |

Figure 3.1 Some effects of the time reversal operator $T$ on state space.

This chapter aims to make good on that, by showing how various expressions of time reversal on a physical state space now follow automatically. I will make this argument case by case, for some common frameworks for dynamical systems in physics. I hope that by illustrating the general technique, it will be clear how to apply it in other physical theories too. In each case, our task will be to determine two things:

1. the *general* character of the time reversal group element in a representation;
2. its *specific* character when it is required to reverse time and 'do nothing else'.

The second task arises because, in general, there are a great number of ways to represent time reversal on state space: we can reverse time and space; we can reverse time and rotate 180 degrees; we can reverse time and exchange matter and antimatter; and so on. Recognising this will be important for our development of the CPT operator in Chapter 8. But, the specific task for us now will be to distinguish 'bare time reversal' from the many other time-reversing transformations.

Our focus in this chapter will be on state space representations. Nearly every theory in modern physics has an expression of this kind; in particular, the framework for analytic mechanics that we discuss in Section 3.3 is so robust that it can be viewed as including both general relativity and quantum theory as special cases.[1] We begin Section 3.1 with some general remarks about state space representations of time translations, which will be applied in all the theories to follow. This includes precise statements of two physical postulates about the experience of time's passage in general: that it is a symmetry of local physics, and that the energy is half-bounded. We then turn to deriving the meaning of time reversal in various physical theories: Newtonian mechanics (Section 3.2), analytic mechanics (Section 3.3), and quantum theories (Section 3.4).

---

[1] Wald (1984, Appendix E) gives a classic introduction to general relativity in the Hamiltonian and Lagrangian frameworks. The fact that quantum theory can be viewed as a special case of analytic mechanics is less well-known, but was pointed out independently by Ashtekar and Schilling (1999), Gibbons (1992), Kibble (1979).

## 3.1  State Space Representations

### *3.1.1  State Space*

State spaces in physics are used to represent possible states of affairs in nature, like the possible locations of the planets in our solar system (Figure 3.2). However, when one spends a little time getting to know state spaces, some patterns among them emerge: many share common structure; and some of them seem perennially useful, while others do not. What makes a structure reasonable, appropriate, or fruitful for use as a state space?

Philosophers of physics have made some comments in this regard, such as the proposal of Albert (2000, p.9) that a state is (at least) a "genuinely instantaneous" and "complete" description of physical facts.[2] Butterfield (2006b,d), inspired by Lewis (1986), characterises this view as one in which states are "temporally intrinsic" facts: when obtaining at a moment, a state does not by itself imply any contingent facts about other times. Thus, physicists sometimes refer to configuration space or phase space in mechanics as an 'instantaneous state space'. However, as Butterfield (2006a,b, 2011) goes on to argue at length, even the state space of classical mechanics cannot possibly be just that; for example, "[m]echanics needs of course to refer to the instantaneous velocity or momentum of a body; and this is temporally extrinsic to the instant in question" (Butterfield 2006a, p.193). I agree. The state spaces to be considered in this chapter are highly structured objects – Hilbert spaces, phase spaces, jet bundles, and the like – and all this structure is needed to understand a symmetry transformation like time reversal.[3]

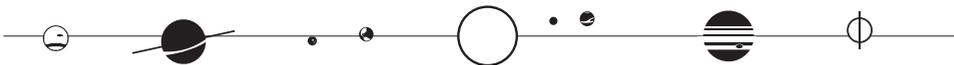

Figure 3.2  A state in the space of the planetary locations.

---

[2] See Section 2.5.1.
[3] Barrett (2018b, 2019, 2020a, 2020b), Dewar (2022, Part IV), Halvorson (2012, 2013, 2019), and Weatherall (2016) have argued on similar grounds that a physical theory can be viewed as a category. I have reservations about some of the remarks in this literature (cf. Roberts 2020), but agree with the general structuralist spirit.

This might seem to contradict some remarks of Albert (2000) about dynamical theories, but I think that it need not. Albert (2000, pp.9–10) distinguishes between states representing complete descriptions of the world at an instant, and the *dynamical conditions* needed to make use of "the full predictive resources" of the laws of physics. In Newtonian mechanics, he takes the states to be particle positions in $\mathbb{R}^{3n}$, while the dynamical conditions also include velocities. I have no problem with this, but will propose a liberal reading of Albert: nothing about his remarks preclude state space from having a great deal of structure beyond this. It can be a manifold, like a tangent bundle of vectors, or be equipped with a symplectic form or a Euclidean metric, among other things.

In the remainder of this section, I will identify two key structural facts associated with general state spaces: that time translations can be viewed as symmetries in isolated systems, and that energy is bounded from below but not from above. This will set the stage for the analysis of time reversal on particular examples of state spaces in the remainder of the chapter.

### *3.1.2 Time Translations Are Automorphisms*

A representation is a homomorphism from a symmetry structure to the automorphisms of a state space. Each state space structure has its own standard of automorphism or 'structure preserving map': for example, a differentiable manifold has diffeomorphisms, while a vector space has linear maps. So, by choosing a state space structure, and therefore a notion of state space automorphisms, we constrain what sorts of representations are possible.

Representing the group of time translations amongst the transformations of a state space is what justifies referring to the representing state space map as 'time evolution' as opposed to something else; this was the thesis of Section 2.3. And, not just any transformations: time translations should be represented among the state space maps that are *automorphisms*, at least for isolated systems. This is because most physical theories are built to capture the repeatability of local experiments, in the following sense.

We seek theories that can be supported by experimental evidence, but also – where possible – theories for which that experimental evidence can be supplied again at a later time. Local dynamical theories are generally designed in this way: the modelling of time evolution is compatible with setting up an isolated experiment today, collecting the results, and then confirming those same results tomorrow when a structurally equivalent experiment is repeated. In more precise terms, a time translation is an automorphism of a dynamical theory. Of course, whether or not this kind of 'homogeneity in time' accurately describes reality is a matter of experience. But as Jauch points out, it turns out to be central to our experience:

> Whether there are systems which are, in this sense, homogeneous in time is of course a matter of experience, and it is indeed one of the fundamental experiences about the physical world that this is the case. (Jauch 1968, p.152)

Thus, we represent time translations amongst the automorphisms of a state space, as postulated by the Representation View. This by itself does not guarantee that the representation can be extended to include time reversal; as we will see in Chapter 4, a representation of time reversal may still fail to arise, giving rise to the failure of temporal symmetry. Our study here will thus be predicated on the assumption that a representation of time reversal exists.

### 3.1.3 Half-Bounded-Energy Representations

A second structural fact about typical representations of time translations, in stark contrast to representations of spatial translations, is that they are associated with a conserved quantity called 'energy' that is bounded from below. To get a more precise sense of what this means, recall that a Lie group (Definition 2.1) is a group with a manifold structure. The continuous transformations in a Lie group, which describe familiar symmetries like rotations or time translations, turn out to be 'locally generated' by an object called a Lie algebra:

> **Definition 3.1** The *Lie algebra* $\mathfrak{g}$ of a Lie group $G$ is the algebra of right-invariant vector fields on $G$, where a 'right-invariant' vector field $X$ is defined by the condition that for all $g, h \in G$, if $\rho_g : G \to G$ is the right-multiplication map on group elements defined by $h \mapsto h \cdot g$, then $d\rho_g(X)|_h = X|_{\rho_g(h)} = X|_{h \cdot g}$.

In more picturesque terms (Figure 3.3), an element of this Lie algebra is any vector field $X$ on the Lie group manifold $G$ with the property that group multiplication 'traces along' the vector field. As this picture suggests, elements of a Lie algebra stand in one-to-one correspondence with the one-parameter subgroups of a Lie group (cf. Olver 1993, Proposition 1.48). In particular, a one-parameter group of time translations $(\mathbb{R}, +)$ is associated with a single Lie algebra element $X$, called the *generator* of the one-parameter group.

A fundamental result in the representation theory of Lie groups is that every homomorphism between Lie groups induces a unique homomor-phism between the Lie algebras of those groups.[4] Thus, every state space representation of the Lie group defines a unique representation of its Lie algebra. For example, in a representation of spacetime symmetries, this group will include a generator for each one-parameter group of time translations $t \mapsto \phi_t$. When this generator $X$ can be viewed as the gradient of a smooth function $h$, then this function $h$ is called the 'Hamiltonian'. Since the Hamiltonian generates time translations, its value is also 'preserved' by time translations, and so in physical terms it can be used to define the conserved quantity known as *energy*. Similarly, each conserved function associated with a one-parameter group of spatial translations is called *three-momentum* or *momentum*, and traditionally denoted $p$.

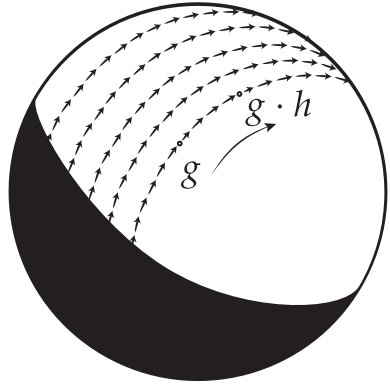

Figure 3.3  A Lie algebra element.

   Although conserved functions $h$ for time translations and $p$ for spatial translations are formally similar at the level of Lie groups and algebras, experience shows that they display a fundamental difference when interpreted as physical quantities. Namely, the possible values of momentum are unbounded above and below, while the possible values of energy are *half-bounded*: they are bounded from below and unbounded from above.[5] This is one of the truly remarkable differences between space and time, although it has received little attention from philosophers. And, it appears to represent an elementary fact about the local structure of our world in everything from the simple harmonic oscillator to quantum electrodynamics. Energy is typically unbounded from above, owing to the fact that velocity can be boosted arbitrarily close to the speed of light, and its lower bound is usually interpreted as defining a 'stable ground state'. The common wisdom about why this lower bound exists has been captured by Malament (1996):

> If it failed, the particle could serve as an infinite energy source (the likes of which we just do not seem to find in nature). Think about it this way. We could first tap the particle to run all the lights in Canada for a week. To be sure, in the process of doing so, we would lower its energy state. Then we could run all the lights for a second week, and lower the energy state of the particle still further. And so on. If the particle had no finite ground state, this process could continue forever. There would never come a stage at which we had extracted all available energy. (Malament 1996, p.5)

[4]  Cf. Hall (2003, Theorem 2.21) or Landsman (2017, Theorem 5.42).
[5]  A well-known exception in the history of physics is the unbounded 'negative energy sea' in Dirac's hole theory of electrodynamics (Dirac 1930). This theory encountered irreparable problems, and half-bounded energy was restored in the modern Fock space formulation of quantum electrodynamics (cf. Duncan 2012, §2.1).

Whatever the reason for this fact, we will use it as the basis for assuming that in a representation of time translations, the generator $h$ must be half-bounded.

This completes our general discussion of state space and representation theory. In the next sections, we will apply it to a number of different theories. In each case, the procedure will be similar:

1. Represent time translations $(\mathbb{R}, +)$ amongst the automorphisms on a state space.
2. Interpret $\tau : t \mapsto -t$ on time translations as time reversal.
3. Extend the representation to include $\tau$ as a group element, interpreted in the representation as an instantaneous time reversal operator $T$.
4. Where possible, use the half-bounded energy constraint on the Hamiltonian $h$ to determine the general character of $T$.
5. When further group structure is available, use it to try to determine the unique definition of $T$.

I hope the result will be a novel, unified approach to the meaning of time reversal: not 'motion reversal', nor a bag of tricks for reversing instants, but as the reversal of time translations in each physical theory where it can be applied.

## 3.2  Newtonian Time Reversal

### 3.2.1  Why Time Translations Are Needed

The world according to atomists like Democritus, Bošković, and Boyle consists of a finite number of indistinguishable particles "variously configured and moved", whose only properties are their locations in smooth three-dimensional Euclidean space (cf. Boyle 1772, p.355). When this is the case, the locations of $n$ particles can be represented by the $C^\infty$ (I will usually say 'smooth') real manifold with Euclidean metric $(\mathbb{R}^{3n}, \cdot)$. Each point $x \in \mathbb{R}^{3n}$ represents a possible particle configuration at an instant, and the motion of the particles through space is represented by a smooth curve $x : \mathbb{R} \to \mathbb{R}^{3n}$ – in my notation, $x = (x_1, \ldots, x_{3n})$ is a coordinate in a manifold representing total configuration, not a spatial coordinate – with $\mathbb{R}$ interpreted as the time axis. So, we write $x(t)$ to represent the configuration $x \in \mathbb{R}^{3n}$ of the system at given time $t \in \mathbb{R}$.

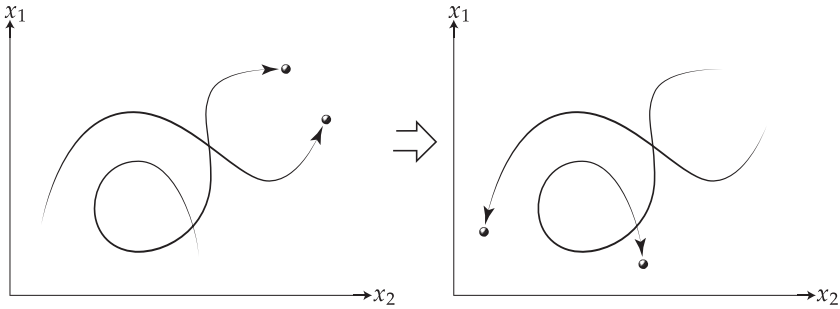

Figure 3.4 The 'folk' view of Newtonian time reversal without time translations.

In this general context, Callender (2000) has set out the common 'folk' view about time reversal:

Relative to a co-ordinisation of spacetime, the time reversal operator takes the objects in spacetime and moves them so that if their old co-ordinates were $t$, their new ones are $-t$, assuming the axis of reflection is the co-ordinate origin. (Callender 2000, p.253)

In other words, the time reverse of a smooth curve $x(t)$ is $x(-t)$, as in Figure 3.4.

I agree with this description of Newtonian time reversal, at least when I put on a substantivalist hat.[6] But it should be viewed as incomplete. We often say, 'Let $x(t)$ represent the trajectories of particles over time'. However, as we have seen in Section 2.5.2, a more complete analysis of time requires including its structural properties: not just the time coordinates, but the time *translations*.

An adequate representation of time in Newtonian mechanics is not possible without this extra structure. In particular, $(\mathbb{R}^{3n}, \cdot)$ cannot be a complete Newtonian state space, since its automorphisms include only the (spatial) transformations of rigid rotation, translation, and reflection. Time translations for this structure cannot even describe an elliptical orbit. So, beware that one's intuitions here can fail: the map $x(t) \mapsto x(-t)$ on smooth curves through a Euclidean manifold is not a complete description of time reversal. It is incomplete because it includes no representation of Newtonian time translations, and so it has no way to say that time translations are reversed. So, let us begin by discussing a more complete state space for Newtonian mechanics.

---

[6] Newtonian substantivalism is traditionally justified using examples like Newton's bucket (Newton 1999, Scholium to the Definitions, pp.58–9). Mach's principle is a standard response (Earman 1989, Chapter 4), but this remains an active research area: see Rynasiewicz (1995a,b, 2014), and the novel recent alternative of Gomes and Gryb (2020) using techniques from Kaluza–Klein theory.

### 3.2.2  *Newtonian Time Translations*

Begin at the level of spacetime. Time translations are usually introduced by extending the Euclidean manifold to include a temporal dimension $\mathbb{R}$, together with some further spacetime structure. Although there are different ways to do this, most[7] admit a group of time translation symmetries isomorphic to $(\mathbb{R}, +)$. So, let me follow my announced practice of postulating the existence of a time translation group, without taking a position on spacetime structure.

To introduce a representation of these time translations, we now need to say what time translation invariance means in the context of Newtonian mechanics. This introduces Newton's second law. However, a rigorous treatment requires some technicalities regarding the symmetries of a differential equation that are not so often discussed. – *Oh dear: Must we complicate a theory as simple as Newtonian mechanics?* – I'm afraid that someone who comes to Newtonian mechanics for its simplicity was misinformed.[8] The situation is as Bill Burke once remarked, and in the spirit of the way Putnam (1962) describes kinetic energy:

Be careful of the naïve view that a physical law is a mathematical relation between previously defined quantities. The situation is, rather, that a certain mathematical structure represents a given physical structure. Thus Newtonian mechanics does not assert that $F = ma$, with $F$, $m$, and $a$ separately defined. Rather, it asserts that the structure of second-order differential equations applies to the motion of masses. (Burke 1985, p.37)

---

[7] This includes Leibnizian, Maxwellian, Galilean, and Newtonian spacetime (Earman 1989, Chapter 2). An exception is Machian spacetime, associated with the work of Barbour (1974); but, insofar as this approach demotes time translations to a 'coordinate relabelling', the question of the meaning of time reversal does not appear to arise.

[8] See Butterfield (2006a,b,c, 2007), Wallace (2022), and Wilson (2013) for some subtleties in the philosophical foundations of classical mechanics. The call-to-arms of Butterfield (2006a,b, 2011) is particularly apt for this discussion: he argues that the properties of classical physics cannot be viewed as defined only at points.

The mathematical structure required for Newtonian mechanics is the following. Observe that, given a smooth curve $x(t)$ through $\mathbb{R}^{3n}$, each point on the curve is associated with a set of values $(x, \dot{x}, \ddot{x}, \dots)$ corresponding to its successively higher derivatives with respect to $t$. These values are elements of the *jet space* of $\mathbb{R}^{3n}$, associated with the smooth manifold $\mathbb{R}^{3n} \times \mathbb{R}^{3n} \times \cdots$ (see Olver 1993; Saunders 1989). In Newtonian mechanics, attention is restricted to the 'two-jet' associated with $M = \mathbb{R}^{3n} \times \mathbb{R}^{3n} \times \mathbb{R}^{3n}$, whose elements have the form $s = (x, \dot{x}, \ddot{x})$. Given a smooth function $F : \mathbb{R} \times \mathbb{R}^{3n} \times \mathbb{R}^{3n} \to \mathbb{R}^{3n}$ of only $(x, \dot{x})$, called a 'force', we say that a *solution to Newton's equation* is a smooth curve $x : \mathbb{R} \to M$ satisfying Newton's second law,[9]

$$\ddot{x} = F(t, x, \dot{x}), \tag{3.1}$$

at each point along the curve. Where are the masses? To simplify the notation, I interpret them as assigned by the function $F$. So, for a harmonic oscillator, $F(x) = -(k/m)x$, where $k$ is called the 'spring constant'.

Viewing the solution $x(t)$ as a set of points $(t, s) \in \mathbb{R} \times M$, it is also convenient to define a function $f : \mathbb{R} \times M \to \mathbb{R}^{3n}$ by

$$f(t, x, \dot{x}, \ddot{x}) := \ddot{x} - F(t, x, \dot{x}). \tag{3.2}$$

Then each point $(t, s) = (t, x, \dot{x}, \ddot{x})$ on a curve is in the solution space if and only if $(t, s) \in \ker f$, meaning that $f(t, s) = \mathbf{0}$. The space of all such curves is called the *solution space* $\mathcal{S}_F$.

On reflection, it should be clear that Newton's equation in this form implies almost nothing about the physical world: without specifying the functional form of $F$, it is just the statement that motion is guided by a second-order differential equation. However, one of the few things it does imply is that the state space of Newtonian mechanics cannot possibly be $\mathbb{R}^3$, nor even $\mathbb{R}^{3n}$. Without an expression of higher derivatives, these spaces do not have enough structure to support Newton's equation. The state space of Newtonian mechanics is rather the state space of a typical differential equation, which has the richer structure of the jet space $\mathbb{R} \times M$. Careful treatments of the philosophy of Newtonian mechanics like that of Wallace (2022, §2) do correctly represent the structure of Newtonian state space, but this is not often discussed in physics or philosophy,[10] and it is a crucial observation for us here.

---

[9] The restriction to smooth force functions of at most first derivatives ensures the existence and uniqueness of solutions (see e.g. Arnol'd 1992, §7.2). This rules out pathological examples of local indeterminism like 'Norton's dome', although the latter is (rightly) the subject of much philosophical debate (cf. Earman 1986; Fletcher 2012; Gyenis 2013; Malament 2008; Norton 2008a; Wilson 2009).

We can now develop a more appropriate understanding of 'symmetry' in Newtonian mechanics: we say that an *automorphism* of a Newtonian system with solution space $\mathcal{S}_F$ is a diffeomorphism $\phi$ on the state space of higher derivatives,

$$\phi : \mathbb{R} \times M \to \mathbb{R} \times M, \tag{3.3}$$

such that $\phi$ preserves the kernel of $f$: if $(t, s_t) \in \ker f$, then $\phi(t, s_t) \in \ker f$. This does what one would expect an automorphism of a solution space to do: if $x(t)$ is a solution to Newton's equation, then so is the curve produced by the transformation $\phi$. As Belot (2013, §3) has emphasised, a Newtonian symmetry is not just a bijection on curves $x(t)$, but a diffeomorphism on $\mathbb{R} \times M$; indeed, a bijection alone does not necessarily preserve all the local structure of a differential equation, and if one exists that does, then it leads to a pathological perspective on symmetries.

This discussion should make clear that the structure of Newtonian state space is not a trivial matter. However, once it is in place, a *representation of time translations* in Newtonian mechanics can be concisely defined: it is a homomorphism $t \mapsto \phi_t$ from the time translations $(\mathbb{R}, +)$ to the automorphism group of $\mathcal{S}_F$, such that for each point $(t_0, s_0) \in \mathbb{R} \times M$ and for each time translation $t$,

$$\phi_t : (t_0, s_0) \mapsto (t_0 + t, s_t), \tag{3.4}$$

for some $s_t \in M$. Writing $\phi_t(s_0) = s_t$ for the restriction of this map to $M$, a representation of $(\mathbb{R}, +)$ thus defines a smooth curve $t \mapsto x(t)$ associated with each initial point $(t_0, s_0)$, which is given by $x(t_0 + t) := \phi_t(s_0)$. Since an automorphism $\phi_t$ of the solution space $\mathcal{S}_F$ by definition preserves $\ker f$, this implies that $x(t)$ is a solution to Newton's equation. As one would expect, each time translation by a duration $t$ maps a point on a curve $x(t_0)$ to a different point on the same curve, $x(t_0) \mapsto x(t_0 + t)$.

Let me emphasise that time translations are represented here by symmetries, as discussed in Section 3.1.2. If the map defined by Eq. (3.4) were not an automorphism of $\mathcal{S}_F$, then it would not provide a representation. This is standard practice in Newtonian mechanics: in locally isolated systems without any hidden degrees of freedom, forces do not seem to display any time dependence, and so time translations are symmetries of the theory; or, in the language of this section, time translations are automorphisms of the solution space.[11] We will return to this assumption in Chapter 4, since it is related to the question of time reversal symmetry and symmetry violation.

---

[10] Recognising the fact that state space must have higher derivatives has implications for much of the philosophical literature on the dimensionality of state space following Albert (1996), such as the contributions in Ney and Albert (2013), which I leave as an invitation to the reader.

### 3.2.3 The Newtonian Time Reversal Operator

Having clarified the structure of Newtonian time translations, we can simply follow the procedure of Chapter 2 to construct the time reversal transformation on Newtonian state space. Here is the fundamental idea, which we will repeat throughout this chapter.

Given a representation $t \mapsto \phi_t$ of the time translation group $(\mathbb{R}, +)$, recall that in Section 2.5.2, we saw that time reversal can be uniquely determined to transform this group as $\tau : t \mapsto -t$. And, in Section 2.6, we showed that it can be 'added in' as a group element to an extension $G$ of those time translations, which satisfies $\tau t \tau^{-1} = -t$. If our representation of $(\mathbb{R}, +)$ can be extended to a representation of $G$, then there will be a state space transformation $T := \phi_\tau$ corresponding to time reversal that reverses each time translation, $T \phi_t T^{-1} = \phi_{-t}$ for all $t \in \mathbb{R}$. There is nothing mysterious about this transformation: it is just the representative of a temporal reflection on state space.

In the previous section, we saw how time translations give rise to curves in Newtonian mechanics. So, by reversing the time translations, time reversal in Newtonian mechanics gives rise to a transformation on curves, whereby the curve $x(t_0 + t) = \phi_t(s_0)$ is replaced with a curve $x^T(t_0 + t) = \phi_{-t}(s_0) = x(t_0 - t)$. In other words, time reversal transforms curves through Newtonian state space as

$$x(t_0 + t) \mapsto x(t_0 - t). \tag{3.5}$$

This makes clear how the 'folk' account of time reversal discussed by Callender (2000) arises! If we parametrise $x(t)$ so that $t_0 = 0$, then we get exactly the transformation $x(t) \mapsto x(-t)$. But, it arises as a reflection of time translations rather than of time coordinates, and so there is no longer any question about 'which coordinate origin' to reflect about of the kind that concerns North (2008).[12] By reversing the whole structure of time, including time translations, that problem is dissolved.

---

[11] Of course, damped systems are often treated heuristically as if they do not satisfy this requirement, but this is generally because they are in reality not isolated; see Section 7.2.

[12] See Section 2.5.2.

Now we can also see the origin of the time reversal transformation $T := \phi_\tau$ on state space, debated by the two camps of Section 2.2: it is just the image of the 'true' time reversal transformation $\tau : t \mapsto -t$ on state space. When a representation of it exists, it is guaranteed to reverse time translations when it acts on them by conjugation:

$$T\phi_t T^{-1} = \phi_{-t}. \qquad (3.6)$$

For example, suppose we adopt a time-independent force $F = F(x, \dot{x})$, which has the property of being independent of the sign of velocity, $F(x, -\dot{x}) = F(x, \dot{x})$. This might be the force $F = (k/m)x$ for the harmonic oscillator, or $F = kq/x$ for a Coulomb force. Let me drop the time parameter $t$ from state space, since this force does not depend on time. Then a representation of the time reversal operator is defined by transforming each state $s = (x, \dot{x}, \ddot{x}) \in M$ in Newtonian state space by

$$T : (x, \dot{x}, \ddot{x}) = (x, -\dot{x}, \ddot{x}). \qquad (3.7)$$

To confirm that this provides a representation of the time reversal group element, we need only check that Eq. (3.6) is satisfied. First, if $(x, \dot{x}, \ddot{x})$ is the initial ($t = 0$) state on a curve $x(t)$, then we have

$$T\phi_t(x, \dot{x}, \ddot{x}) = T(x(t), \dot{x}(t), \ddot{x}(t)) = (x(t), -\dot{x}(t), \ddot{x}(t)). \qquad (3.8)$$

Our assumptions imply that $x(t)$ is a solution only if $x(-t)$ is too,[13] meaning that it evolves from $x(0)$ under the same time translations $\varphi_t$ to $x(-t)$. This curve has initial state $(x, -\dot{x}, \ddot{x})$, which $\varphi_t$ transforms as $\varphi_t(x, -\dot{x}, \ddot{x}) = (x(-t), -\dot{x}(-t), \ddot{x}(-t))$. Therefore, applying the reverse time translation $\varphi_{-t}$ produces $(x(t), -\dot{x}(t), \ddot{x}(t))$, so that we get

$$\phi_{-t}T(x, \dot{x}, \ddot{x}) = \phi_{-t}(x, -\dot{x}, \ddot{x}) = (x(t), -\dot{x}(t), \ddot{x}(t)). \qquad (3.9)$$

Setting these two equations equal, we thus find that $T\phi_t = \phi_{-t}T$, which means that $T$ is a representation of time reversal, $T\phi_t T^{-1} = \phi_{-t}$.

Thus, a Newtonian time reversal operator really does exist! Folk wisdom suggests that it is the identity or non-existent, and there is a grain of truth in that: the construction of $T$ above is indeed the identity when restricted to the submanifold of particle positions $\mathbb{R}^{3n}$. However, time reversal is *not* the identity on true Newtonian state space, the jet space of higher derivatives $(x, \dot{x}, \ddot{x}) \in M$. This follows deductively from the fact that it is a representation of the reversal of time translations. So, when Callender (2000, p.254) proposes that time reversal does nothing but reverse the sign of $t$ and "anything logically supervenient" upon it, we can agree! But, this does not mean that there is no instantaneous time reversal operator, as the Time Reflection Camp seems to suggest: on the contrary, it means that there must be.

---

[13] For $x(t)$ to be a solution means that $\frac{d^2}{dt^2}x(t) = F\left(x(t), \frac{d}{dt}x(t)\right)$ for all $t \in \mathbb{R}$, and so it is true for each $-t \in \mathbb{R}$. Thus, $\frac{d^2}{dt^2}x(-t) = F\left(x(-t), -\frac{d}{dt}x(t)\right) = F\left(x(-t), \frac{d}{dt}x(t)\right)$, where the last equality applies the fact that $F$ does not depend on the sign of velocity. Therefore $x(-t)$ is a solution as well.

How are we to interpret the time reversal transformation $T$ on Newtonian state space? The Instantaneous Camp of Section 2.2 proposed to view it as part of a two-step description of time reversal, in which we first reverse time order and then reverse instants. I would not recommend that it be viewed this way either; at least, not at its foundation. Time reversal is fundamentally the reversal of time translations $\tau t \tau^{-1} = -t$, and the transformation $T$ on state space is just its representative on state space, in that $T \phi_t T^{-1} = \phi_{-t}$. Nor is this somehow really 'motion reversal': since it is an element of a representation, which preserves the essential structure of time, the transformation $T$ has all that is needed to justify viewing it as the reversal of temporal structure.

That said, we can still make sense of the Instantaneous Camp's proposal, as a shorthand way to answer a particular question:

*Given that an initial state $s = (x, \dot{x}, \ddot{x})$ evolves along $x(t)$ according to Newton's equation, what is the solution associated with the 'reversed' initial state $Ts$?*

The answer is: $x(-t)$, which is really $(Tx)(-t)$ together with the fact that $T(x) = x$. This follows immediately from our basic construction: a representation of time reversal must satisfy $T \phi_t T^{-1} = \phi_{-t}$, which is equivalent to $\phi_t T = T \phi_{-t}$; so the evolution $\phi_t(Ts)$ of the initial state $Ts$ by $\phi_t$ can equivalently be written as the evolution $T \phi_{-t}(s)$ of the initial state $s$. Thus, $Ts$ is associated with the solution, $(Tx)(-t) = x(-t)$. The controversy over whether $Ts$ is really a 'time reversed' instantaneous state is, as far as I can tell, only an issue of terminology. One can call it 'motion reversal' or whatever one wishes. But, the basic interpretation of time reversal is still just as the reversal of time translations.

### 3.2.4 Uniqueness of the Newtonian Time Reversal Operator

When a representation of the time reversal operator exists, it is usually not unique. For example, in the case above where force does not depend on the sign of velocity, the transformation $\tilde{T}(x, \dot{x}, \ddot{x}) = (-x, \dot{x}, \ddot{x})$ provides another representation of time reversal, in that it implies that $\tilde{T} \phi_t \tilde{T}^{-1} = \phi_{-t}$.

It is instructive to see this in an example: consider the 'free' particle system for which $F = \mathbf{0}$, and where solutions to Newton's equation are curves of the form $x(t) = \dot{x}t + x$ and time translations can be written as $\varphi_t(x, \dot{x}, \ddot{x}) = (\dot{x}t + x, \dot{x}, \ddot{x})$ for all $t \in \mathbb{R}$. Then,

$$\tilde{T}\phi_t\tilde{T}^{-1}(x, \dot{x}, \ddot{x}) = \tilde{T}\phi_t(-x, \dot{x}, \ddot{x}) = \tilde{T}(\dot{x}t - x, \dot{x}, \ddot{x}) = (-\dot{x}t + x, \dot{x}, \ddot{x})$$
$$= \phi_{-t}(x, \dot{x}, \ddot{x}), \tag{3.10}$$

satisfying our requirement that $\tilde{T}$ is a representation of the temporal reflection $\tau t \tau^{-1} = -t$. So, there are multiple representatives of time reversal in this sense, which have the same effect of transforming curves, as $x(t) \mapsto x(-t)$.

However, the transformation $\tilde{T}$ also reverses 'spatial translations', in that if we define a translation in space $L_a$ by the statement $L_a(x, \dot{x}, \ddot{x}) := (x + a, \dot{x}, \ddot{x})$ for each $a \in \mathbb{R}^{3n}$, then it follows[14] that $\tilde{T}L_a\tilde{T}^{-1} = L_{-a}$. That is what most would expect of 'space and time reversal', but not time reversal alone. So, there is reason to think that not every reversal of a curve $x(t) \mapsto x(-t)$ is 'really' time reversal. But, to distinguish time reversal from space-time reversal, it turns out that we must look beyond the folk wisdom that time reversal just 'reverses little-$t$', and make essential use of the time reversal transformation $T$ on Newtonian state space.

In particular, the Galilei group is a common choice for the symmetries of pre-relativistic spacetime, and so we often suppose that we have a representation of it on Newtonian state space.[15] In this context, the transformation $\tilde{T}$ would normally be defined as a representation of the 'parity and time' transformation $p\tau$. Just as time reversal can be defined as the reversal of time translations, so $p$ can be defined as the reversal of spatial translations, in the sense that $pap^{-1} = -a$ for each spatial translation $s$. In contrast, the time reversal group element $\tau$ commutes with each spatial translation, $\tau a \tau^{-1} = a$. This expresses the 'homogeneity' of time's direction across space: applying time reversal in different spatial locations produces exactly the same result.

Thus, in a representation of the Galilei group with $T = \phi_\tau$ representing time reversal and $L_s = \phi_s$ representing spatial translations, the homomorphism property implies that

$$T L_a T^{-1} = L_a. \tag{3.11}$$

The requirement of homogeneity expressed by Eq. (3.11) is satisfied by the ordinary time reversal transformation $T(x, \dot{x}, \ddot{x}) = (x, -\dot{x}, \ddot{x})$. But, it is not satisfied when $T$ is replaced with the space-time reversal $\tilde{T}$. In this sense, $T$ is the preferred time reversal transformation in Newtonian mechanics.[16] This technique for identifying the time reversal operator turns out to be remarkably general, and we will make use of it in other contexts below.

---

[14] Namely, $\tilde{T}L_a\tilde{T}^{-1}(x, \dot{x}, \ddot{x}) = \tilde{T}L_a(-x, \dot{x}, \ddot{x}) = \tilde{T}(-x + a, \dot{x}, \ddot{x}) = (x - a, \dot{x}, \ddot{x}) = L_{-a}(x, \dot{x}, \ddot{x})$.

[15] For a discussion of the geometry underlying this, see Abraham and Marsden (1978, Theorem 5.4.21 and commentary thereafter) or Marle (1976); a philosophical discussion can be found in Belot (2000) and Earman (1989), or more recently Dewar (2022, Chapter 6).

Here is a general lesson from our discussion: since a state space is a highly structured object, one should expect a representation of time translations to contain significantly more structure than the group of time translations itself. This, in the end, is where the instantaneous time reversal operator gets its 'bells and whistles': a representation $T = \phi_\tau$ of the time reversal group element $\tau$ is a highly structured thing. The same moral applies in nearly every approach to dynamical systems, and so I will appeal to similar arguments in the remainder of this chapter.

What the reader should *not* conclude from this is that Newtonian mechanics is necessarily time reversal invariant. Although we can always extend the time translation group $(\mathbb{R}, +)$ to one that includes time reversal, it is a significant assumption to say that a representation of time translations can be appropriately extended to this larger group too. As we will see in Chapter 4, the latter assumption can fail, even in the context of Newtonian mechanics.[17] Indeed, even more interesting things happen outside the austere world of Democritus, Bošković, and Boyle: if we allow matter to have intrinsic properties besides position (as even Aristotle did[18]), such as a fluid's instantaneous viscosity field in Navier–Stokes theory, then the state space becomes more interesting too. This allows for all sorts of possibilities regarding the failure of time reversal symmetry, even in Newtonian physics.

## 3.3 Analytic Mechanics

There are three great approaches to analytic mechanics: Hamiltonian, Lagrangian, and Hamilton–Jacobi. Folklore has it that these frameworks are all equivalent to Newtonian mechanics, and there are senses in which this is true.[19] But, each adopts a slightly different structure for state space. And, these frameworks are so general as to provide a framework for virtually every area of modern physics. So, to understand the meaning of time reversal in these theories, they should really each be treated independently. As Butterfield observed,[20] the analytic framework,

> helps rebut the false idea that classical mechanics gives us a single matter-in-motion picture. . . . [T]hese equivalences are subtler than is suggested by textbook impressions, and folklore slogans like 'Lagrangian and Newtonian mechanics are equivalent'. (Butterfield 2004, p.29)

---

[16] This suggests that a uniqueness theorem is possible in this context; however, stating this appears to be much simpler in the context of analytic mechanics, and so I reserve this for Section 3.3.

[17] Cf. Roberts (2013b).

[18] In *On Generation and Corruption* Book I, Part 8, 326a.

[19] Philosophical challenges to the folklore are given by Butterfield (2004), Curiel (2013), and North (2009). It has been defended by Barrett (2015, 2019).

Nevertheless, to avoid dragging the reader through the derivation of time reversal in each of these sophisticated frameworks, I hope to be forgiven for analysing just one of them in detail, the symplectic formulation of Hamiltonian mechanics. I will then just give a few brief comments on how the analysis applies to Lagrangian mechanics in Section 3.3.4.[21]

### 3.3.1  State Space for Symplectic Mechanics

With suitable definitions, if $F = -\nabla U$ for some smooth function $U$ of position alone, then with an appropriate definition of a smooth 'Hamiltonian' function $h$, Newton's equation is equivalent to *Hamilton's equations*,[22]

$$\frac{d}{dt} q_i(t) = \frac{\partial h}{\partial p_i} \qquad\qquad \frac{d}{dt} p_i(t) = -\frac{\partial h}{\partial q_i} \qquad (3.12)$$

for each $i = 1, \ldots, n$ and for all $t \in \mathbb{R}$. Hamilton's equations are invariant under time translations, and so they provide a natural context for a representation of time translations. This can be stated most clearly in the language of symplectic mechanics.

One can view symplectic mechanics as a framework for producing Hamilton's equations in 'local' form by elevating time translation invariance to the status of an axiom. The state space is a pair $(M, \omega)$ called a *symplectic manifold*, where $M$ is a smooth $2n$-dimensional real manifold, and $\omega$ is a closed, nondegenerate two-form on $M$ called a *symplectic form*. The central axiom of the theory can be stated as follows:

*Time evolution is along a vector field X that preserves the structure of state space.*

This axiom can be interpreted to mean that $X$ is a smooth vector field with Lie derivative satisfying $\mathcal{L}_X \omega = 0$, called a *symplectic vector field*. That is equivalent[23] to the statement that $\iota_X \omega$ is a closed one-form, which implies that it is 'locally' exact by the Poincaré lemma: in a neighbourhood of every point,

$$\iota_X \omega = dh \tag{3.13}$$

for some smooth function $h : M \to \mathbb{R}$. We can now confirm that Eq. (3.13) is a local geometric expression of Hamilton's equations by an application of Darboux's theorem, which ensures that there is a local coordinate system $(q_1, \ldots, q_n, p_1, \ldots, p_n)$ in which $\omega = \begin{pmatrix} & \mathbb{1} \\ -\mathbb{1} & \end{pmatrix}$, which reproduces[24] the Eqs. (3.12). However, the framework of symplectic mechanics is much more general: $(M, \omega)$ can be any symplectic manifold, and a state $s \in M$ can be used to represent a much more general state of affairs than Boyle or Bošković intended for Newtonian mechanics.

We begin by reviewing the nature of time translations in this framework. A symplectic vector field $X$ is threaded by a one-parameter set of diffeomorphisms called the *symplectic flow* $t \mapsto \phi_t$ along $X$. In general, an automorphism of $(M, \omega)$ is a diffeomorphism that either preserves $\omega$ or reverses its sign; the latter possibility arises because the symplectic form introduces an orientation on $M$, which is arbitrary from a physical perspective. The former is called a *symplectomorphism* and the latter an *anti-symplectomorphism*.

By construction, each $\phi_t$ in a symplectic flow is an symplectomorphism such that $\phi_{t_1} \phi_{t_2} = \phi_{t_1 + t_2}$. Thus, $t \mapsto \phi_t$ forms a representation of the Lie group of time translations. Its local Lie algebra generator in the representation is the smooth function $h$, called the 'Hamiltonian' or 'energy' of the representation. Following the discussion of Section 3.1.3, we will require that the Hamiltonian $h$ be half-bounded, meaning that there exists some fixed lower bound $b \in \mathbb{R}$ such that $h(s) \geq b$ for all $s \in M$, but no such upper bound.

### 3.3.2 Time Reversal in Symplectic Mechanics

Let $t \mapsto \phi_t$ be a representation of time translations amongst the symplectomorphisms of $(M, \omega)$. In other words, $\phi_t$ is a symplectic flow that threads a symplectic vector field $X$. Let $h$ denote a half-bounded local generator of $\phi_t$ in some neighbourhood. Extending the time translations to a group that includes a time reversal element $\tau$, suppose that an extension of the representation $\phi$ exists too. Then, the *time reversal transformation in symplectic mechanics* is just the representative $T := \phi_\tau$ of time reversal on state space. As before, $\tau t \tau^{-1} = -t$ implies that $T\phi_t T^{-1} = \phi_{-t}$, since a representation is a homomorphism.

---

[23] Cartan's 'magic formula' for a two-form $\omega$ and a vector field $X$ states, $\mathcal{L}_X \omega = d\iota_X \omega + \iota_X d\omega$. But $d\omega = 0$, so $\mathcal{L}_X \omega = d\iota_X \omega$. Hence, $\mathcal{L}_X \omega = 0$ if and only if $d\iota_X \omega = 0$, where the latter says $\iota_X \omega$ is closed.

[24] That is, writing an integral curve of $X$ as $q_i(t), p_i(t)$ so that $X = \left( \frac{d}{dt} q_1(t), \ldots, \frac{d}{dt} p_n(t) \right)$. Thus, since for smooth functions $dh = \nabla h = \left( \frac{\partial h}{\partial q_1}, \ldots, \frac{\partial h}{\partial p_n} \right)$, Eq. (3.13) produces Hamilton's equations.

To verify that this notion of time reversal transforms dynamical trajectories in an appropriate way, let $s \in M$ be an initial state, and let us write $t \mapsto s(t) = \phi_t s$ to denote a curve representing the time evolution that begins at $s$. The application of time reversal to each time translation $T\phi_t T^{-1} = \phi_{-t}$ implies that the curve $s(t) = \phi_s(t)$ is transformed to $s(-t) = \phi_{-t}s$, and so time reversal induces a transformation on curves of the form

$$s(t) \mapsto s(-t). \tag{3.14}$$

However, as before, more structure is needed in order to determine what this transformation $T$ is like.

By adopting the (Darboux) coordinates of Hamilton's equations (3.12), one might guess that $T$ should be defined by $T : (q, p) \mapsto (q, -p)$, so as to 'reverse instantaneous momentum'. Writing the symplectic form in terms of the wedge product as $\omega = dp \wedge dq$, this would imply that time reversal also reverses the symplectic form $\omega \mapsto -\omega$, and so is an anti-symplectomorphism.[25] This is indeed the right result. Unfortunately, in symplectic mechanics we do not always have assurance that the coordinates $(q, p)$ represent physical 'position' and 'momentum', respectively. Indeed, if one coordinate system $(q, p)$ happens to represent position and momentum, then there typically remain many coordinate systems $(Q, P)$ related to it by a symplectomorphism, which thus satisfy Hamilton's equations, but which do not represent position and momentum. So, in symplectic mechanics, we will need a more general perspective on the meaning of the time reversal operator.

Happily, we have already done the work of building that general perspective and can prove the general result that time reversal is antisymplectic without this appeal to coordinate systems. Our only substantial assumption will be that the energy associated with time translations is half-bounded. And, to preserve the local structure of symplectic mechanics, we will take our time translations to be associated with a *local* Lie group, defined by the $(\mathbb{R}, +)$ in some neighbourhood of the identity. Since an automorphism of a symplectic manifold $(M, \omega)$ is by definition either symplectic or antisymplectic (as discussed in the end of Section 3.3.1), it will be enough to show that time reversal cannot be symplectic. Here is the result.

---

[25] This is analogous to an antiunitary time reversal operator in quantum theory; see Section 3.4.

**Proposition 3.1** *Let $t \mapsto \phi_t$ be a representation of a local Lie group of $(\mathbb{R}, +)$ in some neighbourhood of the identity, amongst the symplectic and antisymplectic transformations of $(M, \omega)$, with $h$ a half-bounded generator. If the representation extends to one that includes a time reversal operator $T := \phi_\tau$ such that $T\phi_t T^{-1} = \phi_{-t}$, then $T$ is not symplectic, and so it must be antisymplectic.*

*Proof*  Assume for reductio that $T$ is symplectic, and let $s(t) := \phi_t s$ for some $s \in M$. Then $s(t)$ is an integral curve with tangent vector field $X_h$, where $h$ is the half-bounded local generator of time translations. We assumed $T \circ \phi_t \circ T^{-1} = \phi_{-t}$, which is equivalent to $\phi_t \circ T = T \circ \phi_{-t}$ using the fact that $T = T^{-1}$. This implies that $T \circ s(-t) = T \circ \phi_{-t} s = \varphi_t \circ Ts$. So, $(Ts)(-t) = Ts(-t)$ is an integral curve of $X_h$. Moreover, by Hamilton's equations, $s(-t)$ has a Hamiltonian vector field given by $-X_h = X_{-h}$. Combining these two facts implies that

$$X_h = T_* X_{-h}, \tag{3.15}$$

where $T_*$ is the push-forward of $T$ on vector fields. But we have assumed $T$ is symplectic, so Jacobi's theorem can be applied (Abraham and Marsden 1978, Theorem 3.3.19), which says that for symplectic maps, Eq. (3.15) is true if and only if $X_h = X_{-h \circ T}$. Therefore: $h(x) = -h \circ T(x) + c$ for some $c \in \mathbb{R}$ and for all $x \in M$. But $h$ is half-bounded, so we can write $m \leq h \circ T(x)$ for all $x \in M$, which is equivalent to: $-h \circ T(x) + c \leq m + c$. Combining these two thus entails

$$h(x) = -h \circ T(x) + c \leq m + c$$

for all $x \in M$, contradicting the assumption that $h$ is unbounded from above. ∎

One immediate corollary is that, like in Newtonian mechanics, time reversal cannot be the identity transformation: if a representation of time reversal exists, then it must be antisymplectic, whereas the identity is symplectic.

Another interesting application is in electromagnetism. Let $(\mathbb{R}^4, g_{ab})$ be a relativistic spacetime, where $g_{ab}$ is a Lorentzian metric. A two-form $F$ on $\mathbb{R}^4$ that is closed, $dF = 0$, is called a *Maxwell–Faraday field* and is a common way to represent the electromagnetic field. To analyse this system in symplectic mechanics, let $M = T^*\mathbb{R}^4$ be the cotangent bundle over $\mathbb{R}^4$, whose canonical coordinates can be written $(q, p)$ with $q \in \mathbb{R}^4$ representing position in spacetime and $p$ a one-form on $\mathbb{R}^4$ at the point $q$. Let $\theta$ be the canonical one-form on $T^*M$, and let $\omega_0 = d\theta$ be its canonical symplectic form.[26] Using the projection $\pi : T^*M \rightarrow M$, we can pull $F$ back to a two-form on $T^*M$, which we will again denote by $F$. Now, the two-form defined by

$$\omega_F := \omega_0 + eF, \tag{3.16}$$

where $e \in \mathbb{R}$ represents 'electric charge', is again a symplectic form; and, the symplectic flow on $(M, \omega_F)$ generated by the (half-bounded) free Hamiltonian $h(q, p) := \frac{1}{2} g^{ab} p_a p_b$ gives rise to the ordinary equations of motion for a charged particle in an electromagnetic field.[27]

What is the effect of time reversal on this system? From Proposition 3.1, we know that its representative $T$ on the symplectic manifold $(M, \omega_F)$ is antisymplectic, and so reverses the sign of $\omega_F$. If we suppose also that $T : (q, p) \mapsto (q, -p)$, given that we know the coordinates we have adopted represent a particle's position and momentum, then the one-form $\theta$ reverses sign as well, and so $\omega_0 = d\theta$ does too. This implies that $T$ reverses the electromagnetic field, $F \mapsto -F$, since $F = (\omega_F - \omega_0)/e$. Similarly, given an electromagnetic 'four-potential', which is a one-form $A$ satisfying $F = dA$, it follows from this that $A \mapsto -A$. This is a different way of getting to the account of Malament (2004), who instead uses the reversal of a temporal orientation to define time reversal (see Section 2.5.3) and then observes that this induces the transformation $F \mapsto -F$.

### 3.3.3 *Uniqueness of the Hamiltonian Time Reversal Operator*

Like in Newtonian mechanics, there are generally many representations of time reversal on a symplectic manifold: if $T$ satisfies $T\phi_t T^{-1} = \phi_{-t}$, and if $\alpha$ is any symplectomorphism such that $\alpha \varphi_t = \varphi_t \alpha$ for all $t$ (called an 'integral' of time translations), then one can check[28] that this is also satisfied by $\tilde{T} := \alpha \circ T$. But again, the broader context of a symmetry group will often help to determine the meaning of $T$. Let me indicate one scenario in which a uniqueness result can be obtained.

As discussed in Section 3.2.4, the structure of the Galilei group (and the Lorentz group for that matter) guarantees that time reversal, in addition to reversing time translations $T\varphi_t T^{-1} = \varphi_t$, also commutes with spatial translations, $T L_s T^{-1} = L_s$. Moreover, time reversal transforms a velocity boost to the reverse boost, in that $T B_v T^{-1} = B_{-v}$, capturing an intuitive sense in which it reverses velocities. Time reversal similarly reverses rotations, $T R_\theta T^{-1} = R_{-\theta}$. These transformations are the 'generating elements' of the Galilei group, in that all its elements are compositions of them. So, in summary, we have that for each generating element $F$ of the Galilei group, time reversal either commutes with the element, $T F T^{-1} = F$, or maps it to its inverse, $T F T^{-1} = F^{-1}$.

---

[26] The *canonical one-form* $\theta$ on a cotangent bundle $T^*M$ is a standard construction given in coordinates $(q, p)$ by $\sum_i p_i dq^i$. It has the property that $\omega = -d\theta$ is a symplectic form, called the *canonical symplectic form* (cf. Arnol'd 1989, §37).

[27] Cf. Guillemin and Sternberg (1984, p.140).

[28] Namely, $\tilde{T}\phi_t\tilde{T}^{-1} = \alpha(T\phi_t T^{-1})\alpha^{-1} = \alpha(\phi_{-t})\alpha^{-1} = \phi_{-t}\alpha\alpha^{-1} = \phi_{-t}$.

Suppose now that we are concerned with a linear symplectic manifold $(M, \omega)$, meaning that the manifold $M$ is also a vector space, for example in the context of a classical field theory (cf. Section 8.2.2). Suppose further that we have a representation of the Galilei group that is *irreducible*, in the sense that it is a non-trivial representation with no non-trivial subspaces that are also representations. One often uses this condition to characterise 'elementary' systems that cannot be decomposed into further component parts. And, when this is the case, the properties above guarantee that $T$ is unique, up to a multiplicative constant.

This follows immediately from one of the cornerstones of linear representation theory known as *Schur's lemma*, which says that if a transformation $K$ commutes with every element of a linear representation, then it must be a constant multiple of the identity (cf. Blank, Exner, and Havlíček 2008, Theorem 6.7.1). In particular, suppose that given an irreducible representation, there are two transformations $T$ and $\tilde{T}$ that both either commute with each generating element $F$, or map it to its inverse. Then $T\tilde{T}$ commutes with each generating element in the representation. Thus, $T\tilde{T} = c$ for some constant by Schur's lemma, and so applying the fact that $T = T^{-1}$, we get that $\tilde{T} = cT$. We can summarise this general strategy for determining uniqueness in the following.

> **Proposition 3.2** *Let $T = T^{-1}$ be a linear bijection such that, for all the generating elements $F$ of an irreducible linear representation, either $T F T^{-1} = F$ or $T F T^{-1} = F^{-1}$. Then $T$ is the unique transformation with these properties up to a multiplicative constant, in that any other $\tilde{T}$ with these properties satisfies $\tilde{T} = cT$ for some constant $c$.*

As one might expect, this technique finds its most natural home in theories with a built-in linear structure, such as quantum theory. Time reversal in that context is the subject of Section 3.4. But before we turn to the *verdammten Quantenspringerei*, let me offer a few brief remarks on time reversal in Lagrangian mechanics.

### *3.3.4 Lagrangian Mechanics*

Lagrangian mechanics has an elegant geometric formulation due to Klein (1962), which makes its state space structure particularly clear.[29] This formulation reveals that much of Lagrangian mechanics is in fact just a special case of symplectic mechanics. So, let me conclude our discussion of analytic mechanics with a sketch of this fact, which allows one to analyse time reversal using the results above.

Lagrangian mechanics is formulated on the tangent bundle $TM$ of a smooth real manifold $M$. A point in $TM$ is often written $s = (x, \dot{x})$, with $x \in M$ and $\dot{x}$ a vector at $x$, representing the 'configuration' and 'velocity' of a physical system at an instant. Motion in this framework is given by a solution to the *Euler–Lagrange equations*,

$$\frac{d}{dt}\frac{\partial L}{\partial \dot{x}_i} - \frac{\partial L}{\partial x_i} = 0 \tag{3.17}$$

for each $i = 1, 2, \ldots, n$ and for all $t \in \mathbb{R}$.

The state space structure underpinning this framework can often be formulated as a symplectic manifold, and in this sense it is a special case of symplectic mechanics. This applies specifically to systems with a 'regular' Lagrangian, or one for which the Hessian is invertible.[30] For each smooth Lagrangian $L : TM \to \mathbb{R}$, there exists a *canonical closed two-form $\omega_L$* on $TM$, which can be 'pulled back' from the canonical symplectic form $\omega_0$ on $T^*M$, its 'partner' cotangent bundle; this two-form turns out to be a symplectic form if and only if the Lagrangian is regular.[31] There is also a *canonical energy function $h_L$ : $TM \to \mathbb{R}$* defined[32] by $L$. For regular Lagrangians, the Euler–Lagrange equations can then be expressed in the same form as Hamilton's equations, but on the manifold $TM$:

$$\iota_X \omega_L = dh_L. \tag{3.18}$$

In coordinates $(x, \dot{x})$, this reduces to the familiar expression of the Euler–Lagrange equations above (De León and Rodrigues 1989, p.304). And, just as in Hamiltonian mechanics, the time translations here are given by the symplectic flow $t \mapsto \phi_t$ on $TM$ that 'threads' the symplectic vector field $X$.

---

[29] For details on Klein's approach, see Nester (1988) and Curiel (2013), De León and Rodrigues (1989, §7), or Woodhouse (1991, §2).

[30] This condition is required for the existence of the Legendre transformation; for details, see Abraham and Marsden (1978, §3.5). An excellent analysis of the relationship between Hamiltonian and Lagrangian structures can be found in Barrett (2015, 2019).

[31] See Abraham and Marsden (1978, Proposition 3.5.9) or De León and Rodrigues (1989, §7.1).

[32] This makes use of canonical vertical (or 'Liouville') vector field $V$ on $TM$; it is defined in the standard way by $h_L := V(dL) - L$ (cf. Nester 1988).

The state space for 'regular' Lagrangian mechanics can thus be viewed as a symplectic manifold $(TM, \omega_L)$. This means that these models of Lagrangian mechanics are not just equivalent to models of Hamiltonian mechanics: the former are a subset of the latter![33] As a result, the conclusion of Proposition 3.1 applies to this sector of Lagrangian mechanics as well, that time reversal is represented by an antisymplectic transformation $T$. Of course, this conclusion does not apply to the interesting case of non-regular or *singular* Lagrangians, which are ubiquitous in gauge physics and for which the two-form $\omega_L$ in Eq. (3.18) is not symplectic. However, nothing prevents the analysis of time reversal on the Representation View from being carried out in this more general context, as a transformation that reverses time translations associated with singular Lagrangians. I leave this analysis as an invitation to the reader.

## 3.4 Quantum Theories

### 3.4.1 Two Camps on Hilbert Space Quantum Theory

Recall that in Chapter 2, we discussed two camps regarding the meaning of time reversal. The Time Reflection Camp argued that time reversal has the form $t \mapsto -t$, while the Instantaneous Camp argued that time reversal involves this reflection plus a richly structured transformation $T$ on state space (Section 2.2). My thesis was that, once we adopt the Representation View, we find a sense in which both these camps are correct: time reversal does transform time translations as $\tau : t \mapsto -t$; and, its representative on state space is a richly structured operator $T$, of the kind we have now seen in the previous sections. Since a significant portion of this debate has taken place in the context of Hilbert space quantum theory, it is worth briefly reviewing what they say in this context, before we turn to the new derivation of the time reversal operator that the Representation View affords.

Modern quantum theory on Hilbert space was set out by Von Neumann (1932), who took experimental outcomes to be represented by elements of a lattice of closed subspaces of a separable Hilbert space $\mathcal{H}$, or equivalently, by the lattice $L(\mathcal{H})$ of projections onto those subspaces. Given a suitably generalised notion of a 'probability measure' $p$ on a suitable lattice, one can always[34] find a density operator $\rho$ that allows $p$ to be expressed in canonical form (the 'Born rule'):

$$p(F) = \text{Tr}(\rho F) \tag{3.19}$$

for each projection $F \in L(\mathcal{H})$.

---

[33] Compare this to Barrett (2018a).

| **Unitary Operator** $U$ | **Antiunitary Operator** $A$ |
|---|---|
| (U1) $UU^* = U^*U = I$ | (A1) $AA^* = A^*A = I$ |
| (U2) $U(a\psi + b\phi) = aU\psi + bU\phi$ | (A2) $A(a\psi + b\phi) = a^*A\psi + b^*A\phi$ |
| (U3) $\langle U\psi, U\phi \rangle = \langle \psi, \phi \rangle$ | (A3) $\langle A\psi, A\phi \rangle = \langle \psi, \phi \rangle^*$ |

Figure 3.5  Defining properties of unitary and antiunitary operators.

Symmetry transformations in quantum theory are represented by unitary or antiunitary operators (defined by the properties in Figure 3.5), for reasons that I will explain shortly. Members of the Instantaneous Camp, which has included Earman (2002b) and myself (Roberts 2017), defend the standard practice of including an antiunitary time reversal operator, following the definition set out by Wigner (1931).

The statement that time reversal is antiunitary gives rise to the 'complex conjugation' aspect of time reversal: for example, in the Schrödinger representation on $\mathcal{H} = L^2(\mathbb{R})$ with $Q$ defined by $Q\psi := x\psi$ (for all vectors $\psi$ in its domain $D_Q$), with $Q$ interpreted as a 'position observable', the instantaneous time reversal operator $T = K$ is the 'conjugation operator' on wavefunctions, defined by $K\psi := \psi^*$ for all $\psi \in L^2(\mathbb{R})$. One can check that this $T$ is antiunitary.[35] We thus take time reversal to transform a trajectory $\psi(t)$ to $T\psi(-t)$. In contrast, as a member of the Time Reflection Camp, Callender (2000) refers to this $T$ as "Wigner reversal" and argues that 'true' time reversal in quantum theory is just the transformation $\psi(t) \mapsto \psi(-t)$, suggesting that $T$ is actually identity transformation, which is unitary.

---

[34] This is the content of Gleason's theorem. Birkhoff and von Neumann (1936) pointed out that 'probability' in quantum theory cannot be a probability in the ordinary mathematical sense of a bounded measure on a $\sigma$-algebra because the distributive axiom is violated in the lattice $L(\mathcal{H})$. A common response is to propose a more general logic, such as the lattice of Hilbert space projections: see Jauch (cf. 1968) for a classic treatment and Rédei (1996, 1998) for a philosophical perspective; a more general operational perspective can be found in Busch, Grabowski, and Lahti (1995) and in Landsman (2017).

[35] The inner product in the Schrödinger representation is $\langle \psi, \phi \rangle := \int_{\mathbb{R}^n} \psi(x)^*\phi(x)dx$, which implies by the linearity of complex conjugation that $\langle \psi^*, \phi^* \rangle = \int_{\mathbb{R}^n} \psi(x)^{**}\phi(x)^*dx = \int_{\mathbb{R}^n} (\psi(x)^*\phi(x))^*dx = \langle \psi, \phi \rangle^*$.

I have shown elsewhere on the basis of some well-loved adequacy conditions that the time reversal operator must be antiunitary (Roberts 2017, Proposition 1). That result is the following.

**Proposition 3.3** *Let $T$ be a unitary or antiunitary bijection on a separable Hilbert space $\mathcal{H}$. Suppose there is at least one densely-defined self-adjoint operator $H$ on $\mathcal{H}$ that satisfies the following conditions.*

   *i) (positive energy) $0 \leq \langle \psi, H\psi \rangle$ for all $\psi$ in the domain of $H$.*
   *ii) (nontrivial) $H$ is not the zero operator.*
   *iii) (invariance) $Te^{itH}\psi = e^{-itH}T\psi$ for all $\psi$.*

*Then $T$ is antiunitary.*

However, this result did not convince the Time Reflection Camp. Callender (Forthcoming) has responded by rejecting these adequacy conditions. He writes:

Quantum textbooks sometimes address this point and claim that in quantum mechanics time reversal invariance is to be given by *two* operations, a temporal reflection and the operation of complex conjugation $K : \psi \to \psi^*$. This idea can be traced back to Wigner . . . . Roberts (2017) shows that if one assumes that there exists at least one non-trivial time reversal invariant quantum physical system then Wigner's operation follows. But in this context this assumption is a large one for it's up in the air whether quantum mechanics is time reversal invariant. . . . Call it what you like, Wigner's reversal is different from a temporal reflection. Taking the complex conjugation of a state *doesn't follow by logic or definition alone* from a temporal reflection. (Callender Forthcoming, pp.9–10)

Callender has given a reasonable reply. So, let me try to strengthen mine. After a brief primer on symmetries in quantum theory (Section 3.4.2), I will argue that Callender's conclusion that "Wigner's reversal is different from temporal reflection" is too quick: Wigner's antiunitary time reversal transformation is nothing more than the representation of temporal reflection $\tau : t \mapsto -t$, where each $t$ is a time translation. As a result, effects like the conjugation of wavefunctions really do follow from a temporal reflection – indeed, by logic and definition alone.

### 3.4.2 Symmetries of Quantum Theory

As in earlier sections, we begin with the automorphisms of our state space. Experimental outcomes that cannot both occur at once, such as 'z-spin up' and 'z-spin down' in a Stern–Gerlach apparatus, are represented in quantum theory by projections $E, E' \in L(\mathcal{H})$ that are *orthogonal*, $E \perp E' = 0$, meaning that they project onto orthogonal subspaces. An *automorphism* of a Hilbert space lattice $L(\mathcal{H})$ is a bijection **U** on one-dimensional projections that preserves orthogonality:

$$\mathbf{U}(E) \perp \mathbf{U}(E') \text{ if and only if } E \perp E'. \tag{3.20}$$

The group of automorphisms Aut $L(\mathcal{H})$ provides a sensible definition of the symmetries in quantum theory: they preserve the facts about which outcomes can and cannot occur together. They are also strikingly classified by *Uhlhorn's theorem*, which assures us that every automorphism $\mathbf{U} \in$ Aut $L(\mathcal{H})$ implementable by a unique Hilbert space operator $U$ (in that $U\psi \in \mathbf{U}(E)$ if and only if $\psi \in E$) is either unitary or antiunitary (Uhlhorn 1963). Uhlhorn's theorem is a more precise and powerful expression of what is commonly called *Wigner's theorem* in quantum theory (cf. Bargmann 1964).

A representation of time translations in quantum theory is thus a map from the group of time translations $(\mathbb{R}, +)$ to the group of unitary and antiunitary operators on a Hilbert space. In quantum theory, it is entirely standard practice to view time evolution in this way: we begin with a strongly continuous representation of $(\mathbb{R}, +)$ amongst the unitary or antiunitary operators, and from this derive the Schrödinger equation. To show this, we first note that the representation must in fact be entirely unitary, since for each $t \in \mathbb{R}$ we have that $U_t = U_{t/2} U_{t/2}$, which produces a unitary operator regardless of whether $U_{t/2}$ is unitary or antiunitary. We can then apply Stone's theorem (Blank, Exner, and Havlíček 2008, Theorem 5.9.2), which says that for such a representation there exists a unique densely-defined self-adjoint operator $H$ such that $U_t = e^{-itH}$ for all $t \in \mathbb{R}$. This leads immediately to the Schrödinger equation, by defining $\psi(t) := e^{-itH}\psi$ for some $\psi \in \mathcal{H}$ and taking derivatives of both sides.[36] So, on this standard reading of time in quantum theory, the 'little $t$' parameter in the Schrödinger equation does not represent a time coordinate, but a time translation, just as in my general proposal of Chapter 2.

### 3.4.3 Time Reversal in Quantum Theory

We now have the tools to see where Wigner's antiunitary time reversal operator comes from. Let $(\mathbb{R}, +)$ be the group of time translations; as I have shown in Section 2.6, it can always be extended using a semidirect product to a group that includes a time reversal element $\tau$ satisfying $\tau t \tau^{-1} = -t$, for each time translation $t \in \mathbb{R}$. In any representation $\phi$ of this larger group amongst the unitary and antiunitary operators on a Hilbert space, there is no mystery about where Wigner's time reversal operator comes from: we define it to be the representative of the temporal reflection $\tau$, in that,

$$T := \varphi_\tau. \tag{3.21}$$

---

[36] Namely, $\frac{d}{dt}\psi(t) = -iHe^{-itH}\psi = -iH\psi(t)$. A more detailed treatment of the rationale for the Schrödinger equation is Jauch (1968, §10-1 and 10-2) and Landsman (2017, §5.12).

In other words, Wigner's time reversal operator is no different than a temporal reflection: these two have the very same effect on time, guaranteed by the fact that the representation $\phi$ is a ('structure-preserving') homomorphism.

No further assumptions about the nature of time reversal are needed. We will only restrict our attention to a certain very large class of quantum theories, in which energy is bounded from below but not from above. But, this is an assumption about time *translations*, and not about time reversal. On this basis alone, it is possible to prove that temporal reflection, together with logic and definition alone, give rise to Wigner's antiunitary time reversal operator. Let me state the formal proposition first, before turning to its interpretation.

**Proposition 3.4** *Let $t \mapsto U_t$ be a strongly continuous unitary representation from the time translation group $(\mathbb{R}, +)$ to the automorphisms of a separable Hilbert space $\mathcal{H}$, with a half-bounded generator $H$. Let $G$ be the extension of this group to include time reversal $\tau$ satisfying $\tau t \tau^{-1} = -t$. Then:*

1. *the representation of $(\mathbb{R}, +)$ extends to a representation of $G$, with the representative $\tau \mapsto T$ of time reversal satisfying $T U_t T^{-1} = U_{-t}$; and*
2. *in every such representation, $T$ must be antiunitary.*

*Proof* To prove that such an extension exists, let $H$ be the self-adjoint generator satisfying $U_t = e^{-itH}$, and let $\mathrm{sp}(H) = \Lambda \subseteq \mathbb{R}$ be its spectrum. Let $H_s$ be its spectral representation on $L^2(\Lambda)$, meaning that $H_s \psi(x) = x$ for all $\psi$ in its domain, and $V H V^{-1} = H_s$ for some unitary $V : \mathcal{H} \to L^2(\Lambda)$ (cf. Blank, Exner, and Havlíček 2008, §5.8). If $K$ is the conjugation operator on $L^2(\Lambda)$, meaning $K\psi = \psi^*$ for all $\psi \in L^2(\Lambda)$, then $[K, H_s] = 0$, since for all $\psi$ in the domain of $H_s$ we have $K H_s K^{-1} \psi(x) = x \psi(x) = H_s \psi(x)$. Thus, $T := V^{-1} K V$ is the desired antiunitary operator, since our definitions imply that $[T, H] = 0$, and hence $T \mathcal{U}_t T^{-1} = e^{T(-itH)T^{-1}} = e^{itTHT^{-1}} = e^{itH} = U_{-t}$.

To prove that every automorphism $T$ (a unitary or antiunitary by Uhlhorn's theorem) in such a representation must be antiunitarity, suppose for reductio that $T$ is unitary. Since $T U_t T^{-1} = U_{-t}$, we have that,

$$e^{itH} = T e^{-itH} T^{-1} = e^{T(-itH)T^{-1}} = e^{-itTHT^{-1}}, \tag{3.22}$$

where the last equality applies unitarity. By Stone's theorem, the generator of the unitary group is unique, so $-H = THT^{-1}$, and hence $H$ and $-H$ have the same spectrum. But since $H$ is bounded from below, $m \leq \mathrm{sp}(H) = \mathrm{sp}(-H) = -\mathrm{sp}(H) \leq -m$, contradicting the assumption that $\mathrm{sp}(H)$ is unbounded from above.                                                                                                            ∎

I hope this dissolves the remaining mystery surrounding Wigner's antiunitary time reversal operator. Quantum theory is a dynamical theory, and so like any dynamical theory, it admits a representation of time translations with half-bounded energy. Whenever this is the case, Proposition 3.4 shows that time translations extend to include time reversal $\tau : t \mapsto -t$, and, Wigner's time reversal operator is nothing more than its representative on state space, which is always antiunitary. Although a 'unitary time reversal' operator is sometimes associated with the work of Racah (1937), this proposition implies that no such unitary operator reverses time translations in a theory with half-bounded energy.[37]

The Temporal Reflection Camp can also rest reassured that this 'instantaneous' time reversal operator need not be interpreted as 'reversing instants'. Fundamentally, Wigner's time reversal operator is just the representative of a temporal reflection, which has the property that $\tau t \tau^{-1} = -t$, and therefore,

$$TU_t T^{-1} = U_{-t}. \tag{3.23}$$

This $T$ is no more 'instantaneous' than time translations are, in that both are defined as operators on 'instantaneous' state space.

What then of the apparent textbook application of *two* operations, $T$ and $t \mapsto -t$, instead of just $t \mapsto -t$? This is nothing more than a shorthand way to answer the following question:

*Given a solution to the Schrödinger equation $\psi(t) := U_t \psi$ with initial state $\psi$, what is the solution associated with the 'time-reversed' initial state $T\psi$?*

The answer is: $T\psi(-t)$. This follows because Eq. (3.23) implies that $U_t T\psi = TU_{-t}\psi = T\psi(-t)$; so, the unitary dynamics $U_t$ starting with $T\psi$ is given by $T\psi(-t)$. In a misleading sense, this gives the appearance of two operations, $\tau : t \mapsto -t$ and $T : \psi \mapsto T\psi$. But, this is not the origin of the time reversal operator. Time reversal is simply a temporal reflection, which reverses the group of time translations, $U_t \mapsto U_{-t}$. The antiunitary instantaneous time reversal operator $T$ is the operator that implements time reflection, as in Eq. (3.23).

---

[37] This was confirmed explicitly for quantum electrodynamics by Jauch and Rohrlich (1976, pp.88–9). Costa de Beauregard (1980) proposes to restore a unitary reversal of time translations, but at the cost of introducing unbounded negative energy into the theory.

What about Callender's charge that we have assumed time reversal invariance? Here there is a subtlety: Eq. (3.22) does indeed give a standard expression of what it means for a quantum system to be time reversal invariant, which we discuss more in Chapter 4. However, this does not mean that we have assumed time reversal invariance. Although we can always extend time translations to include a time reversal group element, as described in Section 2.6, this does not necessarily mean that we can extend the *representation* of time translations to an appropriate representation of time reversal. If such an extension does exist, then the second part of Proposition 3.4 shows that it must be antiunitary. But, an appropriate representation might fail to exist, in which case we say that a quantum system violates time reversal invariance. I will return to the discussion of time reversal invariance and time reversal violation in Chapter 4.

Note that I am making careful use of the word 'appropriate' when I say that an appropriate representation of time reversal might not exist. Proposition 3.4 is perhaps surprising because the first part shows that, in quantum theory, a representation of time reversal – and hence, of time reversal symmetry – always does exist! This expresses a certain sense in which quantum theory is always symmetric in time, which I will discuss in more detail in Chapter 8. However, it is not always appropriate to call the resulting representative $T$ the 'time reversal operator': there are in general many representatives of time reversal, such as parity–time reversal, just as in the case of classical mechanics (Sections 3.2.4 and 3.3.2).

Fortunately, as in our discussions of classical mechanics, considerations of the more general spacetime symmetry group help to uniquely determine which antiunitary time reversal operator is the 'appropriate' one for time reversal, through the application of Proposition 3.2.

For example, we might adopt the Galilei group, or the Lorentz group, as a more complete group $G$ containing time translations and consider its representation among the automorphisms of a Hilbert space. This representation will be irreducible whenever it describes an 'elementary' system, interpreted as one that cannot be decomposed into component parts. But, time reversal transforms each generating elements of this group to itself or to its inverse, as discussed in Section 3.3.2, and so by Schur's lemma, it follows that any two representatives of time reversal $T$ and $\tilde{T}$ must be related by a multiplicative constant. It is for this reason that in many concrete applications, the choice is uniquely determined as to which antiunitary time reversal operator is appropriate.[38]

### 3.4.4  *The Time Reversal of Spin*

A first topic in many quantum mechanics textbooks is the analysis of *spin*, a degree of freedom associated with a system's total angular momentum.[39] Spin-1/2 systems are often studied on a two-dimensional Hilbert space, with an algebra of observables generated by the Pauli 'spin observables',

$$\sigma_x = \begin{pmatrix} & 1 \\ 1 & \end{pmatrix}, \qquad \sigma_y = \begin{pmatrix} & -i \\ i & \end{pmatrix}, \qquad \sigma_z = \begin{pmatrix} 1 & \\ & -1 \end{pmatrix}.$$

Their eigenvectors are usually interpreted as 'spin-up' and 'spin-down' states of the system with respect to three spatial axes $x, y, z$. The algebraic relations that these observables satisfy are called the 'Pauli relations'.

Time reversal is usually assumed to reverse the sign of the Pauli spin observables, $T\sigma_j T^{-1} = -\sigma_j$ for each $j = x, y, z$. A common explanation of this is that spin is a 'kind' of angular momentum: it is intrinsic angular momentum that does not correspond to rotation in space, but is angular momentum nevertheless. So, the argument goes, since angular momentum is reversed by time reversal, spin must be reversed as well.[40] Of course, one might still wonder *why* a kind of angular momentum that does not actually 'rotate' anything in space must change sign under time reversal.

The Representation View has something to say about this too. To see it, we will first need to get a better grip on how the Pauli spin observables are related to spacetime. Define the 'spin rotations' through $\theta \in (0, 2\pi]$ about each axis $j = x, y, z$ by

$$R_j(\theta) := e^{-(i/2)\theta\sigma_j}. \tag{3.24}$$

Each rotation can be written in explicit form as $R_j(\theta) = \cos(\theta/2)I + \sigma_j \sin(\theta/2)$, where $I$ is the identity operator. The group given by the closure of all these rotations under multiplication is isomorphic to a Lie group called $SU(2)$, and which is *not* isomorphic to the spatial rotation group $SO(3)$. Famously, $SU(2)$ has the unusual property that $R_j(2\pi) = -I$, whereas for $SO(3)$, a rotation through $2\pi$ is the identity. So, viewing 'ordinary' angular momentum as defined by the generators of $SO(3)$, it follows that the spin observables are not ordinary angular momentum.

---

[38] This observation was made explicit in Roberts (2017, Propositions 2 and 3).
[39] See Jauch (1968, Chapter 14) for an elegant introduction.
[40] Cf. Sachs (1987, p.34).

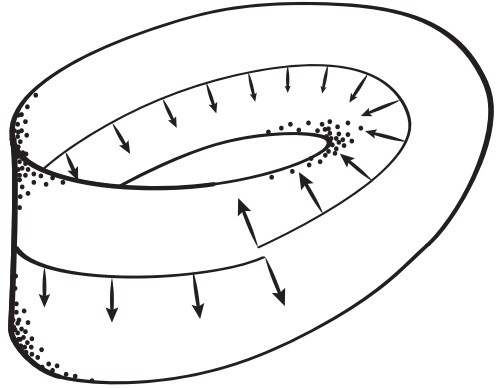

Figure 3.6 On a Möbius strip, a rotation through $2\pi$ reverses the arrow, which is returned to its initial state by another rotation through $2\pi$.

However, we can still think of $SU(2)$ as consisting of spatial rotations in a certain 'degenerate' sense. The group $SU(2)$ is an example of a doubly-degenerate 'covering group' for $SO(3)$. The precise meaning of this is defined in Section 8.3; but, for now, we can visualise a representation of the spin rotation $R_j(\theta)$ using a Möbius strip, as in Figure 3.6. It has the property that, when an arrow is transported through $2\pi$ around the loop of a Möbius strip, it is not returned to its original state, but rather reverses: a second 'copy' of rotations through $2\pi$ is needed to restore it to its original orientation. There are thus two elements of $SU(2)$ corresponding to each 'ordinary' rotation through $\theta$, given by $R_i(\theta)$ and $-R_i(\theta)$. This is what it means to say that the covering group is 'doubly-degenerate'. So, although $SU(2)$ is not isomorphic to $SO(3)$, we can still unambiguously associate each of its elements with a spatial rotation.

Now, recall that as a group element, time reversal generally leaves each spatial rotation intact, $\tau r \tau^{-1} = r$. This is true by definition in the Galilei and Lorentz groups; and, conceptually, it makes sense of the statement that the meaning of time reversal is 'independent of spatial orientation'. These facts can now be carried over to state space using the Representation View: given a group containing both $SU(2)$ and a time reversal group element $\tau$, and which satisfies $\tau r \tau^{-1} = r$, any representation will satisfy

$$T R_j(\theta) T^{-1} = R_j(\theta), \tag{3.25}$$

for each $j = x, y, z$. Using the fact that $T$ is antiunitary, this in turn implies[41] that $Te^{-i\sigma_j}T^{-1} = e^{iT\sigma_j T^{-1}} = e^{-i\sigma_j}$, and hence that $T\sigma_j T^{-1} = -\sigma_j$ for each $j = x, y, z$. In this way, the fact that time reversal does not change spatial rotations can be used to explain why it reverses the Pauli spin observables.

Another interesting fact about the time reversal in a Pauli spin system is that, like a rotation through $2\pi$, applying time reversal twice does not produce the identity: $T^2 = -I$. Only by two more applications of time reversal do we recover: $T^4 = I$. Our discussion above explains this curious fact as well and indeed determines the unique form of the time reversal operator in a Pauli spin system. The proof is an application of the general technique using Schur's lemma, introduced in the discussion of Proposition 3.2.

**Proposition 3.5** *Let $\sigma_x, \sigma_y, \sigma_z$ be an irreducible unitary representation of the Pauli relations, and let $K$ be the conjugation operator in the $\sigma_z$ basis. If $T$ is any antiunitary operator satisfying $T R_j(\theta) T^{-1} = R_j(\theta)$ for $j = x, y, z$, then $T = c\sigma_y K$ for some complex unit $c$, and $T^2 = -I$.*

*Proof*   Our assumptions imply that the antiunitary $T$ reverses each of the spin observables. One can straightforwardly check that $\sigma_y K$ does as well. Since both are antiunitary, the composition $-T\sigma_y K$ is unitary and commutes with $\sigma_x$, $\sigma_y$, and $\sigma_z$. These are the generators of an irreducible representation, and so by Schur's lemma, $-T\sigma_y K = cI$ for some $c \in \mathbb{C}$. This $c$ is a complex unit, $c^*c = 1$, because $-T\sigma_y K$ is unitary. So, multiplying on the right by $\sigma_y K$ and recalling that $(\sigma_y K)^2 = -I$, we find that $T = c\sigma_y K$. Moreover, $T^2 = (c\sigma_2 K)^2 = c^*c(\sigma_2 K)^2 = -I$.                        ∎

Thus, given a reflection of time translations $\tau : t \mapsto -t$ on spacetime, any representation of it on Hilbert space must be antiunitary by Proposition 3.4; and, the form of this antiunitary in a spin-1/2 system is uniquely determined (up to a constant) by Proposition 3.5. Although the concept of 'reversing time translations' at the level of spacetime is relatively simple, its representation in a spin system carries some of the interesting structure of that system.

## 3.5 Summary

This section has been a tour of several different state spaces for modern physics: Newtonian mechanics, analytic mechanics, and quantum theory. All of them include a rich structure for the description of what it means to be a 'solution' to their equation of motion. Although that structure is not always visible in elementary presentations, we have made it visible here through a representation of time translations. This makes it a more straightforward matter to understand what time reversal means: it is nothing more and nothing less than a temporal reflection, represented on a highly structured state space.

---

[41] We use the fact that if $T$ is any antiunitary, then $T(\alpha A) = \alpha^* T A$ for each complex constant $\alpha$. In particular, $Ti\sigma_j = -iT\sigma_j$.

This means that, when it comes to writing down the representative of time reversal on state space, we may find some 'bells and whistles', like the fact that it conjugates wavefunctions. But, the philosopher of time should not be alarmed by this. Time reversal is still ultimately just the automorphism $\tau : t \mapsto -t$ that reverses time translations. But, when it is represented using a highly structured state space, the operator $T$ representing $\tau$ will inevitably pick up some of that novel structure.

# 4

# Philosophy of Symmetry

***Précis.*** *On the Representation View, time reversal has direct empirical significance, breaks the symmetry-to-reality inference, and can be violated in a time-symmetric spacetime.*

Modern physics is firmly rooted in symmetry. Some twentieth-century physicists came reluctantly to appreciate this, with Wigner (1981) recalling how Schrödinger wanted to see the "Gruppenpest" abolished. However, this was largely due to the group theory formalism being viewed at the time as arcane.[1] In more general terms, symmetry has been a cornerstone of physics at least since the eighteenth-century correspondence of Leibniz and Clarke (2000).

Nevertheless, interpreting symmetry is a subtle business. A symmetry is in one sense a relation between two descriptions and in another sense a statement that they are one and the same. Hans Halvorson[2] illustrated the tension using this charming example:

From the perspective of group theory, suppose I ask, "How many groups of order two are there?" You should say: one. But now suppose I ask, "How many groups of order two are there among the subgroups of the Klein four-group?" And you should of course say: three!

---

[1] MIT physicist John Slater seems to have been particularly scandalised: "The authors of the 'Gruppenpest' wrote papers which were incomprehensible to those like me who had not studied group theory. . . . The practical consequences appeared to be negligible, but everyone felt that to be in the mainstream one had to learn about it. Yet there were no good texts from which one could learn group theory. It was a frustrating experience, worthy of the name of a pest. I had what I can only describe as a feeling of outrage at the turn which the subject had taken" (Slater 1975, pp.60–2).

[2] Halvorson (2019, p.259). Halvorson attributes the example to John Burgess, though in a different context.

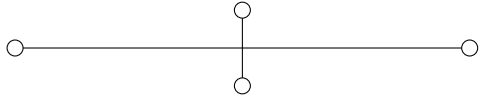

Figure 4.1 The Klein four-group $\mathbb{V}$ describes the symmetries of the figure above: the identity, a flip about each axis, and a flip about both.

This example hearkens back to an old (Fregean) debate.[3] One way to clarify what's going on is to observe that although there is only one group of order two, there are three representations of it on the Klein four-group (Figure 4.1), in the sense of three distinct isomorphisms[4] $\varphi_p, \varphi_t, \varphi_{pt}$ from $\mathbb{Z}_2$ into $\mathbb{V}$. In short, we can dissolve the puzzle by cleanly separating two concepts: the two-element group and its representations in another structure, the Klein four-group.

The Representation View of symmetries developed in Chapter 2 follows a similar strategy. The proposal is to cleanly separate the concept of a symmetry into two parts: a symmetry of spacetime and its representation on a state space. The thesis of this chapter is that a number of issues in the philosophy of symmetry can be clarified and improved by adopting this perspective.

I begin in Section 4.1 by proposing an interpretation of what it means to be a 'dynamical' symmetry, including time reversal symmetry, which recovers its standard definition in the literature as a transformation that 'preserves' the solution space of a dynamical equation. I will then argue that the Representation View has implications for three related philosophical discussions and show how time reversal appears in each.

The first discussion is about epistemology (Section 4.3): if two descriptions related by a dynamical symmetry are empirically indistinguishable, then how can we come to know about that symmetry? Following Kosso (2000) and Brading and Brown (2004), the literature about this question has explored how to answer this question by separating a subsystem with a symmetry from its environment, which does not generally have that symmetry: they call this the 'direct' empirical significance of a symmetry. I will introduce this idea in Section 4.3 and argue that both time translations and time reversal symmetry have direct empirical significance in this sense.

---

[3] Frege (1892) discusses a similar issue in his famous analysis of the meaning of '='. Another influential discussion of mathematical equivalence was given by the Bourbaki group; see Burgess (2015, Chapter 3) for a review.

[4] Writing $\mathbb{Z}_2 = \{0, 1\}$ (with identity 0) and the Klein four-group as $\mathbb{V} = \{1, p, t, pt\}$ (with identity 1), they are defined by $\varphi_p(1) = p$, $\varphi_t(1) = t$, $\varphi_{pt}(1) = pt$, and $\varphi_n(0) = 1$ for each $n = p, t, pt$.

The second discussion is about reference (Section 4.4): if two mathematical models are related by a dynamical symmetry, does it follow that they must refer to the same physical situation? In the debate between Leibniz and Clarke (2000), Leibniz seemed to think so, and this question has recently been the subject of much debate. I will refer to this as the 'symmetry-to-reality inference', following Dasgupta (2016), and argue in Section 4.4 that there are a number of ways in which it can fail.

The final discussion is about whether a dynamical symmetry is necessarily a spacetime symmetry, and vice versa (Section 4.5), as set out in a well-known pair of symmetry principles (SP1) and (SP2) by Earman (1989, §3.4). In Section 4.5, I will argue that on the Representation View, there are senses in which both can fail.

## 4.1  Representing Dynamical Symmetries

### *4.1.1  What Is a Dynamical Symmetry?*

Standard practice has it that a dynamical symmetry is a state space automorphism that preserves the solution space of a dynamical equation.[5] Let me begin by heading off some confusion with a few words about what a dynamical symmetry is not.

First, some discussions define dynamical symmetry as preserving the 'form' of a dynamical equation, which Belot (2003) calls a *nomenclature* symmetry and others call *passive* in analogy to passive and active spacetime symmetries.[6] This deserves the warning rumoured to have been issued by mathematician Graeme Segal: "I defy anybody not to get confused about the difference between passive and active".[7] For example, one must be careful that such symmetries are not viewed as transforming $F = ma$ in such a way that acceleration goes to velocity. To avoid this, the nomenclature approach requires careful analysis of the syntactic 'form' and logical structures of the sentences of a theory.[8] As Belot points out, this often amounts to the same thing in practice. However, these technicalities take one too far afield, and so I will avoid the passive-nomenclature approach.

---

[5] Cf. Belot (2003, 2007, 2013), Butterfield (2006c, 2021), Caulton (2015), De Haro and Butterfield (2019), Earman (1989, §3.4), Møller-Nielsen (2017), Read and Møller-Nielsen (2020), Uffink (2001, p.314), and Wallace (2022), among many others.

[6] The 'passive' usage can be found, for example, in Brown and Holland (1999), Redhead (2003, §3), and Brading and Castellani (2007, §4.1).

[7] I learned this from Jeremy Butterfield (personal correspondence).

[8] The efforts of Barrett and Halvorson (2016a,b), Dewar (2022), Halvorson (2012, 2013, 2019) and collaborators, inspired by Glymour (1970, 1980), provide an enlightened modern expression of this approach, where it is usually referred to as 'translation between theories'.

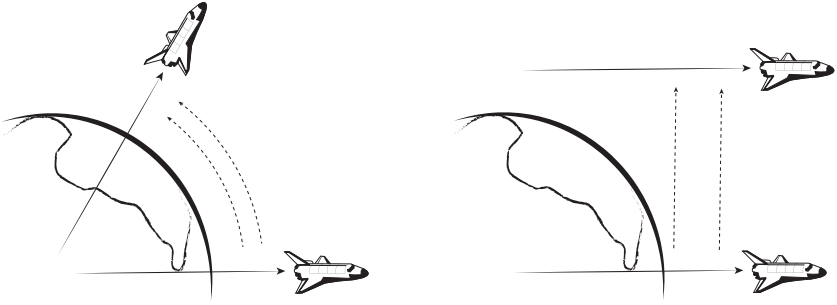

Figure 4.2  In a dynamical equation with a spherically symmetric potential, rotation (left) is a dynamical symmetry while translation (right) is not.

Second, dynamical symmetries are usually more restrictive than 'kinematical' symmetries, by which I mean the automorphisms of a state space. For example, in Hamiltonian mechanics, the kinematical symmetries are the symplectic transformations (see Section 3.3.1). But, when the state space is equipped with a dynamics, such as a Hamiltonian describing a spherically symmetric potential, then only some of the kinematical symmetries preserve the dynamics, such as the rotations about the centre of the potential (see Figure 4.2). Nor do dynamical symmetries necessarily leave an individual solution invariant, as Belot (2003, §4.2) is careful to note: although dynamical symmetries leave the space of all solutions invariant, they often transform each individual solution to a different one, as when rotation transforms a body moving east to a body moving north.

There is a puzzle about the meaning of dynamical symmetries that is not widely appreciated, which is the question of what makes an equation 'dynamical'. One might be tempted to answer that it is the usage of the term that does so: if an equation refers to change over time, then the equation is dynamical. But, what aspect of the formalism distinguishes change with respect to time from change with respect to anything else, like space?

For example, consider a bead on a string in Hamiltonian mechanics, with position and momentum represented by the coordinates $(q, p) \in \mathbb{R} \times \mathbb{R}$. One might wish to declare that motion is described by the solutions to Hamilton's equation, for some smooth function $h(q, p)$:

$$\frac{d}{dt}q(t) = \frac{\partial h}{\partial p}, \qquad\qquad \frac{d}{dt}p(t) = -\frac{\partial h}{\partial q}. \qquad (4.1)$$

But, by choosing the function $h = p$, we find the solutions describe exactly the spatial translations, $q(t) = q_0 + t$ and $p(t) = p_0$, for each initial state $(q_0, p_0)$. In other words, the equation that was declared to be dynamical has the very same structure as one that can be used to refer to non-dynamical translations in space. What distinguishes these two situations?

The problem reappears in field theory as well. In his discussion of classical dynamical symmetry, Earman (1989, pp.45–6) considers any theory with a set of models each of the form $(M, A_1, A_2, \ldots, P_1, P_2, \ldots)$, where $M$ is a smooth manifold, each $A_i$ is an 'absolute object' field on $M$ characterising spacetime structure, and each $P_i$ is a 'dynamic object' field on $M$ characterising physical matter-fields. A dynamical symmetry is then defined to be a transformation of the dynamic objects $P_i$ that preserves the theory's set of models. Here the question becomes the following: What justifies calling a given field 'dynamic' as opposed to 'absolute'?

My answer is to propose the Representation View: a symmetry of a state space can be interpreted as a 'spacetime symmetry' only if it is an element of a *representation* of a spacetime symmetry structure, as argued in Section 2.3. So, in particular, a symmetry can only be viewed as dynamical in the sense of describing 'time translation' if it is a representation of time translations. This is what it means when one says that a curve through state space represents 'time evolution' or that a dynamical field on spacetime is 'dynamical'.[9] Let me now make the consequences of this more precise.

### 4.1.2  Temporal and Dynamical Symmetries

Suppose we have on the one hand a group of symmetries $G$, which could include spacetime or even gauge symmetries, with a preferred subgroup $\mathbb{T} \subseteq G$ interpreted as 'time translations'. In earlier discussions we often focused on $\mathbb{T} = (\mathbb{R}, +)$ to represent time translations; now, I would like to proceed a little more generally and allow $\mathbb{T}$ to be whatever choice of time translations we happen to find appropriate. We will see two alternatives below.

On the other hand, suppose we have a structure $M$ interpreted as 'state space' and which has a group of automorphisms $\text{Aut}(M)$. In earlier discussions, we took $M$ to be a Hilbert space for quantum mechanics, or a symplectic manifold for classical mechanics, but we will leave this open for now. The central postulate of the Representation View is that if the group of symmetries $G$ has meaning in state space, it is through the existence of a representation $\varphi : G \to \text{Aut}(M)$, which is a 'homomorphic copy' of $G$ amongst the symmetries of $M$. This explains why a given state space automorphism $S = \varphi(g)$ is interpreted in the same way as a group element $g$: time translation, rotation, gauge, or any of the like. It is because this is exactly what it represents.

---

[9] A similar view is advocated by Belot (2007, p.171): "Time, in one of its facets, is represented in this scheme by . . . ℝ-actions", where Belot uses 'ℝ-action' to refer to what I call a state space representation of the time translation group $(\mathbb{R}, +)$.

This cleanly separates the concept of a symmetry into two parts, related by a representation: a spacetime part and a state space part. It also allows the concept of a 'dynamical symmetry' to be separated into two parts, corresponding to the two 'sides' of the representation relation. To distinguish them, I will refer to the spacetime version as 'temporal symmetries' and the state space version as 'dynamical symmetries'. The former can be defined in short as transformations that preserve time translations:

> **Definition 4.1** (temporal symmetry) Given a group $G$ with a subgroup $\mathbb{T} \subseteq G$ interpreted as 'time translations', a *temporal symmetry* $g \in G$ is an element that acts invariantly on $\mathbb{T}$ by the automorphism $\alpha_g(\cdot) = g(\cdot)g^{-1}$, in that $\alpha_g(t) \in \mathbb{T}$ for all $t \in \mathbb{T}$. If every element of $G$ is a temporal symmetry, then $G$ is called a *temporal symmetry group*.

The idea is, in short, that a temporal symmetry is a symmetry that preserves time translations, and we refer to a group of these symmetries as a temporal symmetry group. We have seen one example of this already in Section 2.6: let the group of time translations be $\mathbb{T} = (\mathbb{R}, +)$, and let $G :=$ $(\mathbb{R}, +) \rtimes \{\iota, \tau\}$ be its extension to include time reversal. Then, since every element $g \in G$ maps each time translation to another time translation, $gtg^{-1} \in$ $(\mathbb{R}, +)$ for all time translations $t$, we say that $G$ is a temporal symmetry group for $(\mathbb{R}, +)$.

Here is a more interesting example. Let $G = \mathcal{P}$ be the Poincaré group. Then, by definition, each spatial translation $s \in \mathcal{P}$ maps each time translation $t$ to itself via the automorphism $\alpha_s(t) := sts^{-1} = t$. So, the spatial translations preserve $\mathbb{T}$, which means they are temporal symmetries. The Lorentz boosts are a little more subtle: we need to distinguish two cases, illustrated in Figure 4.3.

1. In some applications, we may choose a fixed inertial reference frame, in which time translations are given by a one-parameter group isomorphic to $\mathbb{T} = (\mathbb{R}, +)$. The Lorentz boosts are not temporal symmetries given this definition of time translations, since they result in a different reference frame and therefore a different flow of time translations.

2. In other applications, we might define time translations to be the larger group $\mathbb{T}$ of all the translations along any timelike line (a four-parameter Lie group). Then Lorentz boosts are indeed temporal symmetries, since boosts preserve the set of all timelike vector fields.[10]

---

[10] More formally, in the first case we define the group of time translations $\mathbb{T}$ as the flow along a fixed timelike vector field $\xi^a$ that satisfies the geodesic equation ($\nabla^b \xi_b \xi^a = \mathbf{0}$) in Minkowski spacetime.

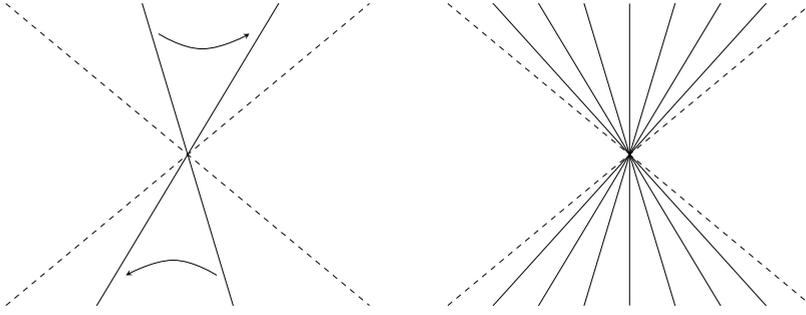

Figure 4.3 Boosts are not a temporal symmetry of translations along a single timelike line (left) but are of the group of all timelike translations (right).

Thus, whether a transformation is a temporal symmetry depends on how we define time translations, as one might expect.

Careful attention to this aspect of the Representation View can help to avoid confusion: for example, as Belot (2013, pp.327,331) has emphasised, boosts do not generally preserve the Euler–Lagrange equations or Hamilton's equations. This led Baker[11] to suggest that Lagrangian and Hamiltonian mechanics leave some puzzlement as to the origin of absolute velocity. But, from the present perspective, Belot's observation is just an instance of the first case above: a preferred frame of reference is needed whenever we choose an Euler–Lagrange or Hamiltonian representation of the time translation group $(\mathbb{R}, +)$. The second case can be recovered by instead representing time translations by a collection of Hamiltonian (or Lagrangian) descriptions related by boosts. Whether boosts are a temporal symmetry in either theory just depends on our definition of time translations.

Temporal symmetries can thus be understood entirely at the level of group transformations. To define dynamical symmetries, we must now consider the other side of the Representation View: a representation on state space. The definition of a dynamical symmetry is then natural, in that it is simply a 'representation' of a temporal symmetry:

---

This vector field (and hence the group of time translations) is transformed non-trivially by a Lorentz boost $\xi^a \mapsto \Lambda^b_a \xi^a \neq \xi^b$, and so it is not a temporal symmetry. The second case defines time translations to be the group $\mathbb{T}$ of flows along *any* of the set of timelike geodesic vector fields $\{\xi^a \mid \xi^a \xi_a > 0, \nabla^b \xi_b \xi^a = \mathbf{0}\}$. This set is preserved by Lorentz boosts, and so boosts are a temporal symmetry of these time translations.

[11] Baker (2020). "What are symmetries?" In: Unpublished Manuscript, http://philsci-archive.pitt.edu/18524/

**Definition 4.2** (dynamical symmetry) Let $\mathbb{T} \subseteq G$ be a group of time translations. Given a representation of time translations $\varphi : \mathbb{T} \to \text{Aut}(M)$, we say that $G$ is a group of *dynamical symmetries* if and only if (1) $G$ is a temporal symmetry group; and (2) there is a non-trivial extension of $\varphi$ to all of $G$, i.e. a representation $\tilde{\varphi} : G \to \text{Aut}(M)$ such that $\tilde{\varphi}(t) = \varphi(t)$ for all $t \in \mathbb{T}$.

The first part of this definition just says that dynamical symmetries are temporal symmetries: they preserve the structure of time translations. The second part expresses a consequence of the Representation View: a dynamical symmetry on state space is just the representative of a spacetime symmetry $g \in G$ in some representation. When a group element $s$ can be represented as a dynamical symmetry in this way, it is common in physics to call this *g-invariance*, using phrases like rotational invariance, translation invariance, time reversal invariance, and so on.

To take an example from quantum theory, let $U : t \mapsto U_t$ be a representation of the time translations $(\mathbb{R}, +)$ amongst the unitary (and antiunitary) operators on a Hilbert space.[12] Let $G$ be the extension of this group to include time reversal $\tau t \tau^{-1} = -t$. As we have seen above, this group $G$ is a temporal symmetry group. Suppose now that we have an extension of $U$ to all of $G$, and let $T := U_\tau$ be the representative of time reversal. Then, $G$ is a dynamical symmetry. In particular, the fact that a representation is a homomorphism implies that $T U_t T^{-1} = U_{-t}$. This is a standard expression of the statement that time reversal is a dynamical symmetry.

Indeed, Definition 4.2 of dynamical symmetries helps explain a well-known property of dynamical symmetries. The solutions of a dynamical equation are associated with a representation $\varphi : t \mapsto \varphi_t$ of time translations: we have agreed that this is what it means for an equation to be 'dynamical'. In particular, any solution $\psi(t)$ with initial condition $\psi(0) = \psi_0$ can be written

$$\psi(t) := \varphi_t \psi_0, \tag{4.2}$$

for all time translations $t$. Let $s \in G$ be a dynamical symmetry, and let $S := \varphi_s$ be its representative on state space. Then, $sts^{-1} = t' \in \mathbb{T}$ implies that $S \varphi_t S^{-1} = \varphi_{t'}$ is a time translation, or equivalently $S^{-1} \varphi_{t'} S = \varphi_t$. Substituting this into Eq. (4.2) gives $\psi(t) = S^{-1} \varphi_{t'} S \psi_0$, which is equivalent to

$$S \psi(t) = \varphi_{t'} S \psi_0. \tag{4.3}$$

---

[12] For a discussion of why unitaries and antiunitaries are the automorphisms of this state space, see Section 3.4.2.

This is the familiar characterisation of dynamical symmetries expressed at the outset of this chapter – that whenever $S$ is a dynamical symmetry, if a curve $\psi(t)$ is a solution to the dynamical equation (Eq. 4.2), then the symmetry-transformed curve $S\psi(t)$ is also a solution (Eq. 4.3). Through an application of the Representation View and our definitions above, we can now offer an explanation of this statement: it follows whenever $S$ represents a symmetry of time translations in spacetime.

### 4.1.3  Time Reversal Symmetry

The discussion above allows us to classify temporal and dynamical symmetries into two basic types. When the time translation group is $\mathbb{T} = (\mathbb{R}, +)$, as in a typical frame of reference, there are only two automorphisms: the identity and the time reversal automorphism $t \mapsto -t$. This was proved in Proposition 2.1. Moreover, the definition of a temporal symmetry implies that it is an automorphism of $\mathbb{T}$. As a result, there are only two kinds of temporal symmetry $t \mapsto t' = sts^{-1}$: the (trivial) identity automorphism $sts^{-1} = t$ and the time reversal automorphism $sts^{-1} = -t$.

We can pull this classification down into a representation $\varphi : G \to \mathrm{Aut}(M)$ of a dynamical symmetry group as well. Defining $S := \varphi_s$, we have from the fact that a representation is a homomorphism that either $S\varphi_t S^{-1} = \varphi_t$ or else $S\varphi_t S^{-1} = \varphi_{-t}$. Thus, a dynamical symmetry preserves solutions in the sense of Eq. (4.3) in one of two ways:

$$S\psi(t) = \varphi_t S\psi, \qquad \text{or} \qquad S\psi(t) = \varphi_{-t} S\psi. \qquad (4.4)$$

In the first case, the symmetry 'commutes' across time translations; in the second, it reverses time. *Time reversal invariance* is nothing more than an example of a symmetry of the second kind. We have seen an example of this at the end of Section 4.1.2.

Recall now from Chapter 3 that the time reversal transformation $T$ on state space is *by definition* the representative of $T := \varphi_\tau$ of a temporal reflection $\tau t \tau^{-1} = -t$, and which thus satisfies $T\varphi_t T^{-1} = \varphi_{-t}$. This means that a representation in which the time reversal element exists is, by definition, one that is time reversal invariant! Be careful though: this does not imply that every dynamical theory is time reversal invariant. Although we may extend time translations to a larger group including time reversal, there is no guarantee that a representation of that larger group exists.

As we will see in the next section, the non-existence of such a representation is precisely how the failure of a symmetry is expressed. This illustrates how deeply a symmetry may be encoded in a dynamical system:

the moment we describe time translations using the group $(\mathbb{R}, +)$, time reversal is a temporal symmetry; and, the moment we have a representation of time reversal on state space, time reversal is a dynamical symmetry. In other words, the meaning of the transformation arises fundamentally from its description as a symmetry. This is not a special or strange result about the meaning of time reversal but a standard, well-motivated feature of the Representation View of symmetry. We discuss it in more detail next.

## 4.2 Invariance and Non-invariance

### 4.2.1 The Symmetry Existence Criterion

The Representation View has novel consequences for the interpretation of dynamical symmetry and asymmetry, which are worth stating clearly. Namely, a representation of a symmetry group plays at least two different semantic roles. The first involves reference: a representation of a spacetime symmetry group allows us to give meaning to those symmetries on state space and refer to them as 'spatial translations', or as 'time translations' or 'time reversal'. The second involves a symmetry property: a dynamical symmetry group is by definition a representation of a temporal symmetry group. So, in the special case of temporal symmetries, a representation plays two roles: it determines the very meaning of those transformations on state space, in addition to determining that they are dynamical symmetries. This second role can be summarised in the following:

(SEC) *Symmetry Existence Criterion.* If a representation of a temporal symmetry group exists, then it is a dynamical symmetry group ('invariance'); if no representation of a group exists, then it is not a dynamical symmetry group ('non-invariance').

A curious consequence is that if a spacetime symmetry group $G$ does happen to be a temporal symmetry group, then the failure of a representation to exist has consequences for reference: not only is there no symmetry, but there is no state space transformation that can meaningfully be interpreted as that spacetime symmetry. When a dynamical symmetry fails, it fails altogether to appear in the state space.

For example, rigid spatial translations often form a temporal symmetry group, as they do in the Poincaré group. This means that our statements about state space automorphisms, like 'the unitary operator $R$ representing rotations', are always made under the implicit assumption that a representation exists. If rotation is not a dynamical symmetry, in that a representation of rotations does not exist, then we have no good reason to view any unitary $R$ as referring to a rotation. In such cases, statements about 'the unitary operator $R$ representing rotations' are not even wrong: they are meaningless.[13]

Because of the dual role that a representation plays in characterising both symmetry and reference, the Symmetry Existence Criterion requires us to reinterpret what it means to *fail* to be a dynamical symmetry. After all, it is common practice to check that a given temporal symmetry is *not* a symmetry of a dynamical system. For example, in the spherically symmetric potential (recall Figure 4.2), a representation of time translations has the property that spatial translations are not generally dynamical symmetries. We often check this by explicitly verifying that an operator resembling spatial translations (such as the unitary $U_s = e^{iaP}$ in quantum theory) fails to preserve the solution space.[14] So, what are we saying when we do this?

When a state space representation of a spacetime symmetry exists, there are often arguments that it is uniquely defined: for example, in Chapter 3, considerations of the Galilei or Poincaré groups allowed us to show that if a representation of time reversal exists, then it is unique. Continuing the language of Section 3.4.3, let me refer to this unique definition as the 'appropriate' representation of a spacetime symmetry. What this means is:

*If* a representation of a temporal symmetry group exists in a dynamical system, *then* it has a particular canonical form that preserves the solution space.

The equivalent contrapositive statement is that: *if* the canonical form of a transformation like rotation or time reversal fails to preserve the solution space, *then* an adequate representation does not exist, and so the dynamical symmetry fails. The difference is subtle: if rotations are not dynamical symmetries, then it is the 'canonical form' of the rotations – but not a representation of them – that fails to preserve the solution space.

As a special case, we have seen above that the very existence of a representation $T$ of the time reversal group element $\tau$ encodes that time reversal is a dynamical symmetry. But, this does not mean that we have somehow illicitly assumed the existence of a symmetry that might not be found in nature. Such constructions should be viewed as conditional statements: that *if* time reversal is a dynamical symmetry *then* its representative $T$ must have a certain canonical form.

---

[13] There is more to say about how this can happen, of course, which will lead us into a discussion of symmetry breaking and symmetry violation: I will return to this in Section 4.5.

[14] Example: the solutions to the Schrödinger equation with spherically symmetric Hamiltonian $H = \frac{1}{2m}P^2 + 1/Q$ are not preserved by the spatial translation operator $U_s = e^{isP}$ when $s \neq 0$.

Thus, when I derive the canonical form of the time reversal operator using an assumption that amounts to time reversal symmetry, Callender (Forthcoming) may rest assured regarding his worry that "this assumption is a large one, for it's up in the air whether quantum mechanics is time reversal invariant". Time reversal invariance is not assumed in this derivation, for there is no guarantee that the time reversal operator exists. If time reversal invariance fails, then the operator $T$ will fail to exist as well. Equivalently, if a representation $T$ of time reversal $\tau$ does exist, then time reversal invariance is sure to hold. This is why we are allowed to assume time reversal invariance in most derivations of the meaning of time reversal, including the original derivation of Wigner (1931, §20).

### 4.2.2 Illustration in the History of Superselection

Although the Symmetry Existence Criterion is not often made explicit in informal discussions of symmetry, it is implicitly adopted by Eugene Wigner and other founders of the Representation View. An illustration of this can be found in one of the interesting physical applications of time reversal. Namely, time reversal appears in the surprising discovery that some pairs of pure quantum states never compose to form a pure state: all of their non-trivial superpositions are mixed states.[15] This phenomenon, known as *superselection*, was first discovered by Wick, Wightman, and Wigner (1952), who derived the so-called fermion or univalence superselection rule: that there is no pure superposition of a pure boson state (of integer-multiple-of-$\hbar$ angular momentum) with a pure fermion state (of half-integer-multiple-of-$\hbar$ angular momentum).[16]

Remarkably, Wick, Wightman, and Wigner made use of the time reversal operator in their argument. So, following the great shock of 1964 when CP symmetry and time reversal symmetry were shown to be experimentally violated, Hegerfeldt, Kraus, and Wigner (1968) wrote a follow-up article entitled "Proof of the Fermion Superselection Rule without the Assumption of Time-Reversal Invariance", which derived the same superselection rule

---

[15] A *state* on a (unital $C^*$) algebra of observables as a positive linear unit-preserving functional $\omega : \mathcal{A} \to \mathbb{C}$. A *mixed* state is one such that $\omega = \lambda\omega_1 + (1 - \lambda)\omega_2$ for some $\lambda \in (0, 1)$ and some states $\omega_1 \neq \omega_2$; otherwise $\omega$ is called *pure*.

[16] According to Wightman (1995, p.752), superselection originated in Wigner's suggestion in the 1940s that some self-adjoint operators are not observables. But as Earman (2008, p.379) points out, all the considerations needed for the fermion superselection rule were already in the Wigner (1931, §20) analysis of time reversal. See in particular Proposition 4.1. See Ruetsche (2004) for a discussion of the subtleties of interpreting mixed states and Earman (2008) for a philosophical appraisal of superselection.

using the spatial rotation operator instead of time reversal. The later authors explained their motivation:

> The fermion or univalence superselection rule, which separates states with integer and half-integer angular momentum, was originally proved under the assumption of time-inversion invariance. Recent experiments on $CP$ violation, combined with the $CPT$ theorem, now seem to question $T$ as a rigorous symmetry. Another proof of the fermion superselection rule without the assumption of $T$ invariance is thus desirable. (Hegerfeldt, Kraus, and Wigner 1968, p.2029)

The very existence of this later article would be puzzling if 'invariance' were taken to mean just 'preservation of the solution space' under an operator with the canonical form of time reversal, since, *the original derivation of fermion superselection did not assume a solution space is preserved.* In the original article, Wick, Wightman, and Wigner (1952, p.103) make no explicit mention of an equation of motion or solution space but simply postulate that "a time inversion operator ... exists" which has the property that $T^2 = -\mathbb{1}$ for a one-fermion system, while $T^2 = \mathbb{1}$ for a one-boson system (see Section 3.4.4). From this one can show that every non-trivial superposition of a boson state $\phi^+$ and a fermion state $\phi^-$ is 'mixed': if $\psi = \alpha\phi^+ + \beta\phi^-$, then $\langle\psi, A\psi\rangle = \lambda\langle\phi^+ A\phi^+\rangle + (1-\lambda)\langle\phi^-, A\phi^-\rangle$ for some $\lambda \in (0,1)$ and for all bounded operators $A \in B(\mathcal{H})$ on the Hilbert space. For completeness, here is a reconstruction of the elegant original argument of Wick, Wightman, and Wigner (1952), in which only the existence of a time reversal operator is assumed.

**Proposition 4.1** (Wick, Wightman, Wigner 1952) *Let $\mathcal{H} = \mathcal{H}^+ \oplus \mathcal{H}^-$ be a direct sum of Hilbert spaces. If there exists a unitary or antiunitary $T : \mathcal{H} \to \mathcal{H}$ such that*

$$T^2\phi^+ = \phi^+ \text{ for all } \phi^+ \in \mathcal{H}^+ \qquad T^2\phi^- = -\phi^- \text{ for all } \phi^- \in \mathcal{H}^-, \qquad (4.5)$$

*then every superposition $\psi = \alpha\phi^+ + \beta\phi^-$ with $|\alpha|^2 + |\beta|^2 = 1$ and $\alpha, \beta \neq 0$ is a mixed state.*

*Proof*  Our assumptions imply that $T^2$ is unitary and commutes with all $A \in B(\mathcal{H})$. Thus,

$$\langle\phi^+, A\phi^-\rangle = \langle T^2\phi^+, T^2A\phi^-\rangle = \langle T^2\phi^+, AT^2\phi^-\rangle = \langle\phi^+, A(-\phi^-)\rangle$$
$$= -\langle\phi^+, A\phi^-\rangle. \qquad (4.6)$$

Hence, $\langle\phi^+, A\phi^-\rangle = 0$. A symmetric argument shows $\langle\phi^-, A\phi^+\rangle = 0$. So, for any superposition $\psi = \alpha\phi^+ + \beta\phi^-$ such that $|\alpha|^2 + |\beta|^2 = 1$,

$$\langle \psi, A\psi \rangle = \langle \alpha\phi^+ + \beta\phi^-, A(\alpha\phi^+ + \beta\phi^-)\rangle,$$
$$= |\alpha|^2\langle\phi^+, A\phi^+\rangle + |\beta|^2\langle\phi^-, A\phi^-\rangle \tag{4.7}$$

for all $A \in B(\mathcal{H})$. Thus, $\psi$ is mixed for all non-trivial ($\alpha, \beta \neq 0$) superpositions. ∎

In this original theorem, there is no mention of an equation of motion, and so there is no solution space to be preserved. Why then does the later article by Hegerfeldt, Kraus, and Wigner (1968) state that the failure of time reversal symmetry invalidates it? I hope you'll agree that these authors, if anyone, are unlikely to be confused about the matter: Wigner in particular is the father of both time reversal and superselection.

The answer, I believe, is that they are implicitly adopting the Representation View on symmetry and are therefore committed to the Symmetry Existence Criterion. On this perspective, although we might not have said anything about the dynamics, the existence of a meaningful representation of the time reversal operator $T$ implicitly assumes that there is one, and so by the Symmetry Existence Criterion it is a dynamical symmetry. Equivalently, the discovery that time reversal invariance fails implies that a representation of the time reversal operator does not exist! This explains how the 1964 discovery of time reversal symmetry violation might have led Wigner and others to rethink the fermion superselection rule.

As the astute reader may note, the assumptions of the operator $T$ in Proposition 4.1 are also satisfied by the CPT operator $\Theta$ constructed by Jost (1957, 1965), which is guaranteed to exist in a very general sense by the CPT theorem.[17] So, although Wick, Wightman, and Wigner (1952) would not have known the possibility of replacing $T$ with $\Theta$, the analysis of rotations by Hegerfeldt, Kraus, and Wigner (1968) is not needed to restore the fermion superselection rule: the CPT operator $\Theta$ could have been used instead. However, the later result derived from rotational symmetry is still an independently interesting study of rotations and superselection. And, this seems to have been the main motivation for the authors of the second article.[18]

---

[17] See Chapter 8.

[18] In personal correspondence, Hegerfeldt tells me that in 1967, he and Kraus had written a paper on why a spin system's change of phase under a global rotation through $2\pi$ is unobservable, but which was rejected by a referee. Wigner was the leading expert on symmetry at the time. Hegerfeldt recalls, "we had never met him, but, young as we were, sent him a copy of our paper and of the referee report and asked for his advice". Wigner wrote back with extensive advice, which led the paper to eventually be published, and their correspondence afterwards led to the publication of Hegerfeldt, Kraus, and Wigner (1968). Charmingly, years later in 1980, Hegerfeldt reports that he

## 4.3  Direct Empirical Significance

With a clear expression of dynamical symmetry and time reversal symmetry in hand, it is possible to add some clarity to the three philosophical discussions of symmetry introduced at the outset of this chapter and to illustrate the role that time reversal plays. We begin in this section with the question of how we come to know about a dynamical symmetry and a review of the literature on direct empirical significance; I will turn to the second and third discussions in Sections 4.4 and 4.5.

Direct empirical significance has been explored as a way of clarifying the epistemology of symmetry: by interpreting symmetries as transformations of a subsystem with respect to its environment. For example, Galileo (1632) famously noticed that experiments done in an isolated cabin below deck on a ship proceed in the same way whether the ship is at rest or in uniform motion. This motivates the Principle of Relativity, that the laws of physics are the same in every inertial reference frame. However, Galileo did not arrive at his conclusion through the impossible task of boosting the universe. He arrived at it by observing a ship, which was in motion with respect to the sea. By breaking the boost symmetry with a comparison to the sea – for example, by looking through the porthole of the cabin – we provide a means of knowing that the two states of motion are actually distinct.

This observation led Kosso (2000) and Brading and Brown (2004) to postulate that two components are needed in order to explain how we know about symmetries: (1) a subsystem (the ship's cabin) on which a transformation is a dynamical symmetry; and (2) an environment (the sea) with respect to which the transformation is not a symmetry. Although two symmetry-related descriptions may be empirically indistinguishable within the subsystem, our ability to distinguish them in relation to the environment explains how we come to know about the symmetry empirically. The existence of these two components is called the *direct empirical significance* of a symmetry or of a symmetry group.

Direct empirical significance has been discussed for rotations, spatial translations, and boosts, as well as for tricky cases like gauge symmetries.[19] But, I am not aware of any comment about the empirical significance of time translations, let alone discrete symmetries like time reversal. The Representation View offers a helpful perspective on both. I will argue here

---

met a physicist on a bus going to a conference in Mexico, and that this physicist revealed that he "had been particularly happy about our paper because he had been the referee on that paper, had rejected it, but had been overruled". (Hegerfeldt, personal correspondence) that there are situations in which both time translation and time reversal have direct empirical significance. However, for discrete transformations like time reversal, the empirical significance piggy-backs on that of time translation symmetry. I will use the Fregean phrase *higher-order* to refer to this kind of direct empirical significance, for reasons that I will soon explain.

[19] Compare Gomes (2019, 2021a,b), Greaves and Wallace (2014), Healey (2007, Chapters 6–7), and Teh (2016).

We begin by adopting the Representation View, in particular on what it means to have a state space representation of time translations.[20] Thus, time translation symmetry is what captures the repeatability of scientific experiments in time. If a description is empirically supported in the lab on Tuesday, then further confirmation may be found on Wednesday by repeating the same experiment. Formally speaking, this is expressed by the postulate that the group of time translations $\mathbb{T}$ can be represented among the automorphisms of a state space $\varphi : \mathbb{T} \to \text{Aut}(M)$. I will adopt the simplest case of the time translation group, $\mathbb{T} = (\mathbb{R}, +)$.

In order to know empirically that an experiment has been repeated at different times, we need a way to compare those times, just as with Galileo's ship. One way to do this is to introduce a 'clock' into the environment. A clock can be a mechanical device that keeps time in the familiar way, or the changing brain state or age of an experimenter that perceives time's passage. This is not just an informal notion: there are formal accounts of clocks and general time-keeping environments, known as 'time observables', which allow the precise modelling of this situation. A time observable is one that keeps track of the parameter $t$ associated with time translations. For example, in quantum systems, a time observable $\mathcal{T}$ is one that tracks the unitary representation of time translations $U_t = e^{-itH}$, in that for any curve $\psi(t) = U_t\psi_0$ with initial state $\psi_0$ and initial expected time $t_0 = \langle \psi_0, \mathcal{T}\psi_0 \rangle$, we have that $\langle \psi(t), \mathcal{T}\psi(t) \rangle = t_0 + t$. There is a great deal of debate over the nature and existence of time observables, which intersects the subtle question of what counts as an 'observable'.[21] However, it is enough for my purposes that in most cases, a time observable system is well-defined.

Given an environment with a clock system in this sense, we can give direct empirical significance to time translations in a way that is analogous to Galileo's ship: instead of comparing the ship to the sea, we compare it to the readings of a clock, as in Figure 4.4. When the initial conditions of the (ship) subsystem are restored from one day to the next, we can confirm that the same behaviour ensues. This can be modelled using a representation of time translations among the automorphisms of the subsystem state space and thus justifies referring to them as symmetries. But, the clock subsystem is not restored: this captures the sense in which an observer can compare the two experiments as distinct. Note that the problem of defining the inverse time translations is not necessarily intractable: they can be defined operationally using a clock running in the reverse direction.[22] Thus, in the same way that the sea environment gives direct empirical significance to boosts, a clock environment gives direct empirical significance to time translations.

---

[20] See Section 3.1.2.

[21] It is well-known that such a time observable $T$ is not generally self-adjoint; however, it can still be modelled as a maximal symmetric or POVM observable. See Busch, Grabowski, and Lahti (1994), Galapon (2009), or Pashby (2014) for an overview, Hegerfeldt and Muga (2010) for a particularly general result, and Roberts (2014, 2018) for my own perspective.

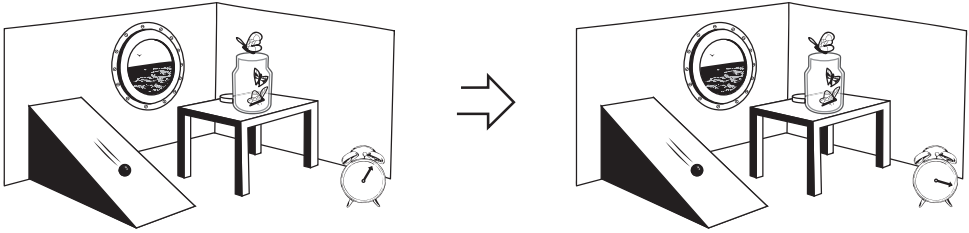

Figure 4.4  A clock system defines the direct empirical significance of time translations.

The direct empirical significance of time reversal is subtly different from this. I have argued that the most natural way to understand discrete symmetries like time reversal is as an automorphism of a continuous group of symmetries (Section 2.6). For example, time reversal is an automorphism of time translation symmetry, spatial reversal is an automorphism of spatial translation symmetry, and so on. In other words, a discrete symmetry is a 'symmetry of a symmetry'. The time translations $\mathbb{T}$ describe a structural property of time; but, since time reversal symmetry describes a structural property of $\mathbb{T}$, it is really a 'property of a property' of time. This is what I mean when I say that it is 'higher-order'.

Being higher-order does not make time reversal any less directly empirically significant. When it exists, the time reversal automorphism is a property of the group of time translations, which themselves have direct empirical significance. For example, if we are able to grasp that the group structure of time translations is $\mathbb{T} = (\mathbb{R}, +)$ in some context, then we can immediately deduce that it has a time reversal automorphism $t \mapsto \tau t \tau = -t$. So, if we have direct empirical access to time translations, then our empirical access to time reversal symmetry is a 'free lunch'. This higher-order property is analogous to Frege's example[23] that 'being unique' is a higher-order property of 'being a moon of the Earth', which is itself directly empirically accessible through the study of 'the Moon'. And, of course, a similar interpretation is available for other discrete symmetries like parity and parity–time reversal as well.

---

[22] See Section 2.2, Figure 2.3. From the perspective of time observables, this corresponds to the fact that for every time observable $\mathcal{T}$, there is a 'reversed' time observable $-\mathcal{T}$.

The argument I am giving here can be summarised as follows.

1. If a group of symmetries has direct empirical significance, then all of its properties have direct empirical significance too.
2. The group of time translations has direct empirical significance, and one of its properties is time reversal symmetry.
3. Therefore, *time reversal symmetry has direct empirical significance too*.

The first premise is a condition of adequacy for a reasonable account of direct empirical significance. The second premise follows from the arguments I have given above: time translation symmetries are directly empirically accessible through the study of a subsystem with respect to a clock environment; and, time reversal is a property of the group of time translations, since it is by definition one of its automorphisms. It follows that time reversal has direct empirical significance. Again, a similar argument can be made about parity and parity–time reversal, too.

Note that I am *not* assuming that a symmetry with 'direct empirical significance' can necessarily be operationally implemented. In most contexts, there is no way to continuously transform a system to its time reverse: in the full Poincaré group, the time reversal operator $\tau$ is not continuously connected to the identity, which makes such a continuous transformation impossible. But, other discrete transformations, like the parity–time transformation, are continuously connected to the identity in the universal covering group of the Poincaré group (discussed in Chapter 8). Moreover, I find it helpful to separate the question of whether we can 'operationally implement' a symmetry from the question of whether we have direct empirical access to it. Direct empirical significance is about whether we can know about a symmetry through the comparison of a subsystem to its environment. In the case of time reversal, we can.

## 4.4 The Symmetry-to-Reality Inference

Leibniz wrote in his third letter to Clarke that the universe would be left unchanged by a transformation that involves only "changing East into West" (Leibniz and Clarke 2000, p.14). This famously inspired Earman and Norton (1987, p.522) to define 'Leibniz Equivalence' in general relativity to mean that, "[d]iffeomorphic models represent the same physical situation", where 'diffeomorphic models' are those related by a symmetry. The *symmetry-to-reality inference* is a generalisation of these principles, which states: if two descriptions are related by a symmetry, then they refer to the same physical situation. The term was coined by Dasgupta (2016, p.840), who formulates it as the statement, "if a putative feature is variant in laws that we have reason to think are true and complete, then this is some reason to think that the feature is not real". More standard usage reformulates this in an equivalent form: if a feature of a theory correctly describes reality, then it is invariant under the symmetries of that theory.

---

[23] Cf. Frege (1884, §53).

Although some commentators have suggested that the symmetry-to-reality inference is always or nearly-always justified, most argue that it is not.[24] In this section I would like to add two additional cautionary remarks about situations in which the symmetry-to-reality inference is either ill-posed or invalid. It is ill-posed in cases where a physical description is mentioned without specifying its reference. And, when the reference of a description is well-specified, there is still a sense in which it is invalid, just as it is invalid to infer that the Klein four-group has only one subgroup of order two. Time reversal will provide one counterexample to the claim.

The first cautionary remark stems from a more general comment about language-to-reality inference, in the influential essay on "Sense and Reference" by Frege (1892). Frege described a language community's common understanding of how to use a proper name as the 'sense' of the name and the object in reality that the name refers to as its 'reference' or 'meaning' (*Bedeutung*). He gave a famous example of this, 'the morning star' and 'the evening star', which he took to have different senses but the same reference: the planet Venus. His cautionary remark about sense and reference was the following.

It may perhaps be granted that every grammatically well-formed expression representing a proper name always has a sense. But this is not to say that to the sense there also corresponds a reference. The words 'the celestial body most distant from the Earth' have a sense, but it is very doubtful if they also have a reference. The expression 'the least rapidly convergent series' has a sense but demonstrably has

---

[24] Something like this inference appears in Baker (2010), Fletcher (2018), and Weatherall (2018); see also Baker (2020, "What are symmetries?" In: Unpublished Manuscript, http://philsci-archive.pitt.edu/18524/, where it is true by definition. In contrast, Belot (2013, p.330) gives a number of examples showing the symmetry-to-reality inference (which he calls 'D2') is "false if understood as a thesis concerning classical symmetries of differential equations". Caution is similarly advised by Butterfield (2021), De Haro and Butterfield (2019), Møller-Nielsen (2017), Read and Møller-Nielsen (2020), and myself (Roberts 2020).

no reference, since for every given convergent series, another convergent, but less rapidly convergent, series can be found. In grasping a sense, one is not certainly assured of a reference. (Frege 1892, p.28)

The same is often true of models in physics. Suppose we are speaking not just of proper names but about whole sentences, and indeed about whole mathematical structures containing sentences, like a model of a quantum system. A model in physics often has sense without reference, which makes any symmetry-to-reality inference impossible. To take just one example: physics students study all manners of facts about the simple harmonic oscillator as a dynamical system – its symmetries, energy levels, basis representations, perturbative approximations, and many other things – without saying anything about which physical system is being referred to. Indeed, part of the power of the harmonic oscillator is that it can refer to countless different things: it is the first-order approximation of any system described by a locally-characterised (meaning, analytic) force.[25] In the use-mention jargon of Quine (1940, §4), the textbook exercises about the harmonic oscillator might thus be viewed as 'mentioning' their use in situations where they have reference, rather than 'using' them as such. De Haro and Butterfield (2018) draw a similar conclusion, pointing out that symmetries (and especially dualities) are often defined in the context of uninterpreted 'bare theories'.

Thus, when the reference of a dynamical system is not fixed, there may be no answer to the question of whether symmetry-related descriptions refer to the same thing: in such cases, the question is not well-posed. So, let us suppose the reference of a given physical description *is* well-posed. I claim that the symmetry-to-reality inference is still not generally valid. This leads to the second remark.

Dasgupta (2016, p.840) formulates the symmetry-to-reality inference as applying to any "putative feature" of reality as it appears in a theory. At this level of generality, there are simple counterexamples to the inference. For example, consider the description of a free Newtonian particle. To avoid formulating an ill-posed question, suppose that each possible solution in state space is identified with some possible trajectory for a physical particle in the absence of forces, each with a different (constant) velocity, as in Figure 4.5. Suppose also that a 'putative feature' of reality is associated with one of the trajectories with non-zero velocity. Is it preserved by dynamical symmetries, as the symmetry-to-reality inference requires?

---

[25] Recall that a harmonic oscillator is a dynamical system with a quadratic potential $U = ax^2$ arising from a linear force $F(x) = -\nabla U = -2ax$. Its widespread usefulness is explained by the fact that if $F'(x)$ is any other force that is 'locally characterised' in the sense of being an analytic function, then $F'(x)$ has a Taylor expansion and is therefore linear in its first-order approximation. In other words: the harmonic oscillator is a first-order approximation of every locally-characterised force.

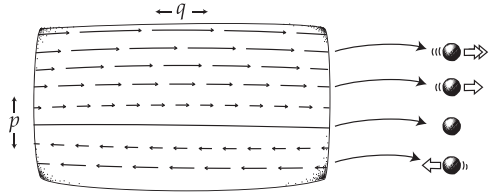

Figure 4.5 Choosing the 'reference' of a dynamical system by associating each curve with a possible motion.

The answer is, 'no'. For example, a representation of parity, which transforms positions and velocities as $\pi : (x, \dot{x}) \mapsto (-x, -\dot{x})$, is a dynamical symmetry of this system.[26] But, it does not preserve each individual curve. A curve is rather transformed to one with opposite velocity, 'reflecting' the phase space diagram in Figure 4.5 about both axes. So, this 'putative feature' of reality is not preserved by a symmetry.

One might respond by observing that, by refining the symmetry-to-reality inference, it is possible to avoid this counterexample. Namely, instead of applying the inference to arbitrary 'putative features' of reality, like individual trajectories, suppose we apply it to the space of all trajectories as a whole. That is, consider the following 'global version' of the symmetry-to-reality inference: that *if two representations of time translations are related by a dynamical symmetry, then they refer to the same physical time translations.*

This formulation of the inference uses 'reference' in a way that would require further analysis, in order to say what it means for a group of time translations (or equivalently, all the possible trajectories at once) to 'refer' to a physical time translation. One might take inspiration from what Dewar (2019) calls *sophistication about symmetries* and view the collection of all symmetry-related trajectories as being what picks out a description of reality, rather than any single trajectory in the solution space.[27] As Dewar writes:

In a slogan, the idea is that we need not insist on finding a theory whose models are invariant under the application of the symmetry transformation, but can rest content with a theory whose models are isomorphic under that transformation. (Dewar 2019, p.298)

---

[26] As a quick check, write the solutions of a classical particle as $x(t) = \dot{x}_0 t + x_0$ for each initial state $(x_0, \dot{x}_0) \in \mathbb{R}^2 \times \mathbb{R}^2$. Then a parity-transformed trajectory has the form $x^{\pi}(t) = -\dot{x}_0 t - x_0$, which is again a solution, with initial state $(-x_0, -\dot{x}_0)$.

[27] For an evaluation of sophistication, see Martens and Read (2020).

This does allow one to avoid the counterexample of parity. For the free particle, parity preserves the representation of time translations as a whole, $\pi \varphi_t \pi^{-1} = \varphi_t$. Spatial translation, spatial rotation, and time translation symmetry do as well. So, whatever the group of time translations $t \mapsto \varphi_t$ refers to, it is preserved by all these symmetries.

Unfortunately, there is a counterexample to this 'global' version of the inference too, which is the case of time reversal symmetry. Recall from Section 4.1.3 that every dynamical symmetry either preserves each individual time translation, $\varphi_t \mapsto \varphi_t$, or reverses it, $\varphi_t \mapsto \varphi_{-t}$. Time reversal is an example of the second case: it does not generally preserve the group of time translations as a whole.

Of course, viewed as a group, the time-reversed representation $t \mapsto \varphi_{-t}$ is isomorphic to the original, since time reversal is, after all, an automorphism. But, this does not mean that the representations of that group are the same. Recall the example of the Klein four-group at the outset: there is just one group of order two but multiple representations of it amongst the subgroups of the Klein four-group. Similarly, when time reversal is a dynamical symmetry, we have two representations of the time translation group $(\mathbb{R}, +)$, given by $t \mapsto \varphi_t$ and $t \mapsto \varphi_{-t}$. Although they are isomorphic, it would be wrong to say that they are one and the same.

One could try to save the symmetry-to-reality inference with yet more sophistication about symmetries. Or, one might replace the time translation group $(\mathbb{R}, +)$ with the quotient group $(\mathbb{R}, +)/\mathbb{Z}_2$ of equivalence classes of time translations of the form $(t, -t)$. If time translations are representations of that group, then time reversal would be what Caulton (2015) calls an 'analytic symmetry' and could not refer to different physical situations. However, the choice of structure to describe time translations has physical content, and so one should be wary of what this commits us to. For example, by adopting the quotient group $(\mathbb{R}, +)/\mathbb{Z}_2$ for time translations, we can only describe physical systems that evolve forwards and backwards in time in exactly the same way. In most dynamical systems in physics, this does not describe any realistic physical system, suggesting the approach is on the wrong track.

## 4.5 The Spacetime–Dynamical Symmetry Relationship

Our last philosophical discussion is about whether every spacetime symmetry is a dynamical symmetry, and vice versa. The tight relationship between spacetime and dynamical symmetries played a pivotal role in the development of spacetime physics, at least since Minkowski (1908, §II) argued that the symmetries of Lorentz's dynamical theories of matter "force this changed conception of space and time upon us". Earman (1989, §3.4), following Stein (1977, p.6), gave an influential discussion of that relationship, as part of his approach to interpreting spacetime theories.[28] His proposal was formulated in terms of two postulates:

(SP1)  A dynamical symmetry is also a spacetime symmetry. And
(SP2)  A spacetime symmetry is also a dynamical symmetry.

Earman himself viewed these principles as "conditions of adequacy on theories of motion" (Earman 1989, p.46). But, an interpretation of spacetime due to Brown and Pooley (2006) and Brown (2005) included an explanation of why this might be so, by arguing that the ghostly concept of 'spacetime structure' actually just encodes (a philosopher might say 'supervenes on') facts about the dynamics of matter fields. Most commentators have emphasised that this is unlikely to be the entire story, with dynamical systems inevitably including some spacetime concepts in their formulation.[29] I agree: as I argued in Section 2.3, a dynamical system must at least admit a representation of time translations, in order to justify being called 'dynamical'. However, many go beyond viewing Earman's principles as adequacy conditions and treat them as truths. Myrvold (2019) even argues that they are analytic truths, or true by the very meaning of the words they contain:[30]

Both connections between dynamical symmetries and spacetime symmetries are analytic. To say that, for test bodies, there is a real distinction between motion along geodesics of the metric and motion that is forced away from geodesics, is to simultaneously refer to a fact about the dynamics of these bodies and their spacetime environment. To say that the field equations do (or do not) require a flat background spacetime is to simultaneously refer to a fact about these field equations and their spacetime setting. (Myrvold 2019, p.142)

The Representation View leads to a different conclusion: both (SP1) and (SP2) can fail, in ways that are important for modern physics. So, these principles are not analytic truths. However, the failure of (SP2) suggests that we should make some interesting adjustments to the mathematical structures we use to describe reality. In this case, Earman was correct: (SP2) is an adequacy principle for good scientific theorising. I will discuss each principle in turn.

---

[28]  That context was the following. We first view a manifold for a spacetime theory as admitting two kinds of smooth (tensor or spinor) fields, the 'absolute' fields describing the structure of spacetime itself and the 'object' fields describing the material contents of spacetime, where the latter evolve according to some dynamical laws. A spacetime symmetry is then a transformation of the absolute fields, while a dynamical symmetry is a transformation of the object fields; both are defined to leave the theory's space of models invariant (Earman 1989, p.45). The approach to dynamical symmetries that I have set out in Section 4.1.2 is compatible with this.

[29]  Cf. Lehmkuhl (2011), Maudlin (2012), Myrvold (2019), Norton (2008b), and Knox (2019).

[30]  Similar conclusions are drawn by Acuña (2016, p.2) and Knox (2019, p.124).

### 4.5.1 Dynamical Symmetries that Are Not Spacetime Symmetries

If a dynamical theory can be formulated on any interesting class of general relativistic spacetimes, then there is a sense in which (SP1) is false. To see this, let me first recall some definitions. For $G$ to be a spacetime symmetry group, its elements must preserve spacetime structure, the way that isometries preserve the structure of a Lorentzian manifold. For $G$ to be a temporal symmetry group, it must preserve the time translations $\mathbb{T}$ when it acts on them by conjugation (Definition 4.1). Finally, for $G$ to be a dynamical symmetry group on some state space, it must both be a temporal symmetry group and admit a representation on state space (Definition 4.2).

Consider now a generic relativistic spacetime with no isometries at all. This is the case for most spacetimes describing realistic gravitational phenomena. Then, of course, 'time translations' cannot be spacetime symmetries, no matter what we take them to be. But, the Representation View still requires that when we formulate a dynamical theory, we must adopt a representation of *some* time translations, although they might be local.[31] Otherwise, the theory is not deserving of the name 'dynamical'.

So, let $\mathbb{T}$ be a structure representing time translations, and suppose that it admits a representation on state space. To make the example simple, assume that the spacetime admits a smooth, complete, timelike vector field, describing the passage of time in some reference frame, and let $\mathbb{T} = (\mathbb{R}, +)$ be the group of time translations that flows along it. Then, $G = \mathbb{T}$ satisfies our definition of a dynamical symmetry group: it is a special case of a temporal symmetry group,[32] and it admits a representation. But, it is not a spacetime symmetry group. So, in any dynamical theory formulated on generic relativistic spacetimes (cf. Wald 1994), there are dynamical symmetries that are not spacetime symmetries.

---

[31] Non-orientable spacetimes by definition do not admit a smooth timelike vector field; in such cases, time translation may still be modelled in local neighbourhoods using a *local* Lie group, such as one associated with $(\mathbb{R}, +)$ in a neighbourhood of the identity; this was our strategy in local symplectic mechanics (Section 3.3.1).

[32] For $G$ to be a temporal symmetry group of $\mathbb{T}$, the elements of $G$ must preserve $\mathbb{T}$ when acting on it by conjugation, $gtg^{-1} \in \mathbb{T}$. This holds trivially when $G = \mathbb{T}$ by the group closure axiom.

### 4.5.2 *Spacetime Symmetries that Are Not Dynamical Symmetries*

The previous symmetry principle fails in the context of gravitation, where spacetimes can be expected to have no non-trivial symmetries. In contrast, (SP2) can fail even for local physics: there are reasonable models in which not every local spacetime symmetry is a dynamical symmetry! So, I do not see any sense in which this principle is analytic. However, when (SP2) fails, working to restore it usually leads to interesting interpretive adjustments to a theory. So, it is not unreasonable to side with Earman and view (SP2) as an adequacy principle for good physical theorising.

There are three ways that (SP2) can fail. Suppose that $G$ is a spacetime symmetry group, and recall that $G$ is a dynamical symmetry group only if it is both a temporal symmetry group and admits a representation on state space. This can fail to occur in three different ways. I will just summarise them here and then discuss their details in the next subsections:

1. *No temporal symmetries*. A spacetime symmetry group $G$ might not be a temporal symmetry group, meaning that it might not preserve the structure of time translations.

2. *No representation of temporal symmetries*. Even if $G$ is a temporal symmetry group, and even if a dynamical theory is given by a representation of time translations, there may be no non-trivial extension of that representation to all of $G$.

3. *No 'adequate' representation of temporal symmetries*. The group $G$ might be a dynamical symmetry group in some representation, but the representation still may not be 'appropriate' given the interpretive constraints associated with a state space, where I use the word 'appropriate' as in Section 4.2.1.

The first two possibilities are already recognised in some corners of the quantum field theory community, where the Representation View is perhaps most well-known. For example, Bogolubov, Logunov, Oksak, and Todorov (1990, p.375) write: "the term 'non-invariance' can mean that either the specified symmetry of the field algebra does not exist, or such a symmetry is not realizable (anti-)unitarily [in a representation]". The third possibility is not a 'strict' failure of (SP2) but is reason enough to doubt it. Let me explain each of these cases in more detail.

### 4.5.3 *No Temporal Symmetries*

A group of spacetime symmetries $G$ does not necessarily preserve the structure of time translations, in which case it is not a dynamical symmetry group. One has some freedom in choosing what we mean by time translations when we formulate a dynamical theory. I have already given one example of this in my discussion of the Lorentz boosts (Section 4.1.2): let $G$ be the group of symmetries of Minkowski spacetime, the Poincaré group. Suppose one chooses a smooth timelike vector field associated with a choice of reference frame and identifies a group of time translations $\mathbb{T} = (\mathbb{R}, +)$ associated with the flow along that reference frame. This is a typical choice in dynamical theories formulated in terms of the Euler–Lagrange equations, and in Hamiltonian's equations as well.

However, on this definition of time translations, Lorentz boosts are spacetime symmetries that are not temporal symmetries. In simple terms, this is because time translations are associated with a fixed reference frame, which is not preserved by velocity boosts. In the context of the Euler–Lagrange equations and Hamilton's equations, this arises in the well-known fact that boosts do not, strictly speaking, preserve the solutions to these dynamical equations. So, boosts are spacetime symmetries, but they are not dynamical symmetries because they fail to even be temporal symmetries.

An appropriate response to this issue is to expand the choice of time translation symmetries in a way that is more appropriate for the Poincaré group. As I showed in Section 4.1.2 (and especially Figure 4.3), choosing time translations to be the larger group that flows along any timelike geodesic restores boosts as temporal symmetries. This temporal symmetry group is preserved by all the spacetime symmetries, and so (SP2) is restored.

In summary, spacetime symmetries can fail to preserve our definition of time translations, in which case they are not dynamical symmetries. This failure of (SP2) is possible because of our freedom to choose what we mean by time translations. We can avoid that failure by introducing a further postulate: that our definition of time translations *must* be preserved by all the spacetime symmetries. However, there remain other ways that (SP1) can fail, which I turn to next.

### 4.5.4 No Representation of Temporal Symmetries

The second way that a spacetime symmetry can fail to be a dynamical symmetry is when no representation exists. Let us set aside the previous concern and assume in what follows that our group of spacetime symmetries preserves time translations, at least at the level of spacetime structure. That is, let us suppose that the spacetime symmetry group $G$ is a temporal symmetry group. The problem is that the structure of state space may still prohibit any non-trivial representation of the spacetime symmetry group, in which case it fails to be a dynamical symmetry group. I will begin with a simple example of this in the case of Hamiltonian mechanics and then argue that the failure of (SP2) is indeed essential to the phenomenon of spontaneous symmetry breaking.

### Failure in Hamiltonian Mechanics

Let me begin with a simple example to illustrate. Suppose spacetime has the symmetries of the complete Galilei group, including time reversal. So, the time translations $\mathbb{T} = (\mathbb{R}, +)$ are spacetime symmetries as well. Suppose our dynamical theory is Hamiltonian mechanics, formulated on a symplectic manifold $(M, \omega)$ as in Section 3.3.1. Strictly speaking, the automorphisms of this state space are the symplectomorphisms, which are the transformations that preserve both $M$ and $\omega$: namely, they are diffeomorphisms, $\phi : M \to M$, that identically preserve the symplectic form, $\phi^* \omega = \omega$, where $\phi^*$ denotes the pull-back of a diffeomorphism $\phi$. So, a representation of time translations in this context is a homomorphism $\varphi : (\mathbb{R}, +) \to \mathrm{Aut}(M, \omega)$, where each time translation $\phi_t \in \mathrm{Aut}(M, \omega)$ is a symplectomorphism.

Now, the time reversal group element $\tau$ is a spacetime symmetry as well. But, if the Hamiltonian for these time translations (representing energy) is bounded from below but not from above, as is usually the case, then it turns out to be impossible to represent it as a symplectomorphism: Proposition 3.1 implies that it must be antisymplectic. The problem, roughly speaking, is that the structure of a symplectic manifold as a state space includes an orientation.[33] This makes it impossible to extend the representation of time translations to one that includes time reversal. So, strictly speaking, time reversal fails to be a dynamical symmetry of symplectic mechanics, even when it is a spacetime symmetry!

One might simply stop there and conclude that classical Hamiltonian mechanics is time reversal symmetry violating. However, a more natural response is to change what we mean by a state space automorphism: by extending the automorphisms to include both symplectic and *anti*symplectic transformations, we can again view time reversal as a dynamical symmetry.[34] In this sense, the failure of (SP2) here is a flag that the structure of state space might need to be reexamined, which leads to an interesting revision of the theory.

This revision is not an analytic truth: it is not *meaningless* to assert that the automorphisms of a state space are symplectomorphisms and also that the symmetries of spacetime are given by the complete Galilei group. If anything, Earman's interpretation of (SP2) as a 'condition of adequacy' is most appropriate: (SP2) gives some indication that anti-symplectic transformations might be automorphisms of classical Hamiltonian mechanics as well.

---

[33] A symplectic form is commonly interpreted as the 'oriented area' of the parallelogram determined by two vectors at a point. See e.g. Arnol'd (1989, §7(C)).
[34] This is the approach taken in Section 3.3.

### Failure for Spontaneous Symmetry Breaking

Another more well-known example in which spacetime symmetries are not dynamical is the case of spontaneous symmetry breaking. Symmetry breaking is a cornerstone of quantum field theory, including the Higgs mechanism, for which degenerate ground states in an electroweak interaction are related by an $SU(2) \times U(1)$ symmetry. But, my point can be made in a simple toy example that is familiar to philosophers of physics: the half-infinite Ising lattice.[35] Let me sketch the structure of that system.

Let the spacetime symmetries be described by the Poincaré group, although any spacetime with rotational symmetries will do. To construct our state space, we will make use of the half-infinite *Ising lattice*, which is an infinite chain of spin-1/2 systems, each imagined to be spatially separated and labelled by the natural numbers, $n = 1, 2, 3, \ldots$, as in Figure 4.6. The algebra of observables describing this system is the Pauli algebra for a countably infinite number of spin systems: that is, for each spin-system $n = 1, 2, 3, \ldots$, we have three self-adjoint operators $\sigma_x^n, \sigma_y^n, \sigma_z^n$ satisfying the Pauli relations, and where any two observables associated with different spin systems on the chain are assumed to commute. The state space for this system is then a representation of the Pauli algebra amongst the operators on a Hilbert space $(\mathcal{A}, \mathcal{H})$. The Hilbert space $\mathcal{H}$ is required to be 'separable', meaning that its dimension is countable; this ensures that some of the basic interpretive structures of quantum theory can be applied.[36]

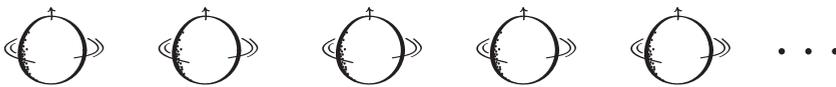

Figure 4.6 A half-infinite Ising lattice.

One of the surprising but well-known features of the half-infinite Ising lattice is that, when we reverse two of the coordinate axes and build a new representation of the Pauli algebra, corresponding to a global rotation in space through the angle $\pi$, the result is not 'unitarily equivalent' to the original. This is what it means to say that spontaneous symmetry breaking occurs. More precisely, calling the first representation $(\mathcal{H}^+, \mathcal{A}^+)$ and the second representation $(\mathcal{H}^-, \mathcal{A}^-)$, unitary inequivalence means that there is no unitary map $R : \mathcal{H}^+ \rightarrow \mathcal{H}^-$ such that $A \mapsto RAR^{-1}$ is a bijection from $\mathcal{A}^+$ to $\mathcal{A}^-$. The structure of this state space prohibits unitarily equivalence.[37]

For our purposes, what is significant about this fact is that, if a representation of the spatial rotation group exists, then it *must* express unitary equivalence: if rotations were not unitary, then the representation would not consist in automorphisms of state space (see Section 3.4.2); and, if it did not relate spatially rotated descriptions, then it would not be homomorphic to the rotation group. So, the half-infinite Ising lattice is a system in which *no representation of the rotation group exists*. As a result, rotations cannot be dynamical symmetries, no matter what we take the dynamics to be, in spite of the fact that they are spacetime symmetries. Thus, we get another failure of (SP2).

A classic response to this situation is to argue that the appropriate state space for quantum field theory is fundamentally the abstract $C^*$ algebra defining the Pauli relations, rather than a Hilbert space representation of this algebra, as is suggested by Segal.[38]

The proper sophistication, based on a mixture of operational and mathematical considerations, gives however a unique and transparent formulation within the framework of the phenomenology described; the canonical variables are fundamentally elements in an abstract algebra of observables, and it is only relative to a particular state of this algebra that they become operators in Hilbert space. (Segal 1959, p.343)

I will not take a position on this view. The lesson that I would like to draw is that, in the presence of spontaneous symmetry breaking, the status of (SP2) can again be used to motivate rethinking the structure of state space. As Segal suggests, the choice is not an analytic truth but requires subtle philosophical as well as 'operational and mathematical' considerations. Again, (SP2) is not analytic but rather an interesting adequacy condition for theorising. Indeed, when we turn to the discussion of dynamical symmetry violation in Section 7.1, I will argue that with a bit of empiricism in mind, it can still be reasonably applied.

---

[37] In this construction, that structure can be viewed as arising from the Hilbert space not having 'enough' basis vectors for an infinite system. One would like to view each spin site as having two orthogonal basis vectors representing 'spin-up' and a 'spin-down' with respect to some axis. But, with a countable set of spin sites, this would mean that there are $2^{\aleph_0}$ basis vectors, and the Hilbert space would not be separable. To avoid this, one can generate a basis set by beginning with the basis vector in which every spin site is 'up' in some direction and then introducing the countable set of basis vectors that are 'down' at only finitely many sites. In contrast, the basis set for a rotated representation, only a finite number of spin sites are 'up'. As a result, global polarisation becomes a representation-dependent property, and unitary equivalence fails. For details, see Sewell (2002, §2.2.3).
[38] See Feintzeig (2015, 2018) for a recent defence of this view.

### 4.5.5 No Adequate Representation

The final example of how (SP2) can fail involves an issue that is commonly referred to as 'symmetry violation'. It might happen that we have a spacetime symmetry that preserves time translations, and even have a representation of that spacetime symmetry. But, our representation might not be 'appropriate' given the interpretive constraints of our theory.

To illustrate what it means to have an 'inappropriate' representation, recall the example from Section 3.2.4 of time reversal for the free particle. For this system, an 'appropriate' representation of the time reversal group element $\tau$ is given by $T(x, \dot{x}) := (x, -\dot{x})$. But, we can also construct an 'inappropriate' transformation $\tilde{T}$ defined by $\tilde{T}(x, \dot{x}) := (-x, \dot{x})$. Our previous calculation revealed[39] that $\tilde{T}$ provides a representation of time reversal. As a result, $\tilde{T}$ reverses each trajectory in the same way that time reversal does, $x(t) \mapsto x(-t)$. But, most would consider this $\tilde{T}$ to be an appropriate representation of space-and-time reversal, and not of time reversal alone. This is what I mean by the statement that $\tilde{T}$ is an inappropriate representative of time reversal.

Potential trouble for (SP2) arises when we have *some* representation of a temporal symmetry but no 'appropriate' representation of it. For example, in the context of quantum theory, it is possible to show that when energy is half-bounded, there is always *some* representation of time reversal: this was the first part of Proposition 3.4 in Chapter 3. In spite of this, it is not generally possible to find a time reversal operator that reverses 'time and only time'. That is what happens when time reversal symmetry violation occurs, and it is known to occur in the theory of electroweak interactions. In such cases, we have no appropriate representation of time reversal, even when time reversal is a symmetry of the background spacetime.

There is a reasonable response to this too. Instead of rejecting (SP2), we might take the discovery of time reversal symmetry violation to imply that we should revise our spacetime structure: since time reversal is a symmetry of Minkowski spacetime, then perhaps Minkowski spacetime by itself does not accurately describe local spacetime structure.

---

[39] See especially Eq. (3.10) of Section 3.2.4.

I agree with this conclusion. However, there remains a gap in the argument, which is to establish why time reversal symmetry violation requires a revision of spacetime structure and not just the rejection of (SP2). Fortunately, this is not a very large gap: I will argue in Section 7.1 that just a little bit of empiricism is enough to choose the former. But, I reserve this argument for Chapter 7, where we discuss the implications of dynamical symmetry violation for the arrow of time in more detail.

### 4.6 Summary

In this chapter I have set out some implications of the Representation View for the philosophy of symmetry. We found that the concept of a symmetry splits into two components: temporal symmetries on spacetime and dynamical symmetries on state space. A representation then plays the dual role by giving meaning to a transformation on the state space of a dynamical theory and also by expressing that it is a symmetry of that theory. This gave rise to what I called the Symmetry Existence Criterion and in particular the fact that if a temporal symmetry fails to be a dynamical symmetry, then there is no representation of that symmetry and so no transformation on state space that can be meaningfully said to correspond to it.

The Representation View also suggested some revisions to three general debates on the philosophy of symmetry: direct empirical significance, the symmetry-to-reality inference, and the relation between spacetime and dynamical symmetries. In the next chapter, I will turn to a fourth: the infamous arrow of time.

# 5

# Arrows That Misfire

*Précis.* *The Representation View provides a strategy for determining whether time itself has an arrow, but most purported arrows fail to establish this.*

The 'arrow of time' is a phrase coined by physicist Arthur S. Eddington, to express that time is asymmetric:[1]

I shall use the phrase 'time's arrow' to express this one-way property of time which has no analogue in space. It is a singularly interesting property from a philosophical standpoint. (Eddington 1928, p.69)

Let us try to get to the bottom of what that 'one-way property' is and determine what evidence there is that time has it. Figure 5.1 summarises some famous responses. And, of course, the list goes on.

   A study of these arrows should proceed with caution, as the possibility of illusion looms. Our naïve human senses often detect phenomena that appear asymmetric in time when they are not. For example, a book will slide to a stop on a tabletop, but never the reverse: it does not spontaneously begin sliding. But, when that experience is carefully described in terms of dynamical systems, we find that the description invariably omits degrees of freedom in a way that hides an underlying temporal symmetry.

---

[1] The Fellowship of Trinity College, Cambridge seem to have been concerned about the arrow of time: Trinity physicist Eddington was preceded 20 years earlier by Trinity philosopher McTaggart (1908, p.474), who wrote at the end of his famous article: "what is that quality, and is it a greater amount of it which determines things to appear as later, and a lesser amount which determines them to appear as earlier, or is the reverse true?" One wonders how much serious concern a denier of the reality of time can muster about its direction.

| Arrow | Example |
|-------|---------|
| Thermodynamic | Dissipating gas |
| Radiation | Expanding electromagnetic waves |
| Quantum measurement | Dynamical state reduction |
| Cosmological | Cosmic expansion |
| T violation | Kaon quark flavour mixing |
| Spacetime | Intrinsic temporal orientation |
| Causal/dependency | Causes leading to effects |

Figure 5.1  Some arrows of time.

This chapter will begin by identifying what is required to have an arrow of time, in the sense that time itself has an asymmetry, through an application of what I call the Representation View.[2] In particular, I will propose that any adequate account of an arrow of time should show both how time itself can be asymmetric and some plausible evidence that supports this. I will argue that, once that background is clarified, most of what is commonly referred to as an 'arrow of time' fails to be a time asymmetry in this sense. The failure can happen in at least three ways: by resorting to *heuristics*; by relying on *boundary conditions*; or by describing a physical system with *missing information*. Those who make use of these techniques may still manage to produce a good explanation of our asymmetric experiences. However, they do not establish an arrow of time.

Section 5.1 will set out what I take to be needed to establish an asymmetry of time itself. The remaining sections will review many of the main contenders for an arrow of time in physics and identify the ways in which each falls short. These include: the radiation arrow (Section 5.2), the arrow of statistical mechanical entropy increase (Section 5.3), cosmological arrows (Section 5.4), quantum collapse (Section 5.5), and causal structure (Section 5.6). I reserve a longer discussion of equilibrium thermodynamics for Chapter 6, where I argue that it falls short as well. Chapter 7 will then present what I take to be the most promising arrow of time: the time reversal symmetry violation that arises from the weak interactions in particle physics.

---

[2]  Recall that the Representation View was introduced in Section 2.3.

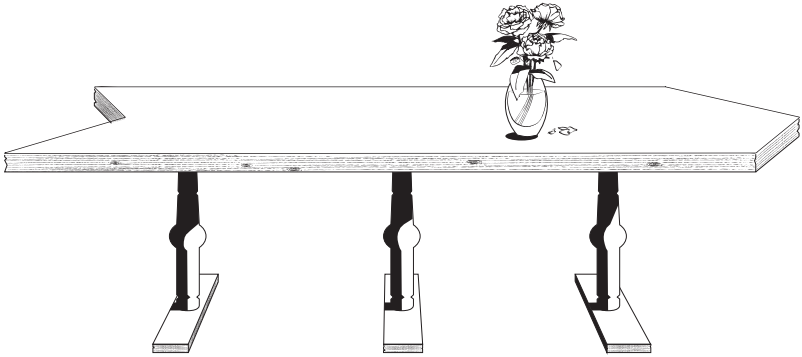

Figure 5.2 Price's table: the asymmetry of the table, like the asymmetry of time, is distinct from the asymmetric placement of items.

## 5.1 Seeking the Arrow of Time Itself

According to Price (2011, p.292), understanding the arrow of time requires answering a number of questions: "Is time anisotropic *at all*, and how could we tell if it is? What could constitute good grounds for taking it to be so, and do we have such grounds?" I will propose answers to these questions in this chapter and in the rest of the book. But first, let me highlight a general principle that they inspire:

*Spacetime–Evidential link:* An account of the asymmetries of time or space must identify both a sense in which time or space *itself* is asymmetric, as well as some plausible empirical evidence that supports that asymmetry.

The difficulty we face in the special case of temporal symmetry is to link it to experience, so that we stand a chance of having reasonable empirical evidence for or against it. In his seminal book, *Time's Arrow and Archimedes' Point*, Price (1996, p.16) characterises the difficulty in a picturesque way: think of time as analogous to a table and the evolution of material systems as analogous to the placement of items on it. Then, the question of whether or not 'time itself' is asymmetric is conceptually different from whether the evolution of a material system is – just as the question of whether the table itself is asymmetric is conceptually different from whether the items on it are asymmetrically placed (Figure 5.2). Our most direct experience is of the evolution of material systems. But, how is this linked to the structure of time itself?

The link that I propose involves a shift of focus in how time is often described by philosophers. Instead of speaking about a temporal 'axis' or coordinate variable, I propose to focus on the structural, functionalist aspects of time and speak instead of time *translations*, following the view motivated in Section 2.4.1. Price himself often makes use of the former, writing for example that, "the contents of the block universe appear to be arranged asymmetrically with respect to the temporal axis" (Price 1996, p.17). But, as I argued in Section 2.4, time is much more than that: it has rich relational, topological, and other structural properties including time translations, with elements of the form $t$ = time shift by two hours, rather than $t$ = two o'clock.

We will correspondingly interpret the statement that 'time has an arrow' to mean that the structure $\mathbb{T}$ used to describe time translations has an asymmetry, in that $\tau : t \mapsto -t$ is not an automorphism of that structure. And, to replace Price's notion of the 'contents' of time, we will adopt the Representation View, developed in Section 2.3: we postulate a representation of time translations $\varphi : \mathbb{T} \to \text{Aut}(M)$ amongst the automorphisms of a state space $M$, in order to give meaning to the notion of time evolution among physical states. According to the Representation View, this is required of any theory that is deserving of the name 'dynamical'. A representation is what encodes the structure of time translations and their symmetries in the context of a dynamical theory.

On this view, Price's table is not quite the right analogy. That table might give the impression that the symmetries of the contents of time are totally independent of the symmetries of time itself. In contrast, on the Representation View, these two are deeply linked: a representation is a homomorphism, which ensures that the symmetries of time translate to a 'homomorphic copy' amongst the symmetries of state space. So, given a representation of time translations, the structure of time is projected down onto state space, like the shadow of a tabletop on the floor (Figure 5.3). This suggests that, by studying dynamical asymmetries in a representation of time translations, it may be possible to infer an asymmetry or arrow of time itself. That is the kind of asymmetry that I will argue for in detail in Chapter 7. But, in this chapter and the next, let me first review of some competing 'arrows' of time.

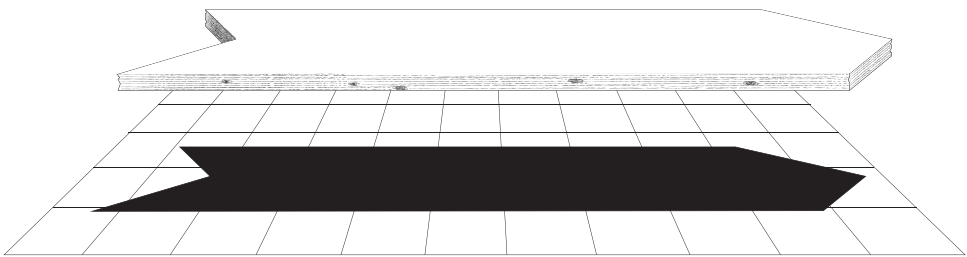

Figure 5.3 The structure of time is projected onto a state space by a representation, the way the structure of a table is projected onto the floor by its shadow.

In the next sections, I will introduce five phenomena that are commonly referred to as 'arrows of time'. I will argue that, at least in their current formulations, these approaches do not yet pass muster as time asymmetries, because they fail to make the Spacetime–Evidential link in a satisfactory way. In each case, this failure happens for one or more of the following reasons:

1. *Heuristic misfire:* making essential use of an informal extra-theoretical judgement, which is not justified by any well-supported physics.
2. *Boundary Condition misfire:* postulating contingent initial or boundary conditions that pick out one particular class of trajectory as special.
3. *Missing Information misfire:* omitting essential information about the time development of a physical system, which if restored would also restore time reversal symmetry.

The first fails to make the Spacetime–Evidential link because of a lack of adequate evidence for a time asymmetry; the second and third fail to make it because they do not establish an asymmetry in the structure of time translations and so cannot be used to infer that time itself is asymmetric.

Let me emphasise going forward that each of these misfires may be associated with an important area of research in the foundations of physics. However, their significance for our purposes is that they do not to establish an arrow of time itself.

## 5.2 The Radiation Arrow

An oscillating charge is associated with an outgoing shell of electromagnetic radiation, which expands with phase velocity equal to the speed of light.[3] The phenomenon is analogous to the circular ripples of a water wave when a stone is dropped in a pond. However, like the ripples in a pond, we never seem to observe this radiation in the reverse form of an inward-collapsing shell, as illustrated in Figure 5.4. Is this an arrow of time?

This question was discussed in a correspondence between Planck (1897a) and Boltzmann (1897) and then more famously in a debate between Ritz (1908) and Einstein (1909) in the journal *Physikalische Zeitschrift*. It has recently generated renewed philosophical interest.[4] The phenomenon is particularly puzzling because the laws of electromagnetism are time reversal invariant: both the outgoing radiation wave and its time reverse are possible! Ritz argued that the observed asymmetry is due to an additional time asymmetric law of nature, while Einstein maintained that it is just a matter of special boundary conditions: roughly speaking, one would have to begin with a ring of charges oscillating in perfect unison in order to produce a collapsing circular wave, just like a water wave. Their conclusions were summarised in a joint statement by Ritz and Einstein (1909).

---

[3] Thomson (1907, p.217) gave an early derivation of this result, which he identified as analogous to the mechanism producing "Röntgen radiation" or X-rays.

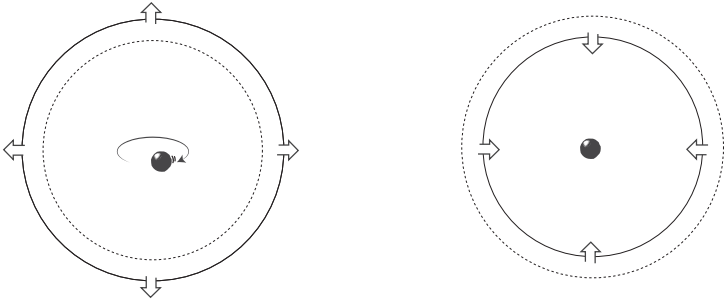

Figure 5.4 An expanding radiation shell (left) is easily produced from an oscillating charge, while a collapsing one (right) is not.

The boundary condition that Einstein identifies – that the experiment begins with a charge oscillation, rather than ending with one – is not evidence for an asymmetry of time itself. Whenever time reversal is a dynamical symmetry, two oppositely-directed representations of time translations are always possible. In the case of the expanding wave, the reverse time development is equally well picked out by the possibility that the experiment *ends* with a charge oscillation and begins with a boundary condition that gives rise to a collapsing wave. Neither is evidence for an asymmetry of the time translations of electromagnetism, any more than specifying a piece at the boundary of a jigsaw puzzle makes its development asymmetric: the piece could be specified either at the beginning or at the end (Figure 5.5). As a result, no link is made to an asymmetry of time itself. In our three categories of failed arrows of time, this is called a Boundary Condition misfire.

---

[4]  Popper (1958) seems to have independently arrived at a similar idea. Davies (1977, Chapter 5), Price (1996, Chapter 3), and Zeh (2007, Chapter 2) give classic discussions. Compare also the Ritz-like position of Frisch (2000, 2005, 2006) to the responses of Earman (2011), North (2003), Norton (2009) and Price (2006).

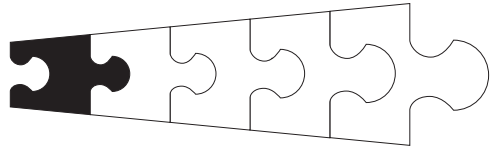

Figure 5.5 Specifying a boundary condition, like specifying a jigsaw puzzle boundary piece, may determine a development from either the beginning or end.

Of course, something might still seem to be missing, at least because most human beings seem to prefer *explanations* that place the oscillating charge at the beginning and the wave ring at the end. But, this by itself is what I have called a Heuristic misfire: it is an extra-theoretic judgement, which is not justified by any well-supported physics.

So, let's try again: one might instead respond by an appeal to statistical facts: in our universe, isolated oscillating charges that produce expanding advanced waves are statistically more likely, whereas coordinated rings of oscillating charges producing collapsing waves are unlikely. However, as Price has convincingly argued, this just moves the heuristic judgement to a different place, through the application of a temporal "double standard":

[T]here would be advanced waves, despite the improbability, if conditions at the center were as they are when we look at the 'normal' case in reverse: in other words, if wave crests were converging, stones being expelled, and so on. The normal case shows that the statistical argument does not exclude these things as we look toward (what we call) the past. To take it to exclude them as we look toward (what we call) the future is thus to apply a double standard, and to fall into a logical circle – to *assume* what we are trying to *prove*. (Price 1996, p.57)

I agree. In our universe, an oscillating charge and a coordinated ring of oscillating charges are equally likely: if the first occurs, the second occurs as well, for example when the waves produced by the single oscillating charge are absorbed into the environment.[5] More generally, the time reversal invariance of electromagnetism guarantees that if one solution is possible, then the time reversed solution is possible as well. One might like to add that an oscillating charge is more likely to occur 'earlier than' a coordinated ring of charges in time. But, in which temporal direction are we looking when we say this? If we respond, "towards (what we normally call) the future", then we have just assumed what we were trying to prove, that there is a preferred direction of time for formulating such statements.[6]

---

[5] Penrose (1979, p.590) points out an exception, that radiation from a star might escape forever into space without absorption; but, likewise, there might also be the reverse phenomenon of source-free radiation from the big bang or a white hole that gets absorbed by a star.

What may still demand explanation is why, when we represent time evolution in either direction, there appear to be unequal numbers of expanding and collapsing waves at every instant. But, this is an example of an inhomogeneity in space, rather than an asymmetry in time. The study of such inhomogeneities in the early universe is an important and active area of research in modern physics, which we will discuss shortly. But, it is not evidence for an arrow of time itself.

So, let us instead turn to the Ritz strategy in this debate, which is to reject that time reversal is a symmetry of the dynamics and instead stipulate a new law of nature that is asymmetric in time. This new law is supposed to be similar to the original laws, except that it restricts solutions to those that satisfy an early-time boundary condition. If such a statement can be made precise, then it might well provide a new dynamical theory in which time reversal symmetry fails. We will discuss theories like that in Chapter 7. But, in the present context, we do not have such a theory. Moreover, most agree that the origin of electromagnetic (EM) asymmetry lies elsewhere. The latter statement might be called, *Earman's Conjecture:*

I will mention a more general conjecture: any EM asymmetry that is clean and pervasive enough to merit promotion to an arrow of time is enslaved to either the cosmological arrow or the same source that grounds the thermodynamic arrow (or a combination of both). But much more work would be needed before I would be willing to make this conjecture with any confidence. (Earman 2011, p.524)

Thus, even on the Ritz approach, the radiation arrow at best reduces to one of the other arrows, either dynamical, cosmological, or thermal. So, let us turn to those other arrows instead.

## 5.3 The Arrow of Statistical Mechanics

A well-known folklore going back to Boltzmann (1896, 1898) says that the increasing entropy of the universe, as observed in processes like melting ice, is what determines time's arrow (Figure 5.6). Reichenbach formulates this as the statement: "positive time is the direction toward higher entropy" (Reichenbach 1956, p.54). This statement can be interpreted in two ways: either in terms of classical thermodynamics, or in terms of statistical mechanics. Classical equilibrium thermodynamics does not aim to describe the fundamental constituents of reality; and, I will give an extended argument in Chapter 6 that it does not establish an arrow of time. So, in this section, I will focus on the case of statistical mechanics. It faces all three of the misfires identified at the outset of this chapter.

---

[6] Price (1991, 1996) argues further that temporal symmetry is supported here by the adoption of the Wheeler and Feynman (1945) absorber theory, where each contribution to an electromagnetic field between a source and an absorber is half-advanced and half-retarded, but adjusting the proposal by dropping the requirement that every source has a perfect absorber. My own sympathies still lie with the photon: as Penrose (1979, p.590) points out, a theory that so tightly constrains the electromagnetic field seems "unfairly biased against the poor photon, not allowing it the degrees of freedom admitted to all massive particles!".

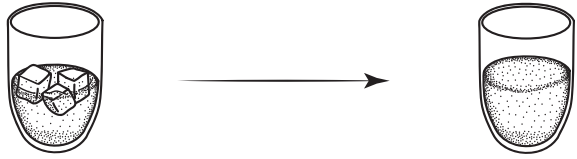

Figure 5.6 The 'arrow' of increasing entropy.

### 5.3.1 *Boundary Condition Misfires*

Statistical mechanics is formulated with either classical or quantum mechanics as its basis. For this reason, it is possible to give a rigorous representation of time translations in this theory, which is usually time reversal invariant (or at least CPT invariant). A variety of well-known strategies can then be used to argue that, on a certain 'coarse grained' level of description, the system can be expected to evolve towards a higher entropy state until it reaches equilibrium. Boltzmann's *Stoßzahlansatz* ('assumption of molecular chaos') in classical statistical mechanics is perhaps the most famous, but a variety of master equations in classical and quantum statistical mechanics also describe the irreversible evolution of probability distributions.[7] Such arguments invariably postulate a special boundary condition in the form of a probability distribution on microstates, or on macrostates, or both, for example to render the state of the universe to be far out of equilibrium in the first microseconds after the big bang.[8]

Thus, like the radiation arrow, this approach to statistical mechanics suffers from a Boundary Condition misfire: a time asymmetry only seems to appear because of special initial or boundary conditions and not because

---

[7] See Zeh (2007): §3.1.1 for a discussion of the former, and §§3.1–3.2 for the latter.
[8] This approach is the subject of a great deal of philosophical debate: see Albert (2000, Chapter 3), Callender (2010), Earman (2006), Penrose (1979, §12.2.3), Penrose (1994), Price (1996, Chapter 2), Price (2004), Wallace (2013, §4.2), and Zeh (2007, §3.1.1).

of an arrow of time itself. Indeed, a Price-style argument can be applied to the statistical mechanical arrow as well, which of course Price gives:

[W]e don't take the *final* microstate to explain the *initial* ordered state. But then by what right do we propose that an *initial* microstate can explain a *final* macrostate? In practice, of course, we are inclined simply to help ourselves to the principle that the past explains the future, but what could possibly justify that inclination here, where the temporal asymmetry of the universe is what we are seeking to explain? (Price 1996, p.42)

One might try to resist Price's conclusion here, by insisting on the logic of Boltzmann's *Stoßzahlansatz*. That is: one should expect all states to produce a maximum entropy state, both to the past and to the future, and so a low entropy past is enough to produce the increasing entropy of our universe. In contrast, a high entropy future is not, since it only produces other high entropy macrostates. Maudlin has responded to Price in this way, suggesting that the process by which macrostates "produce" higher entropy requires a temporal direction.[9]

I won't dwell on the formal issues with the Boltzmann approach, which, as many have noted, does not involve rigorous argumentation.[10] My concern is rather with two overarching concerns about the Boltzmann picture: the origin of the coarse-graining, and the role of time translations. These lead to the next two misfires.

### 5.3.2  *Heuristic Misfires*

A coarse-graining in statistical mechanics is an equivalence relation on microstates, which defines what it means to be in the same macrostate. Without this, one cannot define either equilibrium or entropy in statistical mechanics. A typical approach is to fix a set of 'macroscopic' observables $\{A_i\}$ on the underlying classical or quantum state space and define an equivalence relation on pairs of states by the relation of 'being assigned the same value by all the observables $A_i$'; this partitions the space into macrostates.[11] Intuitively, two states are equivalent if we have no way to observe a distinction between them using our collection of macroscopic observables. But, how do we choose the set of observables on which to base this judgement?

---

[9]  Maudlin (2007, p.134–5) writes: "Even though the laws themselves might run perfectly well in reverse . . . we cannot specify an independent, generic constraint on the final state that will yield (granting the final macrostate is typical) ever decreasing entropy in one direction. . . . This sort of explanation requires that there be a fact about which states produce which. That is provided by a direction of time: earlier states produce later ones".

[10]  See Price (1996, p.40), or Sklar (1993, §7.II.1) for a classic argument that the *Stoßzahlansatz* is usually false. Earman (2006) argues that the whole low-entropy-past approach is "not even false" on the scale of the universe as a whole; and D. Wallace (2011, "The logic of the past hypothesis", Unpublished manuscript, http://philsci-archive.pitt.edu/8894/) and (2013) argues that it is redundant.

Unfortunately, the formalism of statistical mechanics alone does not provide any advice about this question. The problem is, moreover, a serious one, because different choices of observables give rise to different definitions of equilibrium and entropy: thus, the definition of the 'statistical mechanical arrow' depends on it. Indeed, Rovelli (2017) has argued on this basis that the statistical mechanical arrow of time is fundamentally perspectival: formally speaking, there are choices of observables that lead to oppositely-directed arrows of time.

Penrose (1979, p.588) notes that the problem of choosing a set of observables or coarse-graining is "fraught with difficulties", in spite of the fact that by a standard convention, "entropy is a concept that may be bandied about in a totally cavalier fashion!" In the absence of any well-motivated choice for fundamental physics, we face having to embrace this fact: that the approach to equilibrium in statistical mechanics depends on a choice of observables, which is at worst arbitrary and at best a contingent fact about the observations currently available to us. Thus, this arrow suffers from a Heuristic misfire: it does not describe an asymmetry of time itself but rather depends on extra-theoretical judgements that lack an evidential basis in physics.

### 5.3.3 *Missing Information Misfires*

The final misfire arises from a peculiarity of Boltzmann's argument, which is that his *Stoßzahlansatz* makes no mention of the time parameter characterising the approach to equilibrium. As a result, it leaves out some important physical facts about equilibrium, such as the relaxation time required to achieve it. This leaves us with a rather poor explanation of the approach to equilibrium in our universe. But, more importantly for our purposes, if no time translations are associated with the approach to equilibrium, then there is no way to associate it with the structure of time through a representation. Boltzmann's approach by itself lacks the information needed to establish an arrow of time itself.

So, a more appropriate interpretation of Maudlin's "production" process for high entropy states is one that includes a time parameter, like the

---

[11] Conversely, every partition of a state space into macrostates can be associated with a set of observables with this property. See Wills (Forthcoming) for a philosophical discussion of this approach in the context of particle identity debates and Rovelli (2017) for an argument that this makes the statistical mechanical arrow 'perspectival'.

Zwanzig (1960) projection formalism, among others.[12] These approaches to the arrow of time are of a fundamentally different form: they invariably produce a representation of time translations in terms of a dynamical evolution operator on probability distributions, known in differential form as a *master equation*. These approaches provide a much more empirically adequate account of the approach to equilibrium. However, they also generally formalise the fact that we are leaving out information. As Wallace (2013, §3) has pointed out, the fact that such equations 'project out' information over time is what ultimately leads to their time asymmetry. So, this is an example of a Missing Information misfire: an arrow of time only appears to arise because some facts are being ignored.

Statistical mechanics is widely identified as the source of the arrow of time. However, there are devils in the details: this supposed arrow misfires in all three of the ways I have identified above, including heuristic additions, boundary conditions, and missing information. Each of these might be fruitfully used in a physical model. But, none of them provide evidence for an arrow of time.

## 5.4  Cosmological Arrows

The cosmological arrow, in its simplest form, is the observation that the universe on the largest scale is expanding rather than contracting. In a slightly more interesting sense, it is the additional fact that structure formation occurs in an asymmetric way, with great gases clumping together to form galaxies, stars, and planets, but not generally the reverse. Insofar as general relativity is time reversal invariant, these asymmetries are characterised by special boundary conditions: at an early moment, the universe was extremely dense, as well as nearly homogeneous and isotropic, with just enough inhomogeneity to yield clumping and structure formation. At the other end of time, it appears that the universe will continue to expand forever towards an ever-more lifeless and diluted state. One of the great achievements of twentieth-century cosmology is the confirmation of this description of our universe.

### 5.4.1  Boundary Condition Misfire

However, when the cosmological arrow is put this way, the Price-style objection applies, just as it did in electromagnetism and in statistical mechanics.

---

[12] See Zeh (2007, Chapter 3), who argues that Boltzmann's *Stoßzahlansatz* can be viewed in the same spirit as modern statistical mechanical master equations.

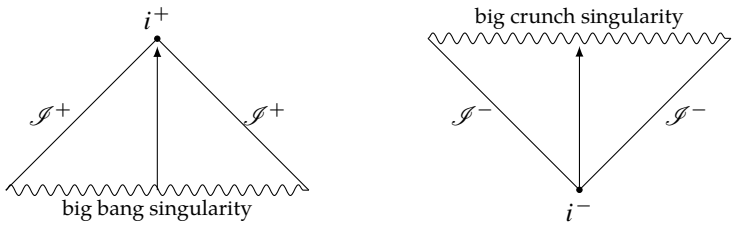

Figure 5.7 Conformal diagrams for a big bang (left) and big crunch (right) universe.

The original model of an expanding universe – called *FLRW spacetime*, for Friedman, Lemaître, Robinson and Walker – is completely time reversible: each of the two available temporal orientations define 'future-directed' developments in opposite directions. This is illustrated in Figure 5.7 using a Penrose conformal diagram, which accurately represents lightcone structure but warps distances so as to allow the representation of infinitely long curves. The first development in the illustration begins with a big bang and then dissipates on its way towards future timelike infinity $i^+$. The second begins dissipated at past timelike infinity $i^-$ and collapses to a big crunch.[13] We cannot say which represents the 'correct' future-directed development without begging the question as to the direction of time. Again, Price says as much:

[U]ntil we find some reason to think otherwise, we should take the view that it is not an objective matter which end of the universe is the 'bang' and which end is the 'crunch'. (Price 1996, p.84)

Let me emphasise that I do not take this to mean that modern cosmologists are confused: far from it. As far as I can see, cosmologists simply use the phrase 'time's arrow' to refer to a completely different issue: not whether cosmic time translations can be represented in opposite directions but whether they couple in an appropriate way to statistical mechanical entropy as we ordinarily understand it. That is the subject of the next subsection.

### 5.4.2 Statistical Mechanical Coupling

The question of how the cosmic expansion couples to statistical physics is tied up with serious foundational issues, as Penrose (1979) has pointed out.

---

[13] Of course, these two descriptions are not generally related by an isometry, i.e. the spacetime is not 'temporally anisotropic' in the sense of Earman (1974, p.29). I will discuss this more in Chapter 7.

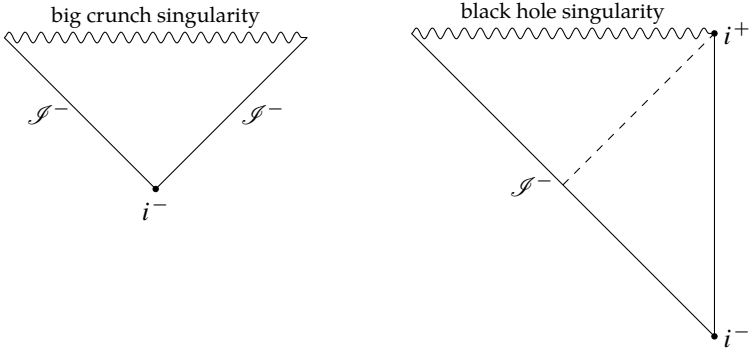

Figure 5.8  The geometry of a big crunch is similar to that of a black hole.

For example, suppose that statistical mechanical entropy were coupled so as to increase with cosmic expansion and therefore to decrease with cosmic contraction. The environment inside a black hole event horizon is formally quite similar to a collapsing universe, as illustrated in the Penrose diagrams of Figure 5.8. So, on this coupling, statistical mechanical entropy should decrease inside the black hole as well. This can even happen near an event horizon where the geometry is not so strange, with water unmelting and glasses unbreaking, among all manner of bizarre things:

Suppose that experiments are performed by the astronaut for a period while he is inside the hole. The behaviour of his apparatus (indeed, of the metabolic processes within his own body) is entirely determined by the conditions at the black hole's singularity (assuming that behaviour is governed by the usual hyperbolic-type differential equations) – as, equally, it is entirely determined by the conditions at the big bang. The situation inside the black hole differs in no essential respect from that at the late stages of a recollapsing universe. If one's viewpoint is to link the local direction of time's arrow directly to the expansion of the universe, then one must surely be driven to expect that our astronaut's experiments will behave in an entropy-decreasing way (with respect to 'normal' time). (Penrose 1979, pp.598–9)

Here Penrose is using the phrase 'time's arrow' as I have suggested, to refer to the arrow of entropy increase. His aim is to state a potential difficulty with assuming it is aligned with the geometry of expansion: this implies the geometry of collapse is aligned with entropy decrease. A great deal of effort has gone into the description of this coupling as part of the search for new physics, which prominently includes the no-boundary proposal of Hartle and Hawking (1983) as well as the Weyl curvature hypothesis of Penrose (1979).

Price has been a prominent critic of the former, charging that Hawking's conjecture does not produce a convincing arrow of time, although it appears that Hawking was never convinced.[14] In my view, there is room for agreement between the two parties: Price is right that Hawking's account is compatible with the possibility of 'no arrow' in the sense of oppositely-directed representations of cosmic time translations; but Hawking's proposal has a different purpose, to establish an 'arrow' in the sense of a plausible thermodynamic coupling.

So, the cosmological arrow as described either suffers from a Boundary Condition misfire or else from all the misfires of thermodynamics, if we view it as fundamentally defined by this coupling. In this form, it does not provide the kind of asymmetry needed to establish an arrow of time itself.

### 5.4.3 Baryogenesis

There is another aspect of the cosmological arrow that may yet provide a plausible source of time asymmetry. A boundary condition by itself does not determine an arrow of time. But, it may be an indication of something that does, like an instance of dynamical symmetry breaking.

For example, if the boundary condition for the big bang is a highly symmetric state, which is symmetric even in the distribution of baryonic (ordinary) matter as compared to antimatter, then some mechanism is needed to explain how the current universe came to be dominated by the former. This is called *baryogenesis*. One obvious proposal is that some symmetry-violating interaction led to more ordinary matter than antimatter. This is called a *charge conjugation symmetry violation*. Even a small amount would help explain baryogenesis, if it could be shown to be magnified by the subsequent cosmic evolution.

Sakharov (1967) pointed out that baryogenesis would have to involve more than just charge conjugation symmetry violation. Its combination with the parity transformation, denoted $CP$, would have to be violated as well, since otherwise baryon symmetry violation could still happen in equal amounts for both matter and antimatter.[15] So, assuming that CPT is a symmetry of whatever this underlying theory is, time reversal symmetry would be violated as well.

How this process works is a matter of active research, since the Sakharov conditions are clearly at best necessary conditions for baryogenesis.[16]

---

[14] Price (1989) published his critique in the journal *Nature* and later expanded on it in his book (Price 1996, Chapter 4). As Brown (2000, p.335) noted in his review of the book, the former article "essentially accused Hawking of sleight of hand".

[15] These transformations are discussed in more detail in our discussion of CPT in Chapter 8.

[16] See White (2016) for a recent introduction.

However, this means that, according to one prominent road to understanding the cosmological arrow of time, it reduces it to a dynamical arrow: namely, a time asymmetry in the dynamical theory governing early universe interactions. I will argue that this kind of time asymmetry really can provide a plausible arrow of time. We will return to the arrow of time in dynamical theories in Chapter 7.

## 5.5 Quantum Collapse

Quantum theory famously has two predictive laws: unitary (Schrödinger) evolution and the statistical Born rule. It is often suggested that the former is usually time reversal invariant, while the latter is not. However, on the most direct reading, this is incorrect: unitary evolution is time symmetric if and only if the Born rule is. Let me begin by clarifying what I mean by that.

### 5.5.1 Harmony between Schrödinger and Born

Suppose that the dynamics of quantum theory is given by a strongly continuous unitary representation $\mathcal{U}_t$ of the time translation group $(\mathbb{R}, +)$ on a Hilbert space. According to the Born rule, if an initial state $\psi$ evolves unitarily for some time $t$, then the probability of finding it in the state $\phi$ is given by

$$\Pr(\psi \xrightarrow{t} \phi) = |\langle \phi, \mathcal{U}_t \psi \rangle|^2. \tag{5.1}$$

Following our discussion of Section 3.4, consider the time reversed representation of this set-up: if the time reversed state $T\phi$ evolves under the same dynamics for a time $t$, then the probability of finding the system in the time reversed state $T\psi$ is given by

$$\Pr(T\phi \xrightarrow{t} T\psi) = |\langle T\psi, \mathcal{U}_t T\phi \rangle|^2 = |\langle \psi, T^* \mathcal{U}_{-t} T\phi \rangle|^2. \tag{5.2}$$

By an elementary theorem of Hilbert space theory (cf. Messiah 1999, §XV.2 Theorem II), the probabilities in Eqs (5.1) and (5.2) are equal for all states $\psi, \phi$ if and only if $T\mathcal{U}_t T^* = e^{i\theta} \mathcal{U}_{-t}$ for some arbitrary phase factor $e^{i\theta}$. But, this phase disappears in the description of true quantum states as 'rays' or one-dimensional subspaces.[17] So, the equality of these equations really holds if and only if $T\mathcal{U}_t T^* = \mathcal{U}_{-t}$. That is just what it means to say that time reversal is a dynamical symmetry in this representation (see Section 4.1.3). In other words, Schrödinger evolution is symmetric in time if and only if the statistical predictions of the Born rule are time symmetric: the two rules are in happy harmony.

---

[17] Another way to look at this is to note that the phase factor $\theta$ can be eliminated by redefining the Hamiltonian generator of $\mathcal{U}_t = e^{-itH}$ by $H \mapsto H - \theta$. Since the Hamiltonian only has physical significance up to an additive constant, this adjustment makes no difference to the predictions of quantum theory.

In Chapter 7, I will argue that quantum theory does provide evidence for a true arrow of time, through the time reversal symmetry violation associated with electroweak theory. In line with what I have just said about the harmony between its two rules, this means that both Schrödinger evolution and the Born rule are time asymmetric in the weak interactions.

However, there are at least three other ways that quantum theory is often said to have an arrow, even when its unitary dynamics and statistical rule are time symmetric. These are: the collapse of the quantum state, decoherence, and branching universes. Although each of these ideas is associated with an important research programme, each of them also misfires as an arrow of time. Let me briefly review why that is.

### *5.5.2 Collapse*

Von Neumann (1955, Chapter 6) postulated that quantum measurement is associated with a 'state reduction' or 'collapse' process. This led to a research programme that supplements or replaces quantum theory with a new dynamical law, which describes how each quantum state will irreversibly evolve to an eigenstate of some observable during the measurement process. The two most prominent approaches to this are the 'GRW' or 'flash' theory, due to Ghirardi, Rimini, and Weber (1986), and its continuous analogue 'CSL' or 'continuous spontaneous localisation', due to Pearle (1989). I will refer to these collectively as *collapse theories*.[18]

Collapse theories are often regarded as being asymmetric in time. To illustrate the idea in simple terms, let me adopt a picturesque example that follows Rovelli (2016). Imagine a quantum state undergoing Schrödinger evolution $\psi(t) = \mathcal{U}_t \psi$, at least approximately and which describes a wavefunction expanding around some point $p$ in space. After evolving this way for a duration of time, the wavefunction collapses to a small shell around a different point $q$, according to the dynamics of some collapse theory. What is the time reverse of this process? It is a system that begins as a small, outward moving shell centred around $p$, which spontaneously

---

[18] For an introduction, see Bassi and Ghirardi (2003); Bassi, Lochan, et al. (2013); Ghirardi and Bassi (2020).

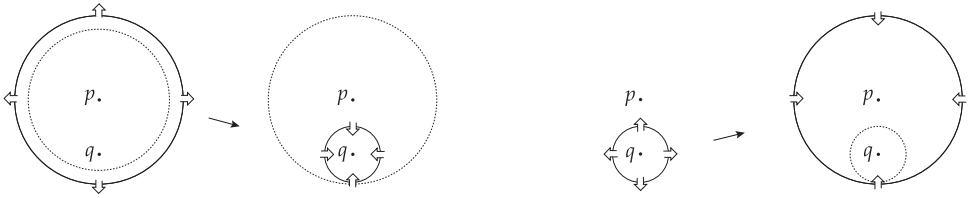

Figure 5.9 A collapse process (left) and the time reversed description (right).

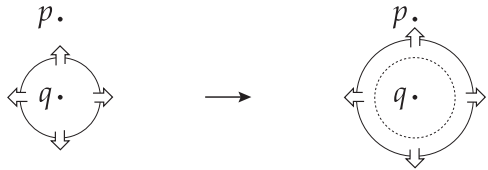

Figure 5.10 Correct evolution from the time reversed initial condition.

'jumps' to a large shell centred at $q$, which it then collapses down towards, as shown in Figure 5.9.

But, no approximation of quantum theory will describe the system this way: an outward moving shell around $q$ will continue to expand around $q$, with no jumping, at least until another collapse event occurs (Figure 5.10). So, this concept of 'collapse' appears to inevitably introduce a time asymmetry. Moreover, the unitary dynamics in this example (and hence the Born rule) are time symmetric! So, the origin of the time asymmetry arises from the addition of a dynamical collapse law.

What does this say about the arrow of time? The answer depends on the status of collapse theories. I will argue in Chapter 7 that, if a dynamical theory is time reversal violating, and if that dynamical theory is accurate, then it provides evidence of a true asymmetry in time. So, if a collapse theory turns out to be both accurate and time asymmetric, then this is good evidence for an arrow of time.

However, neither of these two requirements have been settled in the case of collapse theories. Rovelli argues that this disharmony between the symmetries of the unitary dynamics and collapse is so egregious that we should reject the latter, and any metaphysics that goes with it.[19] Of course, some of this decision might be left up to experiment: collapse theory is associated with some novel empirical predictions, like a slow increase of energy.[20] However, no decisive evidence in its favour has yet been produced, and so its status as a replacement for quantum theory remains unclear. Regarding the time asymmetry of collapse: this claim can be challenged as well. For example, Bedingham and Maroney (2017) propose a framework for collapse theory in which it is time reversal invariant, arguing that the apparent time asymmetry is the result "not of an inherent asymmetry in the dynamics, but of the time asymmetric use of boundary conditions" (Bedingham and Maroney 2017, p.692). Thus, the implications of collapse theories for the arrow of time are at best unsettled and at worst a Boundary Condition misfire.

---

[19] In particular, Rovelli (2016, p.1232) takes this as evidence against a certain kind of realism about the quantum state, writing, "But if the state is taken to be real, the fact that it behaves in a non T-invariant way, when everything we measure about the (classical and quantum) world is T-invariant, sounds illogical".

### 5.5.3 Decoherence

Many of the unusual statistical observations associated with quantum phenomena can be explained using ordinary quantum theory without a collapse postulate. For example, consider the double-slit experiment, in its classic presentation by Feynman (1963b, §1): a beam of electrons is directed at a thin metal plate with two small holes in it, producing an interference pattern on a surface located on the other side. This arises from the fact that the quantum system emerging from the plate is a superposition of two positions. But next, suppose that we 'watch' which hole each electron has passed through, by placing a light source on the side where they emerge, which scatters in different ways depending on which hole the electron passes through. Then the interference pattern disappears, and one observes a concentration of electrons in front of each hole.

The phenomenon of *decoherence* explains the difference between these two observations using ordinary unitary (Schrödinger) evolution. The central idea is to consider how the electron subsystem interacts with its environment. In the first experiment, electrons emerging from the metal plate are more or less isolated from their environment. But, once a macroscopic light source is introduced, the electrons interact with that object in complicated ways, which include a spectacularly large number of degrees of freedom. Decoherence is a body of work showing that, when such interactions are accounted for in the unitary dynamics, the probability of a superposition is suppressed to a vanishingly small value, with respect to some basis.[21] As a result, the probability of observing an interference pattern becomes vanishingly small as well, thus explaining the difference between the two observations.

---

[20] This was pointed out in the original paper of Ghirardi, Rimini, and Weber (1986). See Bassi, Lochan, et al. (2013) for a more recent discussion of the experimental basis for collapse theories.

This suppression of interference is sometimes referred to as the "time arrow of decoherence" (cf. Zeh 2007, p.16). However, this usage of the term should not be understood as referring to an asymmetry in time itself. If the interaction of a subsystem with its environment is time reversal invariant, as most such interactions are, then decoherence is completely symmetric in time: if there is a representation in which interference becomes suppressed towards the future, then there is one in which it becomes suppressed towards the past as well.[22] Of course, the founders of decoherence theory were well-aware of this and of the reason for the apparent time asymmetry:

Environments are notorious for having large numbers of interacting degrees of freedom, making extraction of lost information as difficult as reversing trajectories in a Boltzmann gas. (Zurek 1991, p.40)

In other words, the apparent time asymmetry of decoherence is very similar to that of statistical mechanics. In particular, it is a Missing Information misfire: an arrow only appears when we neglect information about the environment as 'lost'. It is a remarkable fact that so much quantum behaviour can be understood in this way. However, having limited access to information is just a fact about the human condition rather than evidence for a true time asymmetry. Indeed, if those environmental degrees of freedom are restored, then whenever the interactions are time reversal invariant, there is a perfectly adequate representation of time translations in the opposite direction and in which spontaneous coherence occurs.

### 5.5.4 Branching

On the Everett interpretation, also called the 'many worlds' or 'multiverse' approach to quantum theory, the universe is imagined to be a great branching tree, with each path through the tree corresponding to a familiar world of definite measurement outcomes.[23] The strange phenomena of quantum superposition in some basis is then imagined to arise from facts about a multiplicity of branches: for example, in the double-slit experiment described above, each electron travels through both holes in the metal plate, but on different branches of the tree.

---

[21] The modern study of decoherence was launched by the works of Griffiths (1984), Gell-Mann and Hartle (1990), Zeh (1970), and Zurek (1981). For an introduction, see Bacciagaluppi (2020) or Joos et al. (2013).

[22] A similar point is emphasised by Hagar (2012, pp.4603–4).

[23] The many worlds approach is due to Everett III (1957); for a modern defence, see Wallace (2012).

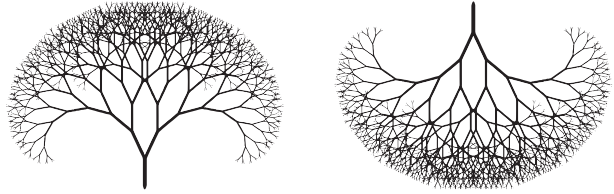

Figure 5.11 Everettian 'branching' happens to the future (left) but not to the past (right).

Since the branches of the Everettian tree appear to 'grow' towards the future but not the past, as in Figure 5.11, this interpretation is sometimes said to be associated with a time asymmetry. Indeed, Rovelli (2016, §) presents this as evidence against the Everett interpretation, given that most quantum interactions are time reversal invariant (or at least CPT invariant).[24] If an Everettian did add a time asymmetric structure to quantum theory of this kind, then in my language it would be a Heuristic misfire: it is an addition to quantum theory that is not justified by any well-supported physics.

However, in modern approaches to the Everett interpretation, such as the one presented by Wallace (2012), Everettian branching is understood in a different way. Branching is not a part of the fundamental quantum dynamics, and so it is not associated with a representation of time translations at the level of fundamental microscopic behaviour. Instead, it is used to help explain the emergence of definite outcomes relative to the experience of a macroscopic observer. Indeed, Wallace (2012, Chapter 3) places decoherence at the foundation of branching: its suppression of superpositions in some basis is taken to imply that there is a preferred basis for measurement outcomes, each one corresponding to a 'branch' in the Everettian tree.[25]

But, as I have argued above, this is a Missing Information misfire when it comes to the arrow of time: it only appears to be an asymmetry from the perspective of an observer who ignores environmental degrees of freedom.

---

[24] CPT invariance is discussed in detail in Chapter 8.
[25] Wallace (2012, p.87) summarises how successive suppression due to decoherence can create a branching structure: "[I]f the state evolves from a basis vector to a superposition of such basis vectors, and if each of those evolves into a superposition of different basis vectors so that no two such superpositions interfere with one another – then we would have branching (relative to that basis, at any rate)".

This is not a problem for Wallace, whose aim here is to help explain the emergence of the macroscopic world. But, neither is it a basis for the arrow of time.

## 5.6  Causal Structure

Our final discussion is about the supposed 'arrow of causation'. Philosophers often locate the arrow of time in the statement that all effects occur later than their causes, or in some more general asymmetric dependence relation such as that of Lewis (1979). However, most philosophers of physics are sceptical of this, in the absence of any well-defined physical property corresponding to 'causation'. This led Russell (1912) to decry causation as "a relic of a bygone age", Norton (2003) to demote it to a "folk science", and Stephen Hawking to report being "very disappointed" by Reichenbach's book, *The Direction of Time*. Hawking's basic concern was that, if causes and effects are states associated with a time reversal invariant dynamics, then they cannot be asymmetric in time:

I was very disappointed. It was rather obscure, and the logic seemed to be circular. It laid great stress on causation, in distinguishing the forward direction of time from the backward direction. But in physics, we believe there are laws that determine the evolution of the universe uniquely. So if state A evolved into state B, one could say that A caused B. But one could equally well look at it in the other direction of time, and say that B caused A. So causality does not define a direction of time. (Hawking 1994, p.346)

This characterisation of Reichenbach's argument is not entirely fair: Reichenbach's idea is, just as Hawking suggests, to begin with a dynamical theory that is time reversal invariant. Reichenbach then uses "causal connections" in a temporally symmetric way as a synonym for this dynamics, applying phrases like "*A* is causally connected to *B*" to refer to states *A* and *B* that lie on the same trajectory of a dynamical theory, without saying which is a 'cause' and which is an 'effect' (Reichenbach 1956, §5). However, to describe the "direction" of such a trajectory, he then goes on to propose we use the increase of statistical mechanical entropy. For our purposes, this approach falters because, as we have seen in Section 5.3, it falls prey to each of the three misfires identified at the outset of this section.

Since the appearance of Reichenbach's book, one of his characterisations of the arrow has become the subject of much discussion, known as the Principle of the Common Cause. In its simplest form, the principle says that every pair of correlated events admit an event in their common past (a 'cause') that increases their probability and screens off the correlation between them.[26] This principle can be used to state the existence of a time asymmetry, so long as it includes the further postulate that there is no common cause in the future. Philosophers sometimes refer to this kind of asymmetry as a 'fork asymmetry'. However, whether it is evidence for an arrow of time depends on how we interpret it. Price (1996, Chapter 6) argues that the Principle of the Common Cause and other purported 'forks' just amount to further boundary conditions.[27] In that case, this approach amounts to a Boundary Condition misfire.

On the other hand, Penrose and Percival (1962, p.616) elevate a version of Reichenbach's principle to the status of a "basic statistical law, which is asymmetric in time". If we view that law as associated with a dynamical theory, which admits a representation of time translations without a reverse representation, then this is a different situation, which I will call a 'dynamical arrow'. This sort of case, I claim, is more promising, and it will be dealt with in Chapter 7.

### 5.7 Summary

This section has presented a brief review of some commonly identified 'arrows of time'. Nearly all of them involve a great deal of interesting physics and philosophy. However, most of them also misfire as arrows of time. They may be Boundary Condition misfires, as when one identifies oscillating charges as to the 'past' of their absorbed radiation; they may be Heuristic misfires, as when one introduces a set of observables to pick out a notion of entropy associated with the statistical mechanical arrow; or they may be Missing Information misfires, as when one ignores environmental degrees of freedom during the decoherence process. All such arrows fail to make the Spacetime–Evidential Link and thus fail to establish an asymmetry of time itself.

There are two more famous arrows that were omitted from our discussion: the arrow of electroweak interactions and the arrow of classical equilibrium thermodynamics. I have left out electroweak theory because, as I will argue in Chapter 7, it provides an arrow that does not misfire! Electroweak interactions provide the most compelling evidence available that time itself has an arrow. In contrast, I have left out classical thermodynamics because my thesis, that there is no thermodynamic arrow in any non-trivial sense, may sound so preposterous to some that it deserves its own discussion. So, if I may be forgiven for this transgression, let me turn to that discussion next.

---

[26] Formally, a common cause $C$ for events $A$ and $B$ has four independent properties: the 'screening-off' conditions $p(A \cap B|C) = p(A|C)p(B|C)$ and $p(A \cap B|C^{\perp}) = p(A|C^{\perp})p(B|C^{\perp})$, and the 'relevance' conditions $p(A|C) > p(A|C^{\perp})$ and $p(B|C) > p(B|C^{\perp})$. This formulation is due to Salmon (1978, 1984). For a recent introduction to this large literature see Hofer-Szabó, Rédei, and Szabó (2013). A classic book on the causal modelling approach is Spirtes, Glymour, and Scheines (2001, Chapter 3).

[27] This includes in particular the "asymmetry of counterfactual dependence" that Lewis (1979) takes to generalise the causal asymmetry.

# 6

# There Is No Thermodynamic Arrow

*Précis.* *The structure of equilibrium thermodynamics, in harmony with statistical mechanics, does not contain a time asymmetry, except when it is trivially supplemented with one.*

Our bodies, the planets and stars, and all traces of our existence will eventually dissipate into the cosmos. According to orthodoxy, this march towards equilibrium is *the* arrow of time.[1] I have tried to defend and clarify orthodoxy in earlier chapters of this book. But, in this case, I think orthodoxy is indefensible: in this chapter, I will argue that there is no non-trivial sense in which the approach to equilibrium in thermodynamics gives rise to a time asymmetry.

The basic argument that I will give, summarised in the next section, is that classical thermodynamics is not so different from its 'partner' theory, statistical mechanics, in how it bears on the arrow of time. It is widely agreed that in statistical mechanics, the approach to equilibrium is symmetric in time, in that the laws predict evolution towards equilibrium with roughly equal likelihood to the future and to the past. Or, in the language developed in this book: the time translations of statistical mechanics admit a representation of time reversal symmetry. The occurrence of statistical mechanical equilibrium in our future rather than in our past is not a fact about temporal symmetry but about contingent initial conditions, contingent boundary conditions, and our special limited perspective as observers.[2] I will argue that classical equilibrium thermodynamics is no different.

---

[1] This is perhaps partly for historical reasons: Eddington (1928, pp.64–9) coined the term 'arrow of time' to describe irreversible thermodynamic processes; see also the remarks at the beginning of Chapter 5.

In contrast, a well-known folklore has it that classical equilibrium thermodynamics, if it were true, would require a robust arrow of time. This folklore was famously captured by Reichenbach:[3]

> If the universe as a whole possesses at every moment a specific entropy, this value is subject to the general law of [thermodynamic] entropy increase . . . . Although this principle leads to the unwelcome consequence that someday our universe will be completely run down and offer no further possibilities of existence to such unequalized systems as living organisms, it at least supplies us with a direction of time: positive time is the direction towards higher entropy. (Reichenbach 1956, p.54)

Reichenbach himself dismissed classical thermodynamics in favour of the more fundamental and accurate tools of statistical mechanics. But, there are good reasons to return to classical thermodynamics and what it says about the arrow of time. In the first place, thermodynamics is a powerful tool in a wide variety of sciences, from biochemistry to stellar astrophysics. It has become an important technique for probing regimes in which the microphysics is not yet known, such as in black hole thermodynamics. And, a key requirement for understanding the reduction of thermodynamics to statistical mechanics is to make sense of the supposed difference between their temporal symmetries.

Uffink (2001, p.307) calls this problem of reconciling temporal symmetries "the most profound problem in the foundations of thermal and statistical physics".[4] Ludwig Boltzmann famously spent most of his career trying to solve the problem. However, not everyone agrees about its significance. Callender (1999, 2001) concludes that we should give up some aspects of the reduction project: just use statistical mechanics to study what is fundamental, and deny that thermodynamics is "universally true and somehow independent of the statistics of the micro-constituents of thermal bodies" (Callender 2001, p.551). Others have recently argued that the difference in temporal symmetries remains a problem that philosophical accounts of reduction must face.[5]

I will argue that this is not a problem in the first place, because there is no non-trivial thermodynamic arrow. Some of my arguments will follow and (I hope) strengthen the work of Uffink (2001), who forcefully demonstrates that the thermodynamic second law stands little chance of rigorously establishing a thermodynamic arrow. My conclusion is more general: I can see no sense in which the structure of thermodynamics contains any basis at all for an arrow of time, except in the trivial sense of inserting it by hand.

---

[2] An argument for this is set out in Section 5.3; see also Price (1996, Chapter 2).
[3] It is echoed by Hawking (1994, p.348), Sklar (1993, §2.I.3) and Zeh (2007, p.5), among others.
[4] The history of the problem is given by Brush (1976b); well-known philosophical discussions can be found in Batterman (2002, Chapter 5), Reichenbach (1956, Chapter III), and Sklar (1993, Chapter 9).
[5] Compare Dizadji-Bahmani, Frigg, and Hartmann (2010); North (2011); Robertson (2021); Valente (2021); Werndl and Frigg (2015).

Section 6.1 summarises the basic idea using the simple example of gas mixtures and addresses a proposal of Brown and Uffink (2001) to locate the thermodynamic arrow in the approach to equilibrium. Section 6.2 then motivates and summarises the mathematical structure of thermodynamics – at least, before a second law is introduced – including how it describes time and change. The remainder of the chapter will argue the temporal symmetry of thermodynamics remains even when the second law is introduced, in any of three classic formulations: that of Clausius (Section 6.3), of Kelvin and Planck (Section 6.4), and of Gibbs (Section 6.5).

## 6.1 Undirected Thermodynamics

*Doesn't the second law say that entropy increases to the future?* Reichenbach (1956, p.49) calls the second law of thermodynamics "a principle that makes possible the mathematical expression of a direction controlling the course of physical occurrences".[6] The idea is to argue from the second law of thermodynamics that entropy is increasing and then define the direction of time as the direction of that increase.

However, virtually all formulations of thermodynamics agree that entropy increase is not an axiom of thermodynamics: it is at most derived as a consequence of the axioms, or in some cases taken to be the 'essence' of a thermodynamic system, all things considered. Of course, one can also skip these derivations entirely and just declare that entropy increases to the future. I fully support this practice when it serves the theoretician. But, when a time asymmetry is inserted in this way, as a bald addition to an otherwise time symmetric theory, I will refer to it as a *trivial* time asymmetry.[7] The thesis of this chapter is that, if thermodynamics has a time asymmetry, then it is trivial. Nothing about the formal structure of thermodynamics requires it, just as in a time symmetric theory like statistical mechanics or Newtonian gravitation. As Uffink (2007, p.938) aptly remarks: "[T]he Second Law has often been understood as demanding continuous monotonic increase of entropy in the course of time . . . . There is, however, no basis for this demand in orthodox thermodynamics".

---

[6] Similar remarks can be found in Davies (1977, §2.1) and Sklar (1993, §2.I.3), among many other places.

[7] In Section 5.1, I called this kind of 'arrow' a *Heuristic misfire:* an informal extra-theoretic judgement, with no justification in terms of any well-supported physics.

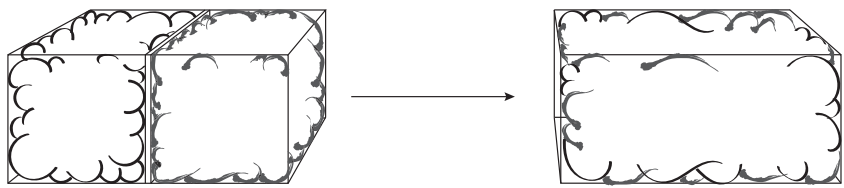

Figure 6.1  Two separated gases and their mixture.

How can thermodynamics be symmetric in time? The basic idea can be illustrated using the simple example of a box that is isolated from any outside influences and which contains two gases in chambers separated by an impenetrable barrier. In a similar system with the barrier removed, where the system is assumed to be 'settled down' into equilibrium, the result is a mixture whose properties can be determined by the properties of the original gases, as shown in Figure 6.1. For example, the mole number of the mixture can be determined on the basis of the component mole numbers; and the entropy of the mixture can be determined on the basis of the component entropies. All predictions in equilibrium thermodynamics have more or less this character: one set of equilibrium systems determines another.[8]

What I would like to point out is that these constraints are fundamentally directionless in time: in this example, as long as the component gases are viewed as 'parts' of the composite, they constrain it in exactly the same way, whether that composite is to the future or to the past. Nothing about the structure of thermodynamics determines that the arrow in Figure 6.1 can only point in one temporal direction. If there is any time asymmetry in thermodynamics, it arise purely from the additional postulate that special initial conditions, or the special perspective of an observer, are associated with one direction in time. But this is just to assume at the outset what we were originally trying to prove.

The situation is strikingly similar to the temporal symmetry that one finds in statistical mechanics.[9] Typical counting arguments in statistical mechanics establish not only that a system will evolve with overwhelming probability to a higher-entropy equilibrium state but also that the system will have evolved from such a state in the past. Like the exploration of a house with a thousand blue rooms and one red room – the next room you will visit is probably blue! One scenario is much more likely than the other, no matter which direction in time one looks in. The predictive mechanism of thermodynamics is similar in spirit: a collection of equilibrium systems determines another equilibrium system (their composite) that maximises entropy, no matter how it is directed in time. In this sense, there is complete harmony between the temporal symmetries of equilibrium thermodynamics and statistical mechanics.

---

[8] This interpretation of thermodynamics has been developed in detail by Callen (1985); see Wills (2022) for philosophical remarks about it.

[9] See Section 5.3.

*But, doesn't thermodynamics still say that equilibrium happens to the future?* A number of careful authors have noticed that the approach to equilibrium itself is not a consequence of the second law, but an independent assumption.[10] For example, in the mixing process for a pair of gases, one must first assume that the two unmixed gases can be associated with a third gas that is an equilibrium mixture of the two; only then can one apply reasoning typically associated with the second law in order to argue that the mixture maximises entropy.

This led Uffink (2001, fn. 93) and Brown and Uffink (2001) to introduce a new postulate into the axioms of thermodynamics, which they call "the Minus First Law":

An isolated system in an arbitrary initial state within a finite fixed volume will spontaneously attain a unique state of equilibrium. (Brown and Uffink 2001)

The Minus First Law is unquestionably an important part of modern physics. In a sense, it is the starting point for a variety of approaches to thermodynamics developed in the twentieth century, often called 'non-equilibrium thermodynamics', which explicitly associate the approach to equilibrium with a continuous time parameter. This approach is used in the study of a variety of physical processes like the relaxation time, mass-flux, and biological cellular processes.[11] But, is the equilibriation process it describes really asymmetric in time?

---

[10] See Uhlenbeck and Ford (1963, §I.3), and especially Brown and Uffink (2001), Marsland III, Brown, and Valente (2015), and Uffink (2001, 2007).

[11] The reciprocity relations of Onsager (1931a,b) provided a framework for studying thermal fluctuations away from equilibrium using techniques pointed out by Einstein (1910) and led to what is known as 'Linear Irreversible Thermodynamics' and 'Extended Irreversible Thermodynamics' (García-Colín and Uribe 1991). An alternative, axiomatic approach sometimes called 'Continuum Thermodynamics' or 'Rational Thermodynamics' was developed by Truesdell (1984), although it has been remarked that, "[i]f there is something 'rational' about Rational Thermodynamics it is certainly well hidden, and interest in it has not stood the test of time". (Lavenda 2010, p.ix). See Haslach Jr. (2011) for a recent overview of non-equilibrium techniques, which are tangential to my purposes here.

Brown and Uffink conclude that it is. Their argument is that when a system approaching equilibrium is time reversed, the result is a system deviating from equilibrium; thus, insofar as this is prohibited by thermodynamics with a Minus First Law, the theory is not time reversal invariant:[12]

> [T]ime reversal ... should then correspond to a world in which processes occur which look like those occurring in a film of our world but played backwards. In particular, the reversal of the spontaneous adiabatic expansion of a gas ... would correspond to a spontaneous adiabatic contraction. But this behaviour is inconsistent with ... the Minus First Law, which as we have seen rules out spontaneous deviations from equilibrium. (Brown and Uffink 2001, p.536)

Time asymmetry here depends on the assumption that a spontaneous decrease of entropy is "inconsistent" with the Minus First Law's statement that equilibrium must be attained. But, inconsistency does not necessarily follow.

The Minus First Law, and any other axiom requiring that a physical system 'attain' equilibrium, is not intrinsically directed in time. The theoretician begins with one description: in equilibrium thermodynamics, it is a collection of equilibrium systems, like the separated gases in Figure 6.1; in non-equilibrium thermodynamics, it is a system out of equilibrium. The Minus First Law only guarantees that this description determines a unique state of equilibrium. This is perfectly consistent with that equilibrium state appearing in both time directions: the initial description is associated with an equilibrium state, which may occur in either the past or the future. The box of separated gases determines an equilibrium state associated with their mixture, whether it occurs in the future or in the past, as in Figure 6.2. Time asymmetry plays no essential role in the requirement that such an equilibrium state exists.

There is no mystery as to why we typically associate one of these scenarios with the future: in practice, one commonly inserts temporally directed language to this description, by referring to an 'intervention' or 'manipulation' that removes the barrier between the gases in a temporally-directed way. When such language is included, one may be inclined to call the mixture the 'effect' or 'future' of the intervention. There is nothing wrong with that. However, it would be misleading to call such an insertion a 'thermodynamic arrow', since there is nothing about the structure of equilibrium thermodynamics that encodes it. Inserting the language of intervention reflects a contingent human perspective, of the kind discussed extensively in Chapter 5, and not an asymmetry of time itself.

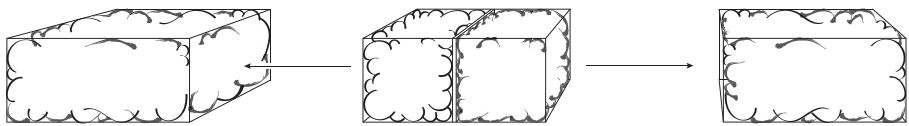

Figure 6.2  An equilibrium state is determined no matter which temporal direction is chosen.

---

[12] This argument has enjoyed some support: compare Marsland III, Brown, and Valente (2015), Myrvold (2020b, Forthcoming), and Valente (2021).

A position very similar to mine has been developed by Myrvold (2020c). Viewing thermodynamics as a theory that relates heat reservoirs to interventions on "manipulable variables", Myrvold develops all the essential structure of thermodynamics, including the zeroth, first, and second laws, in a way that makes no mention of the approach to equilibrium. For my purposes, I prefer to avoid the language of 'manipulation', which might suggest a temporally directed structure that does not exist in the structure of equilibrium states. However, Myrvold is right to identify Brown and Uffink's Minus First Law as lying outside the "province" of thermodynamics, properly conceived. Of course, time asymmetric statements may still be formulated using thermodynamic language.[13] But, these invariably arise from the assumption that the approach to equilibrium is associated with a temporal direction, which is not part of the structure of equilibrium thermodynamics. His diagnosis of the widespread confusion about this is similar to that of Brown and Uffink (2001):

> There is a tendency to conflate the second law of thermodynamics with the tendency of systems to relax to a state of thermal equilibrium, and this has encouraged the idea that the study of equilibration does fall within the scope of thermodynamics. These are not the same thing, however. (Myrvold 2020c, p.1243)

That said, one might still choose to add a time asymmetric axiom to thermodynamics: an axiom which states that systems are associated with not just an equilibrium state but one that occurs in one time direction and not the other. This assumption is implicit in Brown and Uffink's discussion of the Minus First Law. However, this amendment to the 'temporally symmetric Minus First Law' that I have characterised above is no longer a deep structural requirement of thermodynamic theory but the insertion of a time asymmetry of a kind that I call 'trivial'. This is not enough to distinguish the temporal symmetries of thermodynamics from other time symmetric theories.

---

[13] Myrvold (2020c, §9) identifies his statistical version of the Clausius inequality as an example (his Proposition 2) but points out that all the thermodynamic assumptions that go into it are temporally symmetric. Thus, temporal asymmetry enters when a temporal direction is associated with the approach to thermal equilibrium, which lies outside the scope of the theory.

In a nutshell, this is the sense in which I will argue that there is no thermodynamic arrow of time. To make this argument precise, and to square it with the many thermodynamic arrows that have been proposed, a more careful treatment of equilibrium thermodynamics will be needed. In particular, I will aim to show that on a rigorous understanding of the structure of equilibrium thermodynamics, including the first and second laws, there is no thermodynamic arrow of time. So, we begin in the next section with a brief review of the structure of thermodynamics.

## 6.2 Equilibrium Thermodynamics

The Soviet mathematical physicist Vladimir Arnol'd once began an article by declaring, "Every mathematician knows it is impossible to understand an elementary course in thermodynamics". This quote was often passed around by other mathematically-inclined philosophers and physicists when I was a graduate student, as evidence that thermodynamics is a chimera beyond our reach.[14] But, when I finally read the article myself, I found that Arnol'd goes on in the second sentence to say: "The reason is that the thermodynamics is based – as Gibbs has explicitly proclaimed – on a rather complicated mathematical theory, on the contact geometry" (Arnol'd 1990, p.163). In other words, it *is* possible to understand an *advanced* course in thermodynamics!

So, in the remainder of this chapter, I will try to discuss the structure of equilibrium thermodynamics in a mathematically and conceptually rigorous way, at the cost of sometimes introducing a little more differential geometry than is typical.[15] I begin with a brief overview of the structure of thermodynamic state space; I will then discuss two senses in which time and change enter into thermodynamic thinking.

### 6.2.1 Heat, Entropy, and Temperature

In a simple mechanical system, like a mass sitting on top of a hard block of volume $V$, energy $U$ is completely determined by the force of pressure: $dU = -P\,dV$. This situation changes strikingly when the block is replaced with a gas: when we apply pressure to the gas, it gets hotter to our touch! Since this heating effect can be harnessed to produce energy, local energy conservation would require that there be further 'hidden' degrees of freedom into which energy can flow. And of course nowadays we know that there are, such as the microscopic motions of the molecules in the gas.[16] But, never mind the nature of those hidden degrees of freedom: let us embrace our ignorance and refer to those hidden contributions to energy as 'heat'.

---

[14] Cf. Uffink (2001, p.310).
[15] A 'light' version of this approach can be found in the textbook of Callen (1985). Myrvold (2020c) builds a conceptual foundation for thermodynamics that is very close to what I present here. A more detailed development of geometric thermodynamics can be found in Arnol'd (1990), Bravetti (2018), Hermann (1973), or Mrugała (1978); see J. Wills (2021, "Representing mixtures", Unpublished manuscript) and (2022) for philosophical remarks.

Classical equilibrium thermodynamics is the study of the relationship between contributions to energy like pressure and volume, which are directly observable and are called *work*, and those that are not directly observable in this way, which are called *heat*. Burke (1985) calls the latter 'unexamined' degrees of freedom; I will refer to them as 'hidden'. The state space used to describe these degrees of freedom and their structure is what characterises a model of thermodynamics. And, it is this state space that I will argue has no non-trivial time asymmtry.

A *state* of a thermodynamic system is a point in a smooth (for our purposes $C^\infty$) real manifold $N$ of dimension $n$. Each state represents a physical description of the system when it is 'settled down' into stable equilibrium. The *energy* of the system is an exact one-form[17] $dU$, where we view $U$ as a smooth function of the manifold. The various contributions of work to energy, such as pressure and volume, chemical potential and mole number, and so on, all have the form $P_i dX_i$ for some smooth functions $X_i$ and $P_i$. (So, our previous $V$ and $P$ are an example: say, $P_1 dX_1 := P dV$.) We assume at this stage that the functions representing degrees of freedom that are observable or 'not hidden', $(U, X_1, \ldots, X_{n-1})$, form a coordinate system for $N$, with each $P_i : N \to \mathbb{R}$ a smooth function. Since heat is by definition a contribution to energy, the discussion above suggests that it can be represented by a one-form too; a standard convention is to write it with a 'slash', as $\dslash Q$, to emphasise that it might not be exact. We thus arrive at an expression of our conclusion above, that energy can be written as a sum of heat $\dslash Q$ and work $\sum_{i=1} P_i dX_i$, namely,

$$dU = \dslash Q + \sum_{i=1} P_i dX_i. \tag{6.1}$$

Eq. (6.1) is called the *first law of thermodynamics*.[18]

---

[16] Feynman's 'ping pong ball' analogy explains this in simple terms: the compression of a piston into the gas, like a moving ping pong paddle, increases the kinetic energy of the particles that collide with it (Feynman 1963a, §1–2).

[17] A *one-form* on a manifold $N$ is a smooth field of covectors, which at each point maps each vector at that point to a real number. A one-form $\omega$ is *exact* if and only if there exists a smooth function $f : N \to \mathbb{R}$ such that $\omega = df$, where $d$ is the exterior derivative.

[18] If $\dslash Q$ is any one-form on a manifold for which coordinate variables are given by $(U, X_1, \ldots, X_{n-1})$, it can always be expressed as in Eq. (6.1). So, the argumentation above can be viewed as establishing

For our purposes, an important formal question is whether the heat one-form $đQ$ can be expressed in the same way as the other contributions to energy $P_i dX_i$, in the sense that there exist smooth functions $S : N \to \mathbb{R}$ and $T : N \to \mathbb{R}$ such that $đQ = T dS$. The function $T$ is then formally known as an *integrating factor*. The physical significance of this property is that, whatever degrees of freedom give rise to the 'hidden' contributions to energy in the form of heat, they can all be represented by single parameter $S$ called thermodynamic *entropy* and a parameter $T$ called thermodynamic *temperature*. These functions might not exist in general: Myrvold (2020c) presents thermodynamics in a way that does not assume this. However, many applications of thermodynamics apply only to situations in which they do exist.

Carathéodory (1909) gave an influential argument that a one-form representing physical heat should have an integrating factor, given the existence of "adiabatically inaccessible points" in every neighbourhood of each point. His assumption, known as *Carathéodory's principle*, is that in every neighbourhood of every point $p \in N$, there exists a point that is not accessible by any piecewise-smooth curve $\gamma$ satisfying the adiabatic ('no heat exchange') condition $đQ(\bar{\gamma}) = 0$ for every tangent vector $\bar{\gamma}$. Intuitively, this encodes a sense in which processes involving heat exchange are 'always available' in thermodynamics; and, from this it was shown that there are indeed smooth functions $T$ and $S$ satisfying $đQ = T dS$ to characterise temperature and entropy, respectively.[19] This relation of adiabatic accessibility is unexpectedly powerful: in a remarkable series of papers, Lieb and Yngvason (1998, 1999, 2000, 2013) showed that nearly all of thermodynamics can be recovered from axioms characterising adiabatic accessibility, including Carathéodory's principle itself.

Since adiabatic accessibility can be an asymmetric relation, there has been some discussion as to whether it might give rise to the time asymmetry of thermodynamics (Uffink 2001, §9). In short, it does not. One of the central results of Lieb and Yngvason (1999) is a representation theorem, according to which an 'entropy' function $S$ exists that increases with the adiabatic

---

that, whatever the nature of heat is, $N$ is indeed the correct manifold on which to formulate it, as a one-form obeying Eq. (6.1).

[19] The proof of Carathéodory (1909) that $đQ = T dS$ was significantly simplified by Bernstein (1960), who pointed out that Carathéodory only established his conclusion in some local neighbourhood of each point. Conditions for a global integrating factor were obtained by Bernstein (1960) and by Boyling (1968, 1972).

accessibility relation, in that $S(p) \geq S(q)$ if and only if state $p$ is adiabatically accessible from state $q$. But, as Uffink (2001) has emphasised, the axioms used to establish this are completely symmetric in time.[20] Marsland III, Brown, and Valente (2015) have pointed out that as a consequence of this, the representation theorem equally establishes the existence of a decreasing entropy function $-S$ as well.

More importantly for our purposes, there is a different and more perspicuous argument for the existence of an integrating factor, which uses only energy conservation. No asymmetric orderings are needed for entropy and temperature to be defined. This clear-headed approach is due to Jauch (1972) and based on an argument of Tatiana Ehrenfest–Afanassjewa (1925). It can be formulated as follows.

> **Proposition 6.1** (Afanassjewa–Jauch) *Let* $(U, X_1, \ldots, X_{n-1})$ *be a complete set of smooth coordinate functions of a manifold $N$ of dimension $n$, and let $đQ$ be a one-form on $N$, which implies that $dU = đQ + \sum_{i=1}^{n-1} P_i dX_i$ for some smooth functions $(P_1, \ldots, P_n)$ of $N$. Suppose that $W := \sum_{i=1}^{n-1} P_i dX_i$ is 'conserved on adiabats', in that for every closed, piecewise-smooth curve $\gamma$ with tangent vector field $\bar{\gamma}$ satisfying $đQ(\bar{\gamma}) = 0$, we have,*
>
> $$\int_\gamma W = 0. \tag{6.2}$$
>
> *Then there exist smooth functions $S : N \to \mathbb{R}$ and $T : N \to \mathbb{R}$ such that $đQ = T\,dS$ and $T = \partial U / \partial S$.*

The interpretation of the theorem is as follows. Recalling that 'work' degrees of freedom are not enough to capture all the contributions to energy in a thermodynamic system when heat is present, the Afanassjewa–Jauch theorem assumes conversely that, roughly speaking, there are no further contributions to energy besides work and heat. More precisely: when heat exchange is absent, the total work on a closed curve vanishes, as in the classic expression of energy conservation.[21] From this assumption, which I have presented above as the very definition of heat in the formulation of the first law, the theorem ensures that entropy and temperature exist as the formal constituents of that heat. It turns out that Carathéodory's principle of adiabatic inaccessibility follows as a corollary of this theorem.[22]

---

[20] Lieb and Yngvason (1999, §2.6) formulate an additional axiom to guarantee entropy is a concave function, which gives rise to an entropy maximisation principle in the sense of Gibbs (Section 6.5). They also formulate a 'thermal contact' axiom for the description of equilibration (Lieb and Yngvason 1999, §4.1). I conjecture that, insofar as either of these additional axioms are time asymmetric, the time symmetry can be relaxed along the lines suggested in Section 6.1.

[21] See Roberts (2013b) for an interpretation of this statement and its relation to time reversal invariance.

### 6.2.2 Thermodynamics on Contact Manifolds

This completes the conceptual underpinnings of heat in thermodynamic systems. We are now ready to set out the formal structure of thermodynamics, as it might appear in an 'advanced' course. When energy is conserved, the description of a thermodynamic system requires a manifold $M$ of dimension $m = 2n+1$: in local coordinates there is a coordinate function $U$ characterising energy and a set of $2n$ independent coordinate functions $P_i$ and $X_i$, where the special variable $P_0 = T$ is the temperature and $X_0 = S$ is the entropy. Thermodynamics is about situations in which those variables satisfy the first law, $dU = \sum_{i=0}^{n} P_i dX_i$, with each $P_i = \partial U / \partial X_i$ characterising the 'rate of change of energy' with respect to the degree of freedom $X_i$. For example, pressure $P = \partial U / \partial V$ is the rate of change of energy with respect to volume, while temperature $T = \partial U / \partial S$ is the rate of change of energy with respect to entropy.

We will characterise these statements in terms of a particular one-form $\theta$, called the *Gibbs one-form*,

$$\theta := dU - \sum_{i=0}^{n} P_i dX_i. \tag{6.3}$$

At each point of the manifold, there is a special subspace of vectors $v$ of dimension $n$ such that $\theta(v) = 0$, and it is on this subspace that the first law of thermodynamics holds. So, it would be appropriate to formulate thermodynamics on an $n$-dimensional submanifold $N \subset M$ whose vectors $v$ have this property, $\theta(v) = 0$. The contact manifold approach to thermodynamics is built around this idea.

We begin with some background definitions, in order to identify the essential properties of a surface on which the Gibbs one-form vanishes. When each point of $M$ is smoothly assigned a codimension-1 hypersurface, meaning a subspace of dimension $2n$ of the 'total' $(2n + 1)$-dimensional tangent space of that point, the result is called a *field of hypersurfaces*. It can always be locally characterised by a one-form $\eta$ such that $\eta(v) = 0$ for each vector $v$ in a hypersurface. A *contact structure* is a field of hypersurfaces

---

[22] The converse of Carathéodory's principle is that if a heat form $d̸Q$ has an integrating factor $d̸Q = T dS$, then there are adiabatically inaccessible points in every neighbourhood of every point. This well-known statement has a comparatively simple proof (cf. Boyling 1968).

satisfying a non-degeneracy condition called[23] 'maximum non-integrability':
$(d\eta)^n \wedge \eta \neq 0$. The pair $(M, \eta)$, where $M$ is a smooth manifold and $\eta$ is a
contact structure, is called a *contact manifold.*

The physical significance of maximal non-integrability is that it selects a
surface $N$ that is 'as large as it can be', while still satisfying the first law on a
manifold of dimension $2n+1$; and, it turns out that $N$ must have dimension $n$,
capturing an essential property of the first law. If a submanifold $N \subset M$ is in
'contact' with a field of hypersurfaces, in that each of its tangent vectors $v$ is
tangent to a hypersurface $\eta(v) = 0$, then it is called an *integral submanifold* of
the field. An integral submanifold of maximal dimension is called a *Legendre
submanifold*. And, on a $(2n + 1)$-dimensional contact manifold, a Legendre
submanifold always has dimension $n$ (cf. Arnol'd 1989, Appendix 4). So,
this is just what we need to define an $n$-dimensional submanifold on which
the first law holds.

In particular, the Gibbs one-form $\theta$ of Eq. (6.3) is an example of a contact
structure; conversely, every contact structure $\eta$ can be expressed in local
coordinates in the form of the Gibbs one-form (Arnol'd 1990, p.167). So, the
essential structure of the first law really is captured by a contact manifold
of dimension $2n + 1$, together with a Legendre submanifold of dimension $n$,
which characterises a maximal surface on which the first law holds, $\theta = 0$.
The contact manifold $(M, \theta)$ is often referred to as a *thermodynamic phase space*
and the choice of a Legendre submanifold as a 'representation', such as the
energy or entropy representation. This allows for a precise definition of a
model of thermodynamics, originally proposed by Hermann (1973, p.264):

> **Definition 6.1** An 'equilibrium thermodynamic system' is a triple $(M, \theta, N)$,
> where $M$ is a $(2n + 1)$-dimensional manifold, $\theta$ is a contact structure, and $N$ is
> an $n$-dimensional Legendre submanifold.

Writing the contact structure in local coordinates as the Gibbs one-form,
and expressing the energy variable $U$ as a function of the $n$ coordinate
variables on a Legendre submanifold, we now get what Gibbs (1873) called
the 'fundamental relation' of a thermodynamic system:

$$U = f(X_1, \ldots, X_n). \tag{6.4}$$

Taking derivatives, we get $dU = (\partial f/\partial X_1)dX_1 + \cdots + (\partial f/\partial X_n)dX_n$, and
defining $P_i := \partial U/\partial X_i$ for each $i = 1, \ldots n$, we recover the first law.

---

[23] This condition gets its name from the concept of an 'integrable' field of hypersurfaces, characterised
by the condition that $d\eta \wedge \eta = 0$. A manifold can only satisfy maximal non-integrability if it is of odd
dimension; for an introduction, see Arnol'd (1989, Appendix 4).

The $n$ equations for $P_i$ are known as 'equations of state'. For example, a simple gas has thermodynamic phase space given by the five-dimensional manifold $M$, with the Gibbs one-form $\theta = -dU + T\,dS - P\,dV$. The two-dimensional Legendre submanifold is characterised by the coordinate functions $(V, S)$, and the equations of state are determined by calculating $P = \partial U/\partial V$ and $T = \partial U/\partial S$. The statement that $V$ and $S$ are 'extensive' or scale with the energy of the system is captured by a further assumption[24] that $U$ is a first-degree homogeneous function, $f(\lambda V, \lambda S) = \lambda f(V, S) = \lambda U$ for all $\lambda > 0$. The remaining variables $P$ and $T$ are called 'intensive'.

Virtually all aspects of equilibrium thermodynamics can be treated using this geometric foundation as a starting point. Gibbs (1873) himself pointed out that even the description of phase transitions can be captured by the study of the boundaries of convex surfaces, of the kind discussed in Section 6.5.[25] More recent work has approached this through the study of an additional metric and its curvature singularities (cf. Quevedo 2007; Quevedo et al. 2011). But, this is more or less where the consensus about the foundations of thermodynamics ends. We have so far said nothing about dissipation or the second law, or their bearing on the arrow of time. We will review these matters in the remaining sections. But first, let me develop one more piece of background material: the nature of time and time reversal in thermodynamics.

### 6.2.3 Time and Time Reversal in Thermodynamics

It is well-known that the description of time is an awkward part of equilibrium thermodynamics: in spite of the name, the theory does not describe time and change in the way that most dynamical theories do:

In contrast to mechanics, thermodynamics does not possess equations of motion. This, in turn, is due to the fact that thermodynamical processes only take place after an external intervention on the system (such as: removing a partition, establishing thermal contact with a heat bath, pushing a piston, etc.). (Uffink 2001, p.315)

This latter fact led Wallace (2014) to argue that thermodynamics is "misnamed" and better viewed as part of the general study of interventions using control theory. In a similar spirit, Myrvold (2011) formulates it as a theory relative to the "means" available to the experimentalist, referring to thermodynamics as a "resource theory" (Myrvold 2020b, Forthcoming).

---

[24] We will return to this assumption in Section 6.5.
[25] For an introduction, see Wightman (1979).

Other authors refer to large parts of thermodynamics as 'thermostatics' (cf. Tribus 1961). Nevertheless, there are at least two ways that time evolution is routinely represented in thermodynamics, which I will refer to as the *quasistatic process approach* and the *constraint approach*.

My discussion of each approach will adopt the perspective developed in earlier chapters, which I have called the Representation View: instead of studying time as a coordinate variable, we study time translations and view time evolution in a physical theory as a representation of those time translations on a state space.[26] In this way, studying the properties of time translations on state space can be used to determine the symmetries of time itself. As in earlier chapters, we take a representation of time translations to be a homomorphism $\varphi$ from a (possibly local) Lie group of time translations $G$ to the automorphisms of a theory's state space. The set of automorphisms of an equilibrium thermodynamic system $(M, \theta, N)$, which we denote by $\mathrm{Aut}(M, \theta, N)$, consists of *contact transformations*, or smooth maps $\phi : M \to M$ that preserve the contact structure, $\phi^* \theta = \theta$; we assume also that they preserve the Legendre submanifold $N$.

### The Quasistatic Process Approach

The quasistatic process approach, advocated by Carathéodory (1909) and Ehrenfest–Afanassjewa (1925); Ehrenfest-Afanassjewa (1956), views a piecewise-smooth curve through thermodynamic phase space $(M, \theta)$ as an approximate representation of 'quasistatic' change: a process for which the system can at all times be approximated as being in an equilibrium state. This occurs, for example, when the volume of a box of gas is very slowly increased. There is some philosophical work to do in clarifying exactly how this representation is supposed to work (see Lavis 2018; Norton 2014, 2016b).[27] However, that debate is tangential to my purposes, which will aim to determine whether there is any sense of irreversibility in this perspective on time translations.

Viewing time as associated with the Lie group $G = (\mathbb{R}, +)$, a representation of time translations of this kind is given by any map $t \mapsto \varphi_t$ to the contact transformations[28] of $(M, \theta, N)$ that is continuous, preserves $N$, and satisfies $\varphi_{t+t'} = \varphi_t \circ \varphi_{t'}$. This allows us to apply the account of time reversal developed in Chapter 2, as an extension of the group of time translations to include an element that reverses time translations, $t \mapsto -t$. When our representation can be extended to include time reversal, it transforms a curve representing a quasistatic process to one that 'flows in the opposite direction' with the reverse tangent vector field, and we say that the theory is time reversal invariant; otherwise, it is time reversal violating. This recovers one standard approach to time reversal in thermodynamics (cf. Uffink 2001, §3).

---

[26] See Section 2.3.
[27] Norton (2016b) points out that this concern goes back at least to Duhem (1903, §58) and offers an interesting clarification; see Valente (2017) for an alternative view.
[28] Strictly speaking, it is only each smooth part of the flow $\varphi_t$ that defines a (local) representation of time translations; the piecewise-smooth flow $\varphi_t$ may then represent a finite sequence of representations of time translations corresponding to a sequence of successive processes.

Most who argue that thermodynamics is time asymmetric adopt the quasistatic process approach to time and time reversal: this is the case for both Clausius (discussed in Section 6.3) and Planck (discussed in Section 6.4). In the next sections, I will argue that they do not succeed. But first, let me review a very different approach to change in thermodynamics.

### The Constraint Approach

The constraint approach is an alternative view of thermodynamic prediction, which finds its natural home in the work of Gibbs (1876, 1877), and more recently in the textbook by Callen (1985), who writes:

> The single, all-encompassing problem of thermodynamics is the determination of the equilibrium state that eventually results after the removal of internal constraints in a closed, composite system. (Callen 1985, p.26)

This approach is fundamentally about the relationship between two sets of equilibrium systems. The first is a collection of thermodynamic systems, often interpreted as subject to some 'constraint'; the second is a single equilibrium system that is taken to be determined by the first collection, usually by the removal of that constraint. We saw a paradigm example of this above: two gases separated by a barrier can be viewed as determining a mixture when the barrier is removed and once the gases have settled down into an equilibrium state again.

The constraint approach to equilibrium thermodynamics has the advantage of avoiding the awkward appearance of non-equilibrium states. Instead, we describe a process in two stages: one characterising a collection of separate equilibrium systems, and another characterising the single equilibrium system that they determine. We can make predictions by associating these two stages with two moments in time, as when we first observe two separate gases and then later observe a mixture. We can even introduce structure that captures the fact that the entropy of the unconstrained system maximises the entropy of the constrained ones (see Section 6.5). However, this introduction of a temporal direction is entirely independent of the constraint approach because nothing about the relationship it describes between systems requires them to occur in any particular time-order. On the contrary, a collection of equilibrium systems determines another equilibrium system, regardless of whether the latter occurs at an earlier or later time.

Remarkably, the constraint approach provides a 'timeless' view of thermodynamics. We can use it to model changes in time if we wish and can even supplement it with a temporal direction. But, these are at best trivial additions of a thermodynamic arrow, in the sense of ad-hoc additions not required by the otherwise undirected structure of equilibrium states. In the remaining sections, I will argue that even the classic expressions of the second law have this character: when they can be rigorously formulated at all, they are fundamentally undirected in time.

## 6.3 The Clausius Inequality

Most arguments in support of a thermodynamic arrow begin with the second law and then argue that entropy is non-decreasing in just one temporal direction. Perhaps the oldest and most well-known argument of this kind is due to Clausius.[29] The Clausius argument for increasing entropy suffers from conceptual problems, as Uffink (2001) has pointed out. In this section I will give a precise reconstruction of it and argue that its use as evidence for a thermodynamic arrow is completely unfounded.

The Clausius approach adopts less structure than the complete thermodynamic state space introduced in Section 6.2.2. In particular, we begin with a manifold $N$ of dimension $n$ on which the first law holds but without assuming at the outset that there is a global entropy function $S$ that can be viewed as a coordinate variable. Thus, as in Section 6.2.1, we write the first law as $dU = \text{đ}Q + \sum_{i=1}^{n-1} P_i dX_i$, where $(U, X_1, \ldots, X_{n-1})$ are coordinate functions for $N$, each $P_i$ is a smooth functions of $N$, and $\text{đ}Q$ is a one-form representing heat. A central part of this argument is now to assume that it is possibly the case that $\text{đ}Q \neq T dS$ for all smooth functions $T : N \to \mathbb{R}$ and $S : N \to \mathbb{R}$ – and so, by the Afanassjewa–Jauch theorem, energy is not necessarily conserved in the sense that Eq. (6.2) fails! But let us press forward for now. Then, the *Clausius inequality* says: given any 'physically possible' cycle described by a closed curve $c$, together with a function $T$ representing temperature and a one-form $\text{đ}Q$ representing heat, we have that

$$\int_c \text{đ}Q/T \leq 0. \qquad (6.5)$$

---

[29] This argument was introduced by Clausius (1865); see Uffink (2001) for a detailed review and Henderson (2014) and Valente (2021, §4.1) for recent supportive remarks.

Clausius argued that this inequality follows from what is now known as the 'Clausius version' of the second law, that "[h]eat cannot of itself pass from a colder to a hotter body without some other change, connected herewith, occurring at the same time" (see Uffink [2001], p.333).

Two definitions are now needed: we say that a curve $\gamma$ is *thermodynamically reversible* with respect to a one-form $đQ$ if and only if $đQ$ has an integrating factor when it is restricted to that curve, in that there are smooth functions $T$ and $S$ such that $đQ = T\,dS$ at least on every tangent vector to $\gamma$, though not necessarily in general. Otherwise, we will say that $\gamma$ is *thermodynamically irreversible*. If a closed curve $c$ is thermodynamically reversible, then, since the integral of an exact one-form is path-independent, the Clausius inequality becomes identically zero. In short, strict inequality only holds on a closed curve that is thermodynamically irreversible with respect to $đQ$.

The Clausius argument that 'entropy does not decrease' can now be reconstructed as follows:

> **Proposition 6.2** *Let $T$ be a smooth function on a manifold $N$, and let $đQ$ be a one-form. Let $\gamma$ be a smooth curve from $p_i$ to $p_f$ that is adiabatic ($đQ(\bar{\gamma}) = 0$ for all tangent vectors $\bar{\gamma}$) but not necessarily reversible with respect to $đQ$. If the Clausius inequality is satisfied, $\int_c đQ/T \leq 0$ on all closed curves, and if there exists a curve $\gamma'$ from $p_f$ to $p_i$ along which $d(đQ) = 0$, then $đQ = T\,dS$ on $\gamma'$ for some smooth function $S$, and any such function must satisfy $S(p_i) \leq S(p_f)$.*

*Proof*  The fact that $d(đQ/T) = 0$ along $\gamma'$ says that $đQ/T$ is a closed one-form on the contractible manifold $\gamma'$, and so the Poincaré lemma implies that $đQ/T$ is exact on $\gamma'$, in that there exists a smooth function $S$ such that $đQ/T = dS$ for all tangent vectors $\bar{\gamma}'$.

A closed curve $c$ now runs from $p_i$ to $p_f$ along $\gamma$ and then back along $\gamma'$ from $p_f$ to $p_i$, as in Figure 6.3. Therefore, applying our adiabatic assumption and the Clausius inequality,

$$\int_{\gamma'} đQ/T = \underbrace{\int_{\gamma} đQ/T}_{=0} + \int_{\gamma'} đQ/T = \int_c đQ/T \leq 0. \qquad (6.6)$$

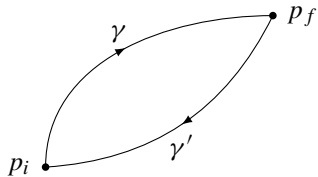

Figure 6.3  The closed curve $c$.

But, we also have by definition that

$$\int_{\gamma'} \dbar Q/T = \int_{S(p_f)}^{S(p_i)} dS = S(p_i) - S(p_f). \tag{6.7}$$

Combining Eqs (6.6) and (6.7) thus establishes that $S(p_i) \leq S(p_f)$.    ■

In summary, the precise sense in which Clausius argues that entropy is non-decreasing is: if thermodynamics is equipped with the structure of a second law in the form of the Clausius inequality, then any (possibly thermodynamically irreversible) adiabatic process $\gamma$ from $p_i$ to $p_f$ can be supplemented with a smooth function $S$, which behaves like entropy with respect to a different, reversible process $\gamma'$ and is such that $S(p_i) \leq S(p_f)$.

As Uffink (2001, p.339) has emphasised, this argument is not perfectly general because of its restriction to adiabatic processes; and, one might be concerned about whether a reversible path $\gamma'$ exists on which entropy is defined or, if it does, whether it represents any physically realistic process. I think the situation for Clausius is much worse: the conceptual problems with this argument are so severe that I do not see any way to justify, on this basis, Clausius' conclusion that entropy tends to increase. Here are three problems with the argument.

### Problem 1

The Clausius argument applies only to systems with a heat form $\dbar Q$ that does not admit global temperature and entropy functions for which $\dbar Q = T\, dS$. By definition, this is required for the existence of an irreversible curve. The problem is that this also implies a violation of local energy conservation, since otherwise we would have $\dbar Q = T\, dS$ by the Afanassjewa–Jauch theorem. That might be taken as evidence that entropy can only increase in 'open' systems, for which local energy conservation can be expected to fail. But, then the often-repeated dictum of Clausius that "[t]he entropy of the universe tends to a maximum" would be false, insofar as the universe is by definition a closed system.[30] This means that it is not an argument that establishes a direction of time for the universe as a whole, but at best for local subsystems.

---

[30] Clausius writes, "Die Entropie der Welt strebt einem Maximum zu" (Clausius 1867, p.44). Followers of Clausius like Planck were aware of difficulties with this statement; see Uffink (2001, pp.338–9). Earman (2006) has argued that such statements about the entropy of the universe are not even false, but meaningless.

## Problem 2

A second, more serious problem is that the Clausius argument only establishes the existence of an increasing entropy function with respect to the curve $\gamma'$: why is the function $S$ associated with this reversible curve chosen, and not some other? Even worse, the smooth function $S$ constructed to represent the 'entropy' of an irreversible curve $\gamma$ in Proposition 6.2 is questionable: it deserves the name at best on the *reversible* curve $\gamma'$ for which $đQ = T\,dS$ but not on the irreversible curve $\gamma$ for which $đQ$ does not generally take this form. When this $đQ$ is not exact, it is not associated with any global entropy function at all. So, the choice to interpret this particular function $S$ as the entropy of a 'possibly irreversible' adiabatic process is totally arbitrary, since a different choice of reversible curve $\gamma$ would have led to a different entropy function. That this arbitrary function is standardly referred to as unqualified 'entropy' is a remarkable sleight of hand.

Let me make the arbitrariness stark: we could equally have chosen the function $S' = -S$ to represent entropy along the (possibly) irreversible curve $\gamma$. This is a function for which entropy is non-*increasing* from $p_i$ to $p_f$. The very same curve $\gamma'$ running in the opposite direction from $p_f$ to $p_i$ undergoes a positive increase in entropy with respect to this new function $S'$. It is hard to see why this reversed path should be impossible from the perspective of the Clausius argument, since $\gamma'$ represents a reversible process by construction. Of course, one could avoid the problem by postulating that $S' = S = 0$. But, then the argument would just result in entropy being constant and not that it 'tends to increase'.

## Problem 3

This last problem suggests that on general symmetry considerations, the Clausius inequality cannot establish a general increase of entropy, even for adiabatic processes. Suppose we say that every 'physically possible' cycle is described by a triple $(đQ, T, c(t))$ that satisfies the Clausius inequality $\int_{c(t)} đQ/T \leq 0$, where I now write the explicit parametrisation of the curve $t \mapsto c(t)$. For each such system, simple properties of integrals[31] imply that $(đQ, -T, c(-t))$, $(-đQ, T, c(-t))$, and $(-đQ, -T, c(t))$ satisfy the Clausius inequality as well, where $t \mapsto c(-t)$ denotes the parametrisation of the curve $c(t)$ in the reverse direction. The former two of these triples describe a process unfolding along the same curve $c$, but in the reverse direction. So, if entropy is increasing along some portion of that curve in the cycle ($đQ, T, c(t)$), then entropy is decreasing in these other two cycles. This problem was pointed out by Ehrenfest–Afanassjewa (1925), who ruled it out by the further postulate of a positive temperature function $T > 0$. However, I do not see any way to rule out any of these systems without assuming by fiat what we were originally trying to prove: that entropy must increase to the future, not to the past.

---

[31] Namely, $\int_{c(t)} đQ/T = -\int_{c(-t)} đQ/T = \int_{c(-t)} dQ/(-T) = \int_{c(-t)} (-dQ)/T$.

I agree that the Clausius inequality does provide a substantial addition to the structure of thermodynamics. However, I can find no physically meaningful sense in which it establishes an increase in entropy, let alone a general arrow of time. Indeed, although Clausius suggested that his inequality implies that the entropy of the universe tends to increase in his earlier papers, it is not surprising that he appears to have deleted any mention of this some years later when those papers were collected into a book.[32]

### 6.4 Planck's Argument

Another popular approach to arguing that entropy is non-decreasing is due to Planck (1897b). Planck produced a series of related arguments in the eleven successive editions of his influential book on thermodynamics, which have been carefully reconstructed and evaluated by Uffink (2001, §7, §10.1). Planck's arguments are even less rigorous than the Clausius argument I have presented above and suffer from similar conceptual problems. It is enough for my purposes to review one of them here. For, as Uffink (2001, p.376) points out, even the final version does not give a substantial improvement on the earlier ones.

Like Clausius, Planck adopts the continuous process approach to describing the passage of time. So, if he can show that a continuous curve from one state to another is such that no physical process exists that can restore the original state, then he will have established an arrow of time. Planck calls such a curve 'irreversible', which differs from the usage of Clausius, who adopts this word specifically to describe curves on which the heat form fails to admit an integrating factor, $đQ \neq T dS$.

Planck's style of argument requires one to say which continuous curves are 'physically impossible', which is not found in the structure of thermodynamics presented in Section 6.2. For this, Planck, adopts an expression of

---

[32] This was pointed out by Kuhn (1987); see Uffink (2001, pp.339–340).

the second law of thermodynamics, which he takes to prohibit the existence of a *perpetuum mobile* of the second kind. This is known as:

*Kelvin's Principle:* No cyclic process exists whose effect on the environment "consists of nothing other than the production of work and the absorption of the equivalent heat."[33]

Here, one considers a description of two thermodynamic systems, one representing the 'target' of investigation and the other the 'environment'. In the environment, since the energy one-form is exact, a closed curve will satisfy $\int_c dU = 0$, and so the first law implies that $-\int_c \dbar Q = \int_c W$, where $\dbar Q$ and $W = \sum_i P_i dX_i$ represent heat and work done on the environment, respectively. Kelvin's principle of no work or heat loss can now be interpreted formally as the statement that $-\int_c \dbar Q = \int_c W \leq 0$. Like in the Clausius argument, we must assume here that the heat one-form $\dbar Q$ is not exact, violating local energy conservation by the Afanassjewa–Jauch theorem, since otherwise the inequality would be identically zero.

Planck's style of argument is now to consider a process along which entropy increases and to present a *reductio ad absurdum* of the hypothesis that some physical process is described by a curve that leads back to the original low-entropy state. To be concrete, I will review just one of Planck's examples. Consider the mixing of two gases of the same temperature, one of which has higher pressure, and which are initially separated by a barrier, as in the example of Section 6.1. Here is a summary of the process Planck (1897b, §118–124) describes and which is illustrated in Figure 6.4.[34]

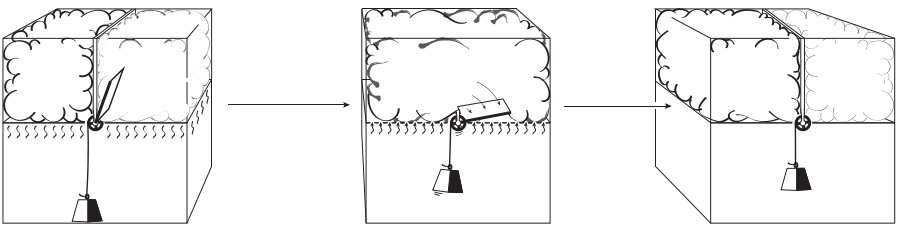

Figure 6.4 Planck's procedure for producing work violating Kelvin's principle from an entropy-decreasing process.

---

- *First stage:* Two unmixed gases are assumed to have the same temperature but one having higher pressure, and the 'environment' is assumed to be associated with some large amount of heat.
- *Second stage:* The gases are slowly (quasistatically) transformed into a mixture, while absorbing heat from the environment and losing work in equal amounts, with the temperature of the gases remaining the same throughout. The environment correspondingly loses heat and gains work in equal amounts, which we interpret as 'heat and work exchange' between the system and its environment.
- *Third stage:* The mixture is now completely isolated from the environment, and the *reductio* hypothesis is applied: we assume there is an entropy-decreasing process that restores the mixture to its original low-entropy state as two unmixed gases. The environment remains in its previous state of having lost heat and gained work.

The result is that the gases complete a cycle, and the environment loses heat and gains work in the sense that $-\int_c đQ = \int_c W > 0$, thus violating Kelvin's principle. The conclusion by *reductio* is that the entropy-decreasing process that restores the original state is impossible, and so time reversal invariance is violated.

As Uffink (2001, §7.5) points out, there are a number of problems with this approach: for example, Planck does not manage to give meaning to the concept of 'entropy' in arbitrary non-equilibrium systems; and, his formulation of the second law is not clearly justified for arbitrary thermodynamic processes.

However, the problems with the Clausius argument discussed in Section 6.3 apply to Planck's argument as well. In the first place, the heat one-form $đQ$ is not exact by assumption, and so Planck's systems do not satisfy local energy conservation. This means that his argument at best establishes a time asymmetry of open systems and not for closed systems or for the universe as a whole. In the second place, the fact that there is not generally a function $S$ such that $đQ = TdS$ means that the association of an entropy function with Planck's proposed irreversible descriptions is questionable at best. Although his discussion is less formal than the one of the previous section, this does not relieve him of the burden of giving meaning to concepts he uses.

Finally, there is a symmetry argument here too, which makes it hard to see how Planck's thinking could ever give rise to a time asymmetry. The key assumption, Kelvin's principle, says that whatever initial thermodynamic system we begin with, we must attain a new system in which there is no net increase in work and heat loss in the environment. Formulated in this way, Kelvin's principle is similar to the Minus First Law's requirement that every system 'attain' a state of equilibrium. This sort of statement is not intrinsically directed in time: it can be consistently applied whether the system attained is in the past or in the future. At its core, the principle only expresses that a target and environment system determine a 'new' target and environment system, in which there is no net increase in work and heat loss in the new environment. But, if this principle can be applied at all, then it can be applied whether that 'new' pair of systems is in the future or in the past.

In other words, the structure of thermodynamics can be retained, even with the addition of Kelvin's principle, without introducing a temporal direction. If one particular temporal direction is chosen, then this appears to be due to a special human understanding of what it means to be an intervention, or to special initial and boundary conditions. But, this is what I have called a 'trivial' introduction of a time asymmetry, which arises from heuristic considerations from outside the theory.

## 6.5 Concavity, Convexity, and Gibbs

Our final discussion of time asymmetry in thermodynamics involves the Gibbs formulation of the second law. Gibbs (1876, 1877) set out his great vision of thermodynamics in a monograph-length work, which spanned over 300 pages in two journal issues. The mathematician Robert Hermann had a positive view of this work and an impressively negative view of the work that followed:

After much reading on the subject, I would say that Gibbs understood, in about 1870, the mathematics of thermodynamics – even in its most 'modern' form – better than almost all of the authors who followed him. Unfortunately, physicists and chemists think that a reasonable statement of the mathematics which they are trying to apply is an 'axiomatization,' and therefore should be avoided as a bad thing which inhibits the creative mind. (Hermann 1973, p.261–2)

In particular, the Gibbs formulation of the second law provides an elegant and conceptually clear addition to the mathematics of thermodynamics, which retains the basic character of the second law as a 'principle of entropy maximisation'. However, the prospects for this approach to establish temporal asymmetry have been succinctly captured by Uffink (2001, p.361): "Obviously, there are no implications for the arrow of time in the second law as formulated by Gibbs". I agree. However, it is helpful to briefly review the Gibbs approach, as a step towards clarifying how it is possible for thermodynamics to have no arrow of time.

A modern view of the Gibbs approach begins with an equilibrium thermodynamic system $(M, \theta, N)$. The Legendre submanifold $N$ is described in terms of a fundamental relation, which I will write in coordinate form as $U = f(S, X_1, \ldots, X_{n-1})$, where $f$ is smooth and is a bijective function of entropy $S$ when all the other variables are held fixed. I will continue to make this argument using coordinate variables, since this will allow me to make the point in relatively simple language. However, deep approaches also exist that are entirely geometrical.[35]

To motivate a second law, we will make just three physical assumptions about a system that has attained equilibrium. I will state these assumptions first and then comment on their interpretation. Perhaps surprisingly, these are all that we need to derive an entropy maximisation principle:

1. *Temperature is positive $T := \partial U / \partial S > 0$.*
2. *Energy is (positively) first-degree homogeneous*, in that when viewed as a function of the other variables $U(S, X_2, \ldots, X_n)$, it satisfies $U(\lambda S, \lambda X_1, \ldots, \lambda X_{n-1}) = \lambda U(S, X_1, \ldots, X_{n-1})$ for all $\lambda > 0$.
3. *Energy has a strongly stable ground state* with a global minimum and no other local minima or maxima; in other words, $U$ is a *convex function*. A convex function of one variable is defined by the property that $U\big(\alpha S_1 + (1 - \alpha)S_2\big) \leq \alpha U(S_1) + (1 - \alpha)U(S_2)$ for all $\alpha \in (0, 1)$, which implies that it has the expected shape illustrated in Figure 6.5.

Let me now interpret these statements. The requirement that temperature $T := \partial U / \partial S$ is positive implies that energy is a monotonically increasing function of entropy. As far as I can see, whether it is increasing or decreasing is conventional. However, the statement that temperature is monotonic is a substantial empirical postulate about the degrees of freedom associated with heat, which by definition are not available to intervention. Postulate 2 says that when energy changes with respect to those degrees of freedom, the entropy function $S$ used to capture them never reaches a local extremum. This need not be true of all physical systems;[36] however, equilibrium thermodynamics can be viewed for our purposes as a theory about those systems for which it is.

---

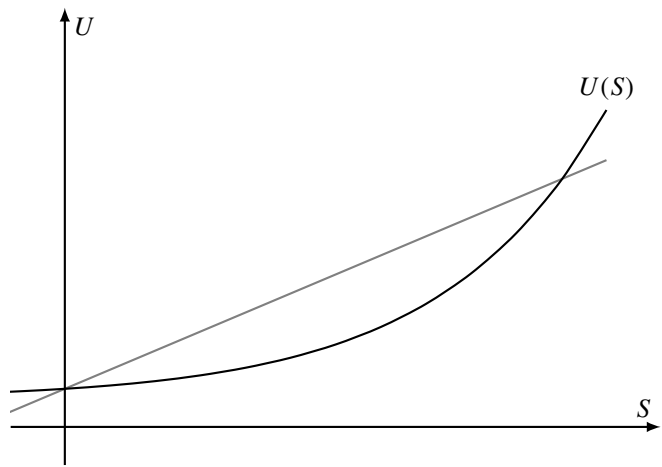

Figure 6.5  A convex function $U(S)$.

The statement that energy is a first degree homogeneous function is also substantial. But it is a natural assumption about thermodynamic systems that have attained equilibrium. Roughly speaking, it expresses the idea that such a system consists of 'similar stuff' in all of its parts, with all inhomogeneities smoothed out. First-degree homogeneity expresses this property in a clever way: we assume that when we scale a thermodynamic system by some factor, such as by doubling the volume and entropy of an ideal gas, the energy is scaled by the same factor.

Finally, we have assumed that energy has a stable ground state in a certain strong sense, expressed by the requirement of convexity. This assumes, as one would typically do, that energy has a global minimum. But, our assumption demands an even stronger sense of stability: that no amount of perturbation will send the system into a 'local' ground state characterised by a local minimum or maximum. Again, there is no requirement that every physical system will satisfy this assumption. But, it is a plausible restriction on the domain of application of thermodynamics; let us restrict our study to systems that are stable enough to satisfy this requirement.

These assumptions, remarkably, are enough to establish an entropy maximisation principle. To manage notation, let me write $X = (U, X_1, \ldots, X_{n-1})$ to denote a set of values of energy and the other non-entropy extensive variables. Then we have the following result.[37]

---

**Proposition 6.3** *Let $U = f(S, X_1, \ldots, X_{n-1})$ be a smooth function, bijective in $S$, such that,*

1. *(positive temperature) $T = \partial U/\partial S > 0$;*
2. *(homogeneity) $U(X) = f(S, X_1, \ldots, X_{n-1})$ is first-degree homogeneous; and*
3. *(strong stability) $U = f(S, X_1, \ldots, X_{n-1})$ is convex.*

*Then $S(X) = S(U, X_1, \ldots, X_{n-1})$ is concave, and for any fixed value of $X = (U, X_1, \ldots, X_{n-1})$,*

$$S(X) = \sup \{ S(\bar{X}) + S(\tilde{X}) \}, \tag{6.8}$$

*where the supremum ranges over all points $\bar{X}$ and $\tilde{X}$ such that $X = \bar{X} + \tilde{X}$.*

*Proof*  The function $U$ is a smooth, convex bijection of $S$, and so it is strictly monotonic in $S$. Since $\partial U/\partial S > 0$, this means that it is strictly increasing, which by an application of the definitions implies that its inverse $S(U, X_1, \ldots, X_{n-1})$ is concave. Since $S$ is the inverse of a homogeneous function, it is also homogeneous.

Now, viewing $S$ as a function of $X := (U, X_1, \ldots, X_{n-1})$, an elementary property of first-degree homogeneous functions $S(X)$ guarantees that $S$ is concave if and only if

$$S(\bar{X} + \tilde{X}) \geq S(\bar{X}) + S(\tilde{X}), \tag{6.9}$$

for all values of $\bar{X}, \tilde{X}$ (Rockafellar 1970, Theorem 4.7). For any fixed value of $X$, this means that $S(X)$ is an upper bound for the set $\{ S(\bar{X}) + S(\tilde{X}) \mid \bar{X} + \tilde{X} = X \}$. Moreover, if $u \leq S(X)$ is any other upper bound, then since $S(X/2) + S(X/2)$ is an element of the set, homogeneity implies

$$u \geq S(X/2) + S(X/2) = S(X)/2 + S(X)/2 = S(X) \tag{6.10}$$

and hence that $u = S(X)$ is a supremum. ∎

Let me comment briefly on the interpretation of this theorem. For a given thermodynamic state with fixed values of energy and the 'work' degrees of freedom, $X = (U, X_1, \ldots, X_n)$, we can consider any pair of states $\bar{X}$ and $\tilde{X}$ that sum to the original ($\bar{X} + \tilde{X} = X$) to be 'possible contributions' to that state. We might now imagine that each of these various possible contributions could in fact occur and note that when we sum the entropies $S(\bar{X}) + S(\tilde{X})$ of each pair of contributing states, we will get a variety of different values. But, the theorem says that when systems achieve equilibrium in the strong sense of our three postulates – positive temperature, homogeneity, and a strong sense of stability – then the only possible contributions that actually occur are those that render the total entropy as large as it can be. The entropy of a given state is the supremum of the possible contributing entropies.

Entropy maximisation in this 'Gibbs sense' has a wide range of applications in equilibrium thermodynamics. It is sufficient, for example, to identify the entropy increase of a pair of mixing gases (cf. Wightman 1979). Wills (2022) has shown that it can be applied in the resolution of a well-known thermodynamic puzzle, the Gibbs paradox.

However, entropy maximisation in the Gibbs sense *has no implications about the direction of time.* The principle might be viewed as describing how the entropy of possible contributions or 'parts' of a system are related to the whole. But, this is a statement about what philosophers call 'mereology', or the nature of parthood relations, and not about the way things change over time.

As a result, when Gibbs entropy maximisation is applied to describe time developments, it remains temporally symmetric. To go back to our paradigm example from Section 6.1: if a mixture in equilibrium is viewed as arising from a pair of separate gases, that mixture will achieve a maximum entropy state, whether it is viewed as occurring in the future or in the past. That is, Gibbs entropy maximisation is most naturally viewed from the 'constraint' perspective on prediction in thermodynamics introduced in Section 6.2.3, that a collection of thermodynamic systems constrains another equilibrium system to achieve maximum entropy, whether that maximum occurs in the future or in the past.

## 6.6 Summary

In spite of the heterodox thesis of this chapter, most of what I have said is already suggested by the work of Uffink (2001) and Brown and Uffink (2001): a direction of time is not encoded in the formal structure of thermodynamics in its first or second laws. Insofar as a time asymmetry is found there, it is 'trivial', in the sense of an ad hoc addition that is not required by the theory. Their argument is only further supported by the presentation I have given of the formal structure of thermodynamics: of the work of Carathéodory, and of the second law as it appears in the work of Clausius, Planck, and Gibbs. When the latter arguments are made precise, any semblance of a thermodynamic arrow of time disappears.

To reach my conclusion, one only needs to take Brown and Uffink's observations about the second law and apply them equally to the 'approach to equilibrium' encoded in their Minus First Law. Like the second law, the association of each collection of thermodynamic systems with a unique equilibrium system does not require a temporal direction. All the formal foundations and predictive structure of thermodynamics can be constructed in its absence. Of course, one can still postulate temporal asymmetry in a trivial way, as in every temporally symmetric theory, by supplementing it with the explicit postulate that change occurs asymmetrically in time. But, if we do not indulge in trivialities, then there is no thermodynamic arrow of time.

# 7

# Time Reversal Violation

---

***Précis.*** *Time reversal symmetry violation in the electroweak interactions provides evidence for an asymmetry of time itself, given the Representation View.*

---

One common way to characterise the arrow of 'time itself' is using the concept of a temporal orientation on a relativistic spacetime.[1] A *temporal orientation* is an equivalence class $[\xi^a]$ of smooth timelike vector fields $\xi^a$ that all 'point' in the same direction: at each spacetime point, the vectors are directed into the same lobe of the light cone, which can then be taken to define the 'future' direction. Thus, given a choice of temporal orientation $[\xi^a]$, we say that a vector field $\chi^a$ is *future-directed* if and only if $\chi^a \in [\xi^a]$. This provides a vivid image of time's arrow, illustrated in Figure 7.1.

There is some debate amongst philosophers about whether the arrow of time given by a temporal orientation is reducible to other kinds of facts.[2] My concern here will be with a related issue, of how we can possibly come to know that the structure of spacetime includes a temporal orientation. This problem was introduced in Section 5.1: the analogy of Price's table showed that, although most of our experiences involve asymmetries in the evolution of material systems, this is not the same as an asymmetry of 'time itself' – just as an asymmetry of the items on the table is not the same as an asymmetry of the table itself. I argued that what is missing is a *Spacetime–Evidential Link:* an account that establishes an asymmetry of time itself, together with some plausible empirical evidence in support of that asymmetry.

---

[1] This standard practice in physics was formulated for the philosophy of time by Earman (1974). Relativistic spacetime was introduced in Section 2.5.3; see Malament (2012) for a systematic treatment of its foundations.

[2] Relationist approaches to time argue that the concept of time can be reduced to other variables: for example, causal theorists like Reichenbach (1928, 1956) and Grünbaum (1963) argue that time can be reduced to causation, while Brown (2005) proposes that spacetime more generally can be reduced to facts about matter-energy fields. In a discussion of such views, Earman (1974, p.20) proposes to consider the "heresy" that time is not reducible to anything else; see Maudlin (2002a) and Maudlin (2007, Chapter 4) for what he calls a "more aggressive" defence of heresy.

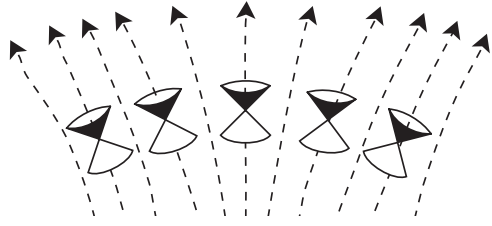

Figure 7.1  A temporal orientation determines a future direction.

In this chapter, I will argue that evidence establishing this link is provided by a dynamical asymmetry that has been established for the laws of nature, which philosophers sometimes called a *lawlike* or *nomic* asymmetry.[3] The example I will discuss is the dynamical asymmetry arising in the electroweak interactions, which are what underpin the phenomenon of radioactivity.

Both philosophers and physicists have proposed that this phenomenon can be used to establish an asymmetry in spacetime structure. For example, Wald writes:

Experiments demonstrating parity violation and (indirectly) demonstrating the failure of time reversal symmetry have shown that two further aspects of spacetime structure can appear in physical laws: the time orientation and space orientation of spacetime. (Wald 1984, p.60)

This idea was explained in more detail by Earman (2002b). Considering a collection of overlapping local neighbourhoods of spacetime, he suggests the following procedure:

Use the failure of time reversal invariance of the laws to pick out the future direction of time in each of these neighbourhoods. (Earman 2002b, p.257)

He then proposes to use the temporal orientation in each neighbourhood to derive a global temporal orientation. Maudlin proposes a view of time asymmetry that is compatible with this, arguing that the discovery of time reversal symmetry violation in the weak interactions implies

---

[3] Cf. Grünbaum (1973, p.273).

that the laws of nature are not Time Reversal Invariant in any sense, and hence that the laws themselves require an intrinsic asymmetry in time directions, and hence that space-time itself, in order to support such laws, must come equipped with an orientation. (Maudlin 2007, p.120)

There is a logical leap in all of these statements: how does one go from a time asymmetry in a dynamical theory – even one confirmed by experimental evidence – to an asymmetry of "space-time itself"? A time orientation is, after all, just a timelike vector field. What would bring us to view it as part of the spacetime geometry, as opposed to its being some property of a matter or radiation field?[4]

One might argue that spacetime structure is entirely reducible to facts about matter and radiation, as in the dynamical approach to relativity proposed by Brown (2005) and Brown and Pooley (2006). But, as I argued in Section 4.5, this radical approach makes it difficult to formulate many modern dynamical theories, which inevitably make use of spacetime concepts in their foundation. Alternatively, one might postulate Earman's symmetry principle (SP2) as an axiom: that every dynamical asymmetry is also a spacetime asymmetry (Earman 1989, p.46). But, following our discussion in Section 4.5, we were left wondering what reason there is to believe this. I will try to offer some support for it in the next section.

What is missing is what I call a Spacetime–Evidential link (Section 5.1), and in particular an account of how symmetry violation in a dynamical theory can provide evidence of an asymmetry in time. So, the current dialectical situation is like squeezing opposite sides of a balloon: a temporal orientation provides a clear picture of time asymmetry but not of how one knows about it, while dynamical asymmetry appears to provide the opposite. Fortunately, a theme developed over the course of this book provides a solution: by adopting the Representation View, we can understand these two as connected through a representation.[5] My argument in this chapter will be that the Representation View, taken together with the experimental evidence for dynamical asymmetry in electroweak interactions, really does provide evidence for a robust time asymmetry. In short, the electroweak interactions establish an arrow of time itself.

Section 7.1 will give my argument that dynamical asymmetry can provide evidence for time asymmetry, as an application of the Representation View.

---

[4] In contexts in which Einstein's equation is satisfied, one might say: when does a timelike vector field belong on the left side, and not the right?

[5] Section 5.1 argued that the Representation View provides an adequate Spacetime–Evidential link. The Representation View itself was introduced in Section 2.3 and described for the specific case of temporal symmetry in Section 2.7.

Section 7.2 will describe a limitation of this approach, associated with the fact that a representation may be 'incomplete'; I will argue that electroweak theory avoids this limitation thanks to evidence from renormalisation group theory. Section 7.3 will then review the discovery of time reversal symmetry violation in the dynamics of electroweak interactions, focusing in particular on how we can come to know such a fact. I emphasise that this is a robust discovery, which is much more general than our current approaches to particle physics. Finally, Section 7.4 will address a critique of Price (1996), which claims to establish that dynamical symmetry violation does not provide evidence for an arrow of time. I will argue that, at least in the case of weak interactions, Price's critique does not succeed.

## 7.1 Representing Time's Arrow

Let me begin with what it might mean to have a temporal asymmetry of spacetime structure. One would like to say that this expresses an asymmetry of 'time itself'. But what does this mean, if not a crude and qualitative expression of substantivalism? And, how can we come to know about it? In this section, I will show how the Representation View developed in Section 2.3 answers both these questions.

The Representation View advises that for a spacetime symmetry to have meaning in the context of a dynamical theory, we must have a representation or 'homomorphic copy' of it amongst the symmetries of that dynamical theory. Otherwise, the dynamical theory would not deserve the name 'dynamical'. This led to my argument in Section 5.1 that Price's table is not quite the right analogy, since it seems to suggest that the structure of time is entirely independent of its contents. On the contrary, according to the Representation View, these two are not independent: a representation 'projects' the structure of time down onto each dynamical theory, just as a shadow is a projection of the structure of a table down onto the floor. This link can be exploited to establish a robust arrow of time, drawing together some threads from earlier chapters. Here is a summary of how this argument goes.

We begin by shifting focus from the concept of a 'time coordinate' to the more structural concept of a 'time translation', motivated by the discussion of Section 2.4. There, I argued that reversing the time translations provides the appropriate definition of time reversal and showed that this generalises the notion of 'reversing temporal orientation'. Namely: given a smooth timelike vector field defining a temporal orientation, one can express time translations as a collection of diffeomorphisms $\varphi_t$ that 'flow' along the integral curves of that vector field, in either direction. If the vector field is complete, then by definition those diffeomorphisms are isomorphic to the Lie group $\mathbb{T} = (\mathbb{R}, +)$. If it is not complete, then the time translations can still be viewed as a *local* Lie group, associated with some neighbourhood of the identity $0$ of $(\mathbb{R}, +)$. As I argued in Section 4.1.2, one can then view temporal symmetry as the statement that time reversal $\tau : t \mapsto -t$ is an automorphism of the time translations, which correspondingly induces a reversal of temporal orientation.

We can use this thinking to encode what it means for 'time itself' to have an asymmetry, similar to the kind afforded by a temporal orientation: it means that $\tau : t \mapsto -t$ is not an automorphism of the time translations. There are various structures with this property that can be used to describe asymmetric time translations.[6] For example, one can add a partial order to our Lie group, replacing $(\mathbb{R}, +)$ with a structure $(\mathbb{R}, +, <)$, called an *ordered Lie group*. Or, one can simply remove all the time translations that flow in one 'direction'. That is, we can choose the time translations $\varphi_t \in (\mathbb{R}, +)$ such that $t \geq 0$, and drop all of their inverses. This turns the time translations into a structure called a *semigroup*, which is roughly speaking a 'group without inverses'. This structure has no time reversing automorphism $t \mapsto -t$, since the inverses have all been removed.[7] Of course, the 'opposite' set of time translations $\varphi_t \in (\mathbb{R}, +)$ such that $t \leq 0$ can be chosen as well. But, this produces the same structure, since it is isomorphic to the first semigroup.

Formulating time asymmetry in terms of the time translations provides a direct, structural perspective on the arrow of 'time itself': namely, as the statement that $t \mapsto -t$ is not an automorphism of the time translations. This is not quite the same as adopting a temporal orientation, for both temporal orientations produce the same asymmetric time translation group; I will discuss this more in Section 7.4.2. But I see this as an advantage: I will argue that it provides an accurate description of the kind of time asymmetry we find in electroweak theory, without requiring any commitment to 'instants', or any other perspective on the metaphysics of what time translations relate.[8]

Given a dynamical theory on a state space $M$, the Representation View ensures that, since this theory is 'dynamical', it must admit a representation of some time translations $\mathbb{T}$ amongst the automorphisms of state space, given by a homomorphism,

$$\varphi : \mathbb{T} \rightarrow \text{Aut}(M), \qquad (7.1)$$

---

[6] The representation theory of both ordered Lie groups and of semigroups is well-studied and so is perfectly amenable to the Representation View advocated in this book; see Neeb (1993) for an introduction.

[7] More precisely, this structure has no non-trivial automorphisms at all; this is a consequence of Proposition 2.1.

[8] This structuralist-functionalist perspective on spacetime was developed in more detail in Section 2.4.1.

When those time translations form a one-parameter set indexed by $t$, the trajectories of the theory are given by a space of curves $\psi(t)$, often expressed as solutions to a differential equation. What makes these trajectories 'dynamical' is the fact that they are determined by a representation of time translations. As I argued in Chapter 3, virtually all dynamical theories can be understood in this way, from Newtonian particle mechanics to quantum theory.

This representation immediately establishes the Spacetime–Evidential link. First, we have an account of what it means for 'time itself' to have an asymmetry: it means that the time translations $\mathbb{T}$ do not admit a time reversal automorphism $t \mapsto -t$. Second, we have an explanation of how one could produce evidence for this time asymmetry: by studying the asymmetries of a dynamical theory, the representation allows us to draw inferences about the asymmetries of time itself, and in particular of the time translations. Like learning about the symmetries of a table by studying its shadow, we can learn about the symmetries of time by studying its representation in a dynamical theory, as I argued in Section 5.1.

There remains a subtlety in establishing such an asymmetry: at least formally speaking, it is possible for the time translations $\mathbb{T}$ to admit a time reversal symmetry $t \mapsto -t$ and for the dynamical theory that represents $\mathbb{T}$ to still be time reversal violating. As I argued in Section 4.5, this happens whenever the structure of state space prevents a representation of time translations from being extended to include time reversal; indeed, we saw that both symplectic mechanics and spontaneous symmetry breaking provide examples of this.[9]

However, *if* a dynamical theory violates time reversal symmetry, while also providing a 'complete' description of the relevant physical information – in a sense to be discussed in Section 7.2 – then this is still a powerful signal. The completeness of the dynamical theory implies that no empirical evidence could ever be produced according to which 'time itself' is temporally symmetric, since no transformation in the dynamical theory can ever represent that symmetry. In other words, if time reversal is not a dynamical symmetry, then no empirical information could ever support the statement that it is a symmetry of time translations.

---

[9] The absence of a representation of the time reversal operator is just what it means for a dynamical theory to be time reversal symmetry violating, by what I have called the 'Symmetry Existence Criterion'; see Section 4.2.1.

Thus, even the slightest hint of empiricism leads to the conclusion proposed by Earman (1989, p.46), that "conditions of adequacy on theories of motion" require us to infer that if a 'complete' dynamical theory violates time reversal symmetry, then spacetime itself must violate this symmetry too. Demanding an empirical basis for temporal symmetry thus leads to the result that *time reversal symmetry violation establishes an arrow of time itself*.

Let me not keep you waiting: in Section 7.3, I will argue that an arrow of time of this kind has indeed been established, through the study of weak interactions and neutral kaon decay. But first, since so much depends on what it means to be a 'complete' theory, let me first discuss what I mean by this in more detail.

### 7.2 Complete Enough Representations

Using time reversal symmetry violation to establish an arrow of time requires the dynamical theory to be 'complete' in a certain sense. I will argue that the Standard Model is indeed complete in the required sense, thanks to insights from renormalisation group theory. However, let me first illustrate why this kind of completeness is needed.

A familiar example of an 'incomplete' dynamical system is the classical damped oscillator, which slows to a stop under the force of friction. To first order approximation, the oscillator follows a curve $x(t)$ satisfying Newton's equation, with

$$m\frac{dx^2}{dt^2} = F(x, \dot{x}) = -kx - c\dot{x}, \tag{7.2}$$

for some constants $c, k > 0$. The first term $-kx$ describes a 'Hooke's law' force, proportional to the oscillator's displacement from some equilibrium position. The second term $-c\dot{x}$ describes a force of damping, such as through friction in a spring or in the air, which is proportional to the system's speed.

This equation manifestly fails to be time reversal invariant: a damped oscillator can slow to a stop, but it cannot spontaneously start oscillating. One can check this by noting that time reversal is not a dynamical symmetry, in the sense of Section 4.1.2: although it is common to justify this by substituting $t \mapsto -t$, a more rigorous check is to observe that there is no transformation of the form $(x, \dot{x}) \mapsto (x, -\dot{x})$ that preserves the solution space. Thus, a representation of time reversal symmetry is impossible for the damped harmonic oscillator.

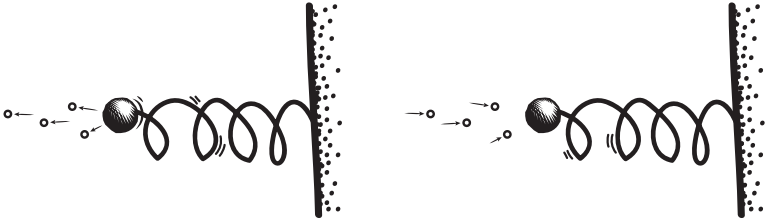

Figure 7.2 A damped oscillator interacts with its microscopic medium (left) in a way that is symmetric under time reversal (right).

However, this is no evidence for an arrow of time, because the description omits relevant degrees of freedom. This includes the energy of the oscillator's motion that is converted into internal energy in the spring, and to a lesser extent into the surrounding medium, jiggling the roughly $10^{26}$ molecules of the spring and in the air. When those missing degrees of freedom are restored, we find a representation of time reversal symmetry: reversing the 'jiggling' produces just the right kind of collision with a rest oscillator so as to start it bouncing, as in Figure 7.2. The time asymmetry of a damped harmonic oscillator is an illusion due to 'losing' information into the environment, of the kind that we saw many times in Chapter 5.

We will soon discuss an example of a more serious time asymmetry in electroweak theory. Unlike the examples above, this theory aims to give a more or less complete description of the interactions involved in electroweak phenomena. But, how can we can be sure that this will not turn out to be another case of missing information? There is an enormous amount of information that is not captured by the Standard Model, especially concerning phenomena occurring at energy scales so high that we have not yet probed them, and perhaps never will. Given the preponderance of false asymmetries that arise from missing information, this should give one pause: Is symmetry violation in electroweak a similar 'illusion' that arises merely from missing information?

Part of the wonder of the human condition is to be continually learning, and so no evidence is perfectly safe. However, in the special case of particle physics, we do have a remarkable framework of ideas that assures us that its dynamics will not be overturned by missing information. In theories that aim to describe the fundamental interactions of matter and energy, the missing information is of a particular kind, corresponding to facts about energy scales that are large enough – or equivalently, about energy variations on distance scales that are small enough – so as to make them inaccessible to any available experiments. However, unlike the dynamics of the damped harmonic oscillator above, the dynamics of modern particle physics is formulated in such a way that this missing information is irrelevant: that is, the dynamics of the theory is decoupled from the details of the missing information. This happy situation is the result of what is known as *renormalisation theory*.

One of the central results about the quantum field theories of the Standard Model is that they are renormalisable. This means, roughly speaking, that their observable predictions remain the same even as one scales the system down to arbitrarily small distances.[10] In particular, the renormalisability of massive Yang–Mills theories, like the electroweak theory that concerns us, was established by 't Hooft (1971). As a result, predictions about accessible energy scales in the these theories *decouple* from the physics of higher energy scales: the fact that we are missing information about the latter is irrelevant. This leaves us in a particularly fortunate situation, characterised here by Huang (1982):

It is fortunate that, at any given stage, we were able to regard certain particles as provisionally fundamental, without jeopardizing the right to change our mind. The reason is that, according to quantum mechanics, it is a good approximation to ignore those quantum states of a system whose excitation energies lie far above the energy range being studied. (Huang 1982, p.1)

Butterfield and Bouatta (2015) have argued that this prediction of renormalisation theory, that high energy scales decouple from accessible scales, helps to explain why a renormalisable theory like the Standard Model is so empirically successful.[11] But, the same result can also be viewed as providing a kind of scientific *confirmation*: the Standard Model implies that high energy scales decouple from accessible energies; so, the spectacular empirical success of the Standard Model is strong evidence *confirming* that this decoupling is true of our world.

That said, the Standard Model does contain conceptual problems related to its very definition. In particular, we have no mathematically rigorous interacting quantum field theory in $3 + 1$ dimensions, let alone a rigorous formulation of electroweak theory.[12] But, these problems are of a different kind. The framework, and in particular electroweak theory, is still 'complete enough' to avoid making mistakes by neglecting higher-energy phenomena. Although it is of course not a complete description of reality, its empirical successes, together with the results of renormalisation theory, provide a remarkable assurance that missing information will not interfere with its predictions. As a result, the discovery of time reversal symmetry violation in this theory is not just a result about an incomplete theory: it holds independently of the missing information.

---

[10] The development of renormalisation theory following Wilson (1971) showed a very broad sense in which this holds; see Duncan (2012, Chapters 16–19) for its conceptual foundations and Rivat (2021) for a historical analysis. Recent discussions of its philosophical significance in the context of scientific realism have been given by Butterfield and Bouatta (2015); Fraser (2018); Rivat (2019); Rivat and Grinbaum (2020); Ruetsche (2018); Williams (2020).

[11] Butterfield and Bouatta (2015, p.439) argue that the generic results following Wilson (1971) explain why quantum field theories formulated at accessible energies are renormalisable, concluding: "It is by analyzing this flow that one deduces that what seemed 'manna from heaven' (that some renormalisable theories are empirically successful) is to be expected: the good fortune we have had is generic".

With this assurance in place, the Representation View now provides a reliable way to test whether the structure of time itself is asymmetric. Suppose a system can be described by a quantum theory whose state space provides a representation of time translations amongst its automorphisms and which is 'complete enough' for the description of the phenomena in the sense described above. Then the test is simple: a dynamical system in which time reversal symmetry is violated provides evidence for an arrow of time itself.

## 7.3 The Discovery

The 'great shock' of electroweak theory is that time reversal is not a dynamical symmetry. I gave a brief history of that discovery in Section 1.5. Let me now discuss in more detail how we came to know about it. I will try to keep the detailed physics relatively light, so as to make the main ideas accessible.

Section 7.3.1 will briefly review some elements of electroweak theory. Section 7.3.2 will then discuss the ideas that led to the 1964 discovery of symmetry violation in this theory. Part of what is so interesting about this story is that the discovery of time reversal symmetry violation was made around a decade before the Standard Model was formulated. So, one naturally wonders: how can we know the symmetries of a theory we don't have? We will soon see that it is thanks to the incredible generality of these experimental discoveries, made possible through the application of symmetry principles that have in any case been of some independent interest to philosophers.

---

[12] For example, Streater (1988, p.144) writes, "To test a model, we must study its predictions. The trouble is that Yukawa theory (and QED when one looked hard at it) is not well defined, and different approximations give quite different predictions, rather than very similar predictions. One cannot tell whether one is testing the theory or the method". See Wallace (2006) for a defence of Standard Model methods.

My discussion will include both the discovery of 'indirect' and 'direct' evidence for time reversal symmetry violation. Taken together, they provide powerful reason to believe that any 'complete enough' dynamical theory of the phenomena at issue will violate time reversal symmetry. By the argument presented in Sections 7.1 and 7.2, this cannot be condemned as just a property of the matter-energy fields involved in the phenomena: it is the robust evidence that time itself has an arrow.

### 7.3.1  *The Standard Model of Electroweak Interactions*

In this discussion, I will make frequent use of three discrete symmetry transformations: parity or 'spatial reversal', described by a unitary transformation P; matter–antimatter exchange, described by a unitary transformation C; and time reversal, described by an antiunitary transformation T.[13] I will also be interested in the combined application of these transformations, such as CP, PT, and CPT. My discussion here will also adopt some standard language from particle physics: when a dynamical theory does not admit an appropriate representation of some transformation X, this is referred to as 'X violation'. Thus, the discussion that concerns us now is T violation in electroweak theory.

The theory of electroweak interactions is a gauge theory: its mathematical formulation in the Standard Model begins with a classical Yang–Mills gauge theory, which is a cousin of electromagnetism when formulated with a $U(1)$ gauge potential. But, we instead choose the non-abelian gauge group $SU(2) \times U(1)$. The classical theory can then be quantised to give a renormalisable quantum field theory, by appeal to a variety of techniques. The result is a theory capable of dealing in a relatively complete and unified way with both the phenomena of electromagnetism and with weak forces, the combination of which explain radioactive decay.

In honour of its founders, electroweak theory is sometimes called the Glashow–Cabibbo–Weinberg–Salam model.[14] The four generators of $SU(2) \times U(1)$ in a Hilbert space representation were associated with four conserved quantities: the first three associated with $SU(2)$ are called 'weak isospin', while the fourth associated with $U(1)$ is called 'weak hypercharge'. Although this model enjoyed some early predictive successes, in the 1960s

---

[13]  I have explained the origin of the time reversal transformation *T* on the Representation View in Chapters 2 and 3. The other transformations have a similar origin in representation theory, which I will introduce in Chapter 8 when we turn to CPT symmetry.

[14]  Its early formulation was given by by Cabibbo (1963), Glashow (1961), Salam (1968), and Weinberg (1967) and was later shown to be renormalisable by 't Hooft (1971).

little was understood about the degrees of freedom associated with it and consequently about the mechanism that could give rise to T violation or CP violation. At the time, only three quark fields were well accepted, with some hints about a fourth 'charm' quark field. At a 1966 conference in Berkeley, soon after the 1964 discovery of CP violation by Cronin and Fitch, Cabibbo is reported to have said that,

At that time, physicists thought that the theory of the weak interactions was in good shape since they 'understood' how to compute radiative correction to $\mu$ decay. In comparison, CP violation was like an atomic bomb blowing up in the background. (Bigi and Sanda 2009, p.145)

A breakthrough came with an application of the Representation View, as presented in Section 2.3. When applied to a global gauge symmetry group such as $SU(2) \times U(1)$, it implies that the gauge transformations only acquire meaning in the context of a state space representation of the group amongst the symmetries of that state space. In an 'elementary' quantum system that cannot be decomposed into further component parts, this would be an irreducible unitary representation.[15]

This led Kobayashi and Maskawa (1973) to analyse the irreducible unitary representations of $SU(2) \times U(1)$. They found that no plausible representations with only four quark fields allowed for the CP and T violation that had been discovered in the previous decade. This led them to propose the six-quark field model, which naturally incorporates CP and T violation: it was this six-quark model that was quickly incorporated into the Standard Model. Kobayashi and Maskawa shared half of the 2008 Nobel Prize in physics for their discovery.

The electroweak contribution to the Lagrangian of the Standard Model does not commute with the 'canonical form' of the antiunitary time reversal operator $T$, in the sense discussed in Section 3.4.3. I will not go into the details of that Lagrangian here, which are not needed for the level of generality of my discussion.[16] What I would like to discuss is the consequence that, in the electroweak theory, since $T$ does not commute with the Lagrangian, it does not commute with the Hamiltonian generator $H$ of the unitary time translations $t \mapsto U_t = e^{-itH}$. This implies[17] that $T$ is not a representation of time reversal, in that $T U_t T^{-1} \neq U_{-t}$. So the representation of time translations $t \mapsto U_t = e^{itH}$ in electroweak theory cannot be extended to include time reversal. For at the group level, time reversal satisfies $\tau t \tau^{-1} = -t$, and so any representation of time reversal $\tau \mapsto T$ must satisfy $T U_t T^{-1} = U_{-t}$. This is the sense in which the electroweak interactions violate time reversal symmetry.

---

[15] This approach, made famous by Wigner (1939) in the analysis of the Lorentz group, had recently been successfully applied by Gell-Mann (1961) to the group $SU(3)$ in the study of strong interactions.

[16] Introductions can be found in Bigi and Sanda (2009, Chapter 8) and Huang (1982, §6.2).

[17] It is easiest to argue for the contrapositive statement: if $U_{-t} = e^{itH} = T e^{-itH} T^{-1}$, then $e^{itH} = e^{T(-itH)T^{-1}} = e^{itTHT^{-1}}$, where the last equality applies the fact that $T$ is antiunitary and thus conjugates complex constants. Since the generator of a continuous one-parameter unitary

Thus, in electroweak theory, time reversal symmetry is violated. Indeed, this same Lagrangian requires the violation of many other discrete symmetries as well. But, before we get too carried away, let me return to the curious puzzle in this story: the representation of time translations that we are considering did not appear in explicit form until after the publication of Kobayashi and Maskawa (1973), long after the discovery of CP violation in 1964. So, the discovery of symmetry violation was not a simple matter of checking whether the dynamics fails to admit a representation of time reversal invariance, as I have sketched above, for the dynamics of electroweak theory had not yet been discovered!

The solution to this puzzle lies in an extraordinary feature of the experimental evidence for symmetry violation. That evidence turns out to be so general that, in almost any reasonable theory that one could produce, the dynamics would have to violate time reversal symmetry, and also all those other symmetries I just alluded to. This is of course what led Kobayashi and Maskawa to search for models that violate time reversal symmetry. I will argue that there is a further lesson from this: that our evidence for time reversal symmetry violation is much more general than the Standard Model alone. These instances of symmetry violation are here to stay, and so they provide remarkably robust evidence for an arrow of time.

This result may still seem a little mysterious: how could one show in such a robust way that the 'reversal of time translations' is not a dynamical symmetry? Understanding this requires a more careful look into the discoveries themselves, which turn out to have some interesting philosophy underpinning them.

### 7.3.2  *The Discovery of CP and T Violation*

Recall that, as recounted in Section 1.5, the discovery of parity violation in 1956 by Chien-Shiung Wu and her collaborators led to the identification of a new class of particle called a K meson or *kaon*. The kaon of their experiment was positively charged. In contrast, the kaon that led to the discovery of CP violation is neutral, called a *neutral kaon*.

---

group is unique, it follows that $T H T^{-1} = H$. Equivalently, if $T H T^{-1} \neq H$, then $T U_t T^{-1} \neq U_{-t}$, and so $T$ is not a representation of time reversal.

At the time, most models of neutral kaons required both CP invariance and T invariance, such as the model of Weinberg (1958). Indeed, CP symmetry is equivalent to T symmetry for a very general class of relativistic quantum theories, by a collection of results known as the CPT theorem, which I will discuss in Chapter 8. So, evidence of CP violation is also good evidence of T violation. When the evidence for T violation relies on the detection of CP violation together with CPT symmetry, it is called *indirect*. In contrast, evidence for T violation that does not rely on CPT symmetry is called *direct*.

The first indirect evidence for T violation was discovered for what is called the 'long-lived' neutral kaon state $K_L$. At the time, these states were best known for their decay into three pions, $K_L \to \pi^0 \pi^+ \pi^-$. The neutral pion $\pi^0$ does not ionise in a spark chamber, and so it is invisible in such a device, but its trajectory can be calculated from the trajectories of the other two charged pions, using the conservation of momentum.

Although the dynamics for these decay events was not yet known, practitioners could still make inferences about the symmetries of whatever the dynamics might turn out to be. These inferences made use of a symmetry principle, which is commonly attributed to Pierre Curie (1894): that every asymmetric effect must originate from an asymmetric cause.[18] More specifically, the form of this principle that is appropriate for dynamical theories was formulated by Belot (2003, §4.2) and Earman (2004): if a final state has an asymmetry, and the initial state lacks that asymmetry, then its only possible origin is in the dynamics.

In fact, this principle does not need to be postulated as an assumption about quantum theory. It is a mathematical fact, which can be formulated for any unitary symmetry and for any unitary representation of time translations.[19] Let me express it here in a form that is particularly relevant for its use in particle physics, using a unitary S-matrix on a Hilbert space.

Recall that an *S-matrix* is a unitary operator, which can be viewed as the 'infinite-time limit' of a unitary group $U_t = e^{-itH}$. What is particularly relevant for our purposes is that if an S-matrix fails to commute with the CP, meaning that $[CP, S] \neq 0$, then the unitary dynamics is CP violating.[20] The possible decay events, like $K_L \to \pi^0 \pi^+ \pi^-$, called *decay*

---

[18] See Belot (2003); Chalmers (1970); Earman (2004); Ismael (1997), and the volume edited by Brading and Castellani (2003), for some classic discussions of Curie's principle; my own view on these matters can be found in Roberts (2013a, 2015a); see also the responses by Castellani and Ismael (2016); Kinney (2021); Norton (2016s).

[19] See Roberts (2015b, p.10, Facts 1 and 2) for two versions and their proofs.

[20] This is the case for most other unitary symmetries as well. For example, suppose the Hamiltonian can be expressed in the form $H = H_0 + V$, with $H_0$ the free Hamiltonian, which satisfies $[CP, H_0] = 0$. Then, writing the S-matrix as a Dyson series $S = \mathcal{T} \exp \left( \int_{\mathbb{R}} V(t) dt \right)$ for some time-ordering operator $\mathcal{T}$, one finds that $[H, CP] = 0$ implies that $[S, CP] = 0$ by the linearity of the integral. Equivalently, if $[S, CP] \neq 0$, then $[H, CP] \neq 0$, which means that the unitary dynamics is CP violating.

*modes*, are associated with non-zero amplitudes of an S-matrix, written $\langle \pi^0 \pi^+ \pi^-, S K_L \rangle \neq 0$. Adopting this general formalism, which applies to an extremely broad class of quantum theories, one can now observe the following fact.

**Proposition 7.1** *Let $CP : \mathcal{H} \to \mathcal{H}$ be any unitary on a Hilbert space (the 'symmetry'), and let $S$ be any bijection (the 'S-matrix') such that $\langle \psi^{out}, S\psi^{in} \rangle \neq 0$. Suppose either of the following conditions hold:*

1. *(in but not out) $CP\psi^{in} = \psi^{in}$ but $CP\psi^{out} = -\psi^{out}$; or*
2. *(out but not in) $CP\psi^{out} = \psi^{out}$ but $CP\psi^{in} = -\psi^{in}$.*

*Then we have CP violation, in the sense that, $(CP)S(CP)^{-1} \neq S$.*

*Proof*   We argue the contrapositive: let $(CP)S = S(CP)$. We will show that $\langle \psi^{out}, S\psi^{in} \rangle = 0$. Since $CP$ is unitary, $\langle \psi^{out}, S\psi^{in} \rangle = \langle (CP)\psi^{out}, (CP)S\psi^{in} \rangle = \langle (CP)\psi^{out}, S(CP)\psi^{in} \rangle$. So, if either the 'in but not out' or the 'out but not in' conditions hold, then,

$$\langle \psi^{out}, S\psi^{in} \rangle = \langle (CP)\psi^{out}, S(CP)\psi^{in} \rangle = -\langle \psi^{out}, S\psi^{in} \rangle, \qquad (7.3)$$

which implies that $\langle \psi^{out}, S\psi^{in} \rangle = 0$.                                    ∎

This means that if a unitary symmetry like CP reverses the sign of an ingoing state but not the outgoing state, or vice versa, then – regardless of what the dynamics describing this process turns out to be – *the dynamics must be CP violating*. Let me illustrate how this was used to show that neutral kaons are CP violating and thus (indirectly) that they are T violating.

When the neutral kaon decays into three pions, $K_L \to \pi^0 \pi^+ \pi^-$, both the ingoing and outgoing states can be viewed as reversing sign under CP:

$$(CP)K_L = -K_L \qquad\qquad (CP)\pi^0 \pi^+ \pi^- = -\pi^0 \pi^+ \pi^-. \qquad (7.4)$$

This decay is compatible with CP symmetry, because Proposition 7.1 does not have any implications for such ingoing and outgoing states. This was the most well-known decay mode associated with the long-lived neutral kaon prior to 1964.

However, a *two*-pion state $\pi^+ \pi^-$ does not reverse sign under CP but is rather left unchanged:

$$(CP)\pi^+ \pi^- = \pi^+ \pi^-. \qquad (7.5)$$

This means that, if one could be sure that a neutral kaon $K_L$ can decay into a two pion state – even once! – then the 'in but not out' criterion would be satisfied, and then Proposition 7.1 guarantees CP violation.

This is exactly what Cronin and Fitch discovered in 1964.[21] In fact, their original intention was to try to show that, to a high degree of accuracy, no two-pion decay events could be found, which would have helped to *confirm* CP violation. As fate would have it, they instead found a small but unmistakable number of CP violating instances in which a neutral kaon decayed into two pions, in about one of every 500 decay events. By the CPT theorem, this also provided the first (indirect) evidence for T violation.

More recently, *direct* evidence for T violation has been produced as well, without appeal to CPT symmetry. This turns out to make use of a different symmetry principle, which is less commonly discussed by philosophers. Curie's principle does not apply to time reversal, or any antiunitary symmetry, due to a subtlety in the proof of Proposition 7.1.[22] However, one can still apply a version of the 'principle of detailed balance', which I call 'Kabir's principle' in this context after its proposal by Kabir (1968, 1970). Here we use the fact that, in just the same way that time reversal symmetry is expressed by $T U_t T^{-1} = U_{-t} = U_t^*$ with $U_t = e^{-itH}$, so also, time reversal symmetry of an S-matrix is expressed by $T S T^{-1} = S^*$. Let me first state the principle and then interpret it.

**Proposition 7.2** *Let $S : \mathcal{H} \to \mathcal{H}$ be a unitary operator on a Hilbert space (the 'S-matrix'), and let $T$ be an antiunitary bijection ('time reversal'). If there exist any $\psi^{in}, \psi^{out} \in \mathcal{H}$ such that*

$$\langle \psi^{out}, S\psi^{in} \rangle \neq \langle T\psi^{in}, ST\psi^{out} \rangle, \tag{7.6}$$

*then $S$ exhibits time reversal violation, $T S T^{-1} \neq S^*$.*

*Proof*  We prove the contrapositive: let $T S T^{-1} = S^*$, and thus $T S = S^* T$. Since $T$ is antiunitary, $\langle \psi^{out}, S\psi^{in} \rangle = \langle T\psi^{out}, T S\psi^{in} \rangle^* = \langle T S\psi^{in}, T\psi^{out} \rangle$. Therefore,

$$\langle \psi^{out}, S\psi^{in} \rangle = \langle T S\psi^{in}, T\psi^{out} \rangle = \langle S^* T\psi^{in}, T\psi^{out} \rangle = \langle T\psi^{in}, ST\psi^{out} \rangle. \tag{7.7}$$

Equivalently, if $\langle \psi^{out}, S\psi^{in} \rangle \neq \langle T\psi^{in}, ST\psi^{out} \rangle$, then $T S T^{-1} \neq S^*$.  ∎

---

[21] See Christenson et al. (1964); Cronin and Greenwood (1982).
[22] This leads to an argument that, in spite of its popularity, Curie's principle is false (Roberts 2013a, 2015b).

The assumption of Proposition 7.2, Kabir's principle, is that the amplitude determining the probability of a certain decay event $\psi^{in} \to \psi^{out}$ is different from the amplitude for the time reverse of that decay, $T\psi^{out} \to T\psi^{in}$. If these two amplitudes are different, then Proposition 7.2 implies that, whatever the dynamics turns out to be, it must be T violating. No appeal to CPT symmetry is needed for this inference, and so such a difference in amplitudes provides direct evidence for T violation.

This kind of evidence has recently been found. Oscillating kaon-antikaon states turned out to be a fruitful place to test for this: a kaon state $K^0$ is known to decay into an 'antikaon' state $\bar{K}^0$, which can then decay back again to $K^0$, creating an oscillation. This is a particularly convenient phenomenon for the application of Kabir's principle, since time reversal $T$ can be viewed as leaving both of these states unchanged: $TK^0 = K^0$ and $T\bar{K}^0 = \bar{K}^0$. So, by checking whether the amplitude for the decay $K^0 \to \bar{K}^0$ is different from its time reverse $\bar{K}^0 \to K^0$, we can check whether this system is T violating. Indeed it is: direct T violation of this kind was discovered by Angelopoulos et al. (1998). Related arguments later showed that $T$ violation also occurs in the $B$ meson sector as well (Lees et al. 2012). Some preliminary recent evidence even suggests a small amount of CP and T violation in the lepton sector through neutrino oscillations (T2K Collaboration 2020). Thus, the experimental evidence for $T$ violation in the Standard Model is becoming ever more secure.

I have formulated the symmetry principles above in the context of S-matrix theory, because this formalism makes the most direct contact with the experimental evidence. However, both have recently been shown to be much more general than that, relying on only a tiny fragment of the structure of quantum theory. In particular, Ashtekar (2015) has shown that both are facts about any dynamical theory that can be formulated in *generalised mechanics*, a framework allowing one to capture a wide variety of theories, from classical and quantum mechanics to gravitation. I will not go into the details of that formalism here; my point is just to say that, from many different perspectives, the experimental evidence for T violation is extremely robust.

In summary, the experimental evidence for T violation is here to stay. The evidence was produced using robust symmetry principles that apply to a wide variety of quantum theories, and even beyond quantum theory. And, when formulated as a renormalisable quantum field theory, the resulting electroweak theory captures this phenomenon in a way that makes missing information from higher energy scales irrelevant.

Let me finally note that, as a result of all this experimental evidence for discrete symmetry violation over the last 60 years, the Standard Model does not admit a representation of any of the following discrete symmetries:[23]

$$C, P, T, CP, CT, \text{ or } PT.$$

This leaves only one combination of these transformations that can be represented in the Standard Model: the combination of all of them, CPT. I will return to the discussion of this transformation in Chapter 8.

## 7.4 The Price Critique

In this chapter, I have argued that time reversal symmetry violation can provide evidence for an arrow of time itself and that the evidence from electroweak theory for such an arrow is extraordinarily strong. In this section, I would like to respond to an important critique of this perspective due to Huw Price (1996, 2011).

Nature is under no obligation to provide us with a physical asymmetry in the structure of time itself. Our world might have had the character that virtually all temporal asymmetries were only apparent. One would just need to explain how this great illusion could happen in a way that is compatible with our experience – and with the results of modern physics. Some of that is explained by our discussion of 'arrows that misfire' in Chapter 5 and Chapter 6. But, another important part of this undertaking has been carried out by Huw Price (1996). Price calls his view the "perspectival view" of time's arrow: that all facts about time asymmetry depend crucially on the standpoint of an observer and thus do not provide evidence for an arrow of time itself.[24] This is a remarkable analysis, and I agree with Price's efforts to dispose of 'misfiring arrows', many of which align with my analysis in previous chapters.

However, I part ways with Price regarding electroweak interactions. About this topic, he writes:

It is true, of course, that the T-asymmetry of the neutral kaon could provide the basis for a universal convention for *labelling* lightcones as 'past' and 'future'. But

---

[23] Both C and P were violated in the original parity violating experiments of Chien-Shiung Wu (Garwin, Lederman, and Weinrich 1957; Wu, Wang, et al. 2015). Given CPT symmetry, this implies that both PT and CT are violated as well. The Cronin and Fitch experiments then settled that both CP and T are violated as well.

[24] Rovelli (2017) similarly identifies a 'standpoint' with the selection of a statistical mechanical coarse-graining, in a related argument that the arrow of time is perspectival; see Section 5.3.2.

this no more requires that time even be anisotropic – let alone objectively oriented – than does our universal signpost for space. (Price 2011, p.294)

This 'universal signpost' will be discussed in detail in the next section. To support his conclusion, Price raises three objections to viewing time asymmetry in a dynamical theory as evidence for a physical arrow of time. They are: that it is independent of the question of time asymmetry; that even if it were not, the time asymmetry can always be dissolved due to a 'gauge' symmetry in its interpretation; and that even without these two objections, existing evidence provides too weak a signal for an arrow of time.

Each of these concerns is important, and I will discuss each of them in turn. As in previous chapters, the response that I will offer draws essentially on the Representation View of Section 2.3. My conclusion, *pace* Price, is that T violation really does imply that time itself is asymmetric – and even 'objectively oriented', if one likes – in a way that avoids his objections.

### 7.4.1  *The Independence Objection*

Price's table analogy provides an indication of his first concern: that the behaviour of matter-energy seems to be independent of the structure of spacetime. If this were strictly true, then an asymmetry of the former would be no evidence for an asymmetry of the latter. Price later makes this concern explicit using a different analogy:

Suppose everything in the universe were to vanish, except a single giant signpost, pointing forlornly to a particular corner of the sky. Or suppose that a universe had always been like this. This spatial asymmetry would not require that space itself be anisotropic, presumably, or that the direction in question be distinguished by anything other than the fact that it happened to be the orientation of the signpost. Similarly in the case of time. The contents of time – that is, the arrangement of physical stuff – might be temporally asymmetric, without time itself having any asymmetry. Accordingly, we need to be cautious in making inferences from observed temporal asymmetries to the anisotropy of time itself. (Price 2011, p.292)

Price is right that an asymmetry in space itself is conceptually different from an asymmetry of a signpost; similarly, an asymmetry in time itself is conceptually different from an asymmetry in the dynamics of its contents. But, like the table analogy, Price's signpost argument only works if those contents are independent of the structure of space and time. Here I disagree.

The very meaning of a dynamical theory requires that it be intimately connected to the structure of time itself: otherwise, it does not deserve the name 'dynamical'. In order to capture this fact, we must postulate the existence of a representation, or a homomorphism from the abstract group of time translations to the symmetries of the state space that describes the "physical stuff". This is what I called the Representation View in Section 2.3. A better metaphor for this view was given in Section 5.1, of learning about a table from its shadow on the floor. In a similar way, the asymmetries of time are projected down onto a dynamical theory by a representation. This is the fundamental problem with Price's signpost analogy: once the link is adequately made between the symmetries of spacetime and the symmetries of a dynamical theory, Price's objection about their independence is dissolved.

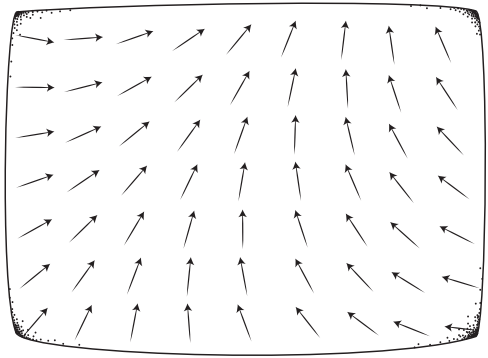

Figure 7.3 A collection of signposts at every spacetime point: we assume that every rigid spatial translation produces a possible collection of signposts as well.

A second problem with Price's signpost example is that it suggests that T violation in electroweak theory is much weaker than it actually is. Price describes a system with a single instance of an asymmetric signpost. But, the evidence from T violation is much stronger and more striking than that because it is much more pervasive. We should rather imagine encountering a great set of signposts at every point in spacetime, as in Figure 7.3 – and also knowing that a rigid translation of the entire set by a spatial translation in any direction will produce another possible set of signposts – but noting that 'spatial reversal' produces a set of signposts that are physically impossible. In other words, we should imagine a system in which there is no way to represent the reflection of spatial transformations $s \mapsto -s$ on the state space of the system. According to the Representation View, the structure of spatial translations on state space is a 'homomorphic copy' or 'shadow' of the structure of space itself. As a result, this discovery would indeed provide evidence that the structure of space has an asymmetry.

The discovery of T violation in electroweak theory is much more like this than it is like Price's solitary signpost. The dynamics of electroweak theory provides a representation of time translations, describing all the possible time evolutions on the state space of electroweak theory. The discovery of T violation shows that on that state space, it is impossible to represent the reversal of those time translations, $t \mapsto -t$. So, since the structure of time translations on state space is a homomorphic copy of the structure of time itself, we have evidence that the time translations lack this symmetry. Thus, although Price's concern is well-motivated, it is avoided once we adopt the Representation View and recognise the full strength of the discovery of T violation.

### 7.4.2 *The 'Gauge' Objection*

Price's second objection concerns whether T violation in a dynamical theory allows the freedom to adopt either of the two temporal orientations on a (temporally orientable) spacetime. Price expresses this in a critique of Maudlin (2002a, 2007), who takes T violation to be evidence for a temporal orientation. Price responds:[25]

What would it mean to say the laws exhibited a specific temporal asymmetry, unless one could orient a choice of temporal coordinate consistently across spacetime, in the manner guaranteed by orientability? This isn't what Maudlin has in mind, however. Orientability is a much weaker condition than existence of an objective distinction between earlier and later – it doesn't imply even that time is anisotropic, much less that it is objectively directed. . . . [W]e should read Maudlin as claiming that the T-violation exhibited by the neutral kaon 'requires' such an orientation. But . . . this is simply not true: a lawlike time asymmetry does not even require temporal anisotropy, let alone the true directionality that Maudlin is after. (Price 2011, p.294)

Price is right that, if T violation in electroweak theory only implied that spacetime admits a smooth timelike vector field, which is to say a temporal orientation, then by definition this would mean that spacetime is *temporally orientable*. Such a vector field does not by itself imply that time has an asymmetry. It does, however, mean that we can choose one temporal orientation and call it 'the future' if we wish. But, says Price, this would only establish a descriptive redundancy, analogous to the descriptive redundancy of gauge theory, which always allows us to call the other temporal orientation 'the future' as well.

However, T violation in electroweak theory again implies a stronger sense of time asymmetry than Price suggests here. T violation establishes that time reversal $t \mapsto -t$ is not an automorphism of the time translations $t$ in spacetime. So, the local structure of spacetime cannot be Minkowski spacetime alone! Nor can it be any temporally orientable spacetime whose time translations admit a time reversal automorphism. This can be captured in the language of orientations by the statement that, in order to accurately describe the symmetries of spacetime, we must choose *some* temporal orientation, although it may not matter which one we choose. One might refer to such a description as a *temporally oriented spacetime*.

---

[25] Price (2011, p.293) gives a similar analysis of a toy time asymmetric model due to Horwich, as does Farr (2020, §2.2.2) for a toy model using monotonic functions.

In a temporally oriented spacetime, Price is right to say that there remains a gauge redundancy as to which temporal orientation we choose. For example, if we say that a decay event like $K^0 \to \bar{K}^0$ is future-directed with respect to the temporal orientation $[\xi^a]$, then we can always say that the opposite decay event $\bar{K}^0 \to K^0$ is future-directed with respect to $[-\xi^a]$. This really is a descriptive redundancy in the theory. However, it is not the same as saying that a spacetime is 'merely orientable'. The symmetries of spacetime are still restricted by an orientation, although it may not matter which.

To see this, let me write $\mathrm{Aut}(M, \eta_{ab})$ to refer to the automorphisms of Minkowski spacetime, called the *complete Poincaré group*. That group is *not* isomorphic to the symmetries of Minkowski spacetime with an orientation, $\mathrm{Aut}(M, \eta_{ab}, [\xi^a])$, which is sometimes called the *orthochronous* Poincaré group and which does *not* include time reversal. However, both temporal orientations $[\xi^a]$ and $[-\xi^a]$ equally give rise to the same orthochronous Poincaré group, in the sense that $\mathrm{Aut}(M, g_{ab}, [\xi^a])$ is isomorphic to $\mathrm{Aut}(M, g_{ab}, [-\xi^a])$. So, both equally capture the temporal asymmetry implied by T violation. It is just an awkwardness of representing asymmetries using a temporal orientation that we must choose one. But, this is avoided when we shift focus to the symmetries of time translations directly, as I have done in the analysis of T violation above.

So, there is a sense in which the comments at the outset of this chapter by Wald, Earman, and Maudlin are correct: the time asymmetry of electroweak theory implies that the spacetime symmetries are characterised by a temporal orientation. And, Price is also right, that we have equal reason to choose either one. But, Price's valid point is no objection to the fact that electroweak theory provides evidence that time itself has an asymmetry, in the sense that I have argued for in the earlier sections of this chapter.

### 7.4.3 *The Weak Signal Objection*

Price's final objection to the claim that T violation establishes an asymmetry of time itself is that it is too rare to matter:

It is true that there appears to be one exception to this general principle [of temporal symmetry] ... the case of the decay of the neutral kaon. Even here the departure from perfect symmetry is tiny, however, and the puzzling character of the existence of this tiny exception serves to highlight the intuitive appeal of the prevailing rule. To a very large extent, then, the laws of physics seem to be blind to the direction of time – they satisfy T-symmetry, as we may say. (Price 1996, p.116)

I agree with Price that the size of time reversal symmetry violation is an important issue for modern physics. For example, it is essential in determining to what extent symmetry violation is responsible for the cosmological asymmetry, and for the baryon asymmetry, if one adopts the Sakharov (1967) conditions discussed in Section 5.4.3. However, one should not conflate concerns about the relationship between different asymmetries with the question of whether time itself has an asymmetry. Although this 'weak signal' objection is relevant to the former, it does not provide any argument against the latter.

It may help to recall what is meant by 'size' in the context of the neutral kaon experiments that Price refers to. Recall from our discussion in Section 7.3.2 that the first evidence for T violation was the decay of the long-lived neutral kaon into two pions, $K_L \to \pi^+\pi^-$. Since $K_L$ and $\pi^+\pi^-$ transform in opposite ways under CP, the discovery of this decay event implies, by our version of Curie's Principle (Proposition 7.1), that the dynamics must be CP violating and therefore by the CPT theorem also T violating. This remarkable decay only occurs in roughly one out of every 500 decays of the neutral kaon. In the modern gauge theory of electroweak interactions, this turns out to be near the maximum amount of T violation that is compatible with the theory (cf. Witten 2018). But it is still a rare occurrence.

The fact that an occurrence is rare is no reason to ignore its implications. T violation implies that a representation of time reversal symmetry does not exist: this rare event is just a witness to the conclusion! Our formulation of Curie's principle (Proposition 7.1) shows that if the decay $K_L \to \pi^+\pi^-$ occurs even once, then time reversal symmetry is violated. Similarly, Kabir's Principle (Proposition 7.2) shows directly that if the probabilities in kaon oscillation differ by even the smallest amount, then time reversal symmetry is violated. Just as one smoking gun out of 500 is still enough to prove a murder, so these rare kaon decays are enough to prove that in our world, time is not symmetric. The time asymmetric dynamics applies to *all* experimental runs, despite the fact that only a small number of them are used to prove the asymmetry itself.

## 7.5 Summary

At last, we have arrived at an asymmetry of 'time itself'. In this chapter I have argued that the time asymmetry established by electroweak theory really does deserve to be called 'the arrow of time'. Perhaps one might like to refer to it more conservatively, as 'one arrow' of time. However, the other asymmetries that are usually called 'arrows' are currently all misfires, as I have argued in Chapters 5 and 6. In this sense, the arrow of time arising from electroweak theory really is unique.

Some subtle argumentation has been needed to establish this. My first step was to move to a more structural perspective, viewing *t* as referring to time translations, rather than to a time coordinate. The Representation View then provides the connection between the structure of time and the trajectories of a dynamical theory: the symmetries of time are 'projected' down onto the state space of a dynamical theory by a representation, just as the symmetries of a table are projected by its shadow down onto the floor. As a result, T violation on the state space of electroweak theory teaches us not only about state space but about the structure of time itself.

Of course, there were a number of subtleties to this story. The evidence for an arrow of time drew in part on the robustness of electroweak theory as a renormalisable quantum field theory as well as on the generality of the experimental evidence for T violation. And, the sense in which 'time itself' has an arrow is not exactly in the sense of a temporal orientation but rather as an asymmetry of time translations. However, I hope now to have shown that, with all these subtleties considered, the arrow of time is secure.

My discussion in this chapter has concerned the 'canonical form' of the time reversal operator, as distinct from other time reversing transformations like CPT. However, there is a remaining doubt about time asymmetry, that there is a sense in which the transformation CPT might be what 'plays the role' of time reversal in relativistic quantum field theory. So, since all such theories are thought to be CPT symmetric, perhaps there is still some sense in which time has a symmetry too. Making sense of this idea requires a careful examination of the meaning of CPT symmetry, which is the subject of the final chapter.

# 8

# Representing CPT

*Précis.* *There are many representations of time reversal symmetry, including PT, CT, and CPT, but only the standard time reversal operator T is associated with an arrow of time itself.*

In an excited phone call, John Wheeler is reported to have told his then-graduate student Richard Feynman at Princeton that he had discovered why all electrons have the same charge and mass: "Because they are all the same electron!"[1] Wheeler proceeded to suggest that a single electron worldline might be wriggling forwards and backwards in time in a great knot, and that when moving backwards we would experience it like its antiparticle, a positron. From the perspective of a future-directed observer like ourselves, the backwards-turning points would appear as electron–positron anni-hilation; the forwards-turning points would appear as electron–positron creation; and, the appearance of distinct particles at a given instant would be explained by the spacelike surface that is the instant cutting through the knot, as in Figure 8.1.

Feynman reported his response:

I did not take the idea that all the electrons were the same one from him as seriously as I took the observation that positrons could simply be represented as electrons going from the future to the past in a back section of their world lines. That, I stole! (Feynman 1972, p.163)

---

[1] As reported by Feynman (1972, p.163). Wheeler appears to have been inspired by Stueckelberg (1942), who had earlier arrived at a similar perspective.

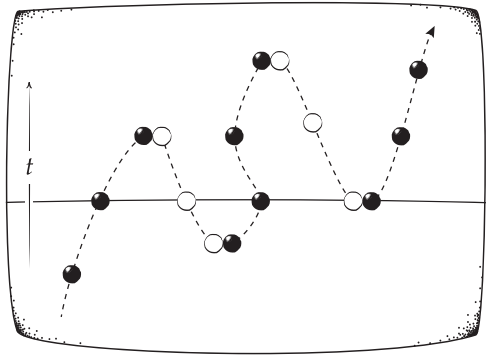

Figure 8.1 Wheeler's knot. On the horizontal spacelike surface there is an electron, a positron, another electron, and an electron–positron creation event.

The idea was fruitful for Feynman: understanding time reversal as including matter–antimatter exchange led him to the absorber theory of Wheeler and Feynman (1945) as well as to the Feynman (1949) theory of positrons.

By inspecting Figure 8.1, one can immediately see that on Feynman's view, inverting the direction of time automatically converts each instantaneous electron state into a positron, and vice versa. Of course, one would like to make this precise in the language of quantum field theory, and many have done so.[2] The proposal has also been defended by philosophers Greaves (2008, 2010) and Arntzenius and Greaves (2009). Writing 'C' to refer to matter–antimatter exchange and 'T' to refer to time reversal, they write,

[T]he operation that ought to be called 'time reversal' – in the sense that it bears the right relation to spatiotemporal structure to deserve that name – is the operation that is usually called TC. (Arntzenius and Greaves 2009, p.584)

Unfortunately, as we have seen in Section 7.3, CT violation in electroweak theory means that there is no representation of CT symmetry in the Standard Model. However, there is a representation of CPT symmetry, which is to say, symmetry under the combination of three transformations: matter–antimatter exchange C, time reversal T, and parity P. This led to another piece of physics lore, that the operation that ought to be called 'time reversal' is the CPT operator, as expressed by Wallace:[3]

---

[2] Early versions appear in Stueckelberg (1942) and Watanabe (1951), and in the original Lüders (1954, p.4) construction of a CPT operator, who refers to CT as time reversal "of the second kind". This phrase is dropped from most later formulations of the CPT theorem, except for that of Bell (1955).

[3] D. Wallace (2011). "The logic of the past hypothesis", Unpublished manuscript, http://philsci-archive.pitt.edu/8894/

[I]n quantum field theory, it is the transformation called CPT, and not the one usually called T, that deserves the name. (Wallace 2011, p.4)[3]

How are we to evaluate these proposals, that time reversal is 'really' CT or CPT? Setting aside labelling conventions, the substantial proposal seems to be that CT or CPT is more appropriate for evaluating whether time has an arrow than T is. The thesis of this chapter is that they are not: the operator that ought to be called 'time reversal', in the sense that it is relevant to the question of whether time itself has an asymmetry, is the standard time reversal operator T.

There are two parts to my argument. In the first, I will draw on the arguments of Chapter 2 to make the case that time reversal must behave appropriately with respect to translations in time and space, in order to deserve the name. Using this behaviour to characterise what it means to reverse 'time alone', I argue in Section 8.1 that only the standard time reversal T is suitable for this purpose. In Section 8.2, I consider the defence of Feynman's proposal due to Greaves (2010), who argues that by following the meaning of classical time reversal through the quantisation procedure, one finds that it is represented by CT. At least when a rigorous approach to quantisation is adopted, I argue that classical time reversal is transformed into an operator T that does not exchange matter and antimatter.

The second part will consider whether there is a more systematic way to view the relationship between matter–antimatter exchange and spacetime symmetry. Both are related to a structure called the 'covering group' of the restricted Lorentz group. Thus, in Section 8.3, I consider the possibility of viewing this structure as a spacetime symmetry group and point out a sense in which it is empirically adequate as a spacetime symmetry group, although a bit eccentric. In Section 8.4, I argue that, even given this spacetime symmetry group, matter–antimatter exchange cannot be a spacetime symmetry on par with parity and time reversal, although it can be viewed in a closely related way, as a symmetry of representations. On this 'spacetime symmetry account' of matter–antimatter exchange, I find there is an interesting relationship between time reversal and matter–antimatter exchange – but still no plausible sense in which T is 'really' CT or CPT. In particular, there is no sense in which the arrow of time established by T violation is erased.

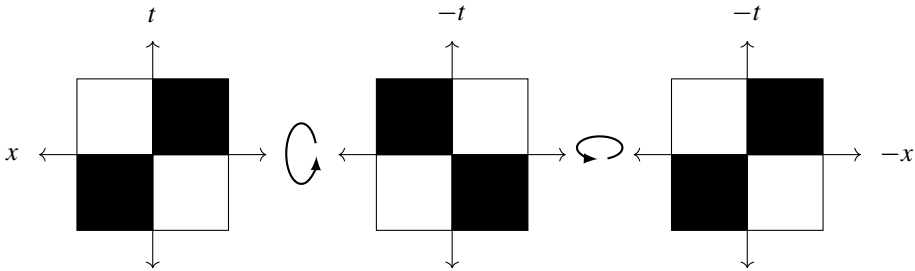

Figure 8.2 Time reversal is not a symmetry, but space-time reversal is.

## 8.1 Against Alternative Routes to Time Symmetry

In Chapter 7, I argued that experimental evidence of T violation from electroweak theory shows that time has an arrow. But, might there be some more liberal perspective on what 'time reversal' means, according to which time is still symmetric? Consider the $2 \times 2$ chequered grid of Figure 8.2, for which neither time reversal nor spatial reversal is a symmetry, but the combination of both is. In this section, I will argue that 'restoring' temporal symmetry in this way is, perhaps surprisingly, always possible – but that this does not undermine the fact that time is asymmetric in the sense established in Chapter 7.

Here is a toy example from Newtonian mechanics: consider a free particle, like a bead that threads a wire, which is constrained by some blocking mechanism to never move to the left, as shown in Figure 8.3. Time translations $\varphi_t$ take their ordinary force-free form on the state space of positions and velocities, $\varphi_t(x, \dot{x}) = (\dot{x}t + x, \dot{x})$, but with the constraint on state space that the only possible states $(x, \dot{x})$ are those that satisfy $\dot{x} \geq 0$. This system is both T violating and P violating, because each would transform a rightward-moving trajectory to an impossible leftward-moving one. In fact, the transformations representing time reversal $T(x, \dot{x}) := (x, -\dot{x})$ and parity $P(x, \dot{x}) := (-x, -\dot{x})$ cannot even be defined on this state space, which includes only states with positive velocities. Nevertheless, one can still represent PT symmetry: if we both turn the string around *and* apply time reversal, by applying $\tilde{T}(x, \dot{x}) := (-x, \dot{x})$, then we do get a dynamical symmetry.[4]

---

[4] To confirm this, observe that for all $(x, \dot{x})$ in the state space, we have, $\tilde{T}\varphi_t\tilde{T}^{-1}(x, \dot{x}) = \tilde{T}\varphi_t(-x, \dot{x}) = \tilde{T}(-x + \dot{x}t, \dot{x}) = (x - \dot{x}t, \dot{x}) = \varphi_{-t}(x, \dot{x})$. Note that this 'PT' transformation, $\tilde{T}$, is *not* the composition of two transformations P and T, neither of which exist. So, it is better notation to introduce a new symbol like $\tilde{T}$. The same turns out to hold of the CPT transformation, as Swanson (2019, p.107) has emphasised.

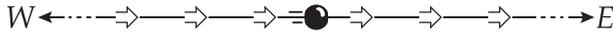

Figure 8.3 Time translations for the asymmetric bead–string system.

Of the two classes of dynamical symmetry described in Section 4.1.3, $S\varphi_t S^{-1} = \varphi_t$ and $S\varphi_t S^{-1} = \varphi_{-t}$, this PT transformation $\tilde{T}$ is an example of the latter: it is a 'time reversing' symmetry,

$$\tilde{T}\varphi_t\tilde{T}^{-1} = \varphi_{-t}. \tag{8.1}$$

Since $\tilde{T}$ reverses all the curves in space, in the sense that $x(t) \mapsto x(-t)$, one might be tempted to call it 'time reversal', especially given the absence of any other candidate. In fact, we have considered this possibility before in Section 3.2.1, where I associated this transformation with a 'folk' view of Newtonian time reversal.

One can make this concern more powerful from the perspective of the Representation View of Section 2.3. On this view, a spacetime symmetry gets its meaning on state space through the presence of a representation. As a spacetime symmetry, time reversal $\tau$ satisfies $\tau t \tau^{-1} = -t$ for all time translations $t$, as I argued in Chapter 2. So, since a representation $\varphi$ of that group is just a homomorphism, Eq. (8.1) says that $\tilde{T}$ provides a representation of time translation reversal! Does this mean that the parity–time reversal transformation $\tilde{T}$ can be understood as referring to the 'true' time reversal operator for this system, or at least an adequate one? If the PT transformation (or some other transformation like it) always restores temporal symmetry, then one might be tempted to conclude that T violation is not enough to establish an arrow of time.

I will make the case that this argument does not work. But, before I turn to that, let me point out a sense in which this more general kind of symmetry is really ubiquitous in any quantum theory, even without appeal to the CPT theorem. If we are willing to do more than reverse 'time alone', then there are always many different ways to obtain a dynamical symmetry that reverses time. Here is the statement of that fact, followed by its interpretation.

**Proposition 8.1** *Let $t \mapsto U_t$ be a strongly continuous unitary representation of $(\mathbb{R}, +)$ on a separable Hilbert space $\mathcal{H}$. Then this representation always extends to a representation of time reversal symmetry, in the sense that there exists some antiunitary $\mathcal{T}$ such that*

$$\mathcal{T}U_t\mathcal{T}^{-1} = U_{-t}. \tag{8.2}$$

*Moreover, for every unitary $U'$ such that $[U_t, U'] = 0$, the operator $\mathcal{T}' := U'\mathcal{T}$ is an antiunitary representation of time reversal symmetry in this sense.*

*Proof*   Our first statement is a corollary of Proposition 3.4. The fact that $\mathcal{T}' := U'\mathcal{T}$ satisfies Eq. (8.2) when $[U_t, U'] = 0$ is thus immediate.   ∎

Here is what this means. First, recall that dynamical systems in fundamental physics are generically associated with half-bounded energy.[5] For quantum systems, it is known that every representation $\mathcal{T}$ of time reversal must be *antiunitary*, which is to say that $\mathcal{T}$ is antilinear, $\mathcal{T}(a\psi + b\phi) = a^*\mathcal{T}\psi + b^*\mathcal{T}\phi$, and that $\mathcal{T}^*\mathcal{T} = I$. This result was proved in the second part of Proposition 3.4. So, if we are willing to accept a transformation as 'time reversal symmetry' whenever it represents an automorphism that reverses time translations, $\mathcal{T}U_t\mathcal{T}^{-1} = U_{-t}$, then time reversal symmetry is always assured. This is true even in the electroweak theory, which is only T violating in the sense that the 'canonical' time reversal transformation $T$ does not provide a representation of time reversal symmetry. In this more general sense, virtually all quantum systems are temporally symmetric.

I find this result fascinating. But, I do not think it provides an argument that time is symmetric after all. Although a representation of time reversal symmetry must reverse time translations, it is also usually part of a larger spacetime symmetry group or gauge group, and this fact constrains it in other ways. For example, in both the Galilei and Lorentz groups, time reversal preserves spatial translations and also reverses velocity boosts. In this context, the meaning of time reversal is constrained by more than just the time translations: it is constrained by its also having appropriate relations to the other symmetry transformations.

For example, returning to the bead on a wire, the parity–time reversal transformation $\tilde{T}$ does not preserve spatial translations; this was shown in the discussion of Newtonian time reversal in Section 3.2.4. Nor does it reverse velocity boosts. Neither of these features are plausible properties of a transformation that reverses 'time alone'. So, once the larger group of spacetime symmetries is taken into consideration, it becomes implausible to identify the parity–time reversal transformation $\tilde{T}$ or the general transformation $\mathcal{T}$ in Proposition 3.4 with time reversal. If anything, the latter can only be viewed as establishing '$UT$ symmetry', where $T$ is time reversal and $U$ is some unitary symmetry.[6]

One can say more: if an irreducible representation of time reversal exists, then it is quite generally unique. This was pointed out in Section 3.3.3 and in particular in Proposition 3.2. For example, consider the case of the Standard

---

[5] For a discussion of this, see Section 3.1.3.
[6] This is always possible because both $T$ and $\mathcal{T}$ are antiunitary: as a result, $U = \mathcal{T}T^{-1}$ is unitary, which implies that $UT = \mathcal{T}$.

Model, which has a representation of all the continuous symmetries of the Poincaré group. Then time reversal must be defined so as to have the correct relations to all those symmetries. And, in an irreducible representation of the Poincaré group, those considerations constrain time reversal so strongly as to make it unique, by Schur's lemma. So, the 'canonical' time reversal operator T is not arbitrary, despite the many available operators given by Proposition 8.1: it is the only representation of time reversal available that behaves appropriately with respect to the entire group of spacetime symmetries. So, the fact that canonical time reversal symmetry is violated in electroweak theory means that no other good options are available to represent temporal symmetry.

Matter–antimatter exchange, or charge conjugation, is no exception to this: in particular, if it is not a spacetime symmetry, then it is not relevant to time asymmetry as I have defined it. That said, some have argued that matter–antimatter exchange actually _can_ be viewed as a spacetime symmetry, with the operator CT behaving the way time reversal should behave, even with respect to the other symmetries. In the next section I will address one such argument, which appeals to the nature of quantisation. In Sections 8.3 and 8.4, I will then propose a different way to try to connect time reversal and charge conjugation, using the universal covering of the Poincaré group. In both cases, I will argue that the standard time reversal operator T prevails and that the argument for time asymmetry developed in Chapter 7 still stands.

## 8.2 On Feynman's View

One precise way to interpret Feynman's proposal is as the claim that CT is the 'correct' time reversal operator. This way of thinking appears to have been smuggled into his famous diagrams, which represent contributions to a scattering amplitude in a perturbative approximation. If the $CT$ transformation exists, then reflecting a Feynman diagram about the vertical 'time' axis produces a $CT$ transformation, in the sense that it both reverses time and exchanges matter and antimatter, as in Figure 8.4.[7]

Even if we follow Feynman and refer to CT as 'time reversal', this does not erase the arrow of time: electroweak theory is CT violating as well, as a consequence of Wu's parity-violating experiment. One can raise further doubts as well: what reason is there to think that Feynman diagrams capture the 'true' nature of time reversal? The fact that reversing a Feynman diagram is similar to an application of CT might be viewed as an artefect of the diagrams which does not adequately capture the symmetries of time itself. In my view, that is exactly what it is. However, let me first review an important argument to the contrary.

---

[7] This was pointed out by Ramakrishnan (1967) and by Feynman (1985, Chapter 3) himself.

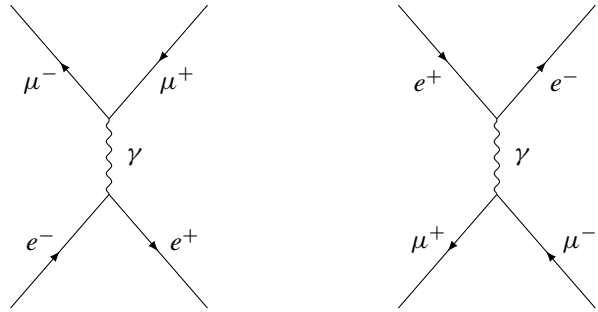

Figure 8.4 A Feynman diagram for an electron–positron decay (left), when vertically reflected, produces a description that both reverses time and exchanges matter and antimatter states (right).

### 8.2.1 The Greaves Quantisation Argument

Arntzenius and Greaves have argued that the Feynman view is needed to explain two puzzles about the CPT theorem.[8] All known relativistic quantum field theories are invariant under CPT, and the 'CPT theorem' is a collection of results that explains this: it shows that a large class of relativistic quantum theories are CPT invariant.[9] However, there is still some room to be puzzled about these results. Greaves (2008, 2010) has put it in the following way.

[O]ne can identify two positive sources of puzzlement:

- How can it come about that one symmetry (e.g., Lorentz invariance) entails another (e.g., CPT) at all?
- How can there be such an intimate relationship between spatiotemporal symmetries (Lorentz invariance, parity reversal, time reversal) on the one hand and charge conjugation, not obviously a spatiotemporal notion at all, on the other? (Greaves 2010, p.28)

---

[8] See Greaves (2008, 2010), Arntzenius and Greaves (2009), and Arntzenius (2011, 2012).
[9] See Bain (2016); Swanson (2018, 2019) for a summary and philosophical perspectives on these results.

Her solution to the second puzzle, echoed by Arntzenius (2011, 2012), is that we must adopt the Feynman view: the transformation normally called CT is what deserves the name 'time reversal'. Thus, both CT and CPT count as 'spatiotemporal', solving the second puzzle. As for the first puzzle, on how continuous relativistic symmetries can give rise to further discrete symmetries, Greaves argues that a "classical PT theorem" shows how this is forced by the geometry of relativistic spacetime.

This section will be concerned with Greaves' solution to the second puzzle.[10] Why is it more 'natural' to say that time reversal is really CT? Greaves (2010, §4) sketches an answer, which draws on her view of quantum field theory as the "quantisation" of a classical field theory. She begins with a classical field theory, making the standard choice for the classical time reversal operator. Standard textbook practice then converts the classical theory into a quantum theory, and Greaves draws on Bell (1955) to determine how classical time reversal is transformed by this procedure. According to Greaves, and perhaps also to Bell, quantisation automatically converts classical time reversal into what is normally called the CT transformation.[11] In this sense, she proposes, CT in quantum field theory is naturally interpreted as time reversal:

[S]tart from a classical field theory, with assumptions about which *classical* transformations deserve the names 'time reversal' and 'parity reversal' already in place (never mind whence!); obtain a QFT by quantization; work out which transformations on QFT states and operators are induced by the already-named transformations on classical fields, and name the former accordingly. . . . when one carries out this *latter* project, with standard names for the classical transformations, *the transformation that is usually called 'TC' receives the name 'T'.* (Greaves 2010, p.39)

If this is right, then the quantisation of classical time reversal is CT, and so the quantisation of classical PT is CPT. That would be quite surprising: time reversal in classical and quantum theory are generally quite similar, in that both can be viewed as representations of time translation reversal.[12] In contrast, Greaves proposes that they are quite different: time reversal in classical theory becomes CT in quantum theory. In the remainder of this section, I will argue that it does not: on a rigorous quantisation procedure, classical time reversal does *not* induce a transformation that exchanges matter and antimatter.

---

[10] Her solution to the first draws on an argument of Bell (1955), which was generalised by Greaves and Thomas (2014) but which has some difficulties. For example, as Swanson (2019, p.120) points out, her PT theorem is too restrictive for the Standard Model, since it holds only for polynomial interactions of tensor fields and thus does not apply to non-polynomial interactions or to those involving spinorial tensors.

[11] Bell summarises the result of quantisation: "Thus the kind of reversal we have been considering implies an automatic change of sign of charge. From the field point of view, such a sign change is no more surprising than the sign change of a velocity, or an angular momentum, with time reversal in particle mechanics" (Bell 1955, p.483).

[12] See Chapter 3.

### 8.2.2 *Quantisation of T Is Not CT*

There are many approaches to 'quantisation', or the practice of converting a classical description into a quantum one. This practice sometimes includes 'eye balling it', and when it does, it is called *canonical quantisation*. Unfortunately, canonical quantisation suffers from a large number of impossibility results.[13] So, it will be helpful to adopt a rigorous alternative. I will argue that, when quantisation is treated in a rigorous way, classical time reversal is not converted into an operator that exchanges matter and antimatter. So, the Greaves quantisation argument does not generally stand up to scrutiny. To illustrate, I will adopt one well-known approach: the *Segal quantisation* of a classical bosonic field.[14]

Begin with a general description of a complex classical field theory, such as a free Klein–Gordon field. Its space of solutions forms a real manifold $S$, which we will assume for simplicity is also a vector space: this holds, for example, in the linear approximation of solutions to a field equation. The manifold has a symplectic structure, which is to say a bilinear map $\omega : S \times S \to \mathbb{R}$ that is skew-symmetric and non-degenerate, arising from the structure of the dynamics.[15] It also has a Riemannian structure, which is to say a bilinear positive symmetric map $g : S \times S \to \mathbb{R}$, arising from the structure of the state space of initial conditions. These will usually satisfy a technical condition of being *compatible*, in the sense that $\frac{1}{2}|\omega(\psi, \phi)| \leq g(\psi, \psi)^{1/2} g(\phi, \phi)^{1/2}$, which helps to guarantee that $g$ provides an inner product. In summary, I will refer to the triple $(S, \omega, g)$ as a *classical field theory* whenever $S$ is a linear manifold, $\omega$ is a symplectic structure, and $g$ is a Riemannian structure that is compatible with it.

The central result of Segal quantisation is that it is always possible to convert a classical field theory of this kind into a quantum theory, called a 'one particle structure', in a way that preserves the essential structure of the original field theory. This structure then forms the basis for a Fock space representation of a quantum field theory. The result is summarised by Kay and Wald (1991, Proposition 3.1):

> **Proposition 8.2** (Segal Quantisation) *For every classical field theory $(S, \omega, g)$ there is a Hilbert space $\mathcal{H}$ and a 'Segal quantisation' map $K : S \to \mathcal{H}$ such that:*
>
> 1. *(adequacy) The range of $K$ is dense in $\mathcal{H}$.*
> 2. *(Riemannian preservation) $\mathrm{Re}\langle K\psi, K\phi \rangle = g(\psi, \phi)$ for all $\psi, \phi \in S$.*
> 3. *(symplectic preservation) $2\,\mathrm{Im}\langle K\psi, K\phi \rangle = \omega(\psi, \phi)$ for all $\psi, \phi \in S$.*
>
> *The pair $(K, \mathcal{H})$ is unique up to unitary equivalence, in the sense that if $(K', \mathcal{H}')$ satisfies (1)–(3), then there is some unitary $U : \mathcal{H} \to \mathcal{H}'$ such that $UK = K'$.*

[13] The canonical quantisation procedure of Dirac (1947), whereby classical observables are 'hatted' to produce quantum observables, suffers from the impossibility results of Groenewold (1946) and of the PhD thesis of Van Hove (1951). A class of similar results is referred to as the *Groenewold–Van Hove Theorem*; for an introduction, see Gotay (2000).

[14] For an introduction, see Kay (1979), Kay and Wald (1991, Appendix A), Segal and Mackey (1963), or Wald (1994). For some alternative approaches see Landsman (1998) or Woodhouse (1991).

[15] For example, if the field theory is formulated with a symplectic manifold as its state space (see Section 3.3), then the solutions' Hamiltonian evolution preserves the symplectic form on state space, which allows one to define a symplectic structure on solutions as well.

The Hilbert space $\mathcal{H}$ built by Segal quantisation can be used to define a symmetric Fock space,

$$\mathcal{F} = \mathbb{C} \oplus \mathcal{H} \oplus \mathcal{S}(\mathcal{H} \otimes \mathcal{H}) \oplus \cdots, \qquad (8.3)$$

where $\mathcal{S}$ is the projection onto the symmetric subspace. One can then follow a standard procedure to define creation, annihilation, and particle number operators of a bosonic quantum field system.[16] Proposition 8.2 ensures that each one-particle structure $\mathcal{H}$ is the unique quantum system capturing the essential properties of the classical bosonic field $(S, g, \omega)$: *adequacy* ensures that every state in the Hilbert space is 'reasonably close' to representing a state in the original classical field theory. And, *Riemannian preservation* and *symplectic preservation* ensure that the metrical and dynamical information are preserved, respectively. In this sense, the resulting Fock space representation is an adequate representation of a bosonic quantum field theory.

The Segal construction moreover guarantees that each one-particle structure can be written as $\mathcal{H} = \mathcal{H}' \oplus \mathcal{H}''$ in a canonical way, with the summands interpreted as 'positive frequency' and 'negative frequency' subspaces. The exchange of these subspaces $(\psi^+ \oplus \psi^-) \mapsto (\psi^- \oplus \psi^+)$ is called *matter–antimatter exchange*. So, to see whether the quantisation of classical time reversal exchanges matter and antimatter, as Greaves proposes it does, we can check whether the Segal quantisation of classical time reversal has this property. As I have suggested, the resulting quantum time reversal operator does *not* exchange matter and antimatter. However, confirming this requires introducing a few details of the Segal construction, which I will now sketch; the remaining details can be found in Kay and Wald (1991, Appendix A).

---

[16] This Fock space construction can be shown to be essentially unique (see Baez, Segal, and Zhou 1992, Theorem 1.10). For an introduction, see Araki (1999, §3.5).

The results of Chapter 3 help to identify the key properties of a time reversal map $T_S : S \to S$ for a classical field theory $(S, \omega, g)$. In summary, time reversal must reverse the symplectic structure, $\omega(T_S\phi, T_S\psi) = -\omega(\phi, \psi)$, and it must preserve the Riemannian structure, $g(T_S\phi, T_S\psi)$ for all $\phi, \psi \in S$. The former follows from the fact that classical time reversal must be antisymplectic (Proposition 3.1). The latter can be motivated from the assumption that time reversal does not alter classical metrical structure. It also follows directly from the fact that any quantum time reversal operator that we end up with must be antiunitary (Proposition 3.4).[17]

How does Segal quantisation convert classical time reversal into a quantum time reversal operator? Given any $T_S : S \to S$, the Segal quantisation map $K$ induces a quantum operator $T_{\mathcal{H}}$ defined by,

$$T_{\mathcal{H}}(K\phi) := K(T_S\phi), \qquad (8.4)$$

for all $K\phi \in \mathcal{H}$. This definition of $T_{\mathcal{H}}$ uniquely extends to all of $\mathcal{H}$, because the range of $K$ is dense. Since classical time reversal $T_S$ reverses $\omega$ and preserves $g$, one can show that this $T_{\mathcal{H}}$ must be antiunitary operator, as we will see shortly.

To determine whether quantum time reversal exchanges matter and antimatter, one can check whether $T_{\mathcal{H}}$ exchanges the canonical positive and negative frequency subspaces in $\mathcal{H} = \mathcal{H} \oplus \mathcal{H}$. In the Segal construction, the quantisation map $K$ is defined so as to transform each classical solution $\phi \in S$ to a quantum state $\psi^+ \oplus \psi^- \in \mathcal{H}$ of the form

$$K(\phi) = \psi^+ \oplus \psi^- := E\phi \oplus FC\phi, \qquad (8.5)$$

where $E$ and $F$ are positive operators, and $C$ is an antilinear map defined on $\mathcal{H}'$, which is a 'complexified copy' of $S$. Omitting some details, here is the main fact about these operators that we will need: whenever $T_S$ preserves $g$ and reverses $\omega$, it commutes with all three of these operators, $E$, $F$, and $C$. This immediately implies our main conclusion, that for any quantum state $\psi^+ \oplus \psi^- = K\phi$,

$$\begin{aligned} T_{\mathcal{H}}(\psi^+ \oplus \psi^-) := K(T_S\phi) &= E(T_S\phi) \oplus FC(T_S\phi) \\ &= T_S(E\phi) \oplus T_S(FC\phi) = (T_{\mathcal{H}}\psi^+) \oplus (T_{\mathcal{H}}\psi^-). \end{aligned} \qquad (8.6)$$

Since the positive and negative frequency states are not exchanged, the quantised time reversal operator is not associated with matter–antimatter exchange.

---

[17] Antiunitarity implies that $\langle T_{\mathcal{H}}\psi, T_{\mathcal{H}}\psi' \rangle = \langle \psi, \psi' \rangle^*$. On the Segal construction of the inner product in Eq. (8.7), this is only possible if $g$ is preserved.

Here are a few more details, which might help to avoid one possible confusion. Segal quantisation begins by constructing a 'complex structure' $J : S \to S$ on the classical solution space, which is a linear map $J : S \to S$ satisfying $J^2 = -I$ and which turns $g$ into a 'Kähler form': $\frac{1}{2}\omega(\psi, \phi) = g(\psi, J\phi)$ and $\omega(J\psi, J\phi) = \omega(\psi, \phi)$ for all $\phi, \psi \in S$. One then converts $S$ into the 'positive frequency' Hilbert space $\mathcal{H}$, by defining $i\psi := -J\psi$ and adopting the inner product,

$$\langle \psi, \phi \rangle := g(\psi, \phi) + \tfrac{i}{2}\omega(\psi, \phi). \tag{8.7}$$

Classical time reversal preserves $g$ and reverses $\omega$, which guarantees it is antiunitary with respect to this inner product,

$$\begin{aligned} \langle T_S\psi, T_S\phi \rangle &= g(T_S\psi, T_S\phi) + \tfrac{i}{2}\omega(T_S\psi, T_S\phi) \\ &= g(\psi, \phi) - \tfrac{i}{2}\omega(\psi, \phi) = \langle \psi, \phi \rangle^*. \end{aligned} \tag{8.8}$$

Using these facts, a short calculation shows[18] that any such $T_S$ must also reverse the complex structure, in that $T_S J = -J T_S$. Since this $J$ intuitively captures 'multiplication by $i$' in the classical field theory, the fact that $T_S J T_S^{-1} = -J$ does capture a sense in which $T_S$ 'conjugates' classical fields.

But, this kind of conjugation is not what captures matter–antimatter exchange in quantum theory and should not be conflated with the conjugation operator $C$ appearing in the Segal quantisation map of Eq. (8.5). The latter is relevant to the exchange of positive and negative frequency subspaces of the quantised field theory, whereas the former is not. On the contrary, the fact that $T_S$ reverses $J$ helps to establish that $T_S$ commutes with $E$, $F$, and $C$ and thus that it does not exchange these subspaces.[19]

Classical time reversal is not converted into a CT transformation by quantisation. However, one might still wonder about Greaves' two puzzles. How can assuming continuous symmetries give rise to more symmetries? And, wouldn't CPT symmetry be a lot easier to explain if charge conjugation were somehow associated with spacetime?

The first puzzle is in fact not so surprising considering the account of discrete symmetries that I have given in Chapter 2. By viewing these symmetries as automorphisms of the continuous symmetries, they arise naturally as further 'higher-order' symmetries, through the semidirect product construction that I have given in Section 2.6. I will discuss this in more detail in Section 8.4. What is more puzzling is how the structure of a given state space can possibly prevent them from being symmetries. My perspective on how this can happen is given in the discussion of T violation in Chapter 7.

---

[18] Irreducibility considerations generally guarantee that $T J T^{-1} = \pm J$ (cf. Wallace 2009, p.218). Moreover, our compatibility assumption for the case that $\phi = \psi$ implies that $g(\phi, \phi) \geq 0$ and hence that $-g(\phi, \phi) = \frac{1}{2}\omega(\phi, J\phi) \leq 0$ for all $\phi \in S$. But this makes $T J T^{-1} = J$ impossible, for then we would have $0 \geq \omega(\phi, J\phi) = -\omega(T\phi, T J\phi) = -\omega(T\phi, J T\phi) \geq 0$ and hence that $\omega(\phi, J\phi) = g(\phi, \phi) = 0$ for all $\phi, \in S$.

[19] Similar remarks have been made by Baker and Halvorson (2010); Wallace (2009).

On the other hand, I find the second question to be tantalising. The ubiquitous symmetry of relativistic quantum field theory under CPT certainly would be a lot less puzzling if C could be viewed as a spacetime symmetry. Quantisation theory is unlikely to make the connection, as we have seen above. But, might there be another way? As Swanson (2019) has pointed out, such a link does exist in algebraic formulations of charge conjugation in quantum field theory. In the next two sections, I will rather consider a somewhat non-standard proposal about how this link can be made. In particular, there is a structure called the 'covering group' of the restricted Poincaré group, which makes a central appearance in the foundations of both spacetime symmetries and matter–antimatter exchange. Section 8.3 proposes a sense in which the spacetime symmetries might be 'enlarged' in a way that goes beyond the Poincaré group, while still remaining empirically adequate. Section 8.4 then considers a possible account of matter–antimatter exchange that makes use of this structure. Even based on this unusual account, I find that time reversal is still just T.

## 8.3 Local Symmetries beyond the Poincaré Group

A 'local' spacetime symmetry is one defined on scales for which gravitation and Planck-scale phenomena can be ignored. These are the scales on which relativistic quantum field theory is formulated. What are the local spacetime symmetries? The success of special relativity provides strong evidence that they consist of at least three things:[20]

1. *Lorentz boosts* describing a change of inertial reference frame;
2. *spatial rotations* describing rigid rotations of the spatial surfaces orthogonal to a timelike line; and
3. *the spacetime translations*, describing rigid translations along a timelike line.

The first two categories can be collected to form a Lie group called the *restricted Lorentz group* $L_+^\uparrow$, written with '$\uparrow$' and '$+$' to indicate that they do not reverse temporal or total orientation, respectively. The inclusion of translations through a semidirect product[21] produces the *restricted Poincaré group* $\mathcal{P}_+^\uparrow = \mathbb{R}^4 \rtimes L^\uparrow$. They are all isometries of Minkowski spacetime, sometimes referred to as the 'continuous symmetries', since as elements of a Lie group they are all continuously connected to the identity. The 'discrete' symmetries are isometries too, consisting of a continuous symmetry composed with either time reversal, parity, or both. Including these produces what is called the *complete Poincaré group* $\mathcal{P}$.

---

[20] For an introduction, see Landsman (1998, esp. §2.2 and Part IV) or Varadarajan (2007, §IX.2).

Of course, the discovery of P, T, and PT violation in electroweak theory indicates that these are not actually spacetime symmetries, as I have argued in Chapter 7. So, the structure of spacetime is not *quite* Minkowski. Instead, common wisdom has it that the local spacetime symmetries are described by the restricted Poincaré group $\mathcal{P}_+^\uparrow$.

Let me take a little inspiration from the proposal of Greaves (2010) discussed above. Is it possible that the spacetime symmetries might include something like matter–antimatter exchange as well? Perhaps common wisdom is not the whole story – the local spacetime symmetries might not be correctly described by the restricted Poincaré group! – if there is an alternative structure that could do the same job and offer a little more.

In this section, I will consider a proposal of this kind, which identifies the spacetime symmetries with the 'universal covering' of the restricted Poincaré group. The elements of the universal covering can all be interpreted as falling into one of the three categories above: boosts, rotations, and translations. But, this group is only locally (and not globally) isomorphic to the restricted Poincaré group. To some extent, this is an example of what philosophers call, 'underdetermination of theory by evidence': two different spacetime symmetry groups are compatible with existing evidence. However, in Section 8.4, I will illustrate a sense in which the universal covering might offer some advantages, through an account of matter–antimatter exchange as a spacetime symmetry.

I will begin in the next section with an example to motivate how the spacetime symmetries might be 'extended' in this way. I will then review the universal covering of the restricted Lorentz group in Section 8.3.2, before suggesting how one might construct a notion of matter–antimatter exchange in Section 8.4. The result might be viewed as a 'spacetime account' of matter–antimatter exchange, or at least the beginning of one. However, even based on this account, I can find no sense in which time reversal requires matter–antimatter exchange.

---

[21] Semidirect products were introduced in Section 2.6. The group multiplication rule for two elements $(x, \Lambda), (x', \Lambda') \in \mathbb{R}^4 \rtimes L^\uparrow$ is given by $(x, \Lambda)(x', \Lambda') = (x + \Lambda x', \Lambda\Lambda')$.

### *8.3.1 What Is the Group of Spatial Rotations?*

Let me begin with an example that is easy to visualise. The continuous symmetries of Euclidean space at a point can be collected together to form a Lie group $SO(3)$ of rigid Euclidean rotations. Those symmetries were given up following the discovery of general relativity. But, in this discussion, let me focus on local regimes in which gravity can be neglected and where one still might wish to postulate that the symmetry group of space at a point is $SO(3)$.

There is still another option. The elements of a different group, $SU(2)$, can also be interpreted as symmetries of space. We have seen this group already in our discussion of spin in Section 3.4.4: its elements are the 'rotations' given by

$$R_x(\theta) = e^{(i/2)\theta\sigma_x}, \qquad R_y(\theta) = e^{(i/2)\theta\sigma_y}, \qquad R_z(\theta) = e^{(i/2)\theta\sigma_z}, \qquad (8.9)$$

where each $\sigma_j$ for $j = x, y, z$ is a Pauli spin observable. As we have seen, this group has the interesting property that a 'rotation' through $2\pi$ does not return a system to where it started. Instead, $R(2\pi) = -I$, and a second rotation through $2\pi$ is needed to restore the identity, $R(4\pi) = I$. Nevertheless, each element of $SU(2)$ can be viewed as a spatial rotation, in the sense that there is a neighbourhood of each element that is isomorphic[22] to a neighbourhood of the ordinary rotation group $SO(3)$. The groups $SU(2)$ and $SO(3)$ have the same local structure but differ in their global description of how the rotations fit together.

The structure of $SU(2)$ obviously does not capture our experience of spatial symmetry on the scale of everyday objects: by rotating a ball through $2\pi$, it appears to go back to where it started.[23] However, the correct spatial symmetry group might still be $SU(2)$, if its strange behaviour is invisible on the scales that are currently accessible to us. Recall the analogy of an arrow on a Möbius strip: when an arrow is transported orthogonally along a loop, a rotation through $4\pi$ is needed to restore it to its original state. However, if we imagine that the arrow is scaled to be very small, as in Figure 8.5, or if it is only detectable in the presence of certain kinds of matter, then it might appear that a rotation through $2\pi$ is enough.

---

[22] Alternatively: these Lie groups have isomorphic Lie algebras at every point; this property will be defined more generally in Section 8.3.2 in the definition of a covering group.

[23] You *can* still experience $SU(2)$ with relative ease, not as a symmetry of space, but through the following procedure: hold your palm upright in front of you and notice that by raising your elbow, it is possible to rotate your hand through 360 degrees, all while keeping your palm upright. Your arm will now be quite twisted; but, by continuing to rotate in the same direction through another 360 degrees, always palm up, you can then untwist your arm into its original state!

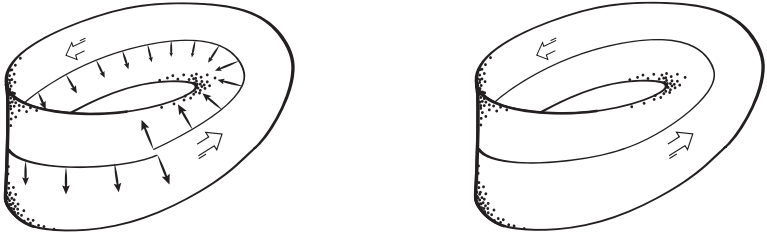

Figure 8.5  The global structure of $SU(2)$ might be hidden from us for scales on which the 'arrows' are small enough to appear invisible.

This is an ordinary case of underdetermination of theory by evidence: either the spatial symmetry group is $SO(3)$, or else it is secretly $SU(2)$ but in a way that gives the illusion of being $SO(3)$ on familiar scales. Of course, we do not need to take $SU(2)$ too seriously in the absence of evidence for it. In contrast, if we had evidence that $SU(2)$ were a more effective model of some phenomenon, in a way that explains the illusion of rotations on ordinary scales, then we might reject $SO(3)$ in favour of $SU(2)$. I do not know of any compelling evidence for this in the case of rotations, at least when considered in isolation. However, I will point out a sense in which this kind of expansion of the Poincaré group does have modelling advantages, in that it might allow one to characterise the matter–antimatter relationship as a spacetime symmetry.

Before I turn to this idea, let me head off a possible source of confusion: the structure of the rotation group for a spin-1/2 system does not provide any evidence for $SU(2)$ over $SO(3)$, as far as I am aware. It is true that the Hilbert space description of a spin-1/2 particle admits an irreducible unitary representation of $SU(2)$, and not of $SO(3)$. However, this does not imply that descriptions related by a rotation of $R(2\pi) = -I$ describe factually different states of affairs.

On the contrary, the statistical predictions of quantum theory are the same for all states $\psi$ on the same *ray*, or set of vectors related by a phase factor. The 'true' state space of quantum theory is thus a ray space; correspondingly, its symmetries are only defined up to a phase factor as well, as discussed in Section 3.4.2. From this perspective, the transformations $R(2\pi) = -I$ and $R(0) = I$ refer to the very same rotation, as Hegerfeldt and Kraus (1968) have pointed out.[24] Indeed, Bargmann (1954) famously showed that the ray space of a spin-1/2 system does admit a representation of $SO(3)$.

---

[24] Their article, written in response to Aharonov and Susskind (1967), is in fact the once-rejected paper that led to their collaboration with Wigner, as reported in Footnote 18.

My point here is rather that we still face a certain amount of underdetermination: whenever we have a representation of $SO(3)$, we always formally have a degenerate representation of $SU(2)$. And, when we turn to the case of the Poincaré group, I will argue that a very similar structure may provide a more effective model of reality, by allowing us to characterise the matter–antimatter relationship.

### *8.3.2  The Covering Group of Spacetime Symmetries*

Let me describe an analogous construction to the one above for the case of the Poincaré group. It will be helpful to first identify the nature of this construction in a more general way.[25]

> **Definition 8.1**  A *covering group* or *cover* over a Lie group $G$ is a Lie group $\hat{G}$, together with a continuous homomorphism or 'covering map' from $\hat{G}$ onto $G$ such that the induced map on Lie algebras is an isomorphism. The *universal covering group* is the unique, simply connected covering group over $G$.

A cover over a Lie group has the same local structure but may not be isomorphic. The 'universal' covering gets its name from the fact that it is a cover for all other covers. For example, $SU(2)$ is a cover for $SO(3)$, and indeed it is simply connected, and so it is the universal covering group over $SO(3)$.

The universal covering group for the restricted Lorentz group $L_+^\uparrow$ is called $SL(2, \mathbb{C})$. It was famously applied by Bargmann (1954) as part of a more general technique for finding unitary representations of Lie groups using their universal coverings. Like the case of $SU(2)$, the group $SL(2, \mathbb{C})$ is a doubly-degenerate cover over $L_+^\uparrow$, and the choice between $L_+^\uparrow$ and $SL(2, \mathbb{C})$ is underdetermined in the same sense that we have discussed in Section 8.3.1. Locally, it behaves just like the restricted Lorentz group. But, it might be that its global structure is somehow hidden from us on ordinary scales. Indeed, $SU(2)$ is a subgroup of $SL(2, \mathbb{C})$, and so this is in fact a more general instance of the problem discussed in the previous section. To understand the symmetries of $SL(2, \mathbb{C})$ and its relationship to Minkowski spacetime, let me introduce a few more details of a classic construction.[26]

The name $SL(2, \mathbb{C})$ stands for the 'special linear group' of $2 \times 2$ complex-valued matrices with unit determinant. To compare it to Minkowski spacetime, we first give a clever expression of the latter using the set $M$ of $2 \times 2$ matrices that are self-adjoint, $A = A^*$. Note that I write $A^*$ to refer to the conjugate-transpose or *adjoint* of $A$; I will later write $\bar{A}$ to refer to conjugation of each matrix element and $A^\top$ to refer to the matrix transpose.

---

[25] See Hochschild (1965, Chapters IV and XII) for an introduction to covering groups.
[26] Classic presentations include Gel'fand, Minlos, and Shapiro (1958, Part II Chapter 1 §1.8) and Naimark (1964, §3.9).

The set $M$ of self-adjoint matrices is closed under matrix addition and multiplication by real scalars. It thus forms a real vector space of four dimensions, with an explicit basis set given by the Pauli matrices:

$$\sigma_0 = \begin{pmatrix} 1 & \\ & 1 \end{pmatrix} \quad \sigma_1 = \begin{pmatrix} & 1 \\ 1 & \end{pmatrix} \quad \sigma_2 = \begin{pmatrix} & -i \\ i & \end{pmatrix} \quad \sigma_3 = \begin{pmatrix} 1 & \\ & -1 \end{pmatrix}. \quad (8.10)$$

We can identify a metric on this space by defining a symmetric bilinear form, $\eta(\sigma_i, \sigma_j) := g_{ij}$, where $g_{ij} = \mathrm{diag}(1, -1, -1, -1)$ is the Minkwoski metric. Then the pair $(M, \eta)$ is isometric to Minkowski spacetime! In particular, since it is a linear space, we can interpret it as the tangent space of Minkowski spacetime at a point. Thus, each element $v \in M$ can be interpreted as a vector at a point in the underlying spacetime manifold, corresponding to an 'instantaneous' translation along spacelike, timelike, or null curves.

It is illuminating to write these elements explicitly in terms of our basis elements and some real numbers $u = (t, x_1, x_2, x_3)$, which can be viewed as representing translations in spacetime. Then the general form of a vector in Minkowski spacetime at a point is

$$v = u \cdot \sigma = \begin{pmatrix} t + x_3 & x_1 - ix_2 \\ x_1 + ix_2 & t - x_3 \end{pmatrix}. \quad (8.11)$$

Writing the Minkowski norm as $|v|^2 := \eta_{\mu\nu} v^\mu v^\nu$ in the Einstein summation convention can now express it in a particularly simple way:

$$\begin{aligned} |v|^2 := \eta_{\mu\nu} v^\mu v^\nu &= t^2 - x_1^2 - x_2^2 - x_3^2 \\ &= (t + x_3)(x_0 - x_3) - (x_1 - ix_2)(x_1 + ix_2) = \det(v), \end{aligned} \quad (8.12)$$

where the determinant is associated with the matrix of Eq. (8.11). The complete Lorentz group $L$ consists of the maps $\Lambda : M \to M$ that preserve the Minkowski norm, $\eta(\Lambda v, \Lambda v) = \eta(v, v)$. By Eq. (8.12), this is equivalent to $\det(\Lambda v) = \det(v)$. The *restricted Lorentz group* $L_+^\uparrow$ then consists of those elements that are continuously connected to the identity.

I have said that $SL(2, \mathbb{C})$ is a cover for $L_+^\uparrow$. The covering map $\Lambda : S \mapsto \Lambda_S$ can be explicitly defined by

$$\Lambda_S v := S v S^*, \quad (8.13)$$

for all $S \in SL(2, \mathbb{C})$ and all $v \in M$. This map is straightforwardly shown[27] to be a continuous, surjective homomorphism $\Lambda : SL(2, \mathbb{C}) \to L_{+}^{\uparrow}$ that is a doubly-degenerate covering, meaning that $\ker \Lambda = \{I, -I\}$. The latter implies the map is a local isomorphism and thus induces an isomorphism of Lie algebras, making this a covering map. Since $SL(2, \mathbb{C})$ is simply connected, it follows that this is in fact the universal covering group over $L_{+}^{\uparrow}$. Each of its elements can be written in polar decomposition as $S = Ue^{A}$, where $U$ is unitary and $A$ is self-adjoint. They are generated by operators of the form $R_j(\theta) = e^{i(\theta/2)\sigma_j}$ and $B_j(s) = e^{(s/2)\sigma_j}$ with $j = x, y, z$, where the former give rise to a subgroup of 'rotations' isomorphic to $SU(2)$, and the latter can be interpreted as Lorentz boosts.

This completes our brief review of $SL(2, \mathbb{C})$.[28] It can be extended to include spacetime translations in a straightforward way, just as they are included in the restricted Poincaré group $\mathcal{P}_{+}^{\uparrow}$, by adopting the semidirect product group $\mathbb{R}^4 \rtimes SL(2, \mathbb{C})$ with multiplication given by $(v, S)(v', S') = (v + Sv'S^*, SS')$. I will denote this group $\overline{\mathcal{P}}_{+}^{\uparrow}$, with a 'bar' covering the letter to remind one that it is the universal covering.

This development of $SL(2, \mathbb{C})$ and its relationship to Minkowski spacetime provides a well-known technique for constructing projective representations of the restricted Lorentz group $L_{+}^{\uparrow}$ on a Hilbert space. However, the interpretation I am considering here goes beyond this usage, in order to give serious consideration to the possibility that C is a spacetime symmetry.

Namely, if physical spacetime were locally described by Minkowski spacetime, then its symmetry group would be the complete Poincaré group. But, spacetime is *not* locally Minkowski, because P, T, and PT symmetry are all violated. So, my plan is now to consider the possibility that it is different in other ways too and in particular that its global structure is given by the universal covering group $\overline{\mathcal{P}}_{+}^{\uparrow}$. This is an unusual way to look at this group. But, in the next section, I will indicate how it suggests an account of matter–antimatter exchange that is closely related to a spacetime symmetry and which allows it to be effectively compared to time reversal.

---

[27] See Varadarajan (2007, p.334–5). To fill in some of the details: $\Lambda_S$ is in the Lorentz group because it preserves the Minkowski norm, $\eta(\Lambda_S v, \Lambda_S v) = \det(\Lambda_S v) = \det(SvS^*) = \det(v) = \eta(v, v)$. It is obviously continuous, and one can check that it is a homomorphism, $\Lambda_{RS}(v) = (RS)v(RS)^* = R(SvS^*)R^* = \Lambda_R \Lambda_S(v)$. To see that it is a twofold covering, let $S \in \ker \Lambda$, so that $v = \Lambda_S v = SvS^*$ for all $v$. It follows by Schur's lemma that $S = cI$ for some $c \in \mathbb{C}$. We thus have that $1 = \det(S) = c^2$ and hence that $S = \pm I$, which is to say that $\ker \Lambda = \{\pm I\}$.

[28] For a more detailed introduction, see Varadarajan (2007, p.334–7), or Naimark (1964, §3.9).

### 8.4 C as a Spacetime Symmetry

I proposed a general strategy for determining the discrete symmetries associated with a collection of continuous spacetime symmetries in Section 2.6. In particular, I showed how time reversal arises as an automorphism of the time translations. In this section, I will argue that matter–antimatter exchange can be understood in a way that is very similar to this: not as an automorphism of spacetime symmetries, but of the space of representations of those symmetries. This brings C, P, T, and their combinations closer to being on par, so long as the symmetries of spacetime are associated with the covering group $\overline{\mathcal{P}}_+^\uparrow$.

Two caveats about this account: first, the concept of 'charge conjugation' is clearly not exhausted by any purely spacetime description. Charges in relativistic field theory are conserved quantities associated with a global gauge group, which in the Standard Model is postulated to be $SU(3) \times SU(2) \times U(1)$. I do not claim that this group is a spacetime structure. What I would like to indicate is how *one aspect* of charge conjugation, which might be called 'matter–antimatter exchange', can be viewed as intimately connected to the spacetime symmetry. I find this to be a natural way to explore the proposal of Greaves (2010). But, as I will argue, its relationship to time reversal does not provide a clean vindication of the Feynman view, and it does not erase the arrow of time associated with T violation.

#### 8.4.1 Extending the Discrete Symmetries

Let me begin by revisiting my account of discrete symmetries. When we restrict attention to the subgroup $(\mathbb{R}, +)$ of time translations in a reference frame, we find that time reversal is the unique non-trivial automorphism, $\tau : t \mapsto -t$ (Proposition 2.1). It is not an element of the original symmetry group but rather a 'symmetry of symmetries', which I referred to as 'higher order' in Section 4.3. However, time reversal can always be 'added' into the group through the construction of the semidirect product $(\mathbb{R}, +) \rtimes \{\iota, \tau\}$. In Section 2.6, I showed how this produces a group element that reverses time translations, $\tau t \tau^{-1} = -t$.

Extending this thinking to the restricted Poincaré group $\mathcal{P}_+^\uparrow$, we find three non-trivial automorphisms: time reversal $\tau$, together with the parity $p$ and their combination $p\tau$. These automorphisms can be defined for each spacetime translation $u = (t, x_1, x_2, x_3) \in \mathbb{R}^4$ in a foliation, as: $u^p :=$ $(t, -x_1, -x_2, -x_3)$ and $u^\tau := (-t, x_1, x_2, x_3)$, and hence $u^{p\tau} = -u$. They can also be defined for each Lorentz boost, as: $p(\Lambda) = \tau(\Lambda) = \pi \Lambda \pi$, where $\pi : (t, x_1, x_2, x_3) = (t, -x_1, -x_2, -x_3)$ is a spatial reversal. The latter arises because both parity and time reversal 'turn around' the spatial direction in which each boost occurs so that the three-velocity is reversed. Thus, the ordinary $p$, $\tau$, and $p\tau$ transformations can be viewed as automorphisms of the Poincaré group. In summary:

$$p : (u, \Lambda) \mapsto (u^p, \pi \Lambda \pi)$$
$$\tau : (u, \Lambda) \mapsto (u^\tau, \pi \Lambda \pi) \qquad (8.14)$$
$$p\tau : (u, \Lambda) \mapsto (-u, \Lambda).$$

These turn out to exhaust the non-trivial automorphisms $\alpha$ of $\mathcal{P}_+^\uparrow$ that are also involutions.[29] So, constructing the 'complete' Poincaré group from these automorphisms produces a group that really is complete.

Similar automorphisms can be found for the covering group $\overline{\mathcal{P}}_+^\uparrow$ as well. Let $v \in M$ be a vector in Minkowski spacetime represented as a $2 \times 2$ self-adjoint matrix, as introduced in Eq. (8.11). The covering map $\Lambda_S v = SvS^*$ defines the restricted Lorentz transformation $\Lambda_S$ associated with each $S \in SL(2, \mathbb{C})$. With a little effort, one can check[30] that $S \mapsto (S^{-1})^*$ is an automorphism of $SL(2, \mathbb{C})$ and that it induces the transformation $\Lambda_S \mapsto \Lambda_{(S^{-1})^*} = \pi \Lambda_S \pi$ on the restricted Lorentz group via the covering map. As a result, $S \mapsto (S^{-1})^*$ induces the same parity and time reversal transformations, as described for the ordinary Poincaré group in Eq. (8.14).[31] We can summarise these as:

$$p : (u, S) \longmapsto (u^p, (S^{-1})^*)$$
$$\tau : (u, S) \longmapsto (u^\tau, (S^{-1})^*) \qquad (8.15)$$
$$p\tau : (u, S) \longmapsto (u^{pt}, S).$$

Can we now introduce matter–antimatter exchange as an automorphism in a similar way? Not quite. There are exactly three non-trivial 'outer' automorphisms of $SL(2, \mathbb{C})$: the inverse-transpose $S \mapsto (S^{-1})^\top$, the complex conjugate $S \mapsto (\bar{S})$, and their combination $S \mapsto (S^{-1})^*$. There are also four antiautomorphisms: the inverse $S \mapsto S^{-1}$ and the transpose $S \mapsto S^\top$, together with the inverse-conjugate and the transpose-conjugate. One can calculate the way that each transforms elements $\Lambda_S$ of the restricted Lorentz group: three of them are summarised in the table of Figure 8.6, from which the effects of the remaining one can be determined. Here, $\pi$ is a total spatial reversal, while $\zeta$ is defined by $\zeta(t, x_1, x_2, x_3) = (t, x_1, -x_2, x_3)$, reversing the spatial translations on just one axis. The problem is that *all* the automorphisms of $SL(2, \mathbb{C})$ induce a non-trivial automorphism of the Lorentz transformations $\Lambda_S$. This makes none of them appropriate for matter–antimatter exchange, which would require a transformation that does not effect any element of the Lorentz group.

---

[29] The spacetime translations admit further automorphisms of the form $a \mapsto \lambda a$ for $\lambda \in \mathbb{R}$; however, the argument of Proposition 2.1 ensures that if they are required to be involutions, then $\lambda = \pm q$.

[30] Conjugation $S \mapsto \bar{S}$ and the inverse-transpose $S \mapsto (S^{-1})^\top$ are both automorphisms of $SL(2, \mathbb{C})$, and thus so is their composition $S \mapsto (S^{-1})^*$. Viewing $S = UA$ in its polar decomposition, one can show that $\Lambda_{S^*} = \Lambda_S^\top$ where $'^\top'$ is the $2 \times 2$ matrix transpose (Varadarajan 2007, pp.335, Eq. (55)). Now, defining $\pi(t, x) = (t, -x)$, one can see by inspection of the form of $v$ in Eq. (8.11) that $v^{-1} = \frac{1}{\det(v)} \pi v$, and that $\det(\pi v) = \det(v)$. This implies that $\Lambda_{S^{-1}} v = S^{-1} v (S^{-1})^* = (S^* v^{-1} S)^{-1} = (S^* \frac{1}{\det(v)} \pi v S)^{-1} = \det(v)(\Lambda_{S^*} \pi v)^{-1} = (\pi \Lambda_S^\top \pi) v$. Combining this with our equation $\Lambda_{S^*} = \Lambda_S^\top$ now establishes that $\Lambda_{(S^{-1})^*} = \pi \Lambda_S \pi$.

[31] This was observed already by Gel'fand, Minlos, and Shapiro (1958, p.160).

| | |
|---|---|
| conjugate: | $\Lambda_S \mapsto \Lambda_{\bar{S}} = \zeta \Lambda_S \zeta$ |
| transpose: | $\Lambda_S \mapsto \Lambda_{S^\top} = \zeta \Lambda_S^\top \zeta$ |
| inverse: | $\Lambda_S \mapsto \Lambda_{S^{-1}} = \pi \Lambda_S^\top \pi$ |

Figure 8.6 Automorphisms and antiautomorphisms of $SL(2, \mathbb{C})$, and their induced effect on the Lorentz group.

As a result, a little more structure is needed to express matter–antimatter exchange. This makes sense, because we clearly do not have enough structure at this abstract level to define something like 'positive frequency' and 'negative frequency' subspaces. However, that kind of structure is afforded at the level of *representations* of $SL(2, \mathbb{C})$. We will now see that, on this space of representations, matter–antimatter exchange takes the form of a symmetry that is very similar to parity and time reversal.

### 8.4.2  Conjugation and Matter–Antimatter Exchange

There are no finite-dimensional unitary representations of $SL(2, \mathbb{C})$, owing to the fact that it is not a compact Lie group. However, the *non*-unitary linear representations of $SL(2, \mathbb{C})$ are the foundation for the theory of spinors, which play an important role in the description of many quantum fields.[32] Weyl (1946, Theorem 8.11.B) showed that these representations of $SL(2, \mathbb{C})$ have a particular canonical form. And, it is this form that can in some contexts be interpreted as distinguishing between matter and antimatter. Let me briefly set out what that is.

---

[32] Spinors appear in the classification of semi-simple Lie groups by Cartan (1913), who in the introduction to his later 1937 *Theory of Spinors* delighted in having 'discovered' spinors before Dirac. Brief modern overviews can be found in R. Geroch (1973, "Special topics in particle physics". In: Unpublished Manunscript of Fall 1973, Version of 25 May 2006, §18) and Wald (1984, Chapter 13); for a detailed introduction, see Carmeli and Malin (2000) or Penrose and Rindler (1984).

Viewing $SL(2,\mathbb{C})$ as a group of $2 \times 2$ matrices, it is by definition a representation amongst the linear transformations of the vector space $V^A :=$ $\mathbb{C}^2$, whose elements are called *spinors* (or sometimes *Weyl spinors*). Note that this concept is not the same as that of a (four-component) Dirac spinor. The tensors built from the vector space $V^A$ are called *spinorial tensors.* I will be concerned with the symmetric spinorial tensors defined on the $n$-fold tensor product $V^A \otimes V^A \otimes \cdots \otimes V^A$ for some $n$. Writing $n = 2j$ for $j = 0, \frac{1}{2}, 1, \frac{3}{2} \cdots$, let $D^j$ denote the subspace of symmetric spinorial tensors defined on this tensor product. Following a typical presentation (cf. Varadarajan 2007, p.336), let $\varphi^j : SL(2,\mathbb{C}) \to D^j$ be the representation

$$\varphi^j : S \mapsto S \otimes S \otimes \cdots \otimes S. \tag{8.16}$$

Since spinors are defined on a complex vector space, there exists another inequivalent representation of $SL(2,\mathbb{C})$ defined by complex conjugation with respect to the first. One writes $V^{A'} = \mathbb{C}^2$ to denote the complex-conjugate vector space[33] and $D^{j'}$ to denote the subspace of symmetric tensors built from this vector space. Then the 'conjugate representation' $\varphi^{j'}$ is given by

$$S \mapsto \bar{S} \otimes \bar{S} \otimes \cdots \bar{S}. \tag{8.17}$$

Weyl's result is that every finite-dimensional irreducible representation of $SL(2,\mathbb{C})$ is equivalent (up to an intertwining) to one of the form

$$\varphi^{(j,j')} := \varphi^j \otimes \varphi^{j'}, \tag{8.18}$$

for some $j, j' = 0, \frac{1}{2}, 1, \frac{3}{2}, \ldots$ . In the description of a quantum field system, one may wish to interpret the first factor as a positive-frequency description, corresponding to an ordinary matter field, while the other is interpreted as a negative-frequency description, corresponding to an antimatter field. Given such an interpretation, the exchange of these factors can be understood as matter–antimatter exchange. Indeed, in the first rigorous statement of a CPT theorem due to Jost (1965), the 'charge conjugation' aspect of CPT is characterised by conjugation of field operators that exchanges these 'undotted' and 'dotted' spinor subspaces.[34]

---

[33] Writing the scalar product in $V^A$ as $av$, where $a \in \mathbb{C}$ and $v \in V^A$, we define $V^{A'}$ to consist of the same set of vectors as $V^A$ but with a scalar product $a \cdot v$ defined by $a \cdot v := a^*v$.

[34] See Haag (1996, §II.5.1) for a modern (and English-language) overview of this theorem.

Of course, to make this connection concretely, a further unitary representation of this structure is needed on a Hilbert space of infinite dimension. This can be carried out by appeal to the representation theory of spinors, which is found in many places, and so I will not develop it here.[35]

What I would like to point out is that these representations of matter and antimatter themselves admit a symmetry, viewed separately from the symmetries of $SL(2, \mathbb{C})$, which exchanges a given representation with its conjugate. Namely, we define the map $c$ on the space of irreducible representations by the definition

$$c\left(\varphi^{(j, j')}\right) := \varphi^{(j', j)}. \tag{8.19}$$

In terms of the canonical direct tensor product $\varphi^j \otimes \varphi^{j'}$, this transformation exchanges the two components. Thus, interpreting one as a 'positive frequency' subspace and the other as a 'negative frequency' subspace, we arrive at a definition of matter–antimatter exchange.

This transformation C is fundamentally different from P and T, which are automorphisms of the spacetime symmetry group $\overline{P}^{\uparrow}_{+}$. In contrast, C can only be defined as a symmetry of its representations. However, the definitions of $\varphi^j$ and $\varphi^{j'}$ allow us to define time reversal and parity as transformations of these representations as well, allowing one to compare all three. Both time reversal and parity take the form $\tau(S) = p(S) = (S^{-1})^*$, since these transformations are only distinct in their transformation of spacetime translations. This induces a transformation of $\varphi^j$ through Eq. (8.16), and on $\varphi^j$ through Eq. (8.17), and therefore on each irreducible representation $\varphi^{(j, j')} = \varphi^j \otimes \varphi^{j'}$.

Moreover, there is a sense in which both P and T are closely connected to matter–antimatter exchange. Both of them involve the conjugation automorphism $S \mapsto \bar{S}$, albeit together with the inverse-transpose as well. Viewed as transformations from one representation $\varphi^{(j, j')}$ of $SL(2, \mathbb{C})$ to another, both are also 'equivalent' to matter–antimatter exchange, in the following sense: consider the element $i\sigma_2 \in SL(2, \mathbb{C})$, where $\sigma_2$ is a Pauli matrix. It can be shown through simple matrix multiplication to transform time reversal $\tau(S) = (S^{-1})^*$ to the conjugate automorphism,

$$(i\sigma_2)\tau(S)(i\sigma_2)^{-1} = \bar{S}. \tag{8.20}$$

---

[35] Comments on this representation theory can be found in Carmeli and Malin (2000, §4.3), Penrose and Rindler (1984), Varadarajan (2007, Chapter IX), or Wald (1984, 357–9).

But, the automorphism $S \mapsto (i\sigma_2)\tau(S)(i\sigma_2)^{-1} = \bar{S}$ transforms a representation in the same way that matter–antimatter exchange does, $c\left(\varphi^{(j,j')}\right) := \varphi^{(j',j)}$. As a result, parity, time reversal, and matter–antimatter exchange are all related by an intertwining map $i\sigma_2$. More precisely, the transformations $p$, $\tau$, and $c$ are related by the property that

$$(i\sigma_2) \circ p = (i\sigma_2) \circ \tau = c \circ (i\sigma_p). \tag{8.21}$$

In the language of representation theory, this means that they are 'equivalent' as symmetries of the space of representations. Although it is more common to propose that matter–antimatter exchange is relevant to time reversal, this suggests a sense in which it is relevant to parity as well. Indeed, Wigner (1957, p.258) himself considered the possibility that "the mirror image of matter is antimatter" in response to Wu's discovery of parity violation.

This provides a neat connection between C, P, and T. Let me now compare it to the remarks about time reversal considered at the beginning of this chapter. In the first place, matter–antimatter exchange is not *quite* a spacetime symmetry – not even adopting the idiosyncratic view of spacetime symmetries as given by $SL(2,\mathbb{C})$ – but rather a symmetry of its representations. So, there is really no sense in which time reversal is 'really' CT or CPT, since neither can be viewed as an automorphism of the continuous spacetime symmetries. This provides a clear sense in which CT and CPT do not behave in an appropriate way to be deserving of the name 'time reversal', following the discussion of Section 8.1. These are not spacetime symmetries but rather symmetries of a representation. As a result, a symmetry involving matter–antimatter exchange cannot 'erase' the time asymmetry established by T violation. Only the latter is relevant for describing the symmetries of 'time alone'.

However, there is still an interesting sense in which the Feynman proposal is perhaps vindicated: time reversal, parity, and matter–antimatter exchange can all be viewed as transforming 'spinor' representations to their conjugates. One might interpret this as the statement that both parity and time reversal 'automatically' include matter–antimatter exchange in a spinor representation. However, exploring this possibility in detail would require considerably more development.

## 8.5 Summary

T violation means that the standard time reversal operator T does not provide a representation of time translation reversal, $t \mapsto -t$. The experimental evidence for this provides strong evidence that time itself has an arrow, as I have argued in Chapter 7. In this chapter, I considered a possible concern about this conclusion: that there is always a large number of representations that 'restore' temporal symmetry, which might include transformations like PT or CPT. One might worry that these would 'erase' the arrow established by T violation and provide an appropriate representation of time reversal symmetry. But, this would be to ignore the broader context of what these transformations mean. The transformation PT reverses spatial translations and so does not deserve the name 'time reversal'. And, when we are considering an irreducible representation, the choice of time reversal operator is generally unique.

There are other arguments that might seek to establish that time reversal is 'really' CT or CPT. Quantisation theory is one; but, on careful mathematical treatments, this transforms classical time reversal to the ordinary quantum time reversal operator T. A more ambitious reinterpretation of C brings us a little closer: by moving up from the ordinary spacetime symmetries to the more exotic universal covering group, I exhibited a close relationship between time reversal and matter–antimatter exchange. However, these transformations are still defined on different spaces. In particular, there is no plausible way to define matter–antimatter exchange as a true spacetime symmetry, even for the universal covering group. Transformations like CT and CPT are certainly important in the foundations of quantum field theory. However, they should not be mistaken for transformations that determine whether time has an arrow.

# Epilogue

There is a story that my former teacher once told me, about a well-known lecturer at a world-class university in the United States. The lecturer decided to spend an entire day in class presenting time reversal. Unfortunately, fifteen minutes into the meeting, the lecturer became so confused that continuing was impossible, and so the class was cancelled. At the next meeting, the lecturer brought a large stack of notes to refer to and continued the lecture as before, but *again* became confused and so again cancelled the class. On the third meeting, when students arrived, they found four chalkboards filled with derivations and the lecturer standing by the door. The lecturer pointed at the chalkboards and said, "*That's* time reversal, and that's all I'm going to say about it!" — and walked out. The students duly copied the blackboards into their notebooks.

I hope this book has filled in some of the details that were missing from that lecture. The study of time asymmetry is a deep, beautiful intertwining of physics and its philosophy, but it is conceptually tricky at times. Fortunately, a simple philosophy helps to clarify this study: time is not just about moments but about the structure of those moments. So, when we study the symmetries of time, that structure must be taken into account.

The passage of time can be viewed as describing the time translations between moments. And, the structure of those time translations provides the basis for a coherent account of temporal symmetry. The account I have presented has two parts. First: time is *symmetric* if and only if reversing the time translations is a symmetry; otherwise it is asymmetric. Second: if a theory has a 'law of motion' describing change over time, then that change must share a common structure with time, through a representation. Like a table casting a shadow on the floor, the asymmetries of time cast a shadow onto our best theories of motion.

By paying close attention to those theories, and the experiments that underpin them, I hope to have clarified some sense in which time has an arrow. It is not just an asymmetry of motion, nor a matter of contingent perspective or illusion. It is a full-fledged asymmetry of time itself. So, indeed: if the structure of time were a canon by Bach, then, at least in our world, it would play differently in one direction than in the reverse.

# Index